\documentclass[11pt]{article}
\usepackage[top=1in, bottom=1in, left=1in, right=1in]{geometry}

\usepackage{color, amsmath, amssymb, amsbsy, amsthm, graphicx, bbm, amsfonts, bm, dsfont, courier, subfig}
\usepackage[toc,page]{appendix}
\usepackage{setspace}
\usepackage{booktabs}
\usepackage[flushleft]{threeparttable}
\usepackage{latexsym}
\usepackage[dvipsnames]{xcolor}
\usepackage{comment} 
\usepackage{wrapfig}
\usepackage{graphicx}
\usepackage{graphics}
\usepackage{multirow}
\usepackage[ruled]{algorithm2e}
\usepackage{xr-hyper}
\usepackage[colorlinks,allcolors=blue]{hyperref}
\usepackage[numbers]{natbib}  
\usepackage{longtable}
\usepackage{enumerate}
 \usepackage{float}

\makeatletter

\newcommand*{\addFileDependency}[1]{
\typeout{(#1)}
\@addtofilelist{#1}
\IfFileExists{#1}{}{\typeout{No file #1.}}
}\makeatother

\usepackage{soul}
\usepackage{threeparttable,booktabs}
\usepackage{etoolbox}
\appto\TPTnoteSettings{\footnotesize}
\usepackage{setspace}
\newtheoremstyle{theoremstyle}
{\topsep} 
{\topsep} 
{\itshape} 
{} 
{} 
{} 
{.5em} 
{\color{black}\ifthenelse{\equal{#3}{}}{{\bfseries #1 #2}}{{\bfseries #1 #2 (#3)}}}
\newtheoremstyle{theoremstylealt}
{\topsep} 
{\topsep} 
{\itshape} 
{} 
{} 
{} 
{.5em} 
{\color{black}\ifthenelse{\equal{#3}{}}{{\bfseries #1 #2$^\prime$}}{{\bfseries #1 #2$^\prime$ (#3)}}}
\newtheoremstyle{examplestyle}
{\topsep} 
{\topsep} 
{} 
{} 
{} 
{} 
{.5em} 
{\color{black}\ifthenelse{\equal{#3}{}}{{\bfseries #1 #2}}{{\bfseries #1 #2 (#3)}}}
\theoremstyle{theoremstyle}\newtheorem{thm}{Theorem}
\theoremstyle{theoremstylealt}
\theoremstyle{theoremstyle}     
\theoremstyle{theoremstyle}\newtheorem{lem}{Lemma}  
\theoremstyle{theoremstyle}        
\theoremstyle{theoremstyle}
\theoremstyle{theoremstyle}\newtheorem{assumption}{Assumption}
\theoremstyle{theoremstylealt}
\theoremstyle{theoremstyle}

\theoremstyle{theoremstyle}\newtheorem{cond}{Condition}

\theoremstyle{theoremstyle}\newtheorem{condition}{Condition}

\theoremstyle{theoremstyle}

\theoremstyle{theoremstyle}

\theoremstyle{examplestyle}
\theoremstyle{examplestyle}
\theoremstyle{examplestyle}

\newcommand{\Expectation}{\mathbb{E}}

\newcommand{\Prob}{\mathbb{P}}

\newcommand{\toProb}{\overset{\mathrm{p}}{\to}}

\newcommand{\diff}{\mathrm{d}}

\renewcommand{\epsilon}{\varepsilon}

\def \hat{\widehat}

\DeclareMathOperator*{\argmax}{argmax} 

\usepackage{setspace}
\usepackage{array}
\newcolumntype{H}{>{\setbox0=\hbox\bgroup}c<{\egroup}@{}}

\begin{document}
	
	\title{Winner's Curse Free Robust Mendelian Randomization with Summary Data}
	
\author{
Zhongming Xie\thanks{Division of Biostatistics, University of California Berkeley.} \and
Wanheng Zhang\thanks{Department of Biostatistics, The University of Texas MD Anderson Cancer Center.} \and
Jingshen Wang\thanks{Division of Biostatistics, University of California Berkeley. Corresponding author.} \and
Chong Wu\thanks{Department of Biostatistics, The University of Texas MD Anderson Cancer Center. Corresponding author. }
} 
	\maketitle

	\begin{abstract}
	In the past decade, the increased availability of genome-wide association studies summary data has popularized Mendelian Randomization (MR) for conducting causal inference. MR analyses, incorporating genetic variants as instrumental variables, are known for their robustness against reverse causation bias and unmeasured confounders. Nevertheless, classical MR analyses utilizing summary data may still produce biased causal effect estimates due to the winner's curse and pleiotropy issues. To address these two issues and establish valid causal conclusions, we propose a unified robust Mendelian Randomization framework with summary data, which systematically removes the winner's curse and screens out invalid genetic instruments with pleiotropic effects. Unlike existing robust MR literature, our framework delivers valid statistical inference on the causal effect without requiring the genetic pleiotropy effects to follow any parametric distribution or relying on perfect instrument screening property. Under appropriate conditions, we demonstrate that our proposed estimator converges to a normal distribution, and its variance can be well estimated. We demonstrate the performance of our proposed estimator through Monte Carlo simulations and two case studies.

\medskip 

 \noindent{\it Keywords}: Bootstrap aggregation; GWAS;  Post-selection inference. 
	\end{abstract}
	\doublespacing
	
	\section{Introduction}

 \subsection{Background and motivation}
	
Drawing inferences about cause and effect lies at the core of uncovering essential scientific principles. In biological and biomedical sciences, causal inference deepens our understanding of underlying etiology and advances developments in disease diagnosis, treatment, and prevention. While observational data present unique opportunities for causal inference by employing large and rich datasets, causal discoveries from observational studies are often susceptible to unmeasured confounding and reverse causation bias issues \citep{imai2011unpacking, flegal2011reverse, gelman2013ask, smith2004mendelian}. As a remedy, Mendelian Randomization (MR) has become a popular research design. Its popularity is not only ascribed to the fact that MR mitigates unmeasured confounding bias by using genetic variants as instrumental variables (IVs) to assess the causal relationship between exposures and outcomes but also credited to the increasing availability of large-scale genome-wide association studies (GWAS) summary data on various complex traits \citep{smith2004mendelian,didelez2007mendelian, lawlor2008mendelian, skrivankova2021strengthening}. 

However, MR with GWAS summary may still produce biased estimates of causal effects due to several sources of bias. These include measurement error in exposure GWAS, winner's curse bias resulting from using the same exposure GWAS for both IV selection and effect estimation, and most crucially, bias from including invalid IVs with pleiotropy \citep{sadreev2021navigating}. Firstly, the effect of IV on exposure is measured by exposure GWAS, which inherently contains measurement error. Ignoring such measurement error can produce biased causal effect estimates, especially when the strength of IVs is weak \citep{ye2021debiased,ma2023breaking}. Secondly, the practice of selecting genetic instruments based on their estimated associations with the exposure variable from GWAS, and using the same data for both instrument selection and estimation, can lead to biased causal effect estimates due to the winner's curse phenomenon \citep{zollner2007overcoming, zhong2010correcting, gkatzionis2019contextualizing}. Lastly, typical MR analyses inevitably involve some invalid IVs that either directly affect the outcome or through unmeasured confounding factors---a phenomenon known as pleiotropy \citep{hemani2018evaluating, watanabe2019global}. The nature of pleiotropy is widespread and usually unknown or complex \citep{watanabe2019global}. Failure to fully account for pleiotropy will also lead to biased causal effect estimates.

A broad literature addresses the biases discussed above to improve the credibility of MR analyses, yet no single approach can simultaneously tackle all these biases. Some methods have made progress in addressing individual issues. For instance, \cite{ye2021debiased} formally tackled the measurement error bias in the popular inverse variance weighted estimator, while \cite{ma2023breaking} proposed a randomized instrument selection and Rao-Blackwellization procedure to address both measurement error bias and winner's curse bias. However, the validity of these methods relies heavily on the assumption that all IVs either have no pleiotropic effects or exhibit balanced pleiotropic effects---an assumption unlikely to hold in practice due to the unknown and complex nature of pleiotropy \citep{watanabe2019global}, potentially leading to biased causal effect estimates.

To account for widespread pleiotropy, many robust MR methods have been proposed. These methods primarily focus on addressing the issue raised by invalid IVs, but often at the expense of neglecting measurement error and winner's curse biases. They can be broadly categorized into two strategies. The first strategy imposes normal mixture model assumptions on the pleiotropic effects. By modeling the observed GWAS summary data within a joint likelihood function, these methods simultaneously estimate the unknown parameters and the desired causal effect.  Such methods include RAPS \citep{zhao2020statistical}, ContMix \citep{burgess2020robust}, MR-APSS \citep{hu2022mendelian},  MRMix \citep{qi2019mendelian}. However, as demonstrated in our simulation studies, when the normal mixture model assumption is violated, these approaches tend to produce false positive findings or have low detection power. Moreover, incorporating procedures to address winner's curse bias, such as that proposed by \cite{ma2023breaking}, is challenging within this framework as it may violate parametric modeling assumptions and result in an incorrect likelihood function. The second strategy avoids imposing parametric modeling assumptions on the pleiotropic effects. Instead, it adopts penalization methods to screen out invalid instruments with pleiotropic effects, using only the selected valid instruments for causal effect estimation. Such methods include, for example, cML \citep{xue2021constrained} and MR-Lasso \citep{luo2008mixture}.  However, these methods either lack rigorous statistical justifications or require that the selected IVs are valid and include all valid IVs (a condition we refer to as ``perfect IV screening"). For example, \cite{xue2021constrained} prove that their procedure can screen out all invalid IVs with a probability tending to one under the asymptotic regime where the number of IVs is fixed, and the sample size tends to infinity. When this is achieved, the resulting causal effect estimate is consistent and asymptotically normal. However, the theoretical results under this asymptotic regime do not account for how the magnitudes of the pleiotropic effects impact the validity of statistical inference. In fact,  perfect IV screening is often unattainable when the pleiotropic effects are small, and the differences between valid and invalid IVs in MR studies are subtle. Notably, two-sample MR is a rapidly evolving field with numerous methodological advancements, such as \citep{morrison2020mendelian,liu2023reciprocal,grant2024bayesian}. For comprehensive reviews of statistical methods in MR, we refer readers to \cite{sanderson2022mendelian} and \cite{boehm2022statistical}.

\subsection{Contribution}

To bridge the aforementioned gaps in the existing literature, we propose a unified MR framework with summary data that simultaneously addresses winner’s curse bias, bias from measurement error in exposure GWAS, and bias from invalid IVs with pleiotropy (Section \ref{section 3}).   Specifically, we propose an \(l_0\) constrained optimization framework that can simultaneously screen out invalid IVs, account for measurement error, and seamlessly integrate with the winner's removal step from \cite{ma2023breaking}. Moreover, we demonstrate that the proposed \(l_0\) constrained optimization framework maintains computational efficiency due to the special form of our objective function. Furthermore, to improve statistical efficiency, we adopt a bootstrap aggregation procedure and use a non-parametric delta method to perform valid inference on the final causal effect.

On the theoretical side, we provide comprehensive theoretical investigations of the proposed method in Section \ref{section 4}. We prove that the final estimator in our proposed method is asymptotically unbiased and converges to a normal distribution even in the presence of directional pleiotropy. Moreover, different from existing theoretical analyses in robust MR, we show that our method can deliver consistent causal effect estimates without perfect invalid IV screening; see detailed discussion in Supplementary Material Section \ref{sup:sec4}. In brief, our theoretical investigation indicates that our proposed method can screen out IVs with large pleiotropic effects, and the resulting causal effect estimator remains consistent even if the selected IVs include some invalid ones with small pleiotropic effects. These theoretical investigations better characterize scenarios where our method performs well and demonstrate its robustness.


Benefiting from the above features in both methodological and theoretical aspects, we demonstrate that our proposed MR framework delivers robust causal effect estimates with improved statistical power in simulated Monte Carlo experiments (Section~\ref{simulation}) and in two case studies (Section~\ref{section 5}).  From our simulated Monte Carlo experiments, we confirm that our proposed method outperforms benchmark methods in terms of type 1 error rates, power, absolute bias, mean squared error, and coverage probability in most scenarios. The results also highlight the importance of simultaneously correcting the winner's curse bias and accounting for measurement error bias and generic pleiotropic effects. From our case study of negative control outcome analyses, in which the population causal effects are believed to be zero by design, we confirm that our approach yields well-controlled Type I error rates (Section \ref{section 5.1}). From our case study to identify causal risk factors for COVID-19 severity, our approach identifies more causal risk factors than the existing approaches, and the identified causal exposures by our proposed method have more supporting evidence.

\section{Framework and challenges}\label{sec2}
In this section, we review the classical two-sample Mendelian Randomization (MR) framework with summary data. We then revisit the pleiotropic effects, measurement error bias, and winner's curse bias within this framework. 

Referring to the causal diagram in Figure \ref{fig:causal-diagram}, we let $X$ denote the exposure, $Y$ the outcome, and $U$ the unmeasured confounder between the exposure and the outcome. The goal of MR analysis is to estimate the causal effect (denoted by $\theta$) of the exposure variable $X$ on the outcome variable $Y$. However, in the presence of unmeasured confounder $U$, it is challenging to directly estimate $\theta$ solely using the information stored in $X$ and $Y$. To overcome this, two-sample MR analyses incorporate $p$ mutually independent SNPs $G_1, \ldots, G_p$ as instrumental variables (IVs) and estimate $\theta$ using the estimated association pairs $\{(\hat{\beta}_{X_j}, \hat{\beta}_{Y_j})\}_{j=1}^p$ collected from two independent GWAS datasets, where $\hat{\beta}_{X_j}$ and $\hat{\beta}_{Y_j}$ are the estimated effect sizes for IV $j$ in exposure and outcome GWAS, respectively. Here, genetic variant $G_j\in\{0,1,2\}$ represents the number of effect alleles of a single-nucleotide polymorphism (SNP) $j$ inherited by an individual. Following the two-sample summary-data MR literature \cite{ye2021debiased,zhao2020statistical}, we assume the following linear structural equation model:
\begin{equation}\label{lsem}
\begin{aligned}
U&=\sum_{j=1}^p \phi_j G_j+E_U, \\
X&=\sum_{j=1}^p \gamma_j G_j+\beta_{XU}U+E_X, \\
Y&=\sum_{j=1}^p \alpha_j G_j+\beta_{YU}U+\theta X+E_Y,
\end{aligned}
\end{equation}
where $E_U$, $E_X$, and $E_Y$ are mutually independent random noises. $E_U$ is independent of $(G_1, \ldots, G_p)$, and $E_X$ and $E_Y$ are independent of $(G_1, \ldots, G_p, U)$. To allow for the valid inference of the causal effect $\theta$, we need $G_j$ ($j=1,\ldots,p$) to be valid IVs in the sense that they satisfy the following three conditions: (1) $\gamma_j\neq 0$, meaning that $G_j$ is associated with $X$ (relevance assumption); (2) $\phi_j=0$, meaning that $G_j$ has no correlated pleiotropic effect with $Y$ (effective random assignment assumption); (3) $\alpha_j = 0$, meaning that $G_j$ has no uncorrelated pleiotropic effect with $Y$  (exclusion restriction assumption).

Provided that all included genetic IVs are valid, two-sample MR analyses can deliver valid inference on $\theta$ by appropriately using information stored in two independent GWAS datasets.  To provide some justifications for this claim, we follow the causal model proposed in \cite{pearl2009causality}. In particular,  in the structural equation models given in Eq~\eqref{lsem}, the total effect of SNP $G_j$ on $Y$ and the total effect of $G_j$ on $X$ are given by:
\begin{align*}
    &\beta_{Y_j}=\mathbb{E}[Y|do(G_j=g_j+1)]-\mathbb{E}[Y|do(G_j=g_j)]=\alpha_j+\beta_{YU}\phi_j+\theta\cdot(\gamma_j+\beta_{XU}\phi_j),
    \\&\beta_{X_j}=\mathbb{E}[X|do(G_j=g_j+1)]-\mathbb{E}[X|do(G_j=g_j)]=\gamma_j+\beta_{XU}\phi_j.
\end{align*}
For a valid IV $G_j$, when $G_j$ satisfies $\phi_j=0$ (effective random assignment assumption) and $\alpha_j=0$ (exclusion restriction assumption), the target causal effect $\theta$ will satisfy $ \beta_{Y_j} = \theta \beta_{X_j}$, where $\beta_{X_j}=\gamma_j$ and $\beta_{Y_j}=\theta\gamma_j$. If the relevance assumption $\gamma_j\neq 0$ is also met, we are then able to use $\beta_{Y_j}$ and $\beta_{X_j}$ to assist valid inference on $\theta$, as they can be well estimated through the estimated association pairs $\{(\hat{\beta}_{X_j}, \hat{\beta}_{Y_j})\}_{j=1}^p$ collected from two independent GWAS dataset in two-sample summary-data MR framework.

However, in practice, due to the widespread pleiotropy in human genetics \citep{hemani2018evaluating, watanabe2019global}, the effective random assignment ($\phi_j = 0$) and exclusion restriction assumptions ($\alpha_j = 0$) are frequently violated, leading to invalid IVs. In the presence of invalid IVs, the total effect of $G_j$ on $Y$ can be expressed as:
\begin{align}\label{eq:sem-causal}
\beta_{Y_j} = \underbrace{\theta \cdot \beta_{X_j}}_{\substack{\text{causal}\\ \text{effect}}} +\underbrace{\alpha_j}_{\substack{\text{uncorrelated}\\ \text{pleiotropy}}} + \underbrace{\beta_{YU} \cdot \phi_j}_{\substack{\text{correlated}\\ \text{pleiotropy}}} \equiv \theta \cdot \beta_{X_j} + r_j.
\end{align}
Here, $\alpha_j$ is the uncorrelated pleiotropic effect that captures the direct effect of $G_j$ on $Y$, and $\beta_{YU} \cdot \phi_j$ is the correlated pleiotropic effect that captures the effect of $G_j$ on $Y$ through the pathway $G_j\rightarrow U \rightarrow Y$. Their combined effect, $r_j = \alpha_j + \beta_{YU} \cdot \phi_j$, represents the total effect of a genetic variant $G_j$ on the outcome $Y$ induced by pleiotropy. These violations make it challenging to accurately estimate $\theta$ using MR. If not appropriately accounted for, genetic pleiotropy can result in biased causal effect estimates in MR analyses (see Section~\ref{simulation} for our simulation results).

On top of the potential bias induced by pleiotropic effects, two additional sources of bias in MR analyses are measurement error bias and winner's curse bias. Measurement error bias arises from the fact that the true effect of an IV on the exposure, $\beta_{X_j}$, is unobserved. Instead, we rely on $\hat{\beta}_{X_j}$, an estimate derived from exposure GWAS (I), which inherently contains measurement error, to conduct MR.  The winner's curse bias, on the other hand, is induced by 
pre-selecting IVs that are strongly associated with the exposure variable to meet the relevance assumption (that is, $\gamma_j\neq 0$). This selection exercise is often based on hard-thresholding measured SNP $z$-scores obtained from GWAS (I): SNP $j$ is selected if $| \hat\beta_{X_j}/ \sigma_{X_j} | > \lambda$, where $\lambda$ is a pre-specified cut-off value, and $\hat\beta_{X_j}$ and $\sigma_{X_j}$ are estimated effect size and its standard error from exposure GWAS dataset, respectively. The selected IVs are then used to construct downstream causal effect estimators. The selected IV-exposure associations tend to overestimate the underlying true association effects $\beta_{X_j}$, as the distribution of any $\hat\beta_{X_j}$ that survives the selection is a truncated Gaussian and the post-selection mean is no longer $\beta_{X_j}$ when commonly used Gaussian assumption on $\hat{\beta}_{X_j}$ is adopted. Subsequently, by doubly using the data in GWAS (I) for IV selection and estimation, classical MR estimators are expected to be biased and have an intractable limiting distribution, making statistical inference problematic. 






%

In the rest of this manuscript, we employ the following model frequently adopted in the Mendelian Randomization literature \citep{zhao2020statistical,qi2019mendelian,xue2021constrained}:
\begin{assumption}[Measurement error model]\label{assumption: measure error}
(i)For any $j\neq j'$, $(\hat{\beta}_{Y_j}, \hat{\beta}_{X_{j}})$ and $(\hat{\beta}_{Y_{j'}}, \hat{\beta}_{X_{j'}})$ are mutually independent.
(ii)For each $j$, the association pair $(\hat{\beta}_{Y_j}, \hat{\beta}_{X_{j}})$ follows
\begin{align*}
		\begin{bmatrix}
		\hat{\beta}_{X_{j}}\\
		\hat{\beta}_{Y_j}
		\end{bmatrix} \sim \mathcal{N}\left(\begin{bmatrix}
		\beta_{X_j} \\
		\theta\beta_{X_j}+r_j
		\end{bmatrix}\ , \begin{bmatrix}
		{\sigma}_{X_j}^2 & 0 \\
		0 & {\sigma}_{Y_j}^2
		\end{bmatrix}\right).
		\end{align*}
    Furthermore, there exists  a positive integer $n\rightarrow\infty$ and positive constants $m$ and $M$ such that $\frac{m}{n}\leq \sigma_{X_j}^2 \leq \frac{M}{n}$,\  $\frac{m}{n}\leq \sigma_{Y_j}^2 \leq \frac{M}{n}$ for $j=1,\ldots,p$. 
\end{assumption}

The assumption of independent SNPs, while seemingly stringent, is grounded in established practice in two-sample MR analyses \citep{ye2021debiased, zhao2020statistical, ma2023breaking}. This approach helps ensure that each selected SNP represents a signal from a unique genetic locus, thereby mitigating potential confounding effects from LD and facilitating clearer interpretation of causal effect estimates. We acknowledge that alternative cis-MR methods such as  Transcriptome-Wide Association Studies (TWAS) \citep{gusev2016integrative, wainberg2019opportunities} and Proteome-Wide Association Studies (PWAS), effectively utilize correlated SNPs, particularly for investigating relationship between omics and complex traits.  However, as the reviewer suggested, when inferring causal relationships between complex traits/diseases (such as the two case studies in Section 6), using independent IVs from the whole genome is typically efficient enough and simple to implement. This strategy is also widely adopted in the literature. Therefore, in line with this common practice, we adopt the independence assumption.   To ensure independent IVs, we apply a sigma-based LD pruning method \citep{ma2023breaking}.

 \section{Methodology}\label{section 3}

\subsection{Measurement error correction and invalid IV screening}

To estimate the causal effect $\theta$, a straightforward approach is to replace the population association effects with their empirical estimates from GWAS in the causal structure equation in \eqref{eq:sem-causal}. Given that all population associations are measured with error in GWAS, 
the sample analogue of the structure equations can be represented as the following two-stage regression model with measurement errors:
\begin{align*}
& \underbrace{\hat{\beta}_{Y_j}}_{\text{response}}   = \underbrace{\theta}_{\substack{\text{target}\\ \text{parameter}}} \cdot \underbrace{\beta_{X_j}}_{\substack{\text{true} \\ \text{covariate}}}  + \underbrace{r_j}_{\substack{\text{unknown}\\ \text{parameter} } } + \underbrace{\nu_j}_{\text{noise}},  \qquad  \underbrace{\hat{\beta}_{X_j} = \beta_{X_j} + u_j}_{\substack{\text{covariates are} \\ \text{measured with error}}},
\end{align*}
where $\nu_j$ and $u_j$ are centered noises. 
 
To operationalize an accurate estimate of $\theta$ using the above two-stage least squares model, we first consider a situation where a set of IVs with $\beta_{X_j}\neq 0 $ (denoted as $\mathcal{S}$) is known. Our method does not require $\mathcal{S}$ to be known, and we will discuss the selection of $\mathcal{S}$ and the practical implementation of our algorithm in the next subsection. With a known $\mathcal{S}$, we propose estimating $\theta$ by solving the following constrained optimization problem: 
\begin{alignat}{2}
    \min_{\theta,r_j} \ &\nonumber\ l\big( \theta, \{ r_j\}_{j\in \mathcal{S}} \big)= \sum_{j\in \mathcal{S}}l_j\big( \theta,  r_j \big) \triangleq  \sum_{j\in \mathcal{S}} \frac{ (\hat{\beta}_{Y_j}- \theta\cdot  \hat{\beta}_{X_j} - r_j  )^2 }{\sigma^2_{Y_j}} -\sum_{j\in \mathcal{S}} \frac{ \theta^2 \cdot {\sigma}_{X_j}^2}{\sigma^2_{Y_j}}\mathds{1}_{(r_j=0)} , \\
   \text{s.t.}\ & \label{eq:optimization}\  \sum_{j\in \mathcal{S}} \mathds{1}_{(r_j=0)}  = v .
\end{alignat}
Intuitively, the objective function above is a bias-corrected least squares function designed to account for measurement error, subject to the constraint that the adopted IVs for estimating $\theta$ are valid.
In the following, we will show that the optimization problem above not only accounts for the measurement errors in $\hat{\beta}_{X_j}$ but also accurately identifies invalid IVs with $r_j \neq 0$. This is achieved with computational efficiency, even when an $l_0$-type constraint is adopted. As a result, the solution of this optimization problem provides an accurate estimate of $\theta$. 

\begin{figure}[!htbp]
	\centering	\includegraphics[width=0.35\linewidth]{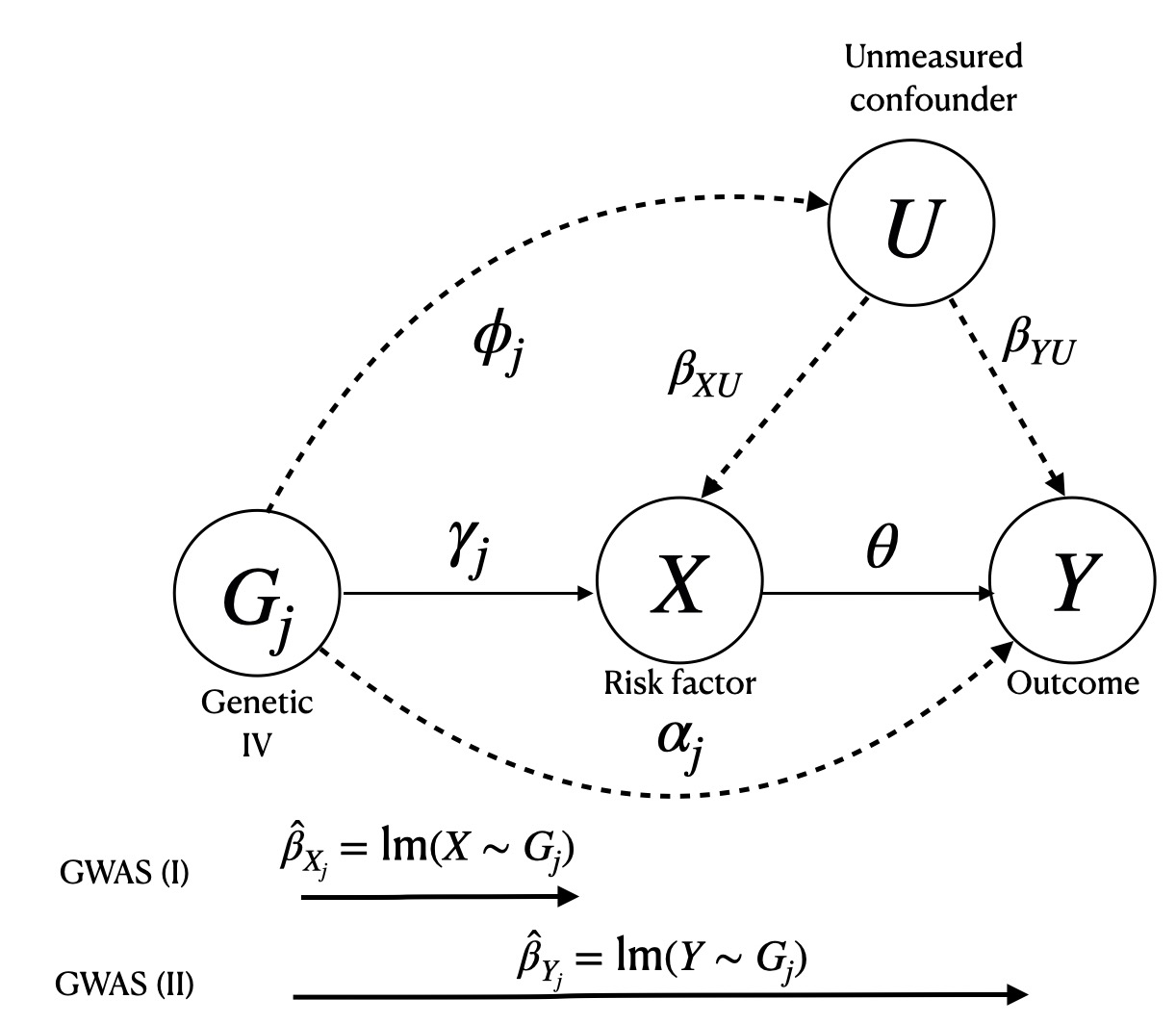}
	\caption{\label{fig:causal-diagram} The causal diagram and GWAS (I) and (II) summary data adopted in the two-sample MR. The corresponding causal effect for each pathway is labeled near the directed edge.  } 
\end{figure}

To start with, when the set of IVs with $r_j = 0$ is known, the solution of the above optimization problem provides an unbiased estimate of $\theta$. As in this case, we have 
\[ 
L(\theta)\triangleq\min_{r_j}l\big( \theta, \{ r_j\}_{j\in \mathcal{S}} \big)= \frac{1}{2}\sum_{j\in \mathcal{V}} \frac{ (\hat{\beta}_{Y_j}- \theta\cdot  \hat{\beta}_{X_j})^2 }{\sigma^2_{Y_j}} -\frac{1}{2}\sum_{j\in \mathcal{V}} \frac{ \theta^2 \cdot {\sigma}_{X_j}^2}{\sigma^2_{Y_j}}. 
 \]
We can verify that $L(\theta)$ is unbiased for the weighted least squares loss function in the 
sense that $\mathbb{E}\big[L(\theta) \big] = \mathbb{E}\big[\sum_{j\in \mathcal{V}}  (\hat{\beta}_{Y_j}- \theta\cdot  {\beta}_{X_j})^2/(2\sigma^2_{Y_j}) \big]$. This suggests that its minimizer is unbiased for the causal effect $\theta$. 

Next, as the set of IVs with $r_j = 0$ is unknown, Problem \eqref{eq:optimization} incorporates an $l_0$-type constraint to screen out invalid IVs. While classical $l_0$-type optimization problems are solved by their convex relaxations, this technique does not apply to our problem due to the inclusion of a measurement error bias correction term in our objective function (that is, the term \(\sum_{j\in \mathcal{S}_{\lambda}} { \theta^2 \cdot {\sigma}_{X_j}^2}/{\sigma^2_{Y_j}}\mathds{1}_{(r_j=0)}\)). To address this issue, we propose an iterative algorithm that mimics block coordinate descent and guarantees the decay of our objective function in Algorithm \ref{alg:optimization}; see justification in the Supplementary Material Section \ref{sup:sec1}.  

Lastly, the number of valid IVs $v$ is unknown and requires tuning. To choose the final set of valid IVs, we propose a generalized Bayesian Information Criteria (GBIC), that is:
\[
    \text{GBIC}(v) = - 2 \hat{l}\big(\hat\theta(v), \{ \hat{r}_{j}(v)\}_{j\in \hat{\mathcal{V}}} \big)+\kappa_n \cdot(s-v), \quad s=|\mathcal{S}|, 
\]where \(\kappa_{n}=\log (n)\), and choose the final set of valid IVs by minimizing the GBIC. The proposed GBIC with $\kappa_n = \log (n)$ is different from the classical BIC criteria that adopts $\kappa_n = \log (s_{\lambda})$. The reason for this choice is that the classical model selection consistency result of the BIC is established in the asymptotic regime with fixed \( s_\lambda \). As we are in an asymptotic regime with $s_\lambda \rightarrow\infty$, our proposed GBIC criteria adjusts $\kappa_n$ accordingly to ensure invalid IV screening consistency. In particular, in Section \ref{sup:sec4} of the Supplemental Material, we demonstrate that our procedure provides a consistent causal effect estimator without requiring the perfect IV screening property under a simplified scenario and Conditions 1-2 and 8-9. One of these conditions imposes a constraint on the penalization coefficient \(\kappa_n\): \(\kappa_{n} \gg \log(s_{\lambda})\). We argue that \(\kappa_n = \log(n)\) is a feasible choice to satisfy this condition, as the order of the sample size is typically larger than the order of the number of selected relevant IVs in a two-sample MR study.

\subsection{Unknown $\mathcal{S}$ and practical implementation}\label{methods: winner's curse removal}

We now consider the realistic scenario where the set \(\mathcal{S}\) is unknown. Because the collection of relevant IVs is not known, practitioners typically perform a pre-selection procedure to identify IVs strongly associated with the exposure. These selected IVs are then used to estimate the causal effect. As discussed in Section \ref{sec2}, selecting genetic instruments based on their estimated associations with the exposure variable from GWAS and using the same data for both instrument selection and estimation can lead to biased causal effect estimates due to the winner's curse phenomenon. To address the issue of winner's curse bias when \(\mathcal{S}\) is unknown, we integrate the proposed method from the previous section with the approach described in \cite{ma2023breaking} to perform Rao-Blackwellized randomized instrument selection.

For each SNP $j=1,2,\ldots,p$, we generate a pseudo SNP-exposure association effect $Z_j\sim \mathcal{N}(0,\eta^2)$, and select SNP $j$ if $\Big| \frac{\hat{\beta}_{X_j}}{ \sigma_{X_j} } + Z_j \Big| >\lambda .$ Define the set of selected SNPs as $\mathcal{S}_{\lambda} = \Big\{j: \Big| \frac{\hat{\beta}_{X_j}}{ \sigma_{X_j} } + Z_j \Big| >\lambda ,\ j = 1, 2,\ldots,p \Big\}$ and its cardinality $|S_{\lambda}|=s_{\lambda}$. For each selected SNP $j\in \mathcal{S}_\lambda$,  we construct an unbiased estimator of $\beta
_{X_j}$ as
\begin{align*}
\hat{\beta}_{X_j,\mathtt{RB}} &= \hat{\beta}_{X_j} - \frac{\sigma_{X_j}}{\eta}\frac{\phi\big(A_{j,+}\big) - \phi\big(A_{j,-}\big)}{1 - \Phi\big(A_{j,+}\big) + \Phi\big(A_{j,-}\big)},\text{ where } A_{j,\pm} =  -  \frac{\hat{\beta}_{X_j} }{ \sigma_{X_j}\eta } \pm\frac{\lambda}{\eta},
\end{align*}

\begin{algorithm}[H]\label{alg:optimization}
\singlespacing\small
\caption{Algorithm to solve the optimization problem in \eqref{eq:optimization winner's curse} }
\KwIn{Data inputs and initial parameters}
\KwOut{Estimated parameters $\hat{\theta}$ and $\hat{r}_j$}
\SetKwBlock{CoordinateDescent}{Block Coordinate Descent}{end}
\SetKwBlock{ValidIVSelection}{Valid IV Selection via GBIC}{end}

\textbf{Initialization } Set $k = 0$, generate $\theta^{(0)} \sim \text{Uniform} \left( \min_{1 \leq j \leq s_\lambda} \frac{\hat{\beta}_{Y_j}}{\hat{\beta}_{X_j}}, \max_{1 \leq j \leq s_\lambda} \frac{\hat{\beta}_{Y_j}}{\hat{\beta}_{X_j}} \right)$\;
\CoordinateDescent{
    \Repeat{$\left| \frac{\theta^{(k+1)} - \theta^{(k)}}{\theta^{(k)}} \right| < 10^{-7}$}{
        Fix $\theta^{(k)}$, update $r_j^{(k+1)}$\;
        Order $\frac{ \left( \hat{\beta}_{Y_j} - \theta^{(k)} \cdot \hat{\beta}_{X_j, \mathtt{RB}} \right)^2 }{\sigma^2_{Y_j}} - \frac{ \theta^2 \cdot \hat{\sigma}_{X_j, \mathtt{RB}}^2}{\sigma^2_{Y_j}}, \quad j = 1, 2, \dots, s_{\lambda} - v$ in decreasing order\;
        Set $r_j^{(k+1)} = \hat{\beta}_{Y_j} - \theta^{(k)} \hat{\beta}_{X_j, \mathtt{RB}}$ for the largest $s_{\lambda} - v$ components, $j = 1, \dots, s_{\lambda} - v$, and $r_j^{(k+1)} = 0$ for $j = s_{\lambda} - v + 1, \dots, s_{\lambda}$\;
        
        Fix $r_j^{(k+1)}$, update $\theta^{(k)}$ by minimizing the following objective function:
        \[
        \theta^{(k+1)} = \underset{\theta \in \mathbb{R}}{\arg\min} \sum_{j \in \mathcal{S}_{\lambda}} \frac{ \left( \hat{\beta}_{Y_j} - \theta \cdot \hat{\beta}_{X_j, \mathtt{RB}} - r_j^{(k+1)} \right)^2 - \theta^2 \cdot \hat{\sigma}_{X_j, \mathtt{RB}}^2 }{\sigma^2_{Y_j}} \mathds{1}_{(r_j^{(k+1)} = 0)}.
        \]       
        \textbf{If} $\left| \frac{\theta^{(k+1)} - \theta^{(k)}}{\theta^{(k)}} \right| < 10^{-7}$ \textbf{then}\ 
            Stop and output $\hat{\theta}(v) = \theta^{(k+1)}$ and $\hat{r}_j(v) = r_j^{(k+1)}$ \;
       \textbf{else} Set $k = k + 1$\;
    }
}

\ValidIVSelection{
    \For{$v = 2, \dots, s_{\lambda}$}{
        Calculate 
        \[
        \text{GBIC}(v) = -2 \hat{l} \left( \hat{\theta}(v), \{ \hat{r}_j(v) \}_{j \in \hat{\mathcal{V}}} \right) + \log(n) \cdot (s_{\lambda} - v);
        \]
    }
    Select $\hat{\mathcal{V}}$ with the smallest $\text{GBIC}(v)$\;
}

\end{algorithm}

$\phi(\cdot)$ and $\Phi(\cdot)$ denote the standard normal density and cumulative distribution functions.   Here, $\eta$ is a pre-specified constant that reflects the noise level of the pseudo SNPs. We recommend using $\eta = 0.5$ as a default value \citep{ma2023breaking}. This choice balances the need for sufficient randomization to address the winner’s curse bias while maintaining the stability of the selection process.  The above procedure only randomizes the IV selection near the cut-off value $\lambda$, which implies that the strong IVs with large $\beta_{X_j}$ are invariably selected.  Here, the choice of the significance cutoff ($\lambda$) for selecting IVs presents a trade-off between including a sufficient number of informative IVs and maintaining the overall strength of the selected IV set. While lowering the cutoff may improve statistical power by incorporating more IVs with moderate effects, setting it too low can introduce weak or null IVs that potentially violate the relevance assumption and compromise the validity of the MR analysis. In our proposed method, we provide a sufficient condition to ensure the asymptotic normality of the estimator, which depends on the average strength of the selected IVs relative to the cutoff value. Specifically, we choose a cutoff of $5\times10^{-5}$, commonly used as a threshold for suggestive significance in GWAS, to strike a balance between including informative IVs and maintaining the validity of the selected IV set. We note that Rao-Blackwellization has also been applied in \cite{bowden2009unbiased} to efficiently combine information from an initial GWAS and a replication study to obtain unbiased estimates of SNP effect sizes. Our approach differs as we do not require a replication study to construct an unbiased estimation for $\beta_{X_j}$ (see Supplement Materials Section 5 for details).  
Benefiting from such randomized IV selection, $\hat{\beta}_{X_j,\mathtt{RB}}$ is free of winner's curse bias, implying that $\mathbb{E}[\hat{\beta}_{X_j,\mathtt{RB}}|j\in\mathcal{S}_{\lambda}]=\beta_{X_j}$. Therefore, our proposed bias-corrected least squares objective function and $l_0$ constraint optimization framework in the previous section can be applied: 
\begin{align} \label{eq:optimization winner's curse}
\min_{\theta\in \mathbb{R}, r_j\in\mathbb{R} } \hat{l}\big( \theta, \{ r_j\}_{j\in \mathcal{S}_{\lambda}} \big) , \text{ s.t. } \ \sum_{j\in \mathcal{S}_{\lambda}} \mathds{1}_{(r_j=0)} = v. 
\end{align}
As one reviewer suggested, we also implemented two $l_1$-type methods and make comparison with our $l_0$ based method through simulations.  Our results demonstrate that while both approaches maintain comparable Type I error control, absolute bias, mean squared error (MSE), and coverage probability across various scenarios, the $l_0$-based CARE method achieves higher statistical power. We have added relevant descriptions, methods, and results in Supplemental Material Section \ref{sup:alg-l1}-\ref{sup:l_1} and Section \ref{sup:l0 and l1}.
where the loss function is defined as 
 \begin{align*}
    &  \hat{l}\big( \theta, \{ r_j\}_{j\in \mathcal{S}_{\lambda}} \big)  \ = \sum_{j\in \mathcal{S}_{\lambda}}\hat{l}_j\big( \theta, r_j \big) = \sum_{j\in \mathcal{S}_{\lambda}} \frac{ (\hat{\beta}_{Y_j}- \theta\cdot  \hat{\beta}_{X_j, \mathtt{RB}} - r_j  )^2 }{\sigma^2_{Y_j}} - \frac{ \theta^2 \cdot {\hat{\sigma}}_{X_j, \mathtt{RB}}^2}{\sigma^2_{Y_j}}\mathds{1}_{(r_j=0)},\\
     &\hat{\sigma}_{X_j,\mathtt{RB}}^{ \mathrm{2} }  =\ \sigma_{X_j}^2\Bigg( 1 - \frac{1}{\eta^2}  \frac{A_{j,+}\phi(A_{j,+}) - A_{j,-}\phi(A_{j,-})}{1 - \Phi(A_{j,+}) + \Phi( A_{j,-} )} +\frac{1}{\eta^2} \Big(\frac{\phi(A_{j,+}) - \phi(A_{j,-}) }{1 - \Phi(A_{j,+}) + \Phi( A_{j, -} )}\Big)^2 \Bigg).
 \end{align*}
\subsection{Bootstrap aggregation and statistical inference}

Since the IV screening step can be rather noisy and we do not expect to perfectly screen out all invalid IVs, we next incorporate bagging (or bootstrap aggregation) \citep{breiman1996bagging} to reduce IV screening variability and to further improve statistical efficiency. Then, we adopt the non-parametric delta method \cite{efron1982jackknife} to construct a confidence interval for our bagged estimator. 

To be specific, we draw bootstrap sample $B$ times from $\mathcal{S}_{\lambda}$. For the $b$-th bootstrap sample (Denoted by $\mathcal{S}_{\lambda,b}^*$), we adjust the loss function as $ \hat{l}_b^*\big( \theta, \{ r_j\}_{j\in \mathcal{S}_{\lambda}} \big)=\sum_{j\in \mathcal{S}_{\lambda}}w_{jb}^*\hat{l}_j\big( \theta,  r_j\big)$,
where $w_{jb}^*$ is the number of occurrences in $\mathcal{S}_{\lambda,b}^*$ for j-th IVs in $\mathcal{S}_{\lambda}$. Then, we conduct the invalid IV screening step for each bootstrap sample $\mathcal{S}_{\lambda,b}^*$ and select $\mathcal{\hat{V}}_b=\left\{j: \hat{r}_{jb}=0 \text{ and } \ j \in \mathcal{S}_{\lambda,b}^*\right\}.$ The downstream causal estimator is derived by aggregating the estimated effects from all bootstrap samples, that is: 
\begin{align}\label{theta definition}\hat{\theta_{b}}=\frac{\sum_{j\in\widehat{\mathcal{V}}_b}  \hat\beta_{Y_j} 
 \hat\beta_{X_j,\mathtt{RB}} / {\sigma}_{Y_j}^{2}}{\sum_{j\in \widehat{\mathcal{V}}_b}  (\hat\beta_{X_j,\mathtt{RB}}^2 - \hat{\sigma}_{X_j,\mathtt{RB}}^{ \mathrm{2} }  ) / {\sigma}_{Y_j}^{2}},\ \ \ \ \widetilde{\theta}=\frac{1}{B}\sum_{b=1}^B \hat{\theta_{b}} , 
\end{align}
where $\hat{\theta}_b$ is obtained by refitting the loss function $\hat{l}\big( \theta, \{ r_j\}_{j\in \hat{\mathcal{V}}_b} \big) $. 

To provide valid statistical inference on the true causal effect $\theta$, we use the non-parametric delta method \citep{efron2014estimation} to estimate the variance of the bagged estimator with $ \hat\sigma_n^2= {\sum}_{j\in \mathcal{S}_{\lambda}} \hat{S}_j^2$, where $\hat{S}_j= B^{-1} \sum_{b=1}^{B}  (w^*_{ib} - B^{-1}\sum_{k=1}^{B} w^*_{ik} )(\widehat{\theta}_b-\widetilde{\theta})$. Then we construct a $(1-\alpha)$-level confidence interval for $\theta$ with $\Big[\widetilde{\theta}-z_{\alpha/2}\cdot\hat\sigma_n, \widetilde{\theta}+z_{\alpha/2}\cdot\hat\sigma_n\Big]$.
Here $\alpha$ is the upper $\alpha/2$-quantile of the standard normal distribution.

In the remainder of this manuscript, we refer to the proposed method as Causal Analysis with Randomized Estimators (CARE). The formalization of our proposed algorithm can be found in Algorithm \ref{alg:care}. We also provide the discussion on the time complexity of this algorithm in Section \ref{sup:sec1} in Supplemental Material.


\section{Theoretical investigations}\label{section 4}

To discuss our theoretical investigations in detail, we begin by revisiting and introducing notations and assumptions. Recall that the set of selected IVs after rerandomization is defined as ${\mathcal{S}}_{\lambda} =\big\{j:|\frac{\hat\beta_{X_j}}{\sigma_{X_j}}+Z_j|>\lambda, j=1,\ldots,p\big\}$ and its cardinality is denoted as $|\mathcal{S_{\lambda}}|=s_{\lambda}$. We next define $ \kappa_{\lambda}$ as the average of squared standardized IV effects to measure the selected IV strength in $\mathcal{S}_{\lambda}$, that is $ \kappa_{\lambda}=\frac{1}{s_{\lambda}}\sum\limits_{j\in  \mathcal{S}_{\lambda}}\frac{\beta_{X_j}^2}{\sigma_{Y_j}^2}. $ 
Among the selected IVs after rerandomization, we denote $\mathcal{V}_{\lambda} = \big\{ j: \ j \in \mathcal{S}_{\lambda} \text{ and } r_j = 0 \big\}$ as the set of valid IVs in ${\mathcal{S}}_{\lambda}$ and denote its cardinality as $|\mathcal{V}_{\lambda} | = v_{\lambda}$.

Considering the dual sources of randomness in our proposed estimator (one from the original GWAS sample, and the other from the bootstrap resampling), we  separate these two sources of randomness by denoting the conditional expectation taken with respect to bootstrap resampling as $   \mathbb{E}^*\big[\cdot\big]= \mathbb{E}\Big[\cdot\big|S_{\lambda},\big\{(\hat{\beta}_{Y_j},\hat{\beta}_{X_{j,\mathtt{RB}}})\big\}_{j \in S_{\lambda}}\Big].
$
Next, we introduce three additional assumptions for our theoretical investigations:

\begin{algorithm}[H]
\singlespacing\small
\SetAlgoLined
\For{$j \leftarrow 1$ \KwTo $p$}{
Generate a pseudo SNP-exposure association effect $Z_j \sim \mathcal{N}(0,\eta^2)$,\\
 \textbf{If} $\left| \frac{\hat{\beta}_{X_j}}{\sigma_{X_j}} + Z_j \right| > \lambda$, \textbf{Then} select SNP $j$.
}
Define the set of selected SNPs as
$\mathcal{S}_{\lambda} = \{j : \left| \frac{\hat{\beta}_{X_j}}{\sigma_{X_j}} + Z_j \right| > \lambda, \ j = 1, 2, \ldots, p \}$ and $|\mathcal{S}_{\lambda}| = s_{\lambda}$, \\
\For{$j \in \mathcal{S}_\lambda$}{
Construct an unbiased estimator of $\hat{\beta}_{X_j,\mathtt{RB}}$ as
    \[
    \hat{\beta}_{j,\mathtt{RB}} = \hat{\beta}_{X_j} - \frac{\sigma_{X_j}}{\eta} \frac{\phi\left(A_{j,+}\right) - \phi\left(A_{j,-}\right)}{1 - \Phi\left(A_{j,+}\right) + \Phi\left(A_{j,-}\right)}, \text{ where } A_{j,\pm} = - \frac{\hat{\beta}_{X_j}}{\sigma_{X_j}\eta} \pm \frac{\lambda}{\eta}
    \]
    and $\phi(\cdot)$ and $\Phi(\cdot)$ denote the standard normal density and cumulative distribution functions.
}
\For{$b = 1$ to $B$}{
    Draw bootstrap sample $\mathcal{S}_{\lambda,b}^*$ from $\mathcal{S}_{\lambda}$,\\
    Conduct the invalid IV screening procedure for $\mathcal{S}_{\lambda,b}^*$
 $$
    \min_{\theta \in \mathbb{R}, r_j \in \mathbb{R}} \left\{ \hat{l}_b^*\left(\theta, \{ r_j \}_{j \in \mathcal{S}_{\lambda}}\right) : \sum_{j \in \mathcal{S}_{\lambda,b}^*} \mathds{1}_{r_j = 0} = v \right\} \quad \Rightarrow \quad \widehat{\mathcal{V}}^*_b(v) = \left\{ j : \hat{r}_j = 0, j \in \mathcal{S}_{\lambda,b}^* \right\},
$$ where $
    \hat{l}_b^*\left(\theta, \{ r_j \}_{j \in \mathcal{S}_{\lambda}}\right) = \sum_{j \in \mathcal{S}_{\lambda}} w_{jb}^* \hat{l}_j\left(\theta, \{ r_j \}_{j \in \mathcal{S}_{\lambda}}\right)$.\\
Select the final estimated set of Valid IVs $\widehat{\mathcal{V}}^*_b$ by GBIC,\\
 Derive the causal estimator for the $b$-th bootstrap 
    \[
    \hat{\theta}_b = A_b^{-1} \sum_{j \in \widehat{\mathcal{V}}_b} \frac{\hat{\beta}_{Y_j} \hat{\beta}_{X_j,\mathtt{RB}}}{\sigma_{Y_j}^{2}}, \quad A_b = \sum_{j \in \widehat{\mathcal{V}}_b} \frac{\hat{\beta}_{X_j,\mathtt{RB}}^2 - \hat{\sigma}_{X_j,\mathtt{RB}}^2}{\sigma_{Y_j}^{2}}.
    \]
    }
 Obtain the final estimator by bootstrap aggregation \(
\widetilde{\theta} = \frac{1}{B} \sum_{b=1}^B \hat{\theta}_b\), \\
Adopt the non-parametric delta method to estimate the variance of the bagged estimator with 
\(
\hat{\sigma}_n^2 = \sum_{j \in \mathcal{S}_{\lambda}} \hat{S}_j^2, \quad \hat{S}_j = \frac{1}{B} \sum_{b=1}^{B} \left(w^*_{ib} - \frac{1}{B} \sum_{k=1}^{B} w^*_{ik}\right)(\widehat{\theta}_b - \widetilde{\theta}),
\)\\
Construct a $(1-\alpha)$-level confidence interval for $\theta$ with 
\(
\left[\widetilde{\theta} - z_{\frac{\alpha}{2}} \cdot \hat{\sigma}_n, \widetilde{\theta} + z_{\frac{\alpha}{2}} \cdot \hat{\sigma}_n \right]
\), here $z_{\frac{\alpha}{2}}$ is the upper $\alpha/2$-quantile of the standard normal distribution.
 \caption{CARE} \label{alg:care}
\end{algorithm}

\begin{assumption}[Variance stabilization]\label{assumption:variance stablization}
There exists a variance stabilizing quantity $a_{\lambda}$ and a vector $\boldsymbol{\tau} \in \mathbb{R}^{s_{\lambda}}$ in which each component is independent of $\left\{(u_j,\nu_j)\right\}_{j \in S_{\lambda}}$ and uniformly bounded away from infinity in probability in the sense that $$
\sup_{j \in S_{\lambda}}\Big|a_\lambda\cdot \mathbb{E}^*\Big[A_b^{-1} \cdot \hat{w}_{jb}\Big]-\tau_j\Big|=o_p(1), 
$$
where $A_b=\sum_{k\in \mathcal{S}_\lambda} \hat{w}_{kb}\cdot(\hat\beta_{X_k,\mathtt{RB}}^2 - \hat{\sigma}_{X_k,\mathtt{RB}}^{ \mathrm{2} }  ) / {\sigma}_{Y_k}^{2}$, and $\hat{w}_{jb} =w_{jb}^* \cdot \mathrm{I}(\hat{r}_{jb} = 0) \cdot \mathrm{I}(w_{jb}^* \geq 1)$. In addition, there is no dominating IV in the sense that $
\frac{\max_{j\in \mathcal{S}_{\lambda}} \beta_{X_j}^2}{\sum_{j\in \mathcal{S}_{\lambda}} \beta_{X_j}^2} \toProb 0$.
\end{assumption}

The first part of the above assumption, intuitively, ensures that our estimator $\tilde{\theta}$ converges to a non-degenerative distribution asymptotically when appropriately scaled by $a_{\lambda}/\sqrt{s_{\lambda}\cdot \kappa_{\lambda}}$. This scaling factor accounts for the number of selected instruments and their average strength, enabling valid statistical inference. The second part of the condition requires that, after selection, no single IV exerts a ``dominating effect" on exposure, which aligns with the biological understanding that complex traits are influenced by many genetic variants with small effects (i.e., the omnigenic model \citep{boyle2017expanded}). To cast more insight into Assumption \ref{assumption:variance stablization}, in Section \ref{sup:sec2.3} of the Supplemental Material, we consider a special case where perfect IV screening is achieved. We show that in this case,  Assumption \ref{assumption:variance stablization} holds for both valid and invalid IVs in $\mathcal{S}_{\lambda}$.

\begin{assumption}[Negligible invalid IV induced bias]\label{assumption: Negligible invalid IV induced bias}
There is negligible bias induced by potential imperfect screening of invalid IVs after bootstrap aggregation in the sense that 
$$
\frac{a_{\lambda}}{\sqrt{s_{\lambda}\cdot \kappa_{\lambda}}}\mathbb{E}^*\big[A_b^{-1} \sum_{j \in S_{\lambda}} \hat{\beta}_{X_{j,\mathtt{RB}}}\cdot r_j \cdot \hat{w}_{jb}/ {\sigma}_{Y_j}^{2}\big]=o_p(1).
$$
\end{assumption}

Our theoretical investigations reveal two sets of sufficient conditions under which Assumption \ref{assumption: Negligible invalid IV induced bias} holds (See Section \ref{sup:sec3} and \ref{sup:sec4} in the Supplemental Material). 
 The first set of sufficient conditions ensures that the selected IVs are ``nearly perfect," meaning they are valid but do not include all possible valid IVs. We show that this nearly perfect IV screening property can be satisfied when there is strong prior knowledge about the trait's genetic architecture or where valid and invalid IVs are easily distinguishable. The second set of sufficient conditions ensures Assumption \ref{assumption: Negligible invalid IV induced bias} holds even if our proposed IV screening procedure does not screen all invalid IVs. In particular, our analysis indicates that when IVs with large $r_j$ values (strong pleiotropic effects) are effectively screened out, our estimator maintains consistency even if the selected set includes some invalid IVs with small $r_j$ values (weak pleiotropic effects). Together, these theoretical investigations suggest that perfect IV screening is not a prerequisite for valid inference in our proposed method. 

\begin{assumption}[Instrument Selection]\label{assumption: instrument selection}
 Define $\underline{\eta}=\min_{1\leq j\leq p}\eta_j$ and $\overline{\eta}= \max_{1\leq j\leq p}\eta_j$, then both $\underline{\eta}$ and $\overline{\eta}$ are bounded and bounded away from zero.
\end{assumption}

The above assumption requires that the parameter $\eta$ should not be too small or too large, as it impacts the concentration behavior and asymptotic normality of our estimator. This assumption can be satisfied by design in our method. We recommend using a default value of $\eta_j = 0.5$ for all $j$ (where $1 \leq j \leq p$), which ensures that both $\underline{\eta}$ and $\overline{\eta}$ are bounded and bounded away from zero. This choice simplifies the implementation while maintaining the theoretical guarantees of our method. Our simulation study also suggests that our method is not sensitive to the choice of $\eta$.

We are now in a position to describe the asymptotic behavior of our bootstrap aggregated estimator. Without loss of generality, we consider a particular form of our estimator in an ideal case where $\tilde{\theta} = \mathbb{E}^*[\hat{\theta}_b]$. 
\begin{thm}\label{thm1:normality}
    Under Assumptions \ref{assumption: measure error}-\ref{assumption: instrument selection}, as $s_\lambda\stackrel{\text{p}}\rightarrow \infty$ and $\frac{\kappa_{\lambda}}{\lambda^2}\stackrel{\text{p}}\rightarrow \infty$, our proposed estimator satisfies the following representation
$$\frac{a_{\lambda}}{\sqrt{s_{\lambda}\cdot \kappa_{\lambda}}}\cdot\big(\tilde{\theta} -\theta\big)=\frac{1}{\sqrt{s_\lambda\cdot\kappa_{\lambda}}}\sum_{j\in S_{\lambda}} \tau_j\cdot\widetilde{u}_j+o_p(1).$$
where $\widetilde{u}_j=\hat{\beta}_{X_{j,\mathtt{RB}}}\big(\theta\cdot\beta_{X_{j}}+\nu_j\big)-\theta\big(\hat{\beta}_{X_{j,\mathtt{RB}}}^2-\hat{\sigma}_{X_{j,\mathtt{RB}}}^2\big)$. Therefore, conditional on the selection event $\mathcal{S}_\lambda$, our estimator converges to a Gaussian distribution, that is 
\begin{align*}
    \tilde{\sigma}^{-1}\big(\tilde{\theta} -\theta\big)\leadsto N(0,1), \text{ where }\tilde{\sigma}^2 =\frac{\sum_{j\in S_{\lambda}} \tau_j^2 \mathbb{V}\big[\widetilde{u}_j|\mathcal{S}_{\lambda}\big]}{a_{\lambda}^2}. 
\end{align*}
\end{thm}

In the theorem above, we consider the asymptotic regime in which both $s_\lambda\stackrel{\text{p}}\rightarrow \infty$ and $\frac{\kappa_{\lambda}}{\lambda^2}\stackrel{\text{p}}\rightarrow \infty$ tend towards infinity. This asymptotic regime is quite natural in the context of MR. On the one hand, $s_\lambda\stackrel{\text{p}}\rightarrow \infty$ requires the number of IVs selected through re-randomization to be large enough, so that our inverse variance weighting-based estimator exhibits concentrated behavior.  On the other hand, the condition $\frac{\kappa_{\lambda}}{\lambda^2} \stackrel{\mathrm{p}}{\rightarrow} \infty$ does not involve the bootstrapping procedure; instead, it pertains to the strength of the selected IVs relative to the threshold $\lambda$ used in the re-randomization step (Step 1). This assumption ensures that, on average, the selected IVs are sufficiently strong compared to the threshold, thereby satisfying the relevance assumption. It is also likely to hold, as it is of the same order as the GWAS sample size $n$ after IV selection through re-randomization. From a theoretical standpoint, both conditions have been rigorously verified in \cite{ma2023breaking} under appropriate conditions.

\section{Simulations studies}\label{simulation}
\begin{figure}[!htbp]
	\centering	\includegraphics[width=0.8\linewidth]{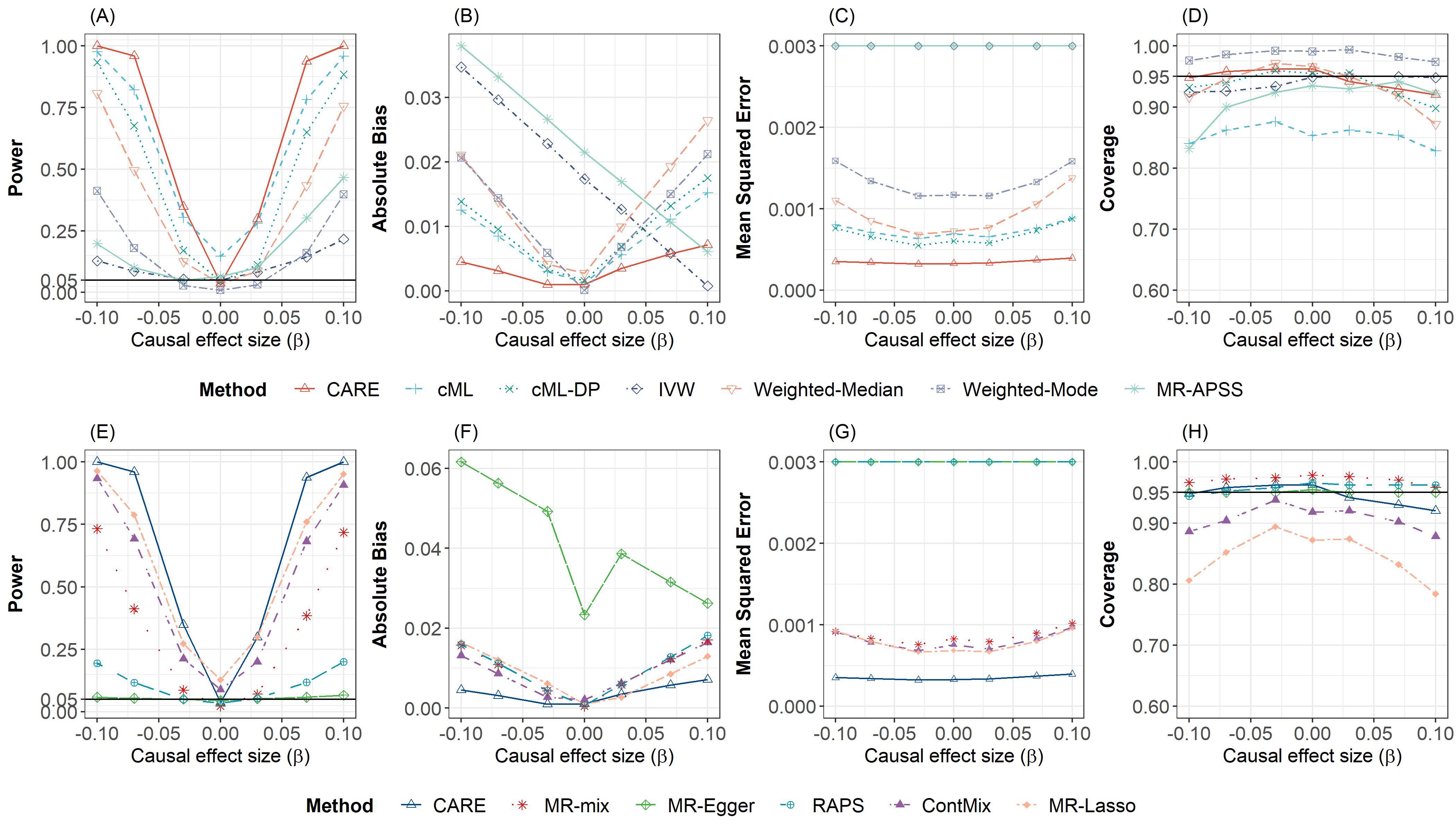}
	\caption{	\label{fig:simulation_normal_main} Power, absolute bias, mean squared error, and coverage of the CARE estimator and several robust MR methods under the main setting with 50\% invalid IVs. Power is the empirical power estimated by the proportion of p-values less than the significance threshold of 0.05.  Coverage is the empirical coverage probability of the 95\% confidence interval.} 
\end{figure}

We generate different simulation settings to evaluate the methods performance. To save space, the simulation settings are put into Supplementary Section \ref{sup:setup}. Figure~\ref{fig:simulation_normal_main} summarizes the performance of various MR methods under the setting of 50\% of the IVs are invalid, which we discuss below.

First, both cML (Type 1 error rate: 0.136) and MR-Lasso (0.112) produce inflated Type 1 error rates. This is because cML and MR-Lasso ignore the randomness in the valid IV selection procedure and assume all invalid IVs have been screened out, which is not the case under this simulation setting. In contrast,  cML-DP (0.042) and CARE (0.042), which explicitly consider the randomness in valid IV selection, yield well-calibrated Type 1 error rates. Furthermore, other benchmark methods, including (random effects) IVW (0.056), MR-Egger (0.050), MRmix (0.020), MR-Median (0.032), MR-mode (0.004), MR-APSS (0.054) and RAPS (0.038) also yield well-controlled Type 1 error rates, though MRmix, MR-Median, MR-mode, and RAPS yield slightly conservative Type 1 error rates. Notably, the winner's curse bias itself does not cause an inflated Type 1 error rate issue \citep{ma2023breaking}, partially explaining the robust performance of many MR methods under the null.


Second, CARE achieves considerably higher statistical power than benchmark methods (Figure~\ref{fig:simulation_normal_main}a). Notably, CARE corrects the winner's curse bias and measurement error bias,  which allows for a more liberal threshold  (say, $p< 5\times 10^{-5}$) for instrument selection, resulting in higher power than other methods that typically use the genome-wide significance level ($p< 5\times 10^{-8}$) as the threshold. Even though MR-APSS, like CARE, allows a liberal threshold ($p< 5\times 10^{-5}$) due to its direct winner's curse bias correction without theoretical guarantee, CARE outperforms MR-APSS,  because of its full correction of the winner's curse bias and meticulous consideration of measurement errors and invalid IVs.

Third, CARE yields smaller absolute bias compared to benchmark methods,  attributable to its comprehensive approach to simultaneously addressing multiple sources of bias (measurement error bias, pleiotropic effects, and winner's curse bias). In comparison, benchmark methods focus on addressing some biases specifically, leading to biased results. For instance, while MR-APSS directly corrects for the winner's curse bias and considers potential invalid IVs, it still presents a larger absolute bias compared to CARE, possibly due to its more limited scope in bias correction and incomplete correction of the winner's curse bias. However, while CARE significantly reduces bias, its estimates are not entirely bias-free. This residual bias likely stems from the subtle differences between valid and invalid IVs. Consequently, the estimates are inevitably influenced by some invalid IVs, albeit to a lesser extent than in other methods. Furthermore, we confirm that ignoring the winner's curse bias and directly applying the measurement error model with $\hat{\beta}_{X_j}$ in CARE generally results in worse performance, particularly concerning the absolute bias (Supplementary Figure~\ref{supfig:simulation_no_correction}). As expected, CARE yields much smaller MSE compared to benchmark methods as CARE has higher power and smaller absolute bias than any benchmark methods.

Fourth, the confidence intervals provided by CARE have coverage probabilities close to the nominal 95\% level. When the absolute causal effect  $|\theta|$ is large (say, 0.1), the absolute bias is relatively large, resulting in slight undercoverage of the true causal effect.

We conduct several additional simulations, including varying proportions of invalid IVs (Supplementary Section \ref{sup:sec_proportion}), uniform-distributed effects in correlated pleiotropy (Supplementary Section \ref{sup:sec_uniform}),
balanced horizontal pleiotropy with InSIDE assumption satisfied (Supplementary Section \ref{sup:sec_bal_pleio}) and directional pleiotropy with InSIDE assumption violated (Supplementary Section \ref{sup:sec_dir_pleio}). The results patterns are similar. 

Furthermore, to validate the results are not sensitive to the specific value of $\eta$ within a reasonable range, we conducted sensitivity analyses using different values of $\eta$ (0.1, 0.3, 0.5, 0.7, 0.9) in our main setting. The results demonstrate that the performance of our method remains stable and consistent for $\eta$ values between 0.3 and 0.9 (Section \ref{sup:sec_eta} in Supplementary Material). As expected, a very small $\eta$ (0.1) led to worse results, likely due to insufficient rerandomization to fully account for the winner's curse bias. Based on these findings, we recommend that practitioners use the default value of $\eta = 0.5$ in most cases without the need for dataset-specific fine-tuning.

 While CARE demonstrates robust performance across various scenarios, it is important to note its limitations. As one reviewer suggested, we consider a simulation scenario that the parameter assumptions of other methods are true (where a three-sample MR design is used and the first GWAS is reserved solely for IV selection based on association strength so that the normality of $\hat{\beta}_{X_j}$ is not distorted).  In this case, some alternative robust MR methods may outperform CARE, indicating that other robust MR methods may outperform CARE  in a three-sample MR desgin (Supplementary Section~\ref{sup:three samples}). Further simulations revealed two situations CARE is suboptimal. Firstly, in settings with non-linear relationships between genetic variants and exposures, CARE showed slightly inflated Type 1 error rates, larger bias, and worse coverage (Section \ref{sup:nonlinear} in Supplementary Material). This limitation stems from the method's underlying assumption of linear relationships, which is common in MR studies and often justified by the predominantly linear or additive nature of genetic effects on complex traits \citep{wainschtein2022assessing}. Unlike our current approach, which exclusively utilizes GWAS summary data to estimate causal effects, recent advancements have addressed the non-linearity issue through methods like DeepMR \cite{malina2022deep}, a deep learning-based approach applicable when individual-level DNA sequence data are available. Secondly, CARE's performance may be compromised when the sample size of the exposure GWAS is small, resulting in a limited number of selected candidate IVs (Section  \ref{sup:sample_size} in Supplementary Material). This issue may also arise due to a relatively small number of independent IVs (Section \ref{sup:sample_size_SNP} in Supplementary Material). Such scenarios can lead to increased sensitivity to violations of IV assumptions and challenge our asymptotic normality results, which require the number of candidate IVs to approach infinity. Users should exercise caution when applying CARE and other MR methods in these scenarios and consider alternative methods or larger sample sizes when possible. 

In the end, it is worth mentioning that the core algorithm in CARE is written in C++ using the R package RcppArmadillo, and each step within the algorithm has a closed-form solution. Consequently, CARE has similar computational efficiency to many other methods, such as cML-DP and MRmix (Supplementary Figure~\ref{supfig:simulation_runtime}), despite utilizing a larger number of IVs and a relatively high number of bootstrap iterations (2,000). Under the main simulation setting (12,000 simulations across 30\%, 50\%, and 70\% invalid IVs), the average computational time of CARE is 12.6 seconds. Notably, the computational time for all methods is less than a minute in most situations when using one single core in a server. Thus, computational time should not be the primary consideration when deciding the method to be used.


\section{Case studies}
\label{section 5}
	In this section, we investigate the performance of proposed CARE in two case studies.  We put the data harmonization details in Supplementary Section \ref{Sec:extension-pruning}. 

\subsection{Negative control outcomes}\label{section 5.1}
To evaluate the Type 1 error rates in real data, we employ negative control outcome analyses, applying CARE and benchmark methods to investigate the causal effect of exposures on outcomes known a priori to have no causal relationship with the exposures.  Briefly, in these negative control outcome analyses, the causal effect size is expected to be $\theta =0$  \citep{sanderson2021use} because negative control outcomes are determined prior to the exposures. However, unmeasured confounding factors may affect the estimates of $\theta$. In particular, following others \citep{sanderson2021use}, we use ease of skin tanning to sun exposures and natural hair color before greying (six outcomes: Ease of skin tanning, Hair color black, Hair color red, Hair color blonde, Hair color light brown, and Hair color dark brown)  as negative control outcomes. These data were downloaded from the IEU OpenGWAS Project \citep{lyon2021variant} with GWAS ID: ukb-b-533 and ukb-d-1747. Notably, both tanning ability and natural hair color before greying are primarily determined at birth (thus, prior to considered exposures) but could be affected by unmeasured confounders \citep{sanderson2021use}. In this setting, the inclusion of invalid IVs due to widespread pleiotropic effects or unmeasured confounding factors (e.g., population stratification) may result in incorrect rejections of the null hypothesis ($\theta = 0$) for MR analyses, leading to inflated Type 1 error rates.




We consider 45 exposures, which include HDL cholesterol, body mass index (BMI), height, Alzheimer's disease, Lung cancer, Type 2 diabetes, stroke, asthma, and many others. All GWAS data are downloaded from the IEU OpenGWAS Project \citep{lyon2021variant}, and details of each exposure are relegated to the Supplementary Table \ref{sup:Table1_Exposure_detail}.  These exposures were selected based on their prevalence in existing literature and relevance to public health. Specifically, traits such as BMI, height, and HDL cholesterol have been extensively studied in genetic epidemiology and are known to be associated with various health outcomes. Disease outcomes like Alzheimer's disease, Type 2 diabetes, and cardiovascular diseases represent major public health concerns and have been the focus of numerous Mendelian randomization studies. This diverse set of exposures covers a wide range of physiological and pathological processes, allowing us to evaluate CARE's performance across various scenarios commonly encountered in Mendelian randomization studies. We apply CARE and benchmark methods to infer causal effects between these 45 exposures and six negative control outcomes (tanning ability and natural hair color before greying), resulting in 270 trait pairs. The corresponding $p$-values should follow a standard uniform distribution, given that the causal effect size $\theta = 0$ under the negative control outcomes analysis. 
\begin{figure}[H]
    \centering
    \begin{minipage}[b]{0.44\textwidth}
        \centering
        \includegraphics[width=\linewidth,height=0.25\textheight]{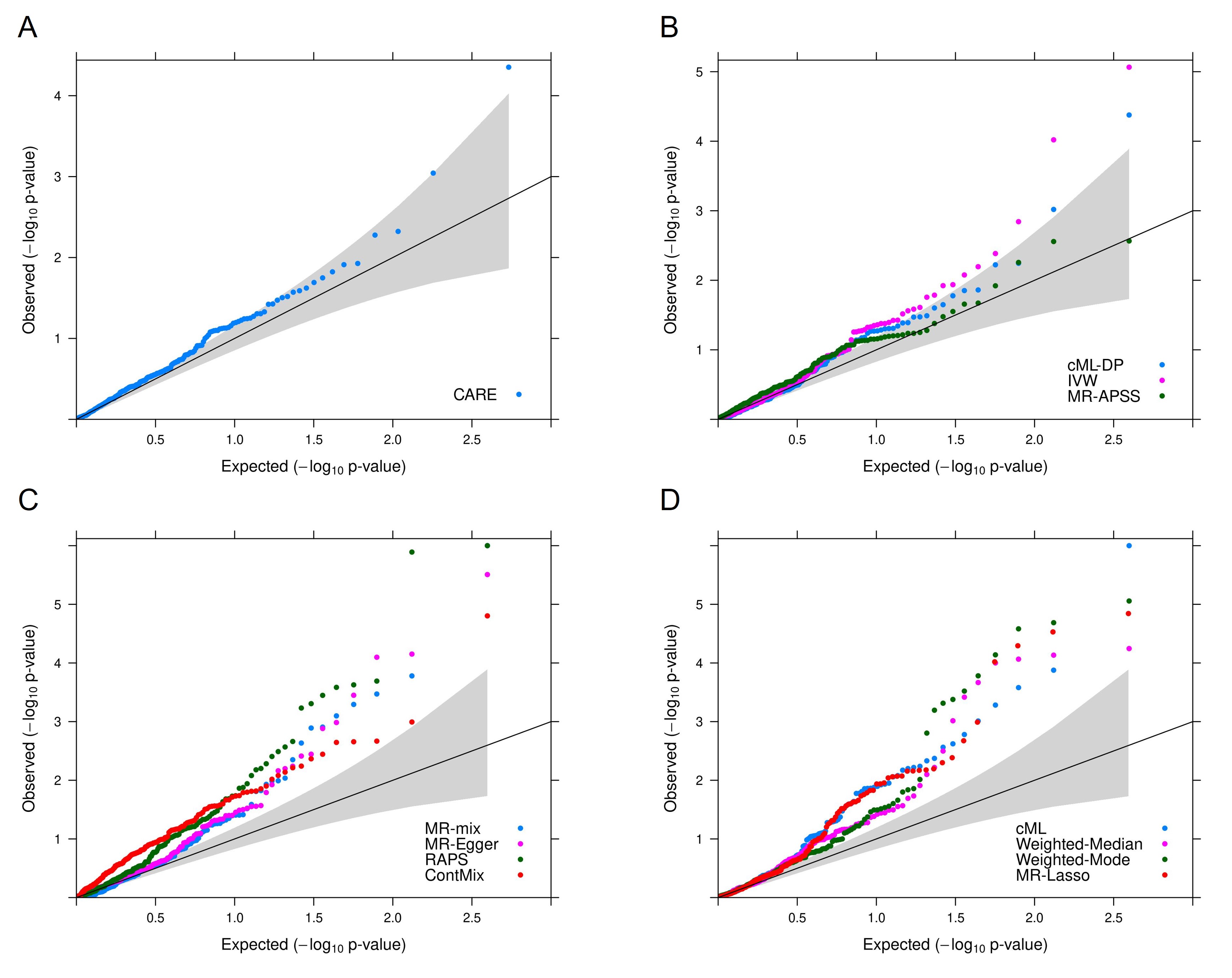}
        \caption{QQ plots of p-values in negative control outcome analysis. The gray-shaded part is 95\% confidence interval.}
        \label{fig:type1error}
    \end{minipage}
    \hfill
    \begin{minipage}[b]{0.44\textwidth}
        \centering
        \includegraphics[width=\linewidth,height=0.15\textheight]{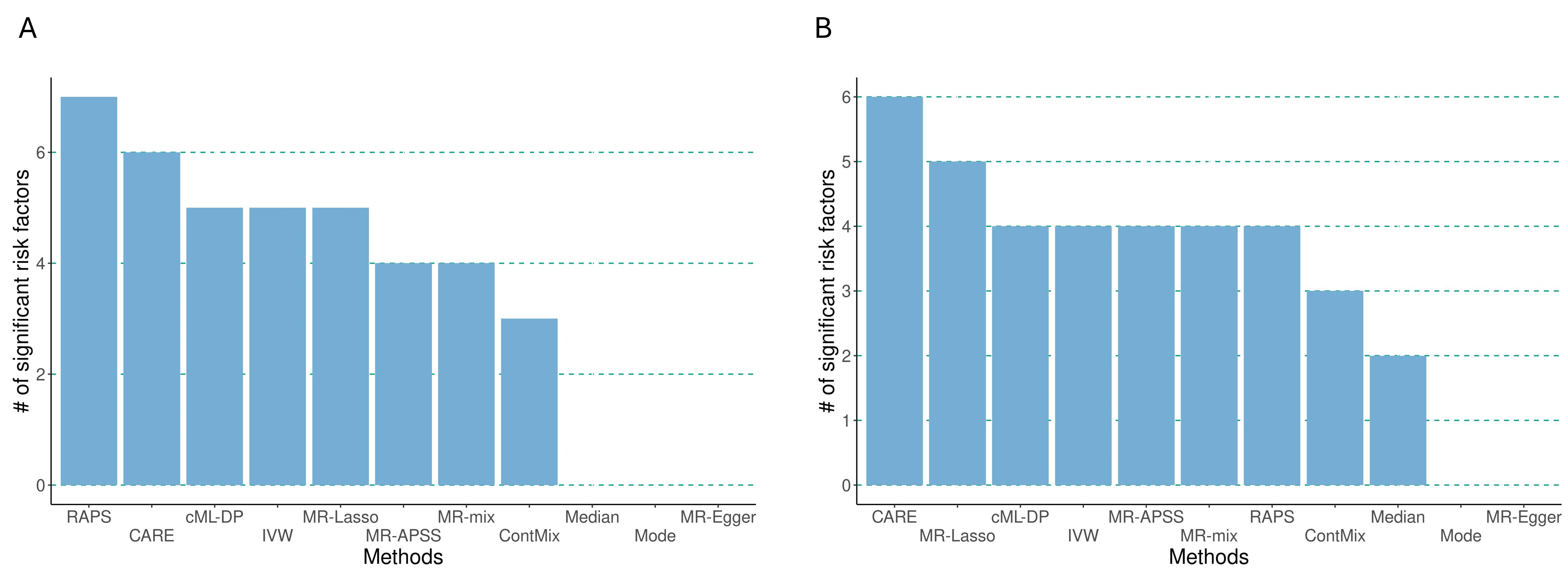}
        \caption{Number of significant causal pairs identified by different methods under Bonferroni-correction threshold $<0.05/45\simeq 10^{-3}$ using (A) 45 exposures used in negative control analysis and (B) 24 exposures that are reported by CDC and existing literature.}
        \label{fig:COVIDhist}
    \end{minipage}
\end{figure}

Figure~\ref{fig:type1error} summarizes the QQ-plots of $-\log_{10}(p)$ values for different methods. First, CARE yields well-calibrated $p$-values, indicating its reliability in controlling type 1 error rates under this negative control outcome analysis (Figure~\ref{fig:type1error}A). Similarly, IVW, cML-DP and MR-APSS also achieve good performance (Figure~\ref{fig:type1error}B). In contrast, MR-mix, MR-Egger, RAPS, ContMix, cML, Weighted-Median, Weighted-Mode, and MR-Lasso yield inflated p-values (Figures~\ref{fig:type1error}C and \ref{fig:type1error}D). One may be surprised that widely used IVW achieves good performance. This is because we make every effort to make a fair comparison between different methods and use the (random effects) IVW to consider pleiotropic effects (i.e., invalid IVs) by allowing over-dispersion in the regression model. As expected, the fixed effects IVW that assumes all used IVs are valid leads to inflated p-values (Supplementary Figure~\ref{supfig:negative_control_qqplot}A). 

To understand why CARE performs well, we highlight two aspects. First, selecting valid IVs can be noisy in real data applications. That explains why cML and MR-Lasso, methods that ignore the screening variability in IV selection, produce inflated p-values (Figure~\ref{fig:type1error}D). Applying bagging reduces the screening variability and thus helps achieve well-calibrated p-values in CARE. Similarly, as cML-DP uses a data perturbation method to account for the screening variability, it also achieves relatively good performance. Second, CARE adopts a rerandomization step to select candidate IVs, accounting for the impact of the winner's curse bias. Breaking the winner's curse bias helps CARE achieve well-calibrated p-values as CARE uses a measurement error model and relies on the unbiasedness estimation of exposure-SNP effect $\beta_{X_j}$. This rerandomization step is crucial for CARE, and we confirm that applying CARE without the rerandomization step leads to inflated p-values (Supplementary Figure~\ref{supfig:negative_control_qqplot}B). 

\subsection{Risk factors identification for COVID-19 severity}\label{section 5.2}

To better understand the underlying causal risk factors for COVID-19 severity and demonstrate the performance of our proposed method CARE, we apply CARE and competing MR methods to systematically identify causal risk factors for COVID-19 severity. Specifically, we investigate the same 45 exposures used in the negative control outcome analysis and use COVID-19 severity (B2) from the covid-19hg (B2, version v7, European ancestry only; \citep{covid2021mapping}) as our outcome data. The dataset includes data from 32,519 hospitalized COVID-19 patients and 2,062,805 population controls. 

First, we compare the number of significant causal exposures identified by CARE and competing methods under the Bonferroni correction ($<0.05/45\simeq 10^{-3}$) (Figure~\ref{fig:COVIDhist}A). CARE identifies 6 causal exposures. In comparison, the competing methods RAPS, cML-DP, IVW, MR-Lasso, MR-APSS, MR-mix, ContMix,  Weighted-Median, Weighted-Mode, MR-Egger identify 7, 5, 5, 5, 4, 4, 3, 0, 0 and 0 causal exposures, respectively. In terms of statistical power, CARE ranks second among all MR methods considered. RAPS achieves the highest power but also yields inflated p-values in our negative control outcome analysis and simulations, primarily due to neglecting variability in valid IV selection step. 


 Second, we compared the risk factors identified by different MR methods to known factors that meet two criteria: (1) they have been reported by the CDC or in peer-reviewed literature, and (2) they overlap with the 45 exposures used in our negative control outcome analyses. Through a comprehensive manual review by two researchers, we identified 24 well-established risk factors for COVID-19 severity (Supplementary Table 1).  Notably, our new method, CARE, demonstrated superior performance by correctly identifying six of these 24 known risk factors: BMI, extreme BMI, HDL cholesterol, obesity class 1, obesity class 2, and overweight. In comparison, benchmark methods showed lower detection rates: MR-LASSO identified 5 risk factors, while cML-DP, IVW, MR-APSS, MR-Mix, and RAPS each identified 4. ContMix detected 3, and Median identified 2. Both Weighted-Mode and MR-Egger failed to identify any risk factors (Figure~\ref{fig:COVIDhist}B).
Importantly, CARE also avoided false positives, i.e., it did not incorrectly identify any factors lacking strong supporting evidence in the literature. In contrast, several benchmark methods produced potential false positives. For example, cML-DP incorrectly identified childhood obesity as a risk factor, while IVW erroneously identified both celiac disease and childhood obesity. Finally, when we focus on four methods with relatively good performance under our negative control outcome analysis, the result patterns are similar (Supplementary Section \ref{sup-sec7.2}). 

In summary, CARE achieves high power in identifying likely causal risk factors for COVID-19 severity, and the identified risk factors can be largely validated by complementary analyses and literature.

\section{Conclusion}
We introduced a unified two-sample Mendelian randomization within the summary data framework, referred to as Causal Analysis with Randomized Estimators (CARE), that accounts for winner's curse, measurement error bias, and genetic pleiotropy simultaneously. Through simulations and biomedical applications, we demonstrate that CARE delivers robust causal effect estimates with improved statistical power. More importantly, the CARE estimator enjoys rigorous theoretical guarantees under mild assumptions, which is often lacking for competing methods.

\bibliographystyle{apalike}
	\bibliography{arxiv}

\pagebreak
 \begin{center}
{\LARGE\bf SUPPLEMENTARY MATERIAL}
\end{center}
\setcounter{section}{0}
\setcounter{equation}{0}
\setcounter{assumption}{0}
\setcounter{figure}{0}
\setcounter{table}{0}
\setcounter{thm}{0}
\setcounter{page}{1}
\makeatletter
\renewcommand{\thesection}{S.\arabic{section}}
\renewcommand{\theequation}{S\arabic{equation}}
\renewcommand{\theassumption}{S\arabic{assumption}}
\renewcommand{\thefigure}{S\arabic{figure}}
\renewcommand{\thethm}{S\arabic{thm}}

    	\tableofcontents 
\section{ Algorithm to solve the optimization problem in \eqref{eq:optimization winner's curse}}\label{sup:sec1}

\subsection{Algorithm to solve the optimization problem in \eqref{eq:optimization winner's curse}}
In the section, we provide an algorithm borrowing ideas from coordinate descent \citep{tseng2001convergence} to solve the optimization problem in \eqref{eq:optimization winner's curse}, that is 
\begin{align*} 
\min_{\theta\in \mathbb{R}, r_j\in\mathbb{R} } \Big\{ \hat{l}\big( \theta, \{ r_j\}_{j\in \mathcal{S}_{\lambda}} \big) : \ \sum_{j\in \mathcal{S}_{\lambda}} \mathbf{1}(r_j = 0) = v \Big\} \quad \Rightarrow \quad\widehat{\mathcal{V}}(v) = \big\{ j: \ \hat{r}_j = 0, j\in \mathcal{S}_{\lambda} \big\},
\end{align*}
This step allows us to screen out invalid IVs and select $\mathcal{V}$.

We note that the proposed algorithm borrows strength from the classical coordinate descent algorithm by iteratively minimizing the objective function by fixing either $\theta$ or $r_j$'s. As our algorithm aims to screen out invalid IVs with $r_j\neq 0$, one difference is that we iteratively search for IVs with large ``residuals" (i.e., $\hat{\beta}_{Yj} -\theta\hat{\beta}_{X_j, \mathtt{RB}}$) in Step 2. (i) so that the objective function can be further minimized. Furthermore, as our optimization problem involves $l_0$ penalty, instead of choosing the model size $v$ based on cross-validation frequently adopted in Lasso-type problems \citep{tibshirani1996regression,zou2006adaptive}, we adopt the Bayesian Information Criterion to select the final set of valid IVs. 

Our proposed algorithm consists of three steps as follows:
\newpage

\begin{algorithm}[H]\label{alg:optimization}
\caption{Algorithm to solve the optimization problem in \eqref{eq:optimization winner's curse} }
\KwIn{Data inputs and initial parameters}
\KwOut{Estimated parameters $\hat{\theta}$ and $\hat{r}_j$}
\SetKwBlock{CoordinateDescent}{Block Coordinate Descent}{end}
\SetKwBlock{ValidIVSelection}{Valid IV Selection via BIC}{end}

\textbf{Initialization } Set $k = 0$, generate $\theta^{(0)} \sim \text{Uniform} \left( \min_{1 \leq j \leq s_\lambda} \frac{\hat{\beta}_{Y_j}}{\hat{\beta}_{X_j}}, \max_{1 \leq j \leq s_\lambda} \frac{\hat{\beta}_{Y_j}}{\hat{\beta}_{X_j}} \right)$\;
\CoordinateDescent{
    \Repeat{$\left| \frac{\theta^{(k+1)} - \theta^{(k)}}{\theta^{(k)}} \right| < 10^{-7}$}{
        Fix $\theta^{(k)}$, update $r_j^{(k+1)}$\;
        Order $\frac{ \left( \hat{\beta}_{Y_j} - \theta^{(k)} \cdot \hat{\beta}_{X_j, \mathtt{RB}} \right)^2 }{\sigma^2_{Y_j}} - \frac{ \theta^2 \cdot \hat{\sigma}_{X_j, \mathtt{RB}}^2}{\sigma^2_{Y_j}}, \quad j = 1, 2, \dots, s_{\lambda} - v$ in decreasing order\;
        Set $r_j^{(k+1)} = \hat{\beta}_{Y_j} - \theta^{(k)} \hat{\beta}_{X_j, \mathtt{RB}}$ for the largest $s_{\lambda} - v$ components, $j = 1, \dots, s_{\lambda} - v$, and $r_j^{(k+1)} = 0$ for $j = s_{\lambda} - v + 1, \dots, s_{\lambda}$\;
        
        Fix $r_j^{(k+1)}$, update $\theta^{(k)}$ by minimizing the following objective function:
        \[
        \theta^{(k+1)} = \underset{\theta \in \mathbb{R}}{\arg\min} \sum_{j \in \mathcal{S}_{\lambda}} \frac{ \left( \hat{\beta}_{Y_j} - \theta \cdot \hat{\beta}_{X_j, \mathtt{RB}} - r_j^{(k+1)} \right)^2 - \theta^2 \cdot \hat{\sigma}_{X_j, \mathtt{RB}}^2 }{\sigma^2_{Y_j}} \mathds{1}_{(r_j^{(k+1)} = 0)}.
        \]       
        \textbf{If} $\left| \frac{\theta^{(k+1)} - \theta^{(k)}}{\theta^{(k)}} \right| < 10^{-7}$ \textbf{then}\ 
            Stop and output $\hat{\theta}(v) = \theta^{(k+1)}$ and $\hat{r}_j(v) = r_j^{(k+1)}$ \;
       \textbf{else} Set $k = k + 1$\;
    }
}

\ValidIVSelection{
    \For{$v = 2, \dots, s_{\lambda}$}{
        Calculate 
        \[
        \text{BIC}(v) = -2 \hat{l} \left( \hat{\theta}(v), \{ \hat{r}_j(v) \}_{j \in \hat{\mathcal{V}}} \right) + \log(n) \cdot (s_{\lambda} - v);
        \]
    }
    Select $\hat{\mathcal{V}}$ with the smallest $\text{BIC}(v)$\;
}

\end{algorithm}
\newpage

\subsection{Justification of unique solution of Problem (4) under fixed $\theta$}\label{sup-optimization}

To cast some insights into the proposed Algorithm \ref{alg:optimization} for solving Problem \eqref{eq:optimization}, we note that in each iteration, our algorithm breaks the optimization into two sub-problems and provides a closed-form global optimal solution for these sub-problems.

In the first sub-problem, we fix $\theta$ and treat Problem \eqref{eq:optimization} as an optimization problem  with respect to $\{r_j\}_{j \in \mathcal{S}}$: 
\[ 
    \min_{\theta,r_j} \ l\big( \theta, \{ r_j\}_{j\in \mathcal{S}} \big),\text{ s.t. }\  \sum_{j\in \mathcal{S}} \mathds{1}_{(r_j=0)}  = v .
 \]
Unlike classical $l_0$ constrained linear regression with an arbitrary design matrix, solving this problem is computationally efficient as we can decompose the original loss function into the sum of $l_j(\theta, r_j)$. Each $l_j(\theta, r_j)$ only depends on a single $r_j$. In this case, a closed-form solution to this optimization problem can be given. 

As we can see that, for invalid IVs with $r_j \neq 0$,  $l_j\big( \theta,  r_j \big)$ reaches its minimum $0$ by setting $r_j=\hat{\beta}_{Y_j}- \theta\cdot  \hat{\beta}_{X_j}$ (See justifications below). While for valid IVs with $r_j = 0$, $l_j\big( \theta,  r_j \big)$ takes a constant value of $
\frac{1}{2}  (\hat{\beta}_{Y_j}- \theta\cdot  \hat{\beta}_{X_j} )^2 /\sigma^2_{Y_j} - \frac{1}{2}  \theta^2 \cdot {\sigma}_{X_j}^2/\sigma^2_{Y_j}$. 

Therefore, to minimize the the loss function $l\big( \theta, \{ r_j\}_{j\in \mathcal{S}} \big)$ for given $\theta$, we only need to find $v$ IVs with the smallest  $
\frac{1}{2}  (\hat{\beta}_{Y_j}- \theta\cdot  \hat{\beta}_{X_j} )^2 /\sigma^2_{Y_j} - \frac{1}{2}  \theta^2 \cdot {\sigma}_{X_j}^2/\sigma^2_{Y_j}
$ and set their $r_j=0$ and the rest of $r_j$ to $\hat{\beta}_{Y_j}- \theta\cdot  \hat{\beta}_{X_j}$. The Block Coordinate Descent Step of our algorithm is indeed providing such a closed-form global optimal solution of the above combinatorial optimization problem. After deriving $\{r_j\}_{j\in \mathcal{S}}$, we then solve our second sub-problem by solving Problem \eqref{eq:optimization} with $\{r_j\}_{j\in \mathcal{S}}$ fixed. 
The alternative minimization of $\theta$ and 
$\{r_j\}_{j\in \mathcal{S}}$ together can ensure the objective function decay.

To justify any given $\theta$, we can give a closed-form solution of the optimization problem 
\[ 
\min\limits_{\{r_j\}_{j \in \mathcal{S}_\lambda}}l\big( \theta, \{ r_j\}_{j\in \mathcal{S}_{\lambda}} \big) \text{ s.t. } \sum_{j\in\mathcal{S}_\lambda}\mathds{1}_{r_j=0}=v, 
 \]
we further investigate $l_j\big( \theta,  r_j \big)$ and 
discuss the solution to this optimization problem in three different situations.

 \[
\frac{\partial l\big( \theta, \{ r_j\}_{j\in \mathcal{S}_{\lambda}} \big)}{\partial r_j}=\frac{\partial l_j\big( \theta, r_j\big)}{\partial r_j} =  - \frac{ \hat{\beta}_{Y_j}- \theta\cdot  \hat{\beta}_{X_j, \mathtt{RB}} - r_j   }{\sigma^2_{Y_j}}  \text{ When } r_j \neq 0.
\]
\[
 l_j\big( \theta,  r_j \big) \triangleq  \frac{1}{2} \frac{ (\hat{\beta}_{Y_j}- \theta\cdot  \hat{\beta}_{X_j, \mathtt{RB}}   )^2 }{\sigma^2_{Y_j}}- \frac{1}{2} \frac{ \theta^2 \cdot {\sigma}_{X_j, \mathtt{RB}}^2}{\sigma^2_{Y_j}}  \text{ When } r_j = 0.
\]

\begin{itemize}
    \item In the case that $\hat{\beta}_{Y_j}- \theta\cdot  \hat{\beta}_{X_j, \mathtt{RB}}>0$, we have  $l_j\big( \theta,  r_j \big)$ reach its local minimum $0$ when $r_j=\hat{\beta}_{Y_j}- \theta\cdot  \hat{\beta}_{X_j, \mathtt{RB}}>0$. 

    When $r_j=0$,
\[
l_j\big( \theta,  r_j \big) =  \frac{1}{2} \frac{ (\hat{\beta}_{Y_j}- \theta\cdot  \hat{\beta}_{X_j, \mathtt{RB}} )^2 }{\sigma^2_{Y_j}} - \frac{1}{2} \frac{ \theta^2 \cdot {\sigma}_{X_j, \mathtt{RB}}^2}{\sigma^2_{Y_j}} 
\]
And when $r_j<0$, we have $\frac{\partial l\big( \theta, \{ r_j\}_{j\in \mathcal{S}_{\lambda}} \big)}{\partial r_j}=\frac{\partial l_j\big( \theta, r_j\big)}{\partial r_j}<0$, and therefore 
\[
l_j\big( \theta,  r_j \big)\geq \lim\limits_{r_j \rightarrow 0^-}\frac{1}{2} \frac{ (\hat{\beta}_{Y_j}- \theta\cdot  \hat{\beta}_{X_j, \mathtt{RB}} -r_j)^2 }{\sigma^2_{Y_j}}> 0.
\]
Therefore we have 
$$\min\limits_{r_j\neq 0} l_j\big( \theta,  r_j \big)= 0 \text{ and } l_j\big( \theta,  r_j=0 \big)=\frac{1}{2} \frac{ (\hat{\beta}_{Y_j}- \theta\cdot  \hat{\beta}_{X_j, \mathtt{RB}} )^2 }{\sigma^2_{Y_j}} - \frac{1}{2} \frac{ \theta^2 \cdot {\sigma}_{X_j, \mathtt{RB}}^2}{\sigma^2_{Y_j}}$$.

    \item In the case that $\hat{\beta}_{Y_j}- \theta\cdot  \hat{\beta}_{X_j, \mathtt{RB}}<0$,  we have  $l_j\big( \theta,  r_j \big)$ reach its local minimum $0$ when $r_j=\hat{\beta}_{Y_j}- \theta\cdot  \hat{\beta}_{X_j, \mathtt{RB}}<0$. 

    When $r_j=0$,
\[l_j\big( \theta,  r_j \big) =  \frac{1}{2} \frac{ (\hat{\beta}_{Y_j}- \theta\cdot  \hat{\beta}_{X_j, \mathtt{RB}} )^2 }{\sigma^2_{Y_j}} - \frac{1}{2} \frac{ \theta^2 \cdot {\sigma}_{X_j, \mathtt{RB}}^2}{\sigma^2_{Y_j}} 
\]
And when $r_j>0$, we have $\frac{\partial l\big( \theta, \{ r_j\}_{j\in \mathcal{S}_{\lambda}} \big)}{\partial r_j}=\frac{\partial l_j\big( \theta, r_j\big)}{\partial r_j}>0$, and therefore 
\[
l_j\big( \theta,  r_j \big)\geq \lim\limits_{r_j \rightarrow 0^+}\frac{1}{2} \frac{ (\hat{\beta}_{Y_j}- \theta\cdot  \hat{\beta}_{X_j, \mathtt{RB}} -r_j)^2 }{\sigma^2_{Y_j}}> 0.
\]
Therefore we have 
$$\min\limits_{r_j\neq 0} l_j\big( \theta,  r_j \big)= 0 \text{ and } l_j\big( \theta,  r_j=0 \big)=\frac{1}{2} \frac{ (\hat{\beta}_{Y_j}- \theta\cdot  \hat{\beta}_{X_j, \mathtt{RB}} )^2 }{\sigma^2_{Y_j}} - \frac{1}{2} \frac{ \theta^2 \cdot {\sigma}_{X_j, \mathtt{RB}}^2}{\sigma^2_{Y_j}}$$.
    \item In the case when $\hat{\beta}_{Y_j}- \theta\cdot  \hat{\beta}_{X_j, \mathtt{RB}}=0$.
    
    When $r_j>0$, we have $\frac{\partial l\big( \theta, \{ r_j\}_{j\in \mathcal{S}_{\lambda}} \big)}{\partial r_j}=\frac{\partial l_j\big( \theta, r_j\big)}{\partial r_j}\geq 0$, and therefore 
\[
l_j\big( \theta,  r_j \big)\geq \lim\limits_{r_j \rightarrow 0^+}\frac{1}{2} \frac{ (\hat{\beta}_{Y_j}- \theta\cdot  \hat{\beta}_{X_j, \mathtt{RB}} -r_j)^2 }{\sigma^2_{Y_j}}\geq 0.
\]
When $r_j<0$, we have $\frac{\partial l\big( \theta, \{ r_j\}_{j\in \mathcal{S}_{\lambda}} \big)}{\partial r_j}=\frac{\partial l_j\big( \theta, r_j\big)}{\partial r_j}<0$, and therefore 
\[
l_j\big( \theta,  r_j \big)\geq \lim\limits_{r_j \rightarrow 0^-}\frac{1}{2} \frac{ (\hat{\beta}_{Y_j}- \theta\cdot  \hat{\beta}_{X_j, \mathtt{RB}} -r_j)^2 }{\sigma^2_{Y_j}}\geq  0.
\]
When $r_j=0$, we have 
\[l_j\big( \theta,  r_j \big) =   - \frac{1}{2} \frac{ \theta^2 \cdot {\sigma}_{X_j, \mathtt{RB}}^2}{\sigma^2_{Y_j}} .
\]

Therefore we have 
$$\min\limits_{r_j\neq 0} l_j\big( \theta,  r_j \big)= 0 \text{ and } l_j\big( \theta,  r_j=0 \big)=\frac{1}{2} \frac{ (\hat{\beta}_{Y_j}- \theta\cdot  \hat{\beta}_{X_j, \mathtt{RB}} )^2 }{\sigma^2_{Y_j}} - \frac{1}{2} \frac{ \theta^2 \cdot {\sigma}_{X_j, \mathtt{RB}}^2}{\sigma^2_{Y_j}}.$$
\end{itemize}

\subsection{Adoption of $l_0$ penalty instead of using Lasso}
We adopted the $l_0$ penalty for three reasons: 
\begin{itemize}
    \item Unlike the classical $l_0$ constrained linear regression, our considered  $l_0$ constrained optimization problem is computationally efficient to solve as closed form solutions of $\left\{r_j\right\}_{j \in \mathcal{S}_\lambda}$ can be derived when $\theta$ is fixed (See previous discussions in Section 1.2). 
    \item Due to the inclusion of a measurement error bias correction term, \(\frac{1}{2}\sum_{j\in \mathcal{S}_{\lambda}} \frac{ \theta^2 \cdot {\sigma}_{X_j, \mathtt{RB}}^2}{\sigma^2_{Y_j}}I(r_j=0)\), in our objective function, adopting a Lasso-type penalty results in an optimization problem with non-differentiable gradients, making the algorithm remains time-consuming to solve. 
    
    \item Empirically, we have actually tested the use of the \( l_1 \) penalty, which was our original idea. There, to enable efficient optimization, we removed the bias correction term for the measurement error.  Our preliminary investigations with the $l_1$ penalty revealed several limitations: i) The number of selected IVs exhibited high sensitivity to small changes in the tuning parameter $\lambda$. ii) The $l_1$ penalty's simultaneous penalization of valid and invalid IVs is suboptimal, given the often subtle differences between these IVs in MR contexts. iii) The convex nature of the $l_1$ penalty resulted in discontinuous jumps in the number of selected IVs as $\lambda$ varied, leading to suboptimal performance. In contrast, the $l_0$ penalty offers several advantages in our specific context: i) It provides a comprehensive set of potential solutions across varying numbers of potential (valid) IVs. ii) It better accommodates the nuanced differences between valid and invalid IVs typically encountered in MR studies.
    
    
\end{itemize}

    These considerations collectively support the use of the $l_0$ penalty as a more suitable approach for our specific optimization problem in the MR framework.

\newpage

\section{Algorithm to solve the optimization problem using $l_1$ penalty}\label{sup:alg-l1}
\begin{itemize}
    \item \textbf{The first approach:} We replace the \( l_0 \) constraint with an \( l_1 \) penalty in the following objective function:
\[
   \hat{l}\big( \theta, \{ r_j\}_{j\in \mathcal{S}},\gamma \big)\triangleq  \sum_{j\in \mathcal{S}}l_j\big( \theta,  r_j \big)
\]
\[
l_j\big( \theta,  r_j ,\gamma\big) =  \frac{1}{2}\frac{ (\hat{\beta}_{Y_j}- \theta\cdot  \hat{\beta}_{X_j} - r_j  )^2 }{\sigma^2_{Y_j}} - \frac{1}{2}\frac{ \theta^2 \cdot \sigma_{X_{j, \mathtt{RB}}}^2}{\sigma^2_{Y_j}}\mathds{1}_{(r_j=0)}+\gamma |r_j|.
\]
For a fixed tuning parameter \(\gamma\), we estimate the parameters by minimizing:
\[
\min_{\theta\in \mathbb{R}, r_j\in\mathbb{R} } \hat{l}\big( \theta, \{ r_j\}_{j\in \mathcal{S}},\gamma \big).
\]
To solve this optimization problem, we alternate between minimizing with respect to \(\theta\) and \(\{ r_j\}_{j\in \mathcal{S}}\). The optimal solution for \(\{ r_j\}_{j\in \mathcal{S}}\) given a fixed \(\theta\) is:
\[
\small
r_j =
\begin{cases} 
\operatorname{sign}(\hat{\beta}_{Y_j} - \theta \cdot \hat{\beta}_{X_j, \mathtt{RB}}) 
\cdot \big(|\hat{\beta}_{Y_j} - \theta \cdot \hat{\beta}_{X_j, \mathtt{RB}}| - \gamma \cdot \sigma^2_{Y_j} \big) 
& \text{if } |\hat{\beta}_{Y_j} - \theta \cdot \hat{\beta}_{X_j, \mathtt{RB}}| > \gamma \cdot \sigma^2_{Y_j} + |\theta| \cdot \sigma_{X_j, \mathtt{RB}}, \\
0 & \text{otherwise}.
\end{cases}
\]
Theoretical justifications for this result can be found in the Supplemental Material Section \ref{sup:l_1}. The optimal solution for \(\theta\), given fixed \(\{ r_j\}_{j\in \mathcal{S}}\), is:
\[
\underset{\theta \in \mathbb{R}}{\arg\min} \sum_{j \in \mathcal{S}} \frac{ \left( \hat{\beta}_{Y_j} - \theta \cdot \hat{\beta}_{X_j, \mathtt{RB}} - r_j \right)^2 - \theta^2 \cdot \hat{\sigma}_{X_j, \mathtt{RB}}^2 }{\sigma^2_{Y_j}} \mathds{1}_{(r_j = 0)}.
\]
We iteratively update \(\theta\) and \(\{ r_j\}_{j\in \mathcal{S}}\) until convergence. The tuning parameter \(\gamma\) is selected via BIC, and the corresponding estimator \(\hat{\theta}(\gamma)\) is used for inference. The full optimization procedure is detailed in \textbf{Algorithm \ref{alg:optimization1}}.
\item \textbf{The second approach:}  We further replace $\mathds{1}_{(r_j= 0)}$ in the measurement error term with $1-|r_j|$ and derive the following objective function,
\begin{align*}
   l\big( \theta, \{ r_j\}_{j\in \mathcal{S}} \big)\triangleq  \sum_{j\in \mathcal{S}}l_j\big( \theta,  r_j \big) 
\end{align*}
\begin{align*}
   l_j\big( \theta,  r_j \big)=\frac{1}{2} \frac{ (\hat{\beta}_{Y_j}- \theta\cdot  \hat{\beta}_{X_j} - r_j  )^2 }{\sigma^2_{Y_j}} -\frac{1}{2}\frac{ \theta^2 \cdot {\sigma}_{X_j}^2}{\sigma^2_{Y_j}}+ (\gamma+\frac{1}{2}\frac{ \theta^2 \cdot {\sigma}_{X_j}^2}{\sigma^2_{Y_j}}) |r_j|.
\end{align*}
For a fixed tuning parameter \(\gamma\), we also estimate the parameters by minimizing:
\[
\min_{\theta\in \mathbb{R}, r_j\in\mathbb{R} } \hat{l}\big( \theta, \{ r_j\}_{j\in \mathcal{S}},\gamma \big).
\]
To solve this optimization problem, we alternate between minimizing with respect to \(\theta\) and \(\{ r_j\}_{j\in \mathcal{S}}\). The optimal solution for \(\{ r_j\}_{j\in \mathcal{S}}\) given a fixed \(\theta\) is:
\[
r_j =
\begin{cases} 
\operatorname{sign}(\hat{\beta}_{Y_j} - \theta \cdot \hat{\beta}_{X_j, \mathtt{RB}}) 
\cdot \big(|\hat{\beta}_{Y_j} - \theta \cdot \hat{\beta}_{X_j, \mathtt{RB}}| - \Tilde{\lambda}_j  \cdot \sigma^2_{Y_j} \big) 
& \text{if } |\hat{\beta}_{Y_j} - \theta \cdot \hat{\beta}_{X_j, \mathtt{RB}}| > \Tilde{\lambda}_j \cdot \sigma^2_{Y_j} , \\
0 & \text{otherwise}.
\end{cases}
\]
where $\Tilde{\lambda}_j=\gamma+\frac{1}{2}\frac{ \theta^2 \cdot {\sigma}_{X_j}^2}{\sigma^2_{Y_j}}$ for all $j\in\mathcal{S}$. Theoretical justifications for this result can be found in the Supplemental Material Section \ref{sup:l_1}. The optimal solution for \(\theta\), given fixed \(\{ r_j\}_{j\in \mathcal{S}}\), is:
\[
\underset{\theta \in \mathbb{R}}{\arg\min} \frac{1}{2}\sum_{j \in \mathcal{S}} \frac{ (\hat{\beta}_{Y_j}- \theta\cdot  \hat{\beta}_{X_j} - r_j  )^2 }{\sigma^2_{Y_j}} -\frac{1}{2}\sum_{j\in \mathcal{S}} \frac{ \theta^2 \cdot {\sigma}_{X_j}^2}{\sigma^2_{Y_j}}+\sum_{j\in \mathcal{S}} (\gamma+\frac{1}{2}\frac{ \theta^2 \cdot {\sigma}_{X_j}^2}{\sigma^2_{Y_j}}) |r_j|.\]
We also iteratively update \(\theta\) and \(\{ r_j\}_{j\in \mathcal{S}}\) until convergence and use BIC to select the tuning parameter \(\gamma\). The corresponding estimator \(\hat{\theta}(\gamma)\) is used for inference. The full optimization procedure is detailed in \textbf{Algorithm \ref{alg:optimization2}}. 

\end{itemize} 
\begin{algorithm}
[H]\label{alg:optimization1}
\small
\caption{Algorithm to solve the optimization problem in \eqref{eq:optimization winner's curse} }
\KwIn{Data inputs and initial parameters}
\KwOut{Estimated parameters $\hat{\theta}$ and $\hat{r}_j$}
\SetKwBlock{CoordinateDescent}{Block Coordinate Descent}{end}
\SetKwBlock{ValidIVSelection}{Valid IV Selection via BIC}{end}

\textbf{Initialization } Set $k = 0$, generate $\theta^{(0)} \sim \text{Uniform} \left( \min_{1 \leq j \leq s_\lambda} \frac{\hat{\beta}_{Y_j}}{\hat{\beta}_{X_j}}, \max_{1 \leq j \leq s_\lambda} \frac{\hat{\beta}_{Y_j}}{\hat{\beta}_{X_j}} \right)$\;
\CoordinateDescent{
    \Repeat{$\left| \frac{\theta^{(k+1)} - \theta^{(k)}}{\theta^{(k)}} \right| < 10^{-7}$}{
        Fix $\theta^{(k)}$, update $r_j^{(k+1)}:$
        
        For $\forall j \in \mathcal{S}$:  If $ |\hat{\beta}_{Y_j} - \theta^{(k)} \cdot \hat{\beta}_{X_j, \mathtt{RB}}| > \gamma \cdot \sigma^2_{Y_j} + |\theta^{(k)}| \cdot \sigma_{X_j, \mathtt{RB}}$, we let $r_j^{(k+1)} =
\operatorname{sign}(\hat{\beta}_{Y_j} - \theta^{(k)}\cdot \hat{\beta}_{X_j, \mathtt{RB}}) 
\cdot \big(|\hat{\beta}_{Y_j} - \theta^{(k)}\cdot \hat{\beta}_{X_j, \mathtt{RB}}| - \gamma \cdot \sigma^2_{Y_j} \big)$.

Otherwise, we set
$r_j^{(k+1)}=0$.
        
        Fix $r_j^{(k+1)}$, update $\theta^{(k)}$ by minimizing the following objective function:
        \[
        \theta^{(k+1)} = \underset{\theta \in \mathbb{R}}{\arg\min} \sum_{j \in \mathcal{S}_{\lambda}} \frac{ \left( \hat{\beta}_{Y_j} - \theta \cdot \hat{\beta}_{X_j, \mathtt{RB}} - r_j^{(k+1)} \right)^2 - \theta^2 \cdot \hat{\sigma}_{X_j, \mathtt{RB}}^2 }{\sigma^2_{Y_j}} \mathds{1}_{(r_j^{(k+1)} = 0)}.
        \]       

           \[
        \theta^{(k+1)} = \frac{\sum_{j \in \mathcal{S}}\frac{\hat{\beta}_{X_j, \mathtt{RB}}\cdot\hat{\beta}_{Y_j}}{\hat{\sigma}_{X_j, \mathtt{RB}}^2}\mathds{1}_{(r_j^{(k+1)} = 0)}}{\sum_{j \in \mathcal{S}}\big(\frac{\hat{\beta}^2_{X_j, \mathtt{RB}}}{\sigma_{Y_j}^2}-\frac{\hat{\sigma}_{X_j, \mathtt{RB}}^2}{\sigma_{Y_j}^2}\big)\mathds{1}_{(r_j^{(k+1)} = 0)}}.
        \]     
        \textbf{If} $\left| \frac{\theta^{(k+1)} - \theta^{(k)}}{\theta^{(k)}} \right| < 10^{-7}$ \textbf{then}\ 
            Stop and output $\hat{\theta}(\gamma) = \theta^{(k+1)}$ and $\hat{r}_j(\gamma) = r_j^{(k+1)}$ \;
       \textbf{else} Set $k = k + 1$\;
    }
}

\ValidIVSelection{
    \For{all candidate $\gamma$ }{
        Calculate 
        \[
        \text{BIC}(\gamma) = -2 \hat{l} \left( \hat{\theta}(\gamma), \{ \hat{r}_j(\gamma) \}_{j \in \hat{\mathcal{V}}_\gamma} \right) + \log(n) \cdot (s_{\lambda} - \hat{v}_\gamma);
        \]
    }
    Select $\hat{\mathcal{V}}_\gamma$ with the smallest $\text{BIC}(\gamma)$\;
}
\end{algorithm}

\newpage

\begin{algorithm}[H]\label{alg:optimization2}
\small
\caption{Algorithm to solve the optimization problem in \eqref{eq:optimization winner's curse} }
\KwIn{Data inputs and initial parameters}
\KwOut{Estimated parameters $\hat{\theta}$ and $\hat{r}_j$}
\SetKwBlock{CoordinateDescent}{Block Coordinate Descent}{end}
\SetKwBlock{ValidIVSelection}{Valid IV Selection via BIC}{end}

\textbf{Initialization } Set $k = 0$, generate $\theta^{(0)} \sim \text{Uniform} \left( \min_{1 \leq j \leq s_\lambda} \frac{\hat{\beta}_{Y_j}}{\hat{\beta}_{X_j}}, \max_{1 \leq j \leq s_\lambda} \frac{\hat{\beta}_{Y_j}}{\hat{\beta}_{X_j}} \right)$\;
\CoordinateDescent{
    \Repeat{$\left| \frac{\theta^{(k+1)} - \theta^{(k)}}{\theta^{(k)}} \right| < 10^{-7}$}{
        Fix $\theta^{(k)}$, update $r_j^{(k+1)}:$
        
        For $\forall j \in \mathcal{S}$:  If $ |\hat{\beta}_{Y_j} - \theta^{(k)} \cdot \hat{\beta}_{X_j, \mathtt{RB}}| > \Tilde{\lambda}_j \cdot \sigma^2_{Y_j}$, we let $r_j^{(k+1)} =
\operatorname{sign}(\hat{\beta}_{Y_j} - \theta^{(k)} \cdot \hat{\beta}_{X_j, \mathtt{RB}}) 
\cdot \big(|\hat{\beta}_{Y_j} - \theta^{(k)} \cdot \hat{\beta}_{X_j, \mathtt{RB}}| - \Tilde{\lambda}_j  \cdot \sigma^2_{Y_j} \big)  \text{ where }\Tilde{\lambda}_j=\gamma+\frac{1}{2}\frac{ \theta^2 \cdot \hat{\sigma}_{X_j, \mathtt{RB}}^2}{\sigma^2_{Y_j}}$.

Otherwise, we set
$r_j^{(k+1)}=0$.
        
        Fix $r_j^{(k+1)}$, update $\theta^{(k)}$ by minimizing the following objective function:
        \[
        \theta^{(k+1)} = \underset{\theta \in \mathbb{R}}{\arg\min} \frac{1}{2} \sum_{j \in \mathcal{S}} \frac{ (\hat{\beta}_{Y_j}- \theta\cdot  \hat{\beta}_{X_j} - r_j^{(k+1)}  )^2 }{\sigma^2_{Y_j}} -\frac{1}{2}\sum_{j\in \mathcal{S}} \frac{ \theta^2 \cdot \hat{\sigma}_{X_j, \mathtt{RB}}^2}{\sigma^2_{Y_j}}+\sum_{j\in \mathcal{S}} (\gamma+\frac{1}{2}\frac{ \theta^2 \cdot \hat{\sigma}_{X_j, \mathtt{RB}}^2}{\sigma^2_{Y_j}}) |r_j^{(k+1)}|.
        \]       
                \[
        \theta^{(k+1)} = \frac{\sum_{j \in \mathcal{S}}\frac{\hat{\beta}_{X_j, \mathtt{RB}}(\hat{\beta}_{Y_j}-r_j^{(k+1)})}{\hat{\sigma}_{X_j, \mathtt{RB}}^2}}{\sum_{j \in \mathcal{S}}\frac{\hat{\beta}^2_{X_j, \mathtt{RB}}}{\sigma_{Y_j}^2}-\frac{\hat{\sigma}_{X_j, \mathtt{RB}}^2}{\sigma_{Y_j}^2}\cdot(1-|r_j^{(k+1)}|)}.
        \]       
        \textbf{If} $\left| \frac{\theta^{(k+1)} - \theta^{(k)}}{\theta^{(k)}} \right| < 10^{-7}$ \textbf{then}\ 
            Stop and output $\hat{\theta}(\gamma) = \theta^{(k+1)}$ and $\hat{r}_j(\gamma) = r_j^{(k+1)}$ \;
       \textbf{else} Set $k = k + 1$\;
    }
}

\ValidIVSelection{
    \For{all candidate $\gamma$ }{
        Calculate 
        \[
        \text{BIC}(\gamma) = -2 \hat{l} \left( \hat{\theta}(\gamma), \{ \hat{r}_j(\gamma) \}_{j \in \hat{\mathcal{V}}_\gamma} \right) + \log(n) \cdot (s_{\lambda} - \hat{v}_\gamma);
        \]
    }
    Select $\hat{\mathcal{V}}_\gamma$ with the smallest $\text{BIC}(\gamma)$\;
}
\end{algorithm}

\clearpage
\section{ Theorectical justifications for two $l_1$ methods} \label{sup:l_1}
\subsection{Method 1}
For a fixed tuning parameter \(\gamma\), we estimate the parameters by minimizing:
\[
\min_{\theta\in \mathbb{R}, r_j\in\mathbb{R} } \hat{l}\big( \theta, \{ r_j\}_{j\in \mathcal{S}},\gamma \big).
\]
To solve this optimization problem, we alternate between minimizing with respect to \(\theta\) and \(\{ r_j\}_{j\in \mathcal{S}}\). The optimal solution for \(\{ r_j\}_{j\in \mathcal{S}}\) given a fixed \(\theta\) is:
\[
r_j =
\begin{cases} 
\operatorname{sign}(\hat{\beta}_{Y_j} - \theta \cdot \hat{\beta}_{X_j, \mathtt{RB}}) 
\cdot \big(|\hat{\beta}_{Y_j} - \theta \cdot \hat{\beta}_{X_j, \mathtt{RB}}| - \gamma \cdot \sigma^2_{Y_j} \big) 
& \text{if } |\hat{\beta}_{Y_j} - \theta \cdot \hat{\beta}_{X_j, \mathtt{RB}}| > \gamma \cdot \sigma^2_{Y_j} + |\theta| \cdot \sigma_{X_j, \mathtt{RB}}, \\
0 & \text{otherwise}.
\end{cases}
\]
To see this, we investigate $l_j\big( \theta,  r_j \big)$ and 
discuss the solution of $r_j$ when fixed $\theta$.
We consider the objective function: 
\begin{align*}
   l\big( \theta, \{ r_j\}_{j\in \mathcal{S}} \big)= \sum_{j\in \mathcal{S}}l_j\big( \theta,  r_j \big) ,\quad
l_j\big( \theta,  r_j \big) \triangleq \frac{1}{2}\frac{ (\hat{\beta}_{Y_j}- \theta\cdot  \hat{\beta}_{X_j} - r_j  )^2 }{\sigma^2_{Y_j}} - \frac{1}{2}\frac{ \theta^2 \cdot {\sigma}_{X_j}^2}{\sigma^2_{Y_j}}\mathds{1}_{(r_j=0)}+\gamma\cdot |r_j|.
\end{align*}
and we have 
\[
\frac{\partial l_j\big( \theta, r_j\big)}{\partial r_j} =  - \frac{ \hat{\beta}_{Y_j}- \theta\cdot  \hat{\beta}_{X_j, \mathtt{RB}} - r_j   }{\sigma^2_{Y_j}}+\gamma  \text{ When } r_j > 0,
\]
\[
\frac{\partial l_j\big( \theta, r_j\big)}{\partial r_j} =  - \frac{ \hat{\beta}_{Y_j}- \theta\cdot  \hat{\beta}_{X_j, \mathtt{RB}} - r_j   }{\sigma^2_{Y_j}}-\gamma  \text{ When } r_j < 0,
\]
\[
 l_j\big( \theta,  r_j \big) \triangleq  \frac{1}{2} \frac{ (\hat{\beta}_{Y_j}- \theta\cdot  \hat{\beta}_{X_j, \mathtt{RB}}   )^2 }{\sigma^2_{Y_j}}- \frac{1}{2} \frac{ \theta^2 \cdot {\sigma}_{X_j, \mathtt{RB}}^2}{\sigma^2_{Y_j}}  \text{ When } r_j = 0,
\]
and consider three different scenarios:
\begin{itemize}
    \item In the case that $\hat{\beta}_{Y_j}- \theta\cdot  \hat{\beta}_{X_j, \mathtt{RB}}>\gamma \cdot\sigma^2_{Y_j}$, 

when $r_j<0$, we have $\frac{\partial l_j\big( \theta, r_j\big)}{\partial r_j}<0$, and therefore 
\[
l_j\big( \theta,  r_j \big)\geq \lim\limits_{r_j \rightarrow 0^-}\frac{1}{2} \frac{ (\hat{\beta}_{Y_j}- \theta\cdot  \hat{\beta}_{X_j, \mathtt{RB}} -r_j)^2 }{\sigma^2_{Y_j}}=\frac{1}{2} \frac{ (\hat{\beta}_{Y_j}- \theta\cdot  \hat{\beta}_{X_j, \mathtt{RB}} )^2 }{\sigma^2_{Y_j}}.
\]
When $r_j=0$,
\[
l_j\big( \theta,  r_j \big) =  \frac{1}{2} \frac{ (\hat{\beta}_{Y_j}- \theta\cdot  \hat{\beta}_{X_j, \mathtt{RB}} )^2 }{\sigma^2_{Y_j}} - \frac{1}{2} \frac{ \theta^2 \cdot {\sigma}_{X_j, \mathtt{RB}}^2}{\sigma^2_{Y_j}}.
\]
When $r_j>0$, we have  $l_j\big( \theta,  r_j \big)$ reach its local minimum when $r_j=\hat{\beta}_{Y_j}- \theta\cdot  \hat{\beta}_{X_j, \mathtt{RB}}-\gamma \cdot\sigma^2_{Y_j}$.
\[
l_j\big( \theta,  r_j \big) =  \frac{1}{2} \frac{ \gamma^2\cdot\sigma^4_{Y_j} }{\sigma^2_{Y_j}} + \gamma\cdot(\hat{\beta}_{Y_j}- \theta\cdot  \hat{\beta}_{X_j, \mathtt{RB}}-\gamma \cdot\sigma^2_{Y_j}).
\]
and
\begin{align*}
   &\quad l_j\big( \theta,  0 \big) -l_j\big( \theta,  r_j \big) \\&=  \frac{1}{2} \frac{ (\hat{\beta}_{Y_j}- \theta\cdot  \hat{\beta}_{X_j, \mathtt{RB}} )^2 }{\sigma^2_{Y_j}} - \frac{1}{2} \frac{ \theta^2 \cdot {\sigma}_{X_j, \mathtt{RB}}^2}{\sigma^2_{Y_j}}- \frac{1}{2} \frac{ \gamma^2\cdot\sigma^4_{Y_j} }{\sigma^2_{Y_j}} - \gamma\cdot(\hat{\beta}_{Y_j}- \theta\cdot  \hat{\beta}_{X_j, \mathtt{RB}}-\gamma \cdot\sigma^2_{Y_j})\\&= \frac{1}{2} \frac{ (\hat{\beta}_{Y_j}- \theta\cdot  \hat{\beta}_{X_j, \mathtt{RB}} )^2 }{\sigma^2_{Y_j}} + \frac{1}{2} \frac{ \gamma^2\cdot\sigma^4_{Y_j} }{\sigma^2_{Y_j}} - \gamma\cdot(\hat{\beta}_{Y_j}- \theta\cdot  \hat{\beta}_{X_j, \mathtt{RB}})- \frac{1}{2} \frac{ \theta^2 \cdot {\sigma}_{X_j, \mathtt{RB}}^2}{\sigma^2_{Y_j}}\\&= \frac{1}{2} \frac{ (\hat{\beta}_{Y_j}- \theta\cdot  \hat{\beta}_{X_j, \mathtt{RB}} -\gamma\cdot\sigma^2_{Y_j})^2 }{\sigma^2_{Y_j}} - \frac{1}{2} \frac{ \theta^2 \cdot {\sigma}_{X_j, \mathtt{RB}}^2}{\sigma^2_{Y_j}}.
\end{align*}

Thus when $\hat{\beta}_{Y_j}- \theta\cdot  \hat{\beta}_{X_j, \mathtt{RB}} -\gamma\cdot\sigma^2_{Y_j}>|\theta|\cdot{\sigma}_{X_j, \mathtt{RB}}$, $l_j\big( \theta,  r_j \big)$ achieves the minimum when 
$$
r_j=\hat{\beta}_{Y_j}- \theta\cdot  \hat{\beta}_{X_j, \mathtt{RB}} -\gamma\cdot\sigma^2_{Y_j}.
$$
Otherwise, $l_j\big( \theta,  r_j \big)$ achieves the minimum when $r_j=0$.

    \item In the case that $\hat{\beta}_{Y_j}- \theta\cdot  \hat{\beta}_{X_j, \mathtt{RB}}<-\gamma \cdot\sigma^2_{Y_j}$, 

when $r_j>0$, we have $\frac{\partial l_j\big( \theta, r_j\big)}{\partial r_j}>0$, and therefore 
\[
l_j\big( \theta,  r_j \big)\geq \lim\limits_{r_j \rightarrow 0^+}\frac{1}{2} \frac{ (\hat{\beta}_{Y_j}- \theta\cdot  \hat{\beta}_{X_j, \mathtt{RB}} -r_j)^2 }{\sigma^2_{Y_j}}=\frac{1}{2} \frac{ (\hat{\beta}_{Y_j}- \theta\cdot  \hat{\beta}_{X_j, \mathtt{RB}} )^2 }{\sigma^2_{Y_j}}.
\]
When $r_j=0$,
\[
l_j\big( \theta,  r_j \big) =  \frac{1}{2} \frac{ (\hat{\beta}_{Y_j}- \theta\cdot  \hat{\beta}_{X_j, \mathtt{RB}} )^2 }{\sigma^2_{Y_j}} - \frac{1}{2} \frac{ \theta^2 \cdot {\sigma}_{X_j, \mathtt{RB}}^2}{\sigma^2_{Y_j}}.
\]
When $r_j<0$, we have  $l_j\big( \theta,  r_j \big)$ reach its local minimum when $r_j=\hat{\beta}_{Y_j}- \theta\cdot  \hat{\beta}_{X_j, \mathtt{RB}}+\gamma \cdot\sigma^2_{Y_j}$.
\[
l_j\big( \theta,  r_j \big) =  \frac{1}{2} \frac{ \gamma^2\cdot\sigma^4_{Y_j} }{\sigma^2_{Y_j}} - \gamma\cdot(\hat{\beta}_{Y_j}- \theta\cdot  \hat{\beta}_{X_j, \mathtt{RB}}+\gamma \cdot\sigma^2_{Y_j}).
\]
and
\begin{align*}
    l_j\big( \theta,  0 \big) -l_j\big( \theta,  r_j \big) &=  \frac{1}{2} \frac{ (\hat{\beta}_{Y_j}- \theta\cdot  \hat{\beta}_{X_j, \mathtt{RB}} )^2 }{\sigma^2_{Y_j}} - \frac{1}{2} \frac{ \theta^2 \cdot {\sigma}_{X_j, \mathtt{RB}}^2}{\sigma^2_{Y_j}}- \frac{1}{2} \frac{ \gamma^2\cdot\sigma^4_{Y_j} }{\sigma^2_{Y_j}} + \gamma\cdot(\hat{\beta}_{Y_j}- \theta\cdot  \hat{\beta}_{X_j, \mathtt{RB}}+\gamma \cdot\sigma^2_{Y_j})\\&= \frac{1}{2} \frac{ (\hat{\beta}_{Y_j}- \theta\cdot  \hat{\beta}_{X_j, \mathtt{RB}} )^2 }{\sigma^2_{Y_j}} + \frac{1}{2} \frac{ \gamma^2\cdot\sigma^4_{Y_j} }{\sigma^2_{Y_j}} + \gamma\cdot(\hat{\beta}_{Y_j}- \theta\cdot  \hat{\beta}_{X_j, \mathtt{RB}})- \frac{1}{2} \frac{ \theta^2 \cdot {\sigma}_{X_j, \mathtt{RB}}^2}{\sigma^2_{Y_j}}\\&= \frac{1}{2} \frac{ (\hat{\beta}_{Y_j}- \theta\cdot  \hat{\beta}_{X_j, \mathtt{RB}} +\gamma\cdot\sigma^2_{Y_j})^2 }{\sigma^2_{Y_j}} - \frac{1}{2} \frac{ \theta^2 \cdot {\sigma}_{X_j, \mathtt{RB}}^2}{\sigma^2_{Y_j}}.
\end{align*}

Thus when $\hat{\beta}_{Y_j}- \theta\cdot  \hat{\beta}_{X_j, \mathtt{RB}} +\gamma\cdot\sigma^2_{Y_j}<-|\theta|\cdot{\sigma}_{X_j, \mathtt{RB}}$, $l_j\big( \theta,  r_j \big)$ achieves the minimum when 
$$
r_j=\hat{\beta}_{Y_j}- \theta\cdot  \hat{\beta}_{X_j, \mathtt{RB}} +\gamma\cdot\sigma^2_{Y_j}.
$$
Otherwise, $l_j\big( \theta,  r_j \big)$ achieves the minimum when $r_j=0$.
    \item In the case that $-\gamma \cdot\sigma^2_{Y_j}\leq \hat{\beta}_{Y_j}- \theta\cdot  \hat{\beta}_{X_j, \mathtt{RB}}\leq \gamma \cdot\sigma^2_{Y_j}$, $l_j\big( \theta,  r_j \big)$ achieves the minimum when $r_j=0$. 
    
\end{itemize}

\newpage

\subsection{Method 2}
We consider the objective function
\begin{align*}
   l\big( \theta, \{ r_j\}_{j\in \mathcal{S}} \big)\triangleq \sum_{j\in \mathcal{S}}l_j\big( \theta,  r_j \big) ,
\end{align*}
where each component loss is given by
\begin{align*}
   l_j\big( \theta,  r_j \big)=\frac{1}{2} \frac{ (\hat{\beta}_{Y_j}- \theta\cdot  \hat{\beta}_{X_j} - r_j  )^2 }{\sigma^2_{Y_j}} -\frac{1}{2}\frac{ \theta^2 \cdot {\sigma}_{X_j}^2}{\sigma^2_{Y_j}}+ (\gamma+\frac{1}{2}\frac{ \theta^2 \cdot {\sigma}_{X_j}^2}{\sigma^2_{Y_j}}) |r_j|.
\end{align*}
When $\theta$ is fixed, the optimization over $r_j$ reduces to minimizing:
$$
\tilde{l}_j\big( \theta,  r_j \big)=\frac{1}{2} \frac{ (\hat{\beta}_{Y_j}- \theta \cdot  \hat{\beta}_{X_j} - r_j  )^2 }{\sigma^2_{Y_j}}+ \tilde{\lambda}_j |r_j|,
$$
with $
\tilde{\lambda}_j = \gamma+\frac{1}{2}\frac{ \theta^2 \cdot {\sigma}_{X_j}^2}{\sigma^2_{Y_j}}.
$

This takes the canonical form of the Lasso problem
$$
\min_{r \in \mathbb{R}} \left\{ \frac{1}{2} (z - r)^2 + \lambda |r| \right\}.
$$
Letting $$
a_j = \hat{\beta}_{Y_j} - \theta \cdot \hat{\beta}_{X_j}, \quad \lambda_j = \tilde{\lambda}_j \cdot \sigma^2_{Y_j},
$$
we have:
$$
\tilde{l}_j(r_j) = \frac{1}{2 \sigma^2_{Y_j}}(a_j - r_j)^2 + \tilde{\lambda}_j |r_j| = \frac{1}{\sigma^2_{Y_j}} \left( \frac{1}{2}(a_j - r_j)^2 + \lambda_j |r_j| \right).
$$

The minimizer of this expression is given by the soft-thresholding operator \citep{donoho1995adapting}, 
$$
r_j^* = S_{\lambda_j}(a_j) = 
\begin{cases}
\operatorname{sign}(a_j) \cdot (|a_j| - \lambda_j), & \text{if } |a_j| > \lambda_j, \\
0, & \text{otherwise}.
\end{cases}
$$
Substituting back, we obtain:
\[
r_j =
\begin{cases} 
\operatorname{sign}(\hat{\beta}_{Y_j} - \theta \cdot \hat{\beta}_{X_j}) 
\cdot \left(|\hat{\beta}_{Y_j} - \theta \cdot \hat{\beta}_{X_j}| - \Tilde{\lambda}_j  \cdot \sigma^2_{Y_j} \right), 
& \text{if } |\hat{\beta}_{Y_j} - \theta \cdot \hat{\beta}_{X_j}| > \Tilde{\lambda}_j \cdot \sigma^2_{Y_j}, \\
0, & \text{otherwise}.
\end{cases}
\]
where $\Tilde{\lambda}_j=\gamma+\frac{1}{2}\frac{ \theta^2 \cdot {\sigma}_{X_j}^2}{\sigma^2_{Y_j}}$ for all $j\in\mathcal{S}$.

\newpage

\section{ Proof of Theorem 1}\label{sup:sec2}

\subsection{Notions and Assumptions}

We first review some notions and assumptions that will be used in our proofs:
\begin{itemize}
    \item The selected set of relevant IVs after randomization: 
    $${\mathcal{S}}_{\lambda} =\left\{j:|\frac{\hat\beta_{X_j}}{\sigma_{X_j}}+Z_j|>\lambda, j=1,\ldots,p\right\}.$$
    \item Cardinality of the set of selected relevant IVs: $s_{\lambda} = |\mathcal{S}_{\lambda} |$.
    \item The average measure of instrument strength after selection:
     $$
     \kappa_{\lambda}=\frac{1}{s_\lambda}\sum\limits_{j\in  \mathcal{S}_\lambda}\frac{\beta_{X_j}^2}{\sigma_{Y_j}^2}.
     $$
     \item In our bagging strategy, we denote the b-th bootstrap sample as $\mathcal{S}_{\lambda,b}^*$ and the number of occurrences in $\mathcal{S}_{\lambda,b}^*$ for j-th IVs of $\mathcal{S}_{\lambda}$ as $w_{jb}^*$. We also denote the selected set of valid IVs as
$$\mathcal{\hat{V}}_b=\left\{j: \hat{r}_{jb}=0 \text{ and } \ j \in \mathcal{S}_{\lambda,b}^*\right\}$$
    and the causal estimator as 
    $$\hat{\theta}_b=A_b^{-1}\sum_{j\in\widehat{\mathcal{V}}_b}  \hat\beta_{Y_j} 
 \hat\beta_{X_j,\mathtt{RB}} / {\sigma}_{Y_j}^{2},$$ 
    where 
     $$A_b=\sum_{j\in \widehat{\mathcal{V}}_b}  (\hat\beta_{X_j,\mathtt{RB}}^2 - \hat{\sigma}_{X_j,\mathtt{RB}}^{ \mathrm{2} }  ) / {\sigma}_{Y_j}^{2}.$$
     \item For convenience, we also denote the conditional expectation taken with respect to bootstrap resampling as 
\begin{align*}
    \mathbb{E}^*\big[\cdot\big]= \mathbb{E}\Big[\cdot\big|S_{\lambda},\big\{(\hat{\beta}_{Y_j},\hat{\beta}_{X_{j,\mathtt{RB}}})\big\}_{j \in S_{\lambda}}\Big].
\end{align*}
\end{itemize}

 Our final estimator is obtained by taking bootstrap aggregation 
 $$\widetilde{\theta}=\frac{1}{B}\sum_{b=1}^B \hat{\theta}_{b}.$$ 

\begin{assumption}[Measurement error model]\label{assumption: measure error}
(i)For any $j\neq j'$, $(\hat{\beta}_{Y_j}, \hat{\beta}_{X_{j}})$ and $(\hat{\beta}_{Y_{j'}}, \hat{\beta}_{X_{j'}})$ are mutually independent.\\
(ii)For each $j$, the association pair $(\hat{\beta}_{Y_j}, \hat{\beta}_{X_{j}})$ follows
\begin{align*}
		\begin{bmatrix}
		\hat{\beta}_{X_{j}}\\
		\hat{\beta}_{Y_j}
		\end{bmatrix} \sim \mathcal{N}\left(\begin{bmatrix}
		\beta_{X_j} \\
		\theta\beta_{X_j}+r_j
		\end{bmatrix}\ , \begin{bmatrix}
		{\sigma}_{X_j}^2 & 0 \\
		0 & {\sigma}_{Y_j}^2
		\end{bmatrix}\right).
		\end{align*}
    Furthermore, there exists positive constants $l$ and $u$ such that $\frac{m}{n}\leq \sigma_{X_j}^2 \leq \frac{M}{n}$,\  $\frac{m}{n}\leq \sigma_{Y_j}^2 \leq \frac{M}{n}$ for $j=1,...,p$. 
\end{assumption}
	
\begin{assumption}[Variance stabilization]\label{sup-assumption:variance stablization}
There exists a variance stabilizing quantity $a_{\lambda}$ and a vector $\boldsymbol{\tau} \in \mathbb{R}^{s_{\lambda}}$ in which each component is independent of $\left\{(u_j,\nu_j)\right\}_{j \in S_{\lambda}}$ and uniformly bounded away from infinity in probability in the sense that $$
\sup_{j \in S_{\lambda}}\Big|a_\lambda\cdot \mathbb{E}^*\Big[A_b^{-1} \cdot \hat{w}_{jb}\Big]-\tau_j\Big|=o_p(1), 
$$
where $A_b=\sum_{k\in \mathcal{S}_\lambda} \hat{w}_{kb}\cdot(\hat\beta_{X_k,\mathtt{RB}}^2 - \hat{\sigma}_{X_k,\mathtt{RB}}^{ \mathrm{2} }  ) / {\sigma}_{Y_k}^{2}$, and 
\begin{align*}
\hat{w}_{jb} =
\begin{cases}
w_{jb}^* \cdot \mathrm{I}(\hat{r}_{jb} = 0) & \text{if } w_{jb}^* \geq 1, \\
0 & \text{if } w_{jb}^* = 0.
\end{cases}
\end{align*}
In addition, there is no dominating instrument in the sense that 
\begin{align*}
\frac{\max_{j\in \mathcal{S}_{\lambda}} \beta_{X_j}^2}{\sum_{j\in \mathcal{S}_{\lambda}} \beta_{X_j}^2} \toProb 0.
\end{align*}
\end{assumption}

\begin{assumption}[Negligible invalid IV induced bias]\label{sup-assumption: Negligible invalid IV induced bias}
There is negligible bias induced by potential imperfect screening of invalid IVs after bootstrap aggregation in the sense that 
$$
\frac{a_{\lambda}}{\sqrt{s_{\lambda}\cdot \kappa_{\lambda}}}\mathbb{E}^*\big[A_b^{-1} \sum_{j \in S_{\lambda}} \hat{\beta}_{X_{j,\mathtt{RB}}}\cdot r_j \cdot \hat{w}_{jb}/ {\sigma}_{Y_j}^{2}\big]=o_p(1).
$$
\end{assumption}

\begin{assumption}[Instrument Selection]\label{sup-assumption: instrument selection}
 Define $\underline{\eta}=\min_{1\leq j\leq p}\eta_j$ and $\overline{\eta}= \max_{1\leq j\leq p}\eta_j$, then both $\underline{\eta}$ and $\overline{\eta}$ are bounded and bounded away from zero.
\end{assumption}

\subsection{Proof}

We begin by decomposing $ \frac{a_\lambda}{\sqrt{s_\lambda\cdot\kappa_\lambda}}(\tilde{\theta}-\theta_0)$ and want to show that there is a leading term in the decomposition converging to a Gaussian distribution. While the remained terms converges to zero in probability. We notice that 
\begin{align*}
     \frac{a_\lambda}{\sqrt{s_\lambda\cdot\kappa_\lambda}}(\tilde{\theta}-\theta_0)= \frac{a_\lambda}{\sqrt{s_\lambda\cdot\kappa_\lambda}}\cdot\mathbb{E}^*\Big[\hat{\theta}_b-\theta_0\Big].
\end{align*}
Here
\begin{align*}
    \hat{\theta}_b-\theta_0&=A_b^{-1} \Big\{ \sum_{j\in\widehat{\mathcal{V}}_b} \tilde{u}_j/ {\sigma}_{Y_j}^{2}+\sum_{j\in\widehat{\mathcal{V}}_b}\hat{\beta}_{X_{j,\mathtt{RB}}}\cdot r_j/ {\sigma}_{Y_j}^{2}\Big\}\\&=A_b^{-1} \Big\{\sum_{j\in \mathcal{S}_\lambda} \hat{w}_{jb}\cdot \tilde{u}_j/ {\sigma}_{Y_j}^{2}+\sum_{j\in \mathcal{S}_\lambda} \hat{w}_{jb}\cdot\hat{\beta}_{X_{j,\mathtt{RB}}}\cdot r_j/ {\sigma}_{Y_j}^{2}\Big\}.
\end{align*}
where $\tilde{u}_j=\beta_{X_{j}}\cdot\big(\nu_j-\theta_0 \cdot u_j\big)+\big(\nu_j\cdot u_j-\theta_0\cdot(u_j^2-\hat{\sigma}_{X_{j,\mathtt{RB}}}^2)\big)$ and 
\begin{align*}
\hat{w}_{jb} =
\begin{cases}
w_{jb}^* \cdot \mathrm{I}(\hat{r}_{jb} = 0) & \text{if } w_{jb}^* \geq 1, \\
0 & \text{if } w_{jb}^* = 0.
\end{cases}
\end{align*}
We then can decompose $\frac{a_\lambda}{\sqrt{s_\lambda\cdot\kappa_\lambda}}(\tilde{\theta}-\theta_0)$ into two terms:
\begin{align*}
    \frac{a_\lambda}{\sqrt{s_\lambda\cdot\kappa_\lambda}}(\tilde{\theta}-\theta_0)&=\underbrace{\frac{a_\lambda}{\sqrt{s_\lambda\cdot\kappa_\lambda}}\sum_{j\in \mathcal{S}_\lambda}\mathbb{E}^*\Big[A_b^{-1} \cdot \hat{w}_{jb}\Big]\cdot\tilde{u}_j/ {\sigma}_{Y_j}^{2}}_{(\mathrm{I})}+\underbrace{\frac{a_\lambda}{\sqrt{s_\lambda\cdot\kappa_\lambda}}\mathbb{E}^*\Big[A_b^{-1}\sum_{j\in \mathcal{S}_\lambda} \hat{w}_{jb}\cdot \hat{\beta}_{X_{j,\mathtt{RB}}}\cdot r_j/ {\sigma}_{Y_j}^{2}\Big]}_{(\mathrm{II})}.
\end{align*}
Assumption \ref{sup-assumption: Negligible invalid IV induced bias} shows that the second term in the above formula satisfies $(\mathrm{II})=o_p(1)$. \\
For the term $\mathrm{(I)}$, we further decompose it as 
\begin{align*}
    &\quad \frac{a_\lambda}{\sqrt{s_\lambda\cdot\kappa_\lambda}}\sum_{j\in \mathcal{S}_\lambda}\mathbb{E}^*\Big[A_b^{-1} \cdot \hat{w}_{jb}\Big]\cdot\tilde{u}_j/ {\sigma}_{Y_j}^{2}\\&=   \underbrace{ \frac{1}{\sqrt{s_\lambda\cdot\kappa_\lambda}}\sum_{j\in \mathcal{S}_\lambda} \tau_j\cdot\tilde{u}_j/ {\sigma}_{Y_j}^{2}}_{(\mathrm{I}.1)}\\&+\underbrace{\frac{1}{\sqrt{s_\lambda\cdot\kappa_\lambda}}\sum_{j\in \mathcal{S}_\lambda}\Big\{a_\lambda\cdot \mathbb{E}^*\Big[A_b^{-1} \cdot \hat{w}_{jb}\Big]-\tau_j\Big\}\cdot\tilde{u}_j/ {\sigma}_{Y_j}^{2}}_{(\mathrm{I}.2)}.
\end{align*}
Here $(\mathrm{I.2})$ has the following upper bounds,
\begin{align*}
    (\mathrm{I}.2)\leq    \sup\limits_{j \in S_\lambda}\Big|a_\lambda\cdot \mathbb{E}^*\Big[A_b^{-1} \cdot \hat{w}_{jb}\Big]-\tau_j\Big|\cdot\frac{1}{\sqrt{s_\lambda\cdot\kappa_\lambda}}\sum_{j\in \mathcal{S}_\lambda}\cdot\tilde{u}_j/ {\sigma}_{Y_j}^{2}.
\end{align*}
Under Assumption \ref{sup-assumption:variance stablization}, we can prove $(\mathrm{I.2})=o_p(1)$.

Combining all the above results, we have 
\begin{align*}
      \frac{a_\lambda}{\sqrt{s_\lambda\cdot\kappa_\lambda}}(\tilde{\theta}-\theta_0)= \frac{1}{\sqrt{s_\lambda\cdot\kappa_\lambda}}\sum_{j\in \mathcal{S}_\lambda} \tau_j\cdot\tilde{u}_j+o_p(1).
\end{align*}
Using the proof of Theorem 1 in \cite{ma2023breaking}, we can show that when Assumption \ref{assumption: measure error} and Assumption \ref{sup-assumption: instrument selection} hold and $\frac{\max_{j\in \mathcal{S}_{\lambda}} \beta_{X_j}^2}{\sum_{j\in \mathcal{S}_{\lambda}} \beta_{X_j}^2} \toProb 0$, conditional on the selection event $\mathcal{S}_\lambda$, $ \frac{1}{\sqrt{s_\lambda\cdot\kappa_\lambda}}\sum_{j\in \mathcal{S}_\lambda} \tau_j\cdot\tilde{u}_j$ converges to a Gaussian distribution as $s_\lambda\stackrel{\text{p}}\rightarrow \infty$ and $\frac{\kappa_{\lambda}}{\lambda^2}\stackrel{\text{p}}\rightarrow \infty$.

Therefore, we can conclude that $ \frac{a_\lambda}{\sqrt{s_\lambda\cdot\kappa_\lambda}}(\tilde{\theta}-\theta_0)$ converges to a Gaussian distribution. 

\subsection{Verifying the Assumption \ref{sup-assumption:variance stablization} in the case with perfect screening property}\label{sup:sec2.3}
To cast more insights into Assumption \ref{sup-assumption:variance stablization}, we next consider a special case where perfect IV screening is achieved. In the case of perfect IV screening, we have 
\begin{align*}
     A_b =\sum_{k\in \mathcal{V}_\lambda}  w_{kb}^*\cdot(\hat\beta_{X_k,\mathtt{RB}}^2 - \hat{\sigma}_{X_k,\mathtt{RB}}^{ \mathrm{2} }  ) / {\sigma}_{Y_j}^{2}. 
  \end{align*}
In what follows, we argue that Assumption \ref{sup-assumption:variance stablization} holds for both valid and invalid IVs in $\mathcal{S}_{\lambda}$:
\begin{itemize}
    \item For valid IVs in $\mathcal{V}_{\lambda}$ ($\mathcal{V}_\lambda$ is the collection of all valid IVs in $\mathcal{S}_\lambda$), we define
\begin{align*}
    \tau_j=a_\lambda\cdot \mathbb{E}^*\Big[\frac{w_{jb}^*}{\sum_{k\in \mathcal{V}_\lambda}  w_{kb}^*\cdot \beta_{X_k}^2  / {\sigma}_{Y_k}^{2}}\Big], \quad a_\lambda=\sum_{k\in \mathcal{V}_\lambda}  \beta_{X_k}^2  / {\sigma}_{Y_k}^{2}, 
\end{align*}
and we have $\tau_j$ independent of $\left\{(u_j,\nu_j)\right\}_{j \in S_{\lambda}}$. In this context, we have $\hat{w}_{jb}  = w_{jb}^*$ and can show that 
$$
\Big|a_\lambda\cdot \mathbb{E}^*\Big[A_b^{-1} \cdot \hat{w}_{jb}\Big]-\tau_j\Big| = \Big|a_\lambda\cdot \mathbb{E}^*\Big[A_b^{-1} \cdot {w}^*_{jb}\Big]-\tau_j\Big|=o_p(1).
$$
For this bound to hold uniformly for \( j \in \mathcal{V}_{\lambda} \) as stated in the assumption, given that \({w}^*_{jb}\) follows a multinomial distribution with an equal mean, we conjecture that this condition is likely to hold as long as \( A_b \) converges to a center that is independent of \( j \). In fact,  under appropriate conditions  (See Section \ref{sec:Ab} 
 in the Supplement Material for full theoretical justifications), we can show that,
\begin{align*}
    A_b=\sum_{k\in \mathcal{V}_\lambda}  \beta_{X_k}^2 / \sigma_{Y_k}^{2}\cdot (1+o_p(1)),
\end{align*}
 which is indeed independent of $j$.

\item For invalid IV $j \in \mathcal{S}_\lambda/\mathcal{V}_\lambda$, under perfect screening property, we have $\hat{r}_{jb}=0$ and therefore $\hat{w}_{jb}=0$. Set $\tau_j=0$ for $j \in \mathcal{S}_\lambda/ \mathcal{V}_\lambda$, we have 
$$
\sup_{j \in \mathcal{S}_\lambda/ \mathcal{V}_\lambda}\Big|a_\lambda\cdot \mathbb{E}^*\Big[A_b^{-1} \cdot \hat{w}_{jb}\Big]-\tau_j\Big|=o_p(1).
$$
\end{itemize}
Combining these two parts of results, we can verify that the Assumption \ref{sup-assumption:variance stablization} is satisfied.

\subsection{The asymptotic analysis of $A_b$ under perfect screening property}\label{sec:Ab}
Notice that 
\begin{align*}
 A_b =\sum_{k\in \mathcal{V}_\lambda}  w_{kb}^*\cdot(\hat\beta_{X_k,\mathtt{RB}}^2 - \hat{\sigma}_{X_k,\mathtt{RB}}^{ \mathrm{2} }  ) / {\sigma}_{Y_k}^{2}. 
\end{align*}
We want to show  $ A_b=\sum_{k\in \mathcal{V}_\lambda} \beta_{X_k}^2 / {\sigma}_{Y_k}^{2}\cdot (1+o_p(1))$ under these two conditions
\begin{align*}
&\frac{\max_{k \in \mathcal{V}_\lambda} \beta_{X_k}^2}{\sum_{k \in \mathcal{V}_\lambda}   \beta_{X_k}^2}  \rightarrow 0 \text{  and  } \frac{v_\lambda\cdot\max_{k \in \mathcal{V}_\lambda}|(\hat\beta_{X_k}^2 - \hat{\sigma}_{X_k,\mathtt{RB}}^{ \mathrm{2} }  ) / {\sigma}_{Y_j}^{2}-\beta_{X_k}^2 / {\sigma}_{Y_j}^{2}|}{\sum_{k \in \mathcal{V}_\lambda}   \beta_{X_k}^2/{\sigma}_{Y_j}^{2}}=o_p(1).
\end{align*}
To prove this result, we begin with the following decomposition,
\begin{align*}
   A_b - \sum_{k\in \mathcal{V}_\lambda} \beta_{X_k}^2 / {\sigma}_{Y_k}^{2}
  & =\sum_{k\in \mathcal{V}_\lambda}  w_{kb}^*\cdot(\hat\beta_{X_k}^2 - \hat{\sigma}_{X_k,\mathtt{RB}}^{ \mathrm{2} }  ) / {\sigma}_{Y_k}^{2}-\sum_{k\in \mathcal{V}_\lambda}  w_{kb}^*\cdot\beta_{X_k}^2 / {\sigma}_{Y_k}^{2}+\sum_{k\in \mathcal{V}_\lambda}  w_{kb}^*\cdot\beta_{X_k}^2 / {\sigma}_{Y_j}^{2}-
  \sum_{k\in \mathcal{V}_\lambda} \beta_{X_k}^2 / {\sigma}_{Y_k}^{2}
  \\& = \sum_{k\in \mathcal{V}_\lambda}  w_{kb}^*\cdot(\hat\beta_{X_k}^2 - \hat{\sigma}_{X_k,\mathtt{RB}}^{ \mathrm{2} }  ) / {\sigma}_{Y_k}^{2}-\sum_{k\in \mathcal{V}_\lambda}  w_{kb}^*\cdot\beta_{X_k}^2 / {\sigma}_{Y_k}^{2}+\sum_{k\in \mathcal{V}_\lambda}  
   (w_{kb}^*-1)\cdot\beta_{X_k}^2 / {\sigma}_{Y_k}^{2}  \\& = \sum_{k\in \mathcal{V}_\lambda}  w_{kb}^*\cdot\Big((\hat\beta_{X_k}^2 - \hat{\sigma}_{X_k,\mathtt{RB}}^{ \mathrm{2} }  ) / {\sigma}_{Y_k}^{2}- \beta_{X_k}^2 / {\sigma}_{Y_k}^{2}\Big)+\sum_{k\in \mathcal{V}_\lambda}  
   (w_{kb}^*-1)\cdot\beta_{X_k}^2 / {\sigma}_{Y_k}^{2}.
\end{align*}
It suffices to prove the two terms on the right-hand side are of the asymptotic order $o_p(\sum_{k\in \mathcal{V}_\lambda} \beta_{X_k}^2 / {\sigma}_{Y_k}^{2})$.\\
Notice that $[w_{1,b}^*, \ldots, w_{s_\lambda,b}^*]$ follows a multinomial distribution with $\mathbb{E}[w_{k,b}^*]=1$ for all $k \in \mathcal{S}_\lambda$ and 
\begin{align*}
    & Var[w_{k,b}^*] =\frac{1}{s_\lambda}\cdot (1-\frac{1}{s_\lambda}),\text{ for all } k \in \mathcal{S}_\lambda,
    \\& Cov(w_{i,b}^*,w_{j,b}^*)= -\frac{1}{s_\lambda} \text{ for all } i,j \in \mathcal{S}_\lambda \text{ such that } i \neq j.
\end{align*}
\begin{itemize}
    \item To show $\frac{\sum_{k\in \mathcal{V}_\lambda}  (w_{kb}^*-1)\cdot\beta_{X_k}^2 / {\sigma}_{Y_k}^{2}}{\sum_{k \in \mathcal{V}_\lambda}   \beta_{X_k}^2/{\sigma}_{Y_k}^{2}}=o_p(1)$,
    
    we have $\sum_{k\in \mathcal{V}_\lambda}  (w_{kb}^*-1)\cdot\beta_{X_k}^2 / {\sigma}_{Y_j}^{2}=O_p(\sqrt{Var[\sum_{k\in \mathcal{V}_\lambda}  (w_{kb}^*-1)\cdot\beta_{X_k}^2 / {\sigma}_{Y_k}^{2}]})$ and 
\begin{align*}
 Var[\sum_{k\in \mathcal{V}_\lambda}  (w_{kb}^*-1)\cdot\beta_{X_k}^2 / {\sigma}_{Y_k}^{2}]&=\sum_{k\in \mathcal{V}_\lambda}   (1-\frac{1}{s_\lambda}) \cdot (\beta_{X_k}^2 / {\sigma}_{Y_k}^{2})^2-\frac{1}{s_\lambda} \sum_{i,j \in \mathcal{V}_\lambda,\ i\neq j}   (\beta_{X_i}^2 / {\sigma}_{Y_i}^{2}) \cdot (\beta_{X_j}^2 / {\sigma}_{Y_k}^{2})\\&=\sum_{k\in \mathcal{V}_\lambda} (\beta_{X_j}^2 / {\sigma}_{Y_k}^{2})^2-\frac{1}{s_\lambda} (\sum_{k \in \mathcal{V}_\lambda}   \beta_{X_k}^2 / {\sigma}_{Y_k}^{2})^2.
\end{align*}
Notice that 
\begin{align*}
\frac{Var[\sum_{k\in \mathcal{V}_\lambda}  (w_{kb}^*-1)\cdot\beta_{X_k}^2 / {\sigma}_{Y_k}^{2}]}{(\sum_{k \in \mathcal{V}_\lambda}   \beta_{X_k}^2 / {\sigma}_{Y_k}^{2})^2} &= \frac{\sum_{k\in \mathcal{V}_\lambda} (\beta_{X_k}^2 / {\sigma}_{Y_k}^{2})^2-\frac{1}{s_\lambda} (\sum_{k \in \mathcal{V}_\lambda}   \beta_{X_k}^2 / {\sigma}_{Y_k}^{2})^2}{(\sum_{k \in \mathcal{V}_\lambda}   \beta_{X_i}^2 / {\sigma}_{Y_k}^{2})^2} \\&= \frac{\sum_{k\in \mathcal{V}_\lambda} (\beta_{X_k}^2 / {\sigma}_{Y_k}^{2})^2}{(\sum_{k \in \mathcal{V}_\lambda}   \beta_{X_k}^2 / {\sigma}_{Y_k}^{2})^2}-\frac{1}{s_\lambda}    \\&\leq \frac{\max_{j \in \mathcal{V}_\lambda} \beta_{X_k}^2/{\sigma}_{Y_k}^{2}\cdot\sum_{k\in \mathcal{V}_\lambda} \beta_{X_k}^2/{\sigma}_{Y_k}^{2}}{(\sum_{k \in \mathcal{V}_\lambda}   \beta_{X_k}^2/{\sigma}_{Y_k}^{2})^2}-\frac{1}{s_\lambda} \\&= \frac{\max_{j \in \mathcal{V}_\lambda} \beta_{X_k}^2/{\sigma}_{Y_k}^{2}}{\sum_{k \in \mathcal{V}_\lambda}   \beta_{X_k}^2/{\sigma}_{Y_k}^{2}}-\frac{1}{s_\lambda},
\end{align*}
if we have 
$$
\frac{\max_{k \in \mathcal{V}_\lambda} \beta_{X_k}^2}{\sum_{k \in \mathcal{V}_\lambda}   \beta_{X_k}^2}\rightarrow 0,
$$
$\frac{\sum_{k\in \mathcal{V}_\lambda}  (w_{kb}^*-1)\cdot\beta_{X_k}^2 / {\sigma}_{Y_k}^{2}}{\sum_{k \in \mathcal{V}_\lambda}   \beta_{X_k}^2/{\sigma}_{Y_k}^{2}}=o_p(1)$ directly follows.

\item To show 
$\frac{\sum_{k\in \mathcal{V}_\lambda}  w_{kb}^*\cdot\big((\hat\beta_{X_k}^2 - \hat{\sigma}_{X_k,\mathtt{RB}}^{ \mathrm{2} }  ) / {\sigma}_{Y_k}^{2}-\beta_{X_k}^2 / {\sigma}_{Y_k}^{2}\big)}{\sum_{k \in \mathcal{V}_\lambda}   \beta_{X_k}^2/{\sigma}_{Y_k}^{2}}=o_p(1)$, we further decompose it into two terms 
$$\frac{\sum_{k\in \mathcal{V}_\lambda}  (w_{kb}^*-1)\cdot\big((\hat\beta_{X_k}^2 - \hat{\sigma}_{X_k,\mathtt{RB}}^{ \mathrm{2} }  ) / {\sigma}_{Y_k}^{2}-\beta_{X_k}^2 / {\sigma}_{Y_k}^{2}\big)}{\sum_{k \in \mathcal{V}_\lambda}   \beta_{X_k}^2/{\sigma}_{Y_k}^{2}}+\frac{\sum_{k\in \mathcal{V}_\lambda}  \big((\hat\beta_{X_k}^2 - \hat{\sigma}_{X_k,\mathtt{RB}}^{ \mathrm{2} }  ) / {\sigma}_{Y_k}^{2}-\beta_{X_k}^2 / {\sigma}_{Y_k}^{2}\big)}{\sum_{k \in \mathcal{V}_\lambda}   \beta_{X_k}^2/{\sigma}_{Y_k}^{2}}.$$

$\frac{\sum_{k\in \mathcal{V}_\lambda}  \big((\hat\beta_{X_k}^2 - \hat{\sigma}_{X_k,\mathtt{RB}}^{ \mathrm{2} }  ) / {\sigma}_{Y_k}^{2}-\beta_{X_k}^2 / {\sigma}_{Y_k}^{2}\big)}{\sum_{k \in \mathcal{V}_\lambda}   \beta_{X_k}^2/{\sigma}_{Y_k}^{2}}=o_p(1)$ can directly follow from Lemma S.13 of the Supplemental Material of \cite{ma2023breaking}.

To prove $\frac{\sum_{k\in \mathcal{V}_\lambda}  (w_{kb}^*-1)\cdot\big((\hat\beta_{X_k}^2 - \hat{\sigma}_{X_k,\mathtt{RB}}^{ \mathrm{2} }  ) / {\sigma}_{Y_k}^{2}-\beta_{X_k}^2 / {\sigma}_{Y_k}^{2}\big)}{\sum_{k \in \mathcal{V}_\lambda}   \beta_{X_k}^2/{\sigma}_{Y_k}^{2}}=o_p(1)$,
we use \begin{align*}
    &\quad \frac{\sum_{k\in \mathcal{V}_\lambda}  (w_{kb}^*-1)\cdot\big((\hat\beta_{X_k}^2 - \hat{\sigma}_{X_k,\mathtt{RB}}^{ \mathrm{2} }  ) / {\sigma}_{Y_k}^{2}-\beta_{X_k}^2 / {\sigma}_{Y_k}^{2}\big)}{\sum_{k \in \mathcal{V}_\lambda}   \beta_{X_k}^2/{\sigma}_{Y_k}^{2}}\\&=O_p(\frac{\mathbb{E}|\sum_{k\in \mathcal{V}_\lambda}  (w_{kb}^*-1)\cdot\big((\hat\beta_{X_k}^2 - \hat{\sigma}_{X_k,\mathtt{RB}}^{ \mathrm{2} }  ) / {\sigma}_{Y_k}^{2}-\beta_{X_k}^2 / {\sigma}_{Y_k}^{2}\big)|}{\sum_{k \in \mathcal{V}_\lambda}   \beta_{X_k}^2/{\sigma}_{Y_k}^{2}}).
\end{align*}
Notice that
\begin{align*}
& \mathbb{E}|\sum_{k\in \mathcal{V}_\lambda}  (w_{kb}^*-1)\cdot\big((\hat\beta_{X_k}^2 - \hat{\sigma}_{X_k,\mathtt{RB}}^{ \mathrm{2} }  ) / {\sigma}_{Y_k}^{2}-\beta_{X_k}^2 / {\sigma}_{Y_k}^{2}\big)|\\&\leq  \max_{j \in \mathcal{V}_\lambda}|(\hat\beta_{X_k}^2 - \hat{\sigma}_{X_k,\mathtt{RB}}^{ \mathrm{2} }  ) / {\sigma}_{Y_k}^{2}-\beta_{X_k}^2 / {\sigma}_{Y_k}^{2}| \cdot \sum_{k\in \mathcal{V}_\lambda} \mathbb{E}|  w_{kb}^*-1|,
\\& \text{ and } \mathbb{E}|w_{kb}^*-1|= \mathbb{E}(w_{kb}^*-1)+2\cdot\mathbb{P}(w_{kb}^*=0)=2\cdot(1-\frac{1}{s_\lambda}).
\end{align*}
If we have
\begin{align*}
\frac{v_\lambda\cdot\max_{k \in \mathcal{V}_\lambda}|(\hat\beta_{X_k}^2 - \hat{\sigma}_{X_k,\mathtt{RB}}^{ \mathrm{2} }  ) / {\sigma}_{Y_j}^{2}-\beta_{X_k}^2 / {\sigma}_{Y_j}^{2}|}{\sum_{k \in \mathcal{V}_\lambda}   \beta_{X_k}^2/{\sigma}_{Y_k}^{2}}=o_p(1).
\end{align*}
Then can show $\frac{\sum_{k\in \mathcal{V}_\lambda}  (w_{kb}^*-1)\cdot\big((\hat\beta_{X_k}^2 - \hat{\sigma}_{X_k,\mathtt{RB}}^{ \mathrm{2} }  ) / {\sigma}_{Y_k}^{2}-\beta_{X_k}^2 / {\sigma}_{Y_k}^{2}\big)}{\sum_{k \in \mathcal{V}_\lambda}   \beta_{X_k}^2/{\sigma}_{Y_k}^{2}}=o_p(1)$ and therefore
$$\frac{\sum_{k\in \mathcal{V}_\lambda}  w_{kb}^*\cdot\big((\hat\beta_{X_k}^2 - \hat{\sigma}_{X_k,\mathtt{RB}}^{ \mathrm{2} }  ) / {\sigma}_{Y_k}^{2}-\beta_{X_k}^2 / {\sigma}_{Y_k}^{2}\big)}{\sum_{k \in \mathcal{V}_\lambda}   \beta_{X_k}^2/{\sigma}_{Y_k}^{2}}=o_p(1).$$
\end{itemize}
Combining all these results, we have $ A_b=\sum_{k\in \mathcal{V}_\lambda} \beta_{X_k}^2 / {\sigma}_{Y_k}^{2}\cdot (1+o_p(1))$.

    \section{ Invalid IV screening consistency}\label{sup:sec3}
	
	In this section, we show that under Conditions \ref{Bound of Orlicz norm}-\ref{high dimension BIC}, the proposed invalid IV screening procedure is ``nearly perfect" as $s_\lambda$ goes to infinity.
	
	\subsection{Notations}\label{notation}

We first introduce notations to be used in the sufficient conditions and our proofs below:
\begin{itemize}
    \item The correct set of valid IVs in $S_\lambda$:
     \begin{align*}
        \mathcal{V}_{\lambda} = \big\{ j \in \mathcal{S}_{\lambda}: \ \beta_{X_j}\neq 0\text{ and }  r_j = 0 \big\}. 
    \end{align*}
    \item Cardinality of the set of valid IVs: $v_{\lambda} = |\mathcal{V}_{\lambda} |$.
    \item The selected set of valid IVs: 
    \begin{align*}
         \hat{\mathcal{V}}_{\lambda} = \big\{ j: \  \hat{r}_{j}= 0 \text{ and } j \in \mathcal{S}_{\lambda} \big\}. 
    \end{align*}
     \item Cardinality of the set of selected valid IVs: $\hat{v}_\lambda = |\hat{\mathcal{V}}_\lambda |$.
    \item For any $\mathcal{V} \subseteq \mathcal{S}_{\lambda}$, we use the following notation:
      \begin{itemize}
          \item Cardinality of the set:  $ v=|\mathcal{V}|.$
          \item 
           The measure of average instrument strength of $\mathcal{V}$:  $\kappa_{\lambda}(\mathcal{V})=\frac{1}{v}\sum\limits_{j\in  \mathcal{V}}\frac{\beta_{X_j}^2}{\sigma_{Y_j}^2}$.
         \item 
         The measure of average pleiotropic effects of $\mathcal{V}$:  
         $r_{\lambda}(\mathcal{V})=\frac{1}{v}\sum\limits_{j\in \mathcal{V}}\frac{r_j^2}{\sigma_{Y_j}^2}.$
         \item Correlation between instrument strength and pleiotropic effects of $\mathcal{V}$ when $\mathcal{V}$ has at least one non-zero $r_j$:  
         $$\rho(\mathcal{V})= \text{Corr}^2\Big( \{\beta_{X_j}\}_{j\in \mathcal{V}},  \{r_j\}_{j\in \mathcal{V}} \Big) = \frac{(\sum\limits_{j\in \mathcal{V}}\frac{
r_j\cdot\beta_{X_j}}{\sigma_{Y_j}^2})^2}{\sum\limits_{j\in \mathcal{V}}\frac{ r_j^2}{\sigma_{Y_j}^2}\cdot \sum\limits_{j\in \mathcal{V}}\frac{ \beta_{X_j}^2}{\sigma_{Y_j}^2}}.$$
      \end{itemize}   
\end{itemize}


To identify invalid IVs, we use the following measurement error models.
$$
\hat \beta_{Y_j}=\theta \cdot \beta_{X_j}+r_j+\nu_j,\quad \hat \beta_{X_{j,\mathtt{RB}}}=\beta_{X_j}+u_{j},\quad j \in \mathcal{S}_\lambda.
$$
and let $n_1=n_2=n$ be the sample sizes of the two GWAS summary datasets for $X$ and $Y$, respectively.

The invalid IV screening is obtained by solving 
$$
\min\limits_{\theta \in \mathbb{R}, r_j } \hat l(\theta,\left\{ r_j\right\}_{j \in \mathcal{S}_\lambda},\left\{\hat \beta_{Y_j},\sigma_{Y_j},\hat \beta_{X_{j,\mathtt{RB}}},\hat{\sigma}_{X_{j,\mathtt{RB}}}\right\}_{j \in \mathcal{S}_\lambda}), \text{ s.t. }\sum\limits_{j \in \mathcal{S}_\lambda}\mathds{1}_{r_j=0}=v.
$$
where 
$$ 
  \hat l( \theta,\left\{ r_j\right\}_{j \in \mathcal{S}_\lambda},\left\{\hat \beta_{Y_j},\sigma_{Y_j},\hat \beta_{X_{j,\mathtt{RB}}},\hat{\sigma}_{X_{j,\mathtt{RB}}}\right\}_{j \in \mathcal{S}_\lambda})=\sum\limits_{j\in \mathcal{S}_\lambda} \frac{(\hat \beta_{Y_j}- \theta \cdot \hat \beta_{X_{j,\mathtt{RB}}}- r_j)^2  }{\sigma_{Y_j}^2}- \sum\limits_{j\in \mathcal{S}_\lambda} \frac{\theta^2 \cdot \hat{\sigma}_{X_{j,\mathtt{RB}}}^2}{\sigma_{Y_j}^2}\cdot \mathds{1}_{r_j=0}.
$$
Here $\mathds{1}(\cdot)$ is the indicator function and $v$ is a tuning parameter representing the unknown number of valid IVs. We propose a generalized Bayesian Information Criterion(GBIC) to select the best $v$:
$$
GBIC(v)=\hat l(\hat \theta,\left\{\hat r_j\right\}_{j \in \mathcal{S}_\lambda},\left\{\hat \beta_{Y_j},\sigma_{Y_j},\hat \beta_{X_{j,\mathtt{RB}}},\hat{\sigma}_{X_{j,\mathtt{RB}}}\right\}_{j \in \mathcal{S}_\lambda})+\kappa_n\cdot (s_\lambda-v).
$$
Then we select $\hat v=\arg\min\limits_{v} GBIC(v)$
and estimate $   \hat{\mathcal{V}}_\lambda = \big\{ j: \  \hat{r}_{j, \hat{v}}= 0 \text{ and } j \in \mathcal{S}_\lambda \big\}$, which is the set of the estimated invalid IVs..

\subsection{Sufficient conditions}

\begin{condition}[Bound of Orlicz norm]\label{Bound of Orlicz norm}
Fix $\lambda$, $\frac{  \hat\sigma_{X_{j,\mathtt{RB}}}^2- \sigma_{X_{j,\mathtt{RB}}}^2}{\sigma_{Y_j}^2}$ is  a sub-exponential random variable for all $j \in \mathcal{S}_\lambda$ and we have 
    $||\frac{\nu_j}{\sigma_{Y_j}}||_{\psi_2}^2$,  $||\frac{u_j}{\sigma_{Y_j}}||_{\psi_2}^2$,  $||\sqrt{\frac{\nu_j^2-\sigma_{Y_j}^2}{\sigma_{Y_j}^2}}||_{\psi_2}$,  $||\sqrt{\frac{  u_j^2- \sigma_{X_{j,\mathtt{RB}}}^2}{\sigma_{Y_j}^2}}||_{\psi_2}$,  $||\sqrt{\frac{  \hat\sigma_{X_{j,\mathtt{RB}}}^2- \sigma_{X_{j,\mathtt{RB}}}^2}{\sigma_{Y_j}^2}}||_{\psi_2}$ bounded away from $\infty$ uniformly for all $j \in \mathcal{S}_\lambda$.
\end{condition}

This condition is a technical condition. It places some restrictions on the tail distributions of the noise terms,  aiming to ensure that they have good concentration behaviors.

\begin{condition}[Orders of the variances and sample sizes]\label{Orders of the variances and sample sizes}
There exist positive constants $m$ and $M$ such that we have $\frac{m}{n}\leq \sigma_{X_j}^2 \text{ , } \sigma_{Y_j}^2 \leq \frac{M}{n}$ for $j=1,...,p$. 
\end{condition}

In this condition, we require the variances of both $\hat \beta_{X_j}$ and $\hat \beta_{Y_j}$ have the orders 
$\frac{1}{n}$ uniformly for all $j \in \mathcal{S}_\lambda$, which is a normal assumption in two-sample summary Mendelian Randomization literature.

\begin{condition}[Plurality and no perfect correlation] \label{Plurality and no perfect correlation}
For all $\mathcal{V}\subseteq \mathcal{S}_{\lambda}$ and $\mathcal{V}$ contains at least one $r_j\neq 0$, whenever $\rho(\mathcal{V}) = 1$, we have $|  \mathcal{V}_{\lambda} | > \big| \mathcal{V}  \big|$; whenever $\rho(\mathcal{V}) <1$, we have the correlation coefficient $\rho(\mathcal{V}) <1$ is upper bounded by a constant $c_0$ smaller than one.

Here $\rho(\mathcal{V})$ measures the correlation between instrument strength and pleiotropic effects of $\mathcal{V}$, which is defined as   
         $$\rho(\mathcal{V})= \text{Corr}^2\Big( \{\beta_{X_j}\}_{j\in \mathcal{V}},  \{r_j\}_{j\in \mathcal{V}} \Big) = \frac{(\sum\limits_{j\in \mathcal{V}}\frac{
r_j\cdot\beta_{X_j}}{\sigma_{Y_j}^2})^2}{\sum\limits_{j\in \mathcal{V}}\frac{ r_j^2}{\sigma_{Y_j}^2}\cdot \sum\limits_{j\in \mathcal{V}}\frac{ \beta_{X_j}^2}{\sigma_{Y_j}^2}}.$$
\end{condition}

This condition is closely related to the plural validity assumption commonly made in two-sample summary data Mendelian Randomization literature \cite{kang2016instrumental,guo2018confidence}, which ensures the uniqueness and identifiability of the causal effect $\theta$. By Cauchy-Schwartz inequality, we can see that $\rho(\mathcal{V})=1$ indicates that there exists a $c \in \mathbb{R}$ such that $\frac{r_j}{\beta_{X_j}}=c$ holds for all $j \in \mathcal{V}$. If the first part of this condition does not hold, there will be a $\mathcal{V^*}$ such that $|\mathcal{V^*}|\geq |\mathcal{V}_\lambda|$ and $\frac{r_j}{\beta_{X_j}}=c>0$ holds for all $j \in \mathcal{V^*}$.  We expect that the invalid IV screening procedure will tend to screen out $\mathcal{S}_\lambda/\mathcal{V^*}$ and leave $\mathcal{V^*}$. Therefore, the sub-sequential causal estimation using $\mathcal{V}^*$ will be centered around $\theta_0+c$ instead of $\theta_0$, where $\theta_0$ is the true causal effect. In this case, we fail to identify the true causal effect. Furthermore, the second part of this condition is to ensure that different clusters of IV set $\mathcal{V}_c=\{j\in \mathcal{S}_\lambda\ |\ \frac{r_j}{\beta_{X_j}}=c\}$ are sufficiently separable, so that there will not be a IV set $\mathcal{V}^* \neq \mathcal{V}_\lambda$ with $\rho(\mathcal{V}^*)\rightarrow 1$ selected by the invalid screening procedure. Without this, we might not be able to distinguish the IV set $\mathcal{V}^*$ and $\mathcal{V}_\lambda$.

\begin{condition}[Boundedness]\label{Boundedness}
For any $\mathcal{V} \in S_{\lambda}$, $|\hat\theta(\mathcal{V})|$ is uniformly bounded away from $\infty$ with probability goes to 1.
\end{condition}

This condition requires that for any subset $\mathcal{V} \subseteq \mathcal{S}_\lambda$, 
the causal estimate $$\hat\theta(\mathcal{V})=\frac{\sum_{j \in \mathcal{V}}\frac{\hat{\beta}_{Y_j}\hat\beta_{X_{j,\mathtt{RB}}}}{\sigma^2_{Y_j}}}{\sum_{j \in \mathcal{V}}\frac{\hat\beta^2_{X_{j,\mathtt{RB}}}-\hat\sigma^2_{X_{j,\mathtt{RB}}}}{\sigma^2_{Y_j}}}$$ based on $\mathcal{V}$ should not be too large. In fact, when the Condition \ref{Bound of Orlicz norm} holds, this condition can be satisfied in the case that $\frac{r_j}{\beta_{X_j}}$ is bounded away from infinity for all $j \in \mathcal{S}_\lambda$ and $\beta_{X_j}$ is sufficiently separated from $0$ for all $j \in \mathcal{S}_\lambda$. To see this, we can decompose $\hat{\theta}(\mathcal{V})$ as follows:
\begin{align*}
\hat\theta(\mathcal{V})&=\frac{\sum_{j \in \mathcal{V}}\frac{(\theta_0\beta_{X_j}+r_j)\beta_{X_j}}{\sigma^2_{Y_j}}+\sum_{j \in \mathcal{V}}\frac{\beta_{X_j}\nu_j}{\sigma^2_{Y_j}}+\sum_{j \in \mathcal{V}}\frac{(\theta_0\beta_{X_j}+r_j)u_j}{\sigma^2_{Y_j}}+\sum_{j \in \mathcal{V}}\frac{u_j\nu_j}{\sigma^2_{Y_j}}}{\sum_{j \in \mathcal{V}}\frac{\beta^2_{X_{j}}}{\sigma^2_{Y_j}}+2\sum_{j \in \mathcal{V}}\frac{\beta_{X_{j}}u_j}{\sigma^2_{Y_j}}+\sum_{j \in \mathcal{V}}\frac{u^2_{j}-\sigma^2_{X_{j,\mathtt{RB}}}}{\sigma^2_{Y_j}}-\sum_{j \in \mathcal{V}}\frac{\hat\sigma^2_{X_{j,\mathtt{RB}}}-\sigma^2_{X_{j,\mathtt{RB}}}}{\sigma^2_{Y_j}}}\\&=\frac{\sum_{j \in \mathcal{V}}(\theta_0+\frac{r_j}{\beta_{X_j}})\cdot\frac{\beta_{X_j}^2}{\sigma^2_{Y_j}}+\sum_{j \in \mathcal{V}}\frac{\beta_{X_j}\nu_j}{\sigma^2_{Y_j}}+\sum_{j \in \mathcal{V}}(\theta_0+\frac{r_j}{\beta_{X_j}})\cdot\frac{\beta_{X_j}u_j}{\sigma^2_{Y_j}}+\sum_{j \in \mathcal{V}}\frac{u_j\nu_j}{\sigma^2_{Y_j}}}{\sum_{j \in \mathcal{V}}\frac{\beta^2_{X_{j}}}{\sigma^2_{Y_j}}+2\sum_{j \in \mathcal{V}}\frac{\beta_{X_{j}}u_j}{\sigma^2_{Y_j}}+\sum_{j \in \mathcal{V}}\frac{u^2_{j}-\sigma^2_{X_{j,\mathtt{RB}}}}{\sigma^2_{Y_j}}-\sum_{j \in \mathcal{V}}\frac{\hat\sigma^2_{X_{j,\mathtt{RB}}}-\sigma^2_{X_{j,\mathtt{RB}}}}{\sigma^2_{Y_j}}}    
\end{align*}

If $\min\limits_{j \in \mathcal{S}_\lambda}|\beta_{X_j}|$ is is sufficiently separated from $0$ , with Condition \ref{Bound of Orlicz norm}, we can verify that there exists a $c>0$ such that the denominator is uniformly larger than $c\cdot \sum_{j \in \mathcal{V}}\frac{\beta_{X_j}^2}{\sigma^2_{Y_j}}$ with probability going to one for any possible $\mathcal{V}$. Similarly, we can also show that when $\max\limits_{j\in \mathcal{S}_\lambda}|\frac{r_j}{\beta_{X_j}}|$ is bounded away from infinity, there exists a $C>0$ such that the numerator is bounded away from $C \cdot \sum_{j \in \mathcal{V}}\frac{\beta_{X_j}^2}{\sigma^2_{Y_j}}$
with probability going to one for any possible $\mathcal{V}$. Therefore, we can verify that $|\hat\theta(\mathcal{V})|$ is uniformly bounded away from $\infty$ with probability going to one.

\begin{condition}[Separation of $r_j\neq0$ and 0]\label{Separation}
\begin{align*}
\min\limits_{j\in\mathcal{S}_\lambda,\ r_j\neq 0} r_j \gg \max \Big\{\frac{\big(s_\lambda\cdot\ln(s_\lambda)\big)^\frac{1}{2}}{n^{\frac{1}{2}}}, \frac{\kappa_{n}^{\frac{1}{2}}}{n^{\frac{1}{2}}}\Big\}.
\end{align*}
\end{condition}

In this condition, we require the pleiotropy effects $r_j \neq 0$ of invalid IVs to be bounded away from $0$. In this case, our invalid IV screening procedure will be able to distinguish invalid IVs and valid IVs. This condition is similar to the "beta-min" condition in the high-dimension linear regression setting. The only difference is that it is made in $r_j$ instead of the parameter of interest $\theta_0$. Without this condition, we will not be able to screen out some invalid IVs with $r_j$ close to 0 in the invalid IV screening procedure, and the perfect screening property will not hold. 

\begin{condition}[The order of $v_\lambda$]\label{Order of the number of valid IVs}
    The number of valid IVs has the same order of $s_\lambda$. In other words, $\frac{v_\lambda}{s_\lambda}$ is bounded away from zero. There exists a constant \( c_1 \) such that \( 0 < c_1 < 1 \). For all \( \mathcal{V} \subseteq \mathcal{S}_{\lambda} \) containing at least one nonzero element \( r_j \neq 0 \), whenever \( \rho(\mathcal{V}) = 1 \), it holds that \( v < c_1 \cdot v_\lambda \).
\end{condition}

The first part of this condition requires that the number of valid IVs in $\mathcal{S}_\lambda$ should be sufficiently large. To be specific, it should be of the same order as the total number of IVs in $\mathcal{S}_\lambda$. This condition can be further weakened by adjusting the penalized coefficient of GBIC. The second part of this condition imposes constraints on the cardinality of the cluster $\{j\in \mathcal{S}_\lambda|\frac{r_j}{\beta_{X_j}}=c\}$. It requires for any $c\neq 0$, the cardinality of the IV clusters $\{j\in \mathcal{S}_\lambda|\frac{r_j}{\beta_{X_j}}=c\}$ should be sufficiently separated from the total number of valid IVs $v_\lambda$ so that the algorithm will not fail to identify the true valid IV set $\mathcal{V}_\lambda$. If there exists a $c_0\neq 0$ such that the cardinality of the IV clusters $\{j\in \mathcal{S}_\lambda|\frac{r_j}{\beta_{X_j}}=c_0\}$ is very close to $v_\lambda$, then the algorithm might fail to distinguish $\{j\in \mathcal{S}_\lambda|\frac{r_j}{\beta_{X_j}}=c_0\}$ and the valid IV set $\mathcal{V}_\lambda$.
\begin{condition}[high dimension BIC]\label{high dimension BIC}
\begin{align*}
\frac{\kappa_{n}}{n\cdot  \min\limits_{j \in \mathcal{S}_{\lambda},\ r_j \neq 0}  r_j^2}\rightarrow 0
\quad  \text{ and } \quad
    \kappa_{n}\gg  ln(s_{\lambda}).
\end{align*}  
\end{condition}

\begin{cond}[high dimension BIC]\label{cond7*}
\begin{align*}
\frac{\kappa_{n}}{n\cdot  \min\limits_{j \in \mathcal{S}_{\lambda},\ r_j \neq 0}  r_j^2}\rightarrow 0
\quad  \text{ and } \quad
    \kappa_{n}\gg  s_\lambda\cdot\ln(s_{\lambda}).
\end{align*}  
\end{cond}

\subsection{Theoretical Results}\label{sup-sec3.3}
Define a collection of set
$$
\bm{\mathcal{V}}_\texttt{valid}=\left\{\mathcal{V}|\mathcal{V}\subseteq \mathcal{S}_\lambda, \ r_j=0 \text{ for all } j \in \mathcal{V} \text{, and } |\mathcal{V}|\geq c_1\cdot |\mathcal{V}_\lambda|\right\}.
$$
Here $c_1$ is the constant that we introduce in Condition \ref{Order of the number of valid IVs}.
\begin{thm}\label{thm s1}
Under Condition \ref{Bound of Orlicz norm}-\ref{high dimension BIC}, our IV screening procedure can consistently select the sets inside $\bm{\mathcal{V}}_\texttt{valid}$. Mathematically, this property is expressed as:
\[
\mathbb{P}(\hat{\mathcal{V}}_\lambda\in \bm{\mathcal{V}}_\texttt{valid})\rightarrow 1, \text{ as } s_\lambda \rightarrow \infty.
\] 
where $\hat{\mathcal{V}}_\lambda$ represents the set of IVs selected by the screening procedure. 
\end{thm}
\begin{thm}\label{thm s2}
Under Condition \ref{Bound of Orlicz norm}-\ref{Order of the number of valid IVs} and Condition \ref{cond7*}, our IV screening procedure can consistently select the complete set of valid IVs $\mathcal{V}_\lambda$. Mathematically, this property is expressed as:
\[
\mathbb{P}(\hat{\mathcal{V}}_\lambda= \mathcal{V}_\lambda)\rightarrow 1, \text{ as } s_\lambda \rightarrow \infty.
\] 
where $\hat{\mathcal{V}}_\lambda$ represents the set of IVs selected by the screening procedure. 
\end{thm}

\subsection{Proof of Theorem \ref{thm s1}}
For any $\mathcal{V} \subseteq \mathcal{{S}}_{\lambda}$, we denote a collection of sparse vectors 
\begin{align*}
    \bm{\mathcal{R}}_{\mbox{v}} =  \left\{ \bm{a} \in \mathbb{R}^{|\mathcal{S}_{\lambda}|\times 1} :  a_j = 0,\ \text{for }j\in \mathcal{V}, \ a_k \neq 0,\ \text{for } k \in \mathcal{V}^c  \right\}
\end{align*}
and a function 
\begin{align*}
h(\mathcal{V}, \theta)&=\min\limits_{ \bm{r}\in   \bm{\mathcal{R}}_{\mbox{v}}} \sum\limits_{j\in{\mathcal{S}}_{\lambda}}  \hat l\left( \theta , \bm{r};   \hat \beta_{Y_j},\sigma_{Y_j},\hat \beta_{X_{j,\mathtt{RB}}},\hat{\sigma}_{X_{j,\mathtt{RB}}}\right)\\
&=\sum\limits_{j\in \mathcal{V}} \frac{(\hat \beta_{Y_j}-\theta \cdot \hat \beta_{X_{j,\mathtt{RB}}})^2-\theta^2 \cdot \hat \sigma_{X_{j,\mathtt{RB}}}^2}{\sigma_{Y_j}^2}.
\end{align*}
Now we want to show $\mathbb{P}(\hat{\mathcal{V}}_{\lambda}\notin \bm{\mathcal{V}}_\texttt{valid}) \rightarrow 0$ as $s_{\lambda} \rightarrow \infty$ by utilizing the following inequality: 
\begin{equation}\label{eq1:selection-inconsistency-inequality}
\begin{aligned}
  \quad \mathbb{P}(\hat{\mathcal{V}}_{\lambda}\notin \bm{\mathcal{V}}_\texttt{valid})
  &= \mathbb{P}\Big(\min\limits_{v\in\mathbb{N}_{+}, v \leq s_\lambda}\big[\min\limits_{|\mathcal{V}|=v, \mathcal{V}\notin \bm{\mathcal{V}}_\texttt{valid}} \min\limits_{\theta \in \mathbb{R}} h(\mathcal{V},  \theta)-\kappa_{n}\cdot v\big] \leq\min\limits_{\theta \in \mathbb{R}} h(\mathcal{V}_{\lambda}, \theta)-\kappa_{n}\cdot v_{\lambda}\Big) \\&\leq \bigcup\limits_{\mathcal{V}\subseteq \mathcal{S}_{\lambda}, \mathcal{V}\notin \bm{\mathcal{V}}_\texttt{valid} }\mathbb{P}\Big( \min\limits_{\theta \in \mathbb{R}} h(\mathcal{V},  \theta)-\kappa_{n}\cdot |\mathcal{V}|\leq\min\limits_{\theta \in \mathbb{R}} h(\mathcal{V}_{\lambda}, \theta)-\kappa_{n}\cdot v_{\lambda}\Big) 
    \\&\leq \sum\limits_{v=1}^{s_\lambda} \tbinom{s_\lambda}{v}\max\limits_{|\mathcal{V}|=v, \mathcal{V}\notin \bm{\mathcal{V}}_\texttt{valid}}\mathbb{P}\Big(\min\limits_{\theta \in \mathbb{R}} h(\mathcal{V},  \theta)-\kappa_{n}\cdot v \leq  h(\mathcal{V}_{\lambda}, \theta_0)-\kappa_{n}\cdot v_{\lambda}\Big) \\&\leq \sum\limits_{v=1}^{s_\lambda} {s_\lambda}^{v}\max\limits_{|\mathcal{V}|=v, \mathcal{V}\notin \bm{\mathcal{V}}_\texttt{valid}}\mathbb{P}\Big(\min\limits_{\theta \in \mathbb{R}} h(\mathcal{V},  \theta)-\kappa_{n}\cdot v \leq  h(\mathcal{V}_{\lambda}, \theta_0)-\kappa_{n}\cdot v_{\lambda}\Big)  \\&\leq  \max\limits_{\mathcal{V}\notin \bm{\mathcal{V}}_\texttt{valid}}e^{(s_{\lambda}+1)\cdot ln(s_{\lambda})}\cdot\mathbb{P}\Big(\min\limits_{\theta \in \mathbb{R}} h(\mathcal{V},  \theta)-\kappa_{n}\cdot |\mathcal{V}| \leq  h(\mathcal{V}_{\lambda}, \theta_0)-\kappa_{n}\cdot v_{\lambda}\Big)\\&=e^{(s_{\lambda}+1)\cdot ln(s_{\lambda})}\cdot\mathbb{P}\Big(\min\limits_{\theta \in \mathbb{R}} h(\mathcal{V}^*,  \theta)-\kappa_{n}\cdot v^* \leq  h(\mathcal{V}_{\lambda}, \theta_0)-\kappa_{n}\cdot v_{\lambda}\Big).
\end{aligned}
\end{equation}
where $\mathcal{V}^*=\argmax\limits_{\mathcal{V}\notin \bm{\mathcal{V}}_\texttt{valid}}e^{(s_{\lambda}+1)\cdot ln(s_{\lambda})}\cdot\mathbb{P}\Big(\min\limits_{\theta \in \mathbb{R}} h(\mathcal{V},  \theta)-\kappa_{n}\cdot |\mathcal{V}| \leq  h(\mathcal{V}_{\lambda}, \theta_0)-\kappa_{n}\cdot v_{\lambda}\Big)$ and $v^*=|\mathcal{V}^*|$.

This is because the above inequality implies that as long as we show that 
\begin{align}
e^{(s_{\lambda}+1)\cdot ln(s_{\lambda})}\cdot\mathbb{P}\Big(\min\limits_{\theta \in \mathbb{R}} h(\mathcal{V}^*,  \theta)-\kappa_{n}\cdot v^* \leq  h(\mathcal{V}_{\lambda}, \theta_0)-\kappa_{n}\cdot v_{\lambda}\Big)\rightarrow 0.  
\end{align} 
then $\mathbb{P}(\hat{\mathcal{V}}_{\lambda}\notin \bm{\mathcal{V}}_\texttt{valid})\rightarrow 0$ holds. Here, we also note that the first equation in Eq \eqref{eq1:selection-inconsistency-inequality} follows from the definition of the optimization problem defined in Equation 2 in the manuscript, the second to the fifth inequalities in Eq \eqref{eq1:selection-inconsistency-inequality} hold following $\min\limits_{ \theta \in \mathbb{R}} h(\mathcal{V}_{\lambda},\theta)\leq h(\mathcal{V}_{\lambda}, \theta_0)$, $\tbinom{s_\lambda}{v}\leq s_\lambda^v $ and some basic calculations. 

To prove Equation (2) goes to 0, we discuss $\mathcal{V}^* \notin \bm{\mathcal{V}}_\texttt{valid}$ in three different cases.
\begin{itemize}
    \item $|\mathcal{V}^*|=v^* \geq v_{\lambda}$ and $\mathcal{V}^* \neq \mathcal{V}_{\lambda}$,
    \item $c_1\cdot v_\lambda \leq |\mathcal{V}^*| < v_{\lambda}$ but $\rho(\mathcal{V}^*)< 1$,
    \item $|\mathcal{V}^*|=v^*< c_1 \cdot v_{\lambda}$.
\end{itemize}
In each case, we show $e^{(s_{\lambda}+1)\cdot ln(s_{\lambda})}\cdot\mathbb{P}\Big(\min\limits_{\theta \in \mathbb{R}} h(\mathcal{V}^*,  \theta)-\kappa_{n}\cdot v^* \leq  h(\mathcal{V}_{\lambda}, \theta_0)-\kappa_{n}\cdot v_{\lambda}\Big)\rightarrow 0 $ 
and therefore (2) holds uniformly for all $\mathcal{V}^*\subseteq \mathcal{S}_\lambda$ and $\mathcal{V}^* \neq \mathcal{V}_{\lambda}$.

\subsubsection{Case 1: $|\mathcal{V}^*|=v^* \geq v_{\lambda}$ and $\mathcal{V}^* \neq \mathcal{V}_{\lambda}$}
To show the above results, we analyze the asymptotic properties of $ h(\mathcal{V}_{\lambda}, \theta_0)$, $ \min\limits_{\theta \in \mathbb{R}} h(\mathcal{V}^*,  \theta)$ and $\kappa_{n}$.
We start with $h(\mathcal{V}_{\lambda}, \theta_0)$ and decompose it below following our notation defined in Section \ref{notation}.
\begin{align}
   h(\mathcal{V}_{\lambda}, \theta_0)&\nonumber=\sum\limits_{j\in \mathcal{V}_{\lambda}} \frac{(\hat \beta_{Y_j}-\theta_0 \cdot \beta_{X_j})^2  }{\sigma_{Y_j}^2}+\theta_0^2 \cdot \sum\limits_{j\in \mathcal{V}_{\lambda}} \frac{ ( \hat \beta_{X_{j,\mathtt{RB}}}-\beta_{X_j})^2- \hat\sigma_{X_{j,\mathtt{RB}}}^2}{\sigma_{Y_j}^2}\\
   &\nonumber-2\theta_0\cdot \sum\limits_{j\in  \mathcal{V}_{\lambda}} \frac{(\hat \beta_{Y_j}-\theta_0 \cdot \beta_{X_j})( \hat \beta_{X_{j,\mathtt{RB}}}-\beta_{X_j})}{\sigma_{Y_j}^2} \\
   & \label{eq:expansion-of-h}= \sum\limits_{j\in \mathcal{V}_{\lambda}} \frac{\nu_j^2  }{\sigma_{Y_j}^2}+\theta_0^2 \cdot \sum\limits_{j\in \mathcal{V}_{\lambda}} \frac{  u_j^2- \hat\sigma_{X_{j,\mathtt{RB}}}^2}{\sigma_{Y_j}^2}-2\theta_0\cdot \sum\limits_{j\in  \mathcal{V}_{\lambda}} \frac{\nu_ju_j}{\sigma_{Y_j}^2}.
\end{align}
Next, we study the asymptotic property of $ \min\limits_{\theta \in \mathbb{R}} h(\mathcal{V}^*,  \theta)$.
We denote $\hat \theta(\mathcal{V}^*)=\arg\min\limits_{\theta\in \mathbb{R}}h(\mathcal{V}^*,\theta)$ and decompose $h(\mathcal{V}^*,\hat{\theta}(\mathcal{V}^*))$ in a similar way as $h(\mathcal{V}_{\lambda},\theta_0)$, following our notation defined in Section \ref{notation}.
\begin{align*}
    h(\mathcal{V}^*, \hat {\theta}(\mathcal{V}^*))&=\nonumber\sum\limits_{j\in \mathcal{V}^*} \frac{(\hat \beta_{Y_j}-\hat{\theta}(\mathcal{V}^*) \cdot \beta_{X_j})^2  }{\sigma_{Y_j}^2}+\hat\theta(\mathcal{V}^*)^2 \sum\limits_{j\in \mathcal{V}^*} \frac{( \hat \beta_{X_{j,\mathtt{RB}}}-\beta_{X_j})^2- \hat\sigma_{X_{j,\mathtt{RB}}}^2}{\sigma_{Y_j}^2}\\
    &+2\hat{\theta}(\mathcal{V}^*) \sum\limits_{j\in  \mathcal{V}^*} \frac{(\hat{\theta}(\mathcal{V}^*)\cdot \beta_{X_j}-\theta_0\cdot \beta_{X_j}-r_j)\cdot( \hat \beta_{X_{j,\mathtt{RB}}}-\beta_{X_j})}{\sigma_{Y_j}^2}\\
    &-2\hat{\theta}(\mathcal{V}^*)  \sum\limits_{j\in \mathcal{V}^*} \frac{(\hat \beta_{Y_j}-\theta_0 \cdot \beta_{X_j}-r_j)( \hat \beta_{X_{j,\mathtt{RB}}}-\beta_{X_j})}{\sigma_{Y_j}^2}\\
    &=\sum\limits_{j\in \mathcal{V}^*} \frac{(\theta_0 \cdot \beta_{X_j}+r_j+\nu_j -\hat{\theta}(\mathcal{V}^*) \cdot \beta_{X_j})^2  }{\sigma_{Y_j}^2}+\hat\theta(\mathcal{V}^*)^2  \sum\limits_{j\in \mathcal{V}^*} \frac{u_j^2- \hat\sigma_{X_{j,\mathtt{RB}}}^2}{\sigma_{Y_j}^2}\\&+2\hat{\theta}(\mathcal{V}^*) \sum\limits_{j\in  \mathcal{V}^*} \frac{(\hat{\theta}(\mathcal{V}^*)\cdot \beta_{X_j}-\theta_0\cdot \beta_{X_j}-r_j)\cdot u_j}{\sigma_{Y_j}^2}-2\hat{\theta}(\mathcal{V}^*)  \sum\limits_{j\in \mathcal{V}^*} \frac{\nu_j u_j}{\sigma_{Y_j}^2}\\
    &=\sum\limits_{j\in \mathcal{V}^*} \frac{(\theta_0 \cdot \beta_{X_j}+r_j-\hat{\theta}(\mathcal{V}^*) \cdot \beta_{X_j})^2  }{\sigma_{Y_j}^2}+\sum\limits_{j\in \mathcal{V}^*} \frac{\nu_j ^2}{\sigma_{Y_j}^2}+2\sum\limits_{j\in \mathcal{V}^*} \frac{(\theta_0 \cdot \beta_{X_j}+r_j -\hat{\theta}(\mathcal{V}^*)\cdot \beta_{X_j}) \cdot \nu_j }{\sigma_{Y_j}^2}\\&+\hat\theta(\mathcal{V}^*)^2  \sum\limits_{j\in \mathcal{V}^*} \frac{u_j^2- \hat\sigma_{X_{j,\mathtt{RB}}}^2}{\sigma_{Y_j}^2}+2\hat{\theta}(\mathcal{V}^*) \sum\limits_{j\in  \mathcal{V}^*} \frac{(\hat{\theta}(\mathcal{V}^*)\cdot \beta_{X_j}-\theta_0\cdot \beta_{X_j}-r_j)\cdot u_j}{\sigma_{Y_j}^2}-2\hat{\theta}(\mathcal{V}^*)  \sum\limits_{j\in \mathcal{V}^*} \frac{\nu_j u_j}{\sigma_{Y_j}^2}.
\end{align*}
With these decomposition, the probability in Equation (2) can be rewritten as 
\begin{align*}
 &\quad\mathbb{P}\Big(\min\limits_{\theta \in \mathbb{R}} h(\mathcal{V}^*,  \theta)-\kappa_{n}\cdot v^* \leq  h(\mathcal{V}_{\lambda}, \theta_0)-\kappa_{n}\cdot v_{\lambda}\Big)
 \\&=\mathbb{P}\Big(\sum\limits_{j\in \mathcal{V}^*} \frac{(\theta_0 \cdot \beta_{X_j}+r_j-\hat{\theta}(\mathcal{V}^*) \cdot \beta_{X_j})^2  }{\sigma_{Y_j}^2}+\kappa_{n}\cdot 
(v_{\lambda}- v^*)
 \leq -\sum\limits_{j\in \mathcal{V}^*} \frac{\nu_j ^2}{\sigma_{Y_j}^2}-2\sum\limits_{j\in \mathcal{V}^*} \frac{(\theta_0 \cdot \beta_{X_j}+r_j -\hat{\theta}(\mathcal{V}^*)\cdot \beta_{X_j}) \cdot \nu_j }{\sigma_{Y_j}^2}
 \\&\quad-\hat\theta(\mathcal{V}^*)^2  \sum\limits_{j\in \mathcal{V}^*} \frac{u_j^2- \sigma_{X_{j,\mathtt{RB}}}^2}{\sigma_{Y_j}^2}+\hat\theta(\mathcal{V}^*)^2  \sum\limits_{j\in \mathcal{V}^*} \frac{\hat\sigma_{X_{j,\mathtt{RB}}}^2- \sigma_{X_{j,\mathtt{RB}}}^2}{\sigma_{Y_j}^2}-2\hat{\theta}(\mathcal{V}^*) \sum\limits_{j\in  \mathcal{V}^*} \frac{(\hat{\theta}(\mathcal{V}^*)\cdot \beta_{X_j}-\theta_0\cdot \beta_{X_j}-r_j)\cdot u_j}{\sigma_{Y_j}^2}
 \\&\quad+2\hat{\theta}(\mathcal{V}^*)  \sum\limits_{j\in \mathcal{V}^*} \frac{\nu_j u_j}{\sigma_{Y_j}^2}+\sum\limits_{j\in \mathcal{V}_{\lambda}} \frac{\nu_j^2  }{\sigma_{Y_j}^2}+\theta_0^2   \sum\limits_{j\in \mathcal{V}_{\lambda}} \frac{  u_j^2- \sigma_{X_{j,\mathtt{RB}}}^2}{\sigma_{Y_j}^2}-\theta_0^2   \sum\limits_{j\in \mathcal{V}_{\lambda}} \frac{  \hat\sigma_{X_{j,\mathtt{RB}}}^2- \sigma_{X_{j,\mathtt{RB}}}^2}{\sigma_{Y_j}^2} -2\theta_0  \sum\limits_{j\in  \mathcal{V}_{\lambda}} \frac{\nu_ju_j}{\sigma_{Y_j}^2}\Big)  .
 \end{align*}
When $|\mathcal{V}^*|=v^* \geq v_{\lambda}$ and $\mathcal{V}^* \neq \mathcal{V}_{\lambda}$, we know that there is at least one $r_j \in \mathcal{V}^*$ such that $r_j \neq 0$. So $\rho(\mathcal{V}^*)$ is well-defined and by Condition \ref{Plurality and no perfect correlation} we have $
\rho(\mathcal{V}^*)\leq c_0$.
By some calculations,  we can see that 
\begin{align*}
    \sum\limits_{j\in \mathcal{V}^*} \frac{(\theta_0 \cdot \beta_{X_j}+r_j-\hat{\theta}(\mathcal{V}^*) \cdot \beta_{X_j})^2  }{\sigma_{Y_j}^2}\geq \min\limits_{\theta \in \mathbb{R}}\sum\limits_{j\in \mathcal{V}^*} \frac{(\theta_0 \cdot \beta_{X_j}+r_j-\theta \cdot \beta_{X_j})^2 }{\sigma_{Y_j}^2} =\sum\limits_{j\in \mathcal{V}^*}\frac{ r_j^2}{\sigma_{Y_j}^2}\cdot(1-\rho(\mathcal{V}^*)).
\end{align*}
with probability 1.
Therefore,
\begin{align*}
        \sum\limits_{j\in \mathcal{V}^*} \frac{(\theta_0 \cdot \beta_{X_j}+r_j-\hat{\theta}(\mathcal{V}^*) \cdot \beta_{X_j})^2  }{\sigma_{Y_j}^2}\geq \sum\limits_{j\in \mathcal{V}^*}\frac{ r_j^2}{\sigma_{Y_j}^2}\cdot(1-c_0).
\end{align*}
with probability 1.

Then under Condition \ref{Boundedness}, there exists a $C_0>0$ such that 

{\small
\begin{align*}
 &\quad\mathbb{P}\Big(\min\limits_{\theta \in \mathbb{R}} h(\mathcal{V}^*,  \theta)-\kappa_{n}\cdot v^* \leq  h(\mathcal{V}_{\lambda}, \theta_0)-\kappa_{n}\cdot v_{\lambda}\Big)
 \\&\leq \mathbb{P}\Big(        \sum\limits_{j\in \mathcal{V}^*} \frac{(\theta_0 \cdot \beta_{X_j}+r_j-\hat{\theta}(\mathcal{V}^*) \cdot \beta_{X_j})^2  }{\sigma_{Y_j}^2}-\kappa_{n}\cdot (v^*-v_\lambda)
 \leq -\sum\limits_{j\in \mathcal{V}^*} \frac{\nu_j ^2-\sigma_{Y_j}^2}{\sigma_{Y_j}^2}-\hat\theta(\mathcal{V}^*)^2  \sum\limits_{j\in \mathcal{V}^*} \frac{u_j^2- \sigma_{X_{j,\mathtt{RB}}}^2}{\sigma_{Y_j}^2}
 \\&\quad +\hat\theta(\mathcal{V}^*)^2  \sum\limits_{j\in \mathcal{V}^*} \frac{\hat\sigma_{X_{j,\mathtt{RB}}}^2- \sigma_{X_{j,\mathtt{RB}}}^2}{\sigma_{Y_j}^2}-2 \sum\limits_{j\in  \mathcal{V}^*} \frac{(\hat{\theta}(\mathcal{V}^*)\cdot \beta_{X_j}-\theta_0\cdot \beta_{X_j}-r_j)\cdot \nu_j}{\sigma_{Y_j}^2}-2\hat{\theta}(\mathcal{V}^*) \sum\limits_{j\in  \mathcal{V}^*} \frac{(\hat{\theta}(\mathcal{V}^*)\cdot \beta_{X_j}-\theta_0\cdot \beta_{X_j}-r_j)\cdot u_j}{\sigma_{Y_j}^2}
 \\&\quad+2\hat{\theta}(\mathcal{V}^*)  \sum\limits_{j\in \mathcal{V}^*} \frac{\nu_j u_j}{\sigma_{Y_j}^2}+\sum\limits_{j\in \mathcal{V}_{\lambda}} \frac{\nu_j^2-\sigma_{Y_j}^2  }{\sigma_{Y_j}^2}+\theta_0^2   \sum\limits_{j\in \mathcal{V}_{\lambda}} \frac{  u_j^2- \sigma_{X_{j,\mathtt{RB}}}^2}{\sigma_{Y_j}^2}-\theta_0^2   \sum\limits_{j\in \mathcal{V}_{\lambda}} \frac{  \hat\sigma_{X_{j,\mathtt{RB}}}^2- \sigma_{X_{j,\mathtt{RB}}}^2}{\sigma_{Y_j}^2} -2\theta_0  \sum\limits_{j\in  \mathcal{V}_{\lambda}} \frac{\nu_ju_j}{\sigma_{Y_j}^2}\Big) 
  \\&\leq \mathbb{P}\Big(\sum\limits_{j\in \mathcal{V}^*} \frac{(\theta_0 \cdot \beta_{X_j}+r_j-\hat{\theta}(\mathcal{V}^*) \cdot \beta_{X_j})^2  }{\sigma_{Y_j}^2}-\kappa_{n}\cdot (v^*-v_\lambda)
 \leq \big|\sum\limits_{j\in \mathcal{V}^*} \frac{\nu_j ^2-\sigma_{Y_j}^2}{\sigma_{Y_j}^2}\big|+ \hat\theta(\mathcal{V}^*)^2 \big|\sum\limits_{j\in \mathcal{V}^*} \frac{u_j^2- \sigma_{X_{j,\mathtt{RB}}}^2}{\sigma_{Y_j}^2}\big|
 \\&\quad +2\big|\hat{\theta}(\mathcal{V}^*) \sum\limits_{j\in  \mathcal{V}^*} \frac{(\hat{\theta}(\mathcal{V}^*)\cdot \beta_{X_j}-\theta_0\cdot \beta_{X_j}-r_j)\cdot u_j}{\sigma_{Y_j}^2}\big|+2\big| \sum\limits_{j\in  \mathcal{V}^*} \frac{(\hat{\theta}(\mathcal{V}^*)\cdot \beta_{X_j}-\theta_0\cdot \beta_{X_j}-r_j)\cdot \nu_j}{\sigma_{Y_j}^2}\big|+2\big|\hat{\theta}(\mathcal{V}^*) \sum\limits_{j\in \mathcal{V}^*} \frac{\nu_j u_j}{\sigma_{Y_j}^2} \big|
 \\&\quad+\hat\theta(\mathcal{V}^*)^2 \big|\sum\limits_{j\in \mathcal{V}^*} \frac{\hat\sigma_{X_{j,\mathtt{RB}}}^2- \sigma_{X_{j,\mathtt{RB}}}^2}{\sigma_{Y_j}^2}\big|+\big|\sum\limits_{j\in \mathcal{V}_{\lambda}} \frac{\nu_j^2-\sigma_{Y_j}^2  }{\sigma_{Y_j}^2}\big|+\theta_0^2  \big|\sum\limits_{j\in \mathcal{V}_{\lambda}} \frac{  u_j^2- \sigma_{X_{j,\mathtt{RB}}}^2}{\sigma_{Y_j}^2}\big|+\theta_0^2  \big|\sum\limits_{j\in \mathcal{V}_{\lambda}} \frac{  \hat\sigma_{X_{j,\mathtt{RB}}}^2- \sigma_{X_{j,\mathtt{RB}}}^2}{\sigma_{Y_j}^2}\big|+2\theta_0 \big|\sum\limits_{j\in  \mathcal{V}_{\lambda}} \frac{\nu_ju_j}{\sigma_{Y_j}^2}\big|\Big) 
   \\&\leq \mathbb{P}\Big(\sum\limits_{j\in \mathcal{V}^*} \frac{(\theta_0 \cdot \beta_{X_j}+r_j-\hat{\theta}(\mathcal{V}^*) \cdot \beta_{X_j})^2  }{\sigma_{Y_j}^2}-\kappa_{n}\cdot (v^*-v_\lambda)
 \leq \big|\sum\limits_{j\in \mathcal{V}^*} \frac{\nu_j ^2-\sigma_{Y_j}^2}{\sigma_{Y_j}^2}\big|+C_0^2 \big|\sum\limits_{j\in \mathcal{V}^*} \frac{u_j^2- \sigma_{X_{j,\mathtt{RB}}}^2}{\sigma_{Y_j}^2}\big|
 \\&\quad+C_0^2 \big|\sum\limits_{j\in \mathcal{V}^*} \frac{\hat\sigma_{X_{j,\mathtt{RB}}}^2- \sigma_{X_{j,\mathtt{RB}}}^2}{\sigma_{Y_j}^2}\big|+2C_0\big| \sum\limits_{j\in  \mathcal{V}^*} \frac{(\hat{\theta}(\mathcal{V}^*)\cdot \beta_{X_j}-\theta_0\cdot \beta_{X_j}-r_j)\cdot u_j}{\sigma_{Y_j}^2}\big|+2\big| \sum\limits_{j\in  \mathcal{V}^*} \frac{(\hat{\theta}(\mathcal{V}^*)\cdot \beta_{X_j}-\theta_0\cdot \beta_{X_j}-r_j)\cdot \nu_j}{\sigma_{Y_j}^2}\big|\\&\quad+2C_0\big| \sum\limits_{j\in \mathcal{V}^*} \frac{\nu_j u_j}{\sigma_{Y_j}^2} \big|+\big|\sum\limits_{j\in \mathcal{V}_{\lambda}} \frac{\nu_j^2-\sigma_{Y_j}^2  }{\sigma_{Y_j}^2}\big|+\theta_0^2  \big|\sum\limits_{j\in \mathcal{V}_{\lambda}} \frac{  u_j^2- \sigma_{X_{j,\mathtt{RB}}}^2}{\sigma_{Y_j}^2}\big|+\theta_0^2  \big|\sum\limits_{j\in \mathcal{V}_{\lambda}} \frac{  \hat\sigma_{X_{j,\mathtt{RB}}}^2- \sigma_{X_{j,\mathtt{RB}}}^2}{\sigma_{Y_j}^2}\big|+2\theta_0 \big|\sum\limits_{j\in  \mathcal{V}_{\lambda}} \frac{\nu_ju_j}{\sigma_{Y_j}^2}\big|\Big).
\end{align*}
Denote the event
\begin{align*}
& \quad \sum\limits_{j\in \mathcal{V}^*} \frac{(\theta_0 \cdot \beta_{X_j}+r_j-\hat{\theta}(\mathcal{V}^*) \cdot \beta_{X_j})^2  }{\sigma_{Y_j}^2}-\kappa_{n}\cdot (v^*-v_\lambda)
 \leq \big|\sum\limits_{j\in \mathcal{V}^*} \frac{\nu_j ^2-\sigma_{Y_j}^2}{\sigma_{Y_j}^2}\big|+C_0^2 \big|\sum\limits_{j\in \mathcal{V}^*} \frac{u_j^2- \sigma_{X_{j,\mathtt{RB}}}^2}{\sigma_{Y_j}^2}\big|
 \\&\quad+C_0^2 \big|\sum\limits_{j\in \mathcal{V}^*} \frac{\hat\sigma_{X_{j,\mathtt{RB}}}^2- \sigma_{X_{j,\mathtt{RB}}}^2}{\sigma_{Y_j}^2}\big|+2C_0\big| \sum\limits_{j\in  \mathcal{V}^*} \frac{(\hat{\theta}(\mathcal{V}^*)\cdot \beta_{X_j}-\theta_0\cdot \beta_{X_j}-r_j)\cdot u_j}{\sigma_{Y_j}^2}\big|+2\big| \sum\limits_{j\in  \mathcal{V}^*} \frac{(\hat{\theta}(\mathcal{V}^*)\cdot \beta_{X_j}-\theta_0\cdot \beta_{X_j}-r_j)\cdot \nu_j}{\sigma_{Y_j}^2}\big|\\&\quad+2C_0\big| \sum\limits_{j\in \mathcal{V}^*} \frac{\nu_j u_j}{\sigma_{Y_j}^2} \big|+\big|\sum\limits_{j\in \mathcal{V}_{\lambda}} \frac{\nu_j^2-\sigma_{Y_j}^2  }{\sigma_{Y_j}^2}\big|+\theta_0^2  \big|\sum\limits_{j\in \mathcal{V}_{\lambda}} \frac{  u_j^2- \sigma_{X_{j,\mathtt{RB}}}^2}{\sigma_{Y_j}^2}\big|+\theta_0^2  \big|\sum\limits_{j\in \mathcal{V}_{\lambda}} \frac{  \hat\sigma_{X_{j,\mathtt{RB}}}^2- \sigma_{X_{j,\mathtt{RB}}}^2}{\sigma_{Y_j}^2}\big|+2\theta_0 \big|\sum\limits_{j\in  \mathcal{V}_{\lambda}} \frac{\nu_ju_j}{\sigma_{Y_j}^2}|
\end{align*}
as $\mathcal{C}(\mathcal{V}^*)$ and 
\begin{align*}
\delta(\mathcal{V}^*)=\sum\limits_{j\in \mathcal{V}^*} \frac{(\theta_0 \cdot \beta_{X_j}+r_j-\hat{\theta}(\mathcal{V}^*) \cdot \beta_{X_j})^2  }{\sigma_{Y_j}^2}-\kappa_{n}\cdot (v^*-v_\lambda),
\end{align*}
we have
\begin{align*}
\mathcal{C}(\mathcal{V}^*) \subseteq       &\ \left\{C_0^2\cdot
\big|\sum\limits_{j\in \mathcal{V}^*} \frac{u_j^2- \sigma_{X_{j,\mathtt{RB}}}^2}{\sigma_{Y_j}^2}\big|\geq \frac{\delta(\mathcal{V}^*)}{10}\right\}\bigcup \left\{\theta_0^2\cdot\big|\sum\limits_{j\in \mathcal{V}_{\lambda}} \frac{  u_j^2- \sigma_{X_{j,\mathtt{RB}}}^2}{\sigma_{Y_j}^2}\big|\geq \frac{\delta(\mathcal{V}^*)}{10}\right\}\\&\bigcup\left\{\big|\sum\limits_{j\in \mathcal{V}^*} \frac{\nu_j ^2-\sigma_{Y_j}^2}{\sigma_{Y_j}^2}\big|\geq \frac{\delta(\mathcal{V}^*)}{10}\right\}\bigcup\left\{\big|\sum\limits_{j\in \mathcal{V}_\lambda} \frac{\nu_j ^2-\sigma_{Y_j}^2}{\sigma_{Y_j}^2}\big|\geq \frac{\delta(\mathcal{V}^*)}{10}\right\} \bigcup\left\{\theta_0^2\cdot\big|\sum\limits_{j\in \mathcal{V}_{\lambda}} \frac{  \hat\sigma_{X_{j,\mathtt{RB}}}^2- \sigma_{X_{j,\mathtt{RB}}}^2}{\sigma_{Y_j}^2}\big|\geq \frac{\delta(\mathcal{V}^*)}{10}\right\}\\& \bigcup
 \left\{C_0^2\cdot\big|\sum\limits_{j\in \mathcal{V}^*} \frac{\hat\sigma_{X_{j,\mathtt{RB}}}^2- \sigma_{X_{j,\mathtt{RB}}}^2}{\sigma_{Y_j}^2}\big|\geq 
\frac{\delta(\mathcal{V}^*)}{10}\right\}\bigcup\left\{2C_0\cdot\big|\sum\limits_{j\in  \mathcal{V}_{\lambda}} \frac{\nu_ju_j}{\sigma_{Y_j}^2}\big|\geq \frac{\delta(\mathcal{V}^*)}{10}\right\}
\bigcup\left\{2\theta_0\cdot\big| \sum\limits_{j\in \mathcal{V}^*} \frac{\nu_j u_j}{\sigma_{Y_j}^2} \big|\geq \frac{\delta(\mathcal{V}^*)}{10}\right\}\\&\bigcup\left\{2\big|\sum\limits_{j\in  \mathcal{V}^*} \frac{(\hat{\theta}(\mathcal{V}^*)\cdot \beta_{X_j}-\theta_0\cdot \beta_{X_j}-r_j)\cdot \nu_j}{\sigma_{Y_j}^2}\big|\geq \frac{\delta(\mathcal{V}^*)}{10}\right\}\\&\bigcup\left\{2C_0\cdot\big|\sum\limits_{j\in  \mathcal{V}^*} \frac{(\hat{\theta}(\mathcal{V}^*)\cdot \beta_{X_j}-\theta_0\cdot \beta_{X_j}-r_j)\cdot u_j}{\sigma_{Y_j}^2}\big|\geq \frac{\delta(\mathcal{V}^*)}{10}\right\}.
\end{align*}
}

When $|\mathcal{V}^*|=v^*\geq v_{\lambda}$, we know that the number of $r_j$ that is not equal to zero for $j \in \mathcal{V}^*$ is at least $v^*-v_{\lambda}$. Then if $\frac{\kappa_{n}}{  \min\limits_{j \in \mathcal{S}_{\lambda},\ r_j \neq 0} \frac{ r_j^2}{\sigma_{Y_j}^2}}\rightarrow 0$ (Condition \ref{high dimension BIC}), we have $\frac{\kappa_{n}\cdot (v^*-v_{\lambda})}{ \sum\limits_{j\in \mathcal{V}^*}\frac{ r_j^2}{\sigma_{Y_j}^2}}\rightarrow 0$. So there exists a $c>0$ such that 
$$
\delta(\mathcal{V}^*)=\sum\limits_{j\in \mathcal{V}^*} \frac{(\theta_0 \cdot \beta_{X_j}+r_j-\hat{\theta}(\mathcal{V}^*) \cdot \beta_{X_j})^2  }{\sigma_{Y_j}^2}-\kappa_{n}\cdot (v^*-v_\lambda)\geq \sum\limits_{j\in \mathcal{V}^*}\frac{ r_j^2}{\sigma_{Y_j}^2}\cdot c.
$$
uniformly holds for all $|\mathcal{V}^*|=v^* \geq v_{\lambda}$ and $\mathcal{V}^* \neq \mathcal{V}_{\lambda}$.
Then $\mathbb{P}( \mathcal{C}(\mathcal{V}^*))$ is bounded by
\begin{align*}
&\ \mathbb{P}\Big(
C_0^2 \cdot\big|\sum\limits_{j\in \mathcal{V}^*} \frac{u_j^2- \sigma_{X_{j,\mathtt{RB}}}^2}{\sigma_{Y_j}^2}\big|\geq \frac{c}{10}\sum\limits_{j\in \mathcal{V}^*}\frac{ r_j^2}{\sigma_{Y_j}^2}\Big)+\mathbb{P}\Big(\theta_0^2 \cdot\big|\sum\limits_{j\in \mathcal{V}_{\lambda}} \frac{  u_j^2- \sigma_{X_{j,\mathtt{RB}}}^2}{\sigma_{Y_j}^2}\big|\geq \frac{c}{10}\sum\limits_{j\in \mathcal{V}^*}\frac{ r_j^2}{\sigma_{Y_j}^2}\Big)
\\& +\mathbb{P}\Big(C_0^2 \cdot\big|\sum\limits_{j\in \mathcal{V}^*} \frac{\hat\sigma_{X_{j,\mathtt{RB}}}^2- \sigma_{X_{j,\mathtt{RB}}}^2}{\sigma_{Y_j}^2}\big|\geq 
\frac{c}{10}\sum\limits_{j\in \mathcal{V}^*}\frac{ r_j^2}{\sigma_{Y_j}^2}\Big)+\mathbb{P}\Big(\theta_0^2 \cdot\big|\sum\limits_{j\in \mathcal{V}_{\lambda}} \frac{  \hat\sigma_{X_{j,\mathtt{RB}}}^2- \sigma_{X_{j,\mathtt{RB}}}^2}{\sigma_{Y_j}^2}\big|\geq \frac{c}{10}\sum\limits_{j\in \mathcal{V}^*}\frac{ r_j^2}{\sigma_{Y_j}^2}\Big)\\&+\mathbb{P}\Big(\big|\sum\limits_{j\in \mathcal{V}^*} \frac{\nu_j ^2-\sigma_{Y_j}^2}{\sigma_{Y_j}^2}\big|\geq \frac{c}{10}\sum\limits_{j\in \mathcal{V}^*}\frac{ r_j^2}{\sigma_{Y_j}^2}\Big)+\mathbb{P}\Big(\big|\sum\limits_{j\in \mathcal{V}_\lambda} \frac{\nu_j ^2-\sigma_{Y_j}^2}{\sigma_{Y_j}^2}\big|\geq \frac{c}{10}\sum\limits_{j\in \mathcal{V}^*}\frac{ r_j^2}{\sigma_{Y_j}^2}\Big) \\& +\mathbb{P}\Big(2 \theta_0\cdot\big|\sum\limits_{j\in  \mathcal{V}_{\lambda}} \frac{\nu_ju_j}{\sigma_{Y_j}^2}\big|\geq \frac{c}{10}\sum\limits_{j\in \mathcal{V}^*}\frac{ r_j^2}{\sigma_{Y_j}^2}\Big)
+\mathbb{P}\Big(2 C_0\cdot\big| \sum\limits_{j\in \mathcal{V}^*} \frac{\nu_j u_j}{\sigma_{Y_j}^2} \big|\geq \frac{c}{10}\sum\limits_{j\in \mathcal{V}^*}\frac{ r_j^2}{\sigma_{Y_j}^2}\Big)\\& +\mathbb{P}\Big(2\big|\sum\limits_{j\in  \mathcal{V}^*} \frac{(\hat{\theta}(\mathcal{V}^*)\cdot \beta_{X_j}-\theta_0\cdot \beta_{X_j}-r_j)\cdot \nu_j}{\sigma_{Y_j}^2}\big|\geq \frac{1}{10}\big(\sum\limits_{j\in \mathcal{V}^*} \frac{(\theta_0 \cdot \beta_{X_j}+r_j-\hat{\theta}(\mathcal{V}^*) \cdot \beta_{X_j})^2  }{\sigma_{Y_j}^2}-\kappa_{n}\cdot (v^*-v_\lambda)\big)\Big)\\&+\mathbb{P}\Big(2C_0\cdot\big|\sum\limits_{j\in  \mathcal{V}^*} \frac{(\hat{\theta}(\mathcal{V}^*)\cdot \beta_{X_j}-\theta_0\cdot \beta_{X_j}-r_j)\cdot u_j}{\sigma_{Y_j}^2}\big|\geq \frac{1}{10}\big(\sum\limits_{j\in \mathcal{V}^*} \frac{(\theta_0 \cdot \beta_{X_j}+r_j-\hat{\theta}(\mathcal{V}^*) \cdot \beta_{X_j})^2  }{\sigma_{Y_j}^2}-\kappa_{n}\cdot (v^*-v_\lambda)\big)\Big).
\end{align*}
Using Lemma \ref{lem1}, we know that there exists a $c'>0$ such that the first eight terms are bounded by
$
 2\cdot e^{-c' \cdot \min\big\{\frac{ {v^*}^2\cdot r_{\lambda}^2(\mathcal{V}^*)}{v_\lambda},\ v^*\cdot r_{\lambda}^2(\mathcal{V}^*), \ v^* \cdot r_{\lambda}(\mathcal{V}^*)\big\}}.
$
We also have 
\begin{align*}   
&\big|\sum\limits_{j\in  \mathcal{V}^*} \frac{(\hat{\theta}(\mathcal{V}^*)\cdot \beta_{X_j}-\theta_0\cdot \beta_{X_j}-r_j)\cdot u_j}{\sigma_{Y_j}^2}\big|\leq \sqrt{\sum\limits_{j\in  \mathcal{V}^*} \frac{(\hat{\theta}(\mathcal{V}^*)\cdot \beta_{X_j}-\theta_0\cdot \beta_{X_j}-r_j)^2}{\sigma_{Y_j}^2}}\sqrt{\sum\limits_{j\in  \mathcal{V}^*} \frac{u_j^2}{\sigma_{Y_j}^2}} \text{ and }
\\&\big|\sum\limits_{j\in  \mathcal{V}^*} \frac{(\hat{\theta}(\mathcal{V}^*)\cdot \beta_{X_j}-\theta_0\cdot \beta_{X_j}-r_j)\cdot \nu_j}{\sigma_{Y_j}^2}\big|\leq \sqrt{\sum\limits_{j\in  \mathcal{V}^*} \frac{(\hat{\theta}(\mathcal{V}^*)\cdot \beta_{X_j}-\theta_0\cdot \beta_{X_j}-r_j)^2}{\sigma_{Y_j}^2}}\sqrt{\sum\limits_{j\in  \mathcal{V}^*} \frac{\nu_j^2}{\sigma_{Y_j}^2}}.
\end{align*}
That means there exists a $c''>0$ such that $\mathbb{P}\Big(2\big|\sum\limits_{j\in  \mathcal{V}^*} \frac{(\hat{\theta}(\mathcal{V}^*)\cdot \beta_{X_j}-\theta_0\cdot \beta_{X_j}-r_j)\cdot \nu_j}{\sigma_{Y_j}^2}\big|\geq \frac{\delta(\mathcal{V}^*)}{10}\Big)$ is further bounded by $\mathbb{P}\Big(\sum\limits_{j\in  \mathcal{V}^*}\frac{\nu_j^2}{\sigma_{Y_j}^2}\geq c''\sum\limits_{j\in \mathcal{V}^*} \frac{r_j^2}{\sigma_{Y_j}^2}\Big)$ through
\begin{align*}
    &\quad \mathbb{P}\Big(2\big|\sum\limits_{j\in  \mathcal{V}^*} \frac{(\hat{\theta}(\mathcal{V}^*)\cdot \beta_{X_j}-\theta_0\cdot \beta_{X_j}-r_j)\cdot \nu_j}{\sigma_{Y_j}^2}\big|\geq \frac{1}{10}\big(\sum\limits_{j\in \mathcal{V}^*} \frac{(\theta_0 \cdot \beta_{X_j}+r_j-\hat{\theta}(\mathcal{V}^*) \cdot \beta_{X_j})^2  }{\sigma_{Y_j}^2}-\kappa_{n}\cdot (v^*-v_\lambda)\big)\Big)\\&\leq    \mathbb{P}\Big(2\sqrt{\sum\limits_{j\in  \mathcal{V}^*}\frac{\nu_j^2}{\sigma_{Y_j}^2}}\geq \frac{\frac{1}{10}\big(\sum\limits_{j\in \mathcal{V}^*} \frac{(\theta_0 \cdot \beta_{X_j}+r_j-\hat{\theta}(\mathcal{V}^*) \cdot \beta_{X_j})^2  }{\sigma_{Y_j}^2}-\kappa_{n}\cdot (v^*-v_\lambda)\big)}{\sqrt{\sum\limits_{j\in \mathcal{V}^*} \frac{(\theta_0 \cdot \beta_{X_j}+r_j-\hat{\theta}(\mathcal{V}^*) \cdot \beta_{X_j})^2  }{\sigma_{Y_j}^2}}}\Big)\\& \leq \mathbb{P}\Big(\sum\limits_{j\in  \mathcal{V}^*}\frac{\nu_j^2}{\sigma_{Y_j}^2}\geq c''\sum\limits_{j\in \mathcal{V}^*} \frac{r_j^2}{\sigma_{Y_j}^2}\Big).
\end{align*}
Similarly, 
\begin{align*}
    &\quad \mathbb{P}\Big(2C_0\cdot\big|\sum\limits_{j\in  \mathcal{V}^*} \frac{(\hat{\theta}(\mathcal{V}^*)\cdot \beta_{X_j}-\theta_0\cdot \beta_{X_j}-r_j)\cdot u_j}{\sigma_{Y_j}^2}\big|\geq \frac{1}{10}\big(\sum\limits_{j\in \mathcal{V}^*} \frac{(\theta_0 \cdot \beta_{X_j}+r_j-\hat{\theta}(\mathcal{V}^*) \cdot \beta_{X_j})^2  }{\sigma_{Y_j}^2}-\kappa_{n}\cdot (v^*-v_\lambda)\big)\Big)\\& \leq \mathbb{P}\Big(\sum\limits_{j\in  \mathcal{V}^*}\frac{u_j^2}{\sigma_{Y_j}^2}\geq c''\sum\limits_{j\in \mathcal{V}^*} \frac{r_j^2  }{\sigma_{Y_j}^2}\Big).
\end{align*}
Using Lemma \ref{lem1}, we can also show that these two terms are bounded by
$
 2\cdot e^{-c' \cdot \min\big\{\frac{ {v^*}^2\cdot r_{\lambda}^2(\mathcal{V}^*)}{v_\lambda},\ v^*\cdot r_{\lambda}^2(\mathcal{V}^*), \ v^* \cdot r_{\lambda}(\mathcal{V}^*)\big\}}.
$
To prove $ e^{(s_{\lambda}+1)\cdot ln(s_{\lambda})}\cdot\mathbb{P}\Big(\min\limits_{\theta \in \mathbb{R}} h(\mathcal{V}^*,  \theta)-\kappa_{n}\cdot v^* \leq  h(\mathcal{V}_{\lambda}, \theta_0)-\kappa_{n}\cdot v_{\lambda}\Big) \rightarrow 0$ for any $\mathcal{V}^* \subseteq \mathcal{S}_\lambda$ such that $|\mathcal{V}^*|=v^*\geq v_{\lambda}$, we only need to show
\begin{align*}
    2e^{(s_{\lambda}+1)\cdot ln(s_{\lambda})}\cdot e^{-c'\cdot min\left\{ v^*\cdot r_{\lambda}^2(\mathcal{V}^*), \ v^* \cdot r_{\lambda}(\mathcal{V}^*), \  \frac{{v^*}^2\cdot r_{\lambda}^2(\mathcal{V}^*)}{v_{\lambda}}  \right\} }\rightarrow 0.
\end{align*}
Using $v^* \geq v_\lambda$, it suffices to show
\begin{align*}
    2e^{(s_{\lambda}+1)\cdot ln(s_{\lambda})}\cdot e^{-c'\cdot min\left\{ v^*\cdot r_{\lambda}^2(\mathcal{V}^*), \ v^* \cdot r_{\lambda}(\mathcal{V}^*) \right\} }\rightarrow 0.
\end{align*}
We prove this formula in Lemma \ref{lem2} and conclude that 
\begin{align*}
    e^{(s_{\lambda}+1)\cdot ln(s_{\lambda})}\cdot\mathbb{P}\Big(\min\limits_{\theta \in \mathbb{R}} h(\mathcal{V}^*,  \theta)-\kappa_{n}\cdot v^* \leq  h(\mathcal{V}_{\lambda}, \theta_0)-\kappa_{n}\cdot v_{\lambda}\Big) \rightarrow 0.
\end{align*}
uniformly for all $\mathcal{V}^* \subseteq \mathcal{S}_\lambda$ such that $|\mathcal{V}^*|=v^*\geq v_{\lambda}$ and $\mathcal{V}^* \neq \mathcal{V}_{\lambda}$.
\medskip

\subsubsection{Case 2: When $c_1\cdot v_\lambda \leq |\mathcal{V}^*| < v_{\lambda}$ but $\rho(\mathcal{V}^*)< 1$ }
When $c_1\cdot v_\lambda \leq |\mathcal{V}^*| < v_{\lambda}$ but $\rho(\mathcal{V}^*)< 1$, we can know from Condition \ref{Plurality and no perfect correlation} that $
\rho(\mathcal{V}^*)\leq c_0$.
Therefore,
\begin{align*}
        \sum\limits_{j\in \mathcal{V}^*} \frac{(\theta_0 \cdot \beta_{X_j}+r_j-\hat{\theta}(\mathcal{V}^*) \cdot \beta_{X_j})^2  }{\sigma_{Y_j}^2}\geq \sum\limits_{j\in \mathcal{V}^*}\frac{ r_j^2}{\sigma_{Y_j}^2}\cdot(1-c_0).
\end{align*}
with probability 1.
Similarly, under Condition \ref{Boundedness}, as $n \rightarrow \infty$, there exists a $C_0>0$ such that 
{\small
\begin{align*}
 &\quad\mathbb{P}\Big(\min\limits_{\theta \in \mathbb{R}} h(\mathcal{V}^*,  \theta)-\kappa_{n}\cdot v^* \leq  h(\mathcal{V}_{\lambda}, \theta_0)-\kappa_{n}\cdot v_{\lambda}\Big)
   \\&\leq \mathbb{P}\Big(\sum\limits_{j\in \mathcal{V}^*} \frac{(\theta_0 \cdot \beta_{X_j}+r_j-\hat{\theta}(\mathcal{V}^*) \cdot \beta_{X_j})^2  }{\sigma_{Y_j}^2}-\kappa_{n}\cdot (v^*-v_\lambda)
 \leq \big|\sum\limits_{j\in \mathcal{V}^*} \frac{\nu_j ^2-\sigma_{Y_j}^2}{\sigma_{Y_j}^2}\big|+C_0^2 \big|\sum\limits_{j\in \mathcal{V}^*} \frac{u_j^2- \sigma_{X_{j,\mathtt{RB}}}^2}{\sigma_{Y_j}^2}\big|
 \\&\quad+C_0^2 \big|\sum\limits_{j\in \mathcal{V}^*} \frac{\hat\sigma_{X_{j,\mathtt{RB}}}^2- \sigma_{X_{j,\mathtt{RB}}}^2}{\sigma_{Y_j}^2}\big|+2C_0\big| \sum\limits_{j\in  \mathcal{V}^*} \frac{(\hat{\theta}(\mathcal{V}^*)\cdot \beta_{X_j}-\theta_0\cdot \beta_{X_j}-r_j)\cdot u_j}{\sigma_{Y_j}^2}\big|+2\big| \sum\limits_{j\in  \mathcal{V}^*} \frac{(\hat{\theta}(\mathcal{V}^*)\cdot \beta_{X_j}-\theta_0\cdot \beta_{X_j}-r_j)\cdot \nu_j}{\sigma_{Y_j}^2}\big|\\&\quad+2C_0\big| \sum\limits_{j\in \mathcal{V}^*} \frac{\nu_j u_j}{\sigma_{Y_j}^2} \big|+\big|\sum\limits_{j\in \mathcal{V}_{\lambda}} \frac{\nu_j^2-\sigma_{Y_j}^2  }{\sigma_{Y_j}^2}\big|+\theta_0^2  \big|\sum\limits_{j\in \mathcal{V}_{\lambda}} \frac{  u_j^2- \sigma_{X_{j,\mathtt{RB}}}^2}{\sigma_{Y_j}^2}\big|+\theta_0^2  \big|\sum\limits_{j\in \mathcal{V}_{\lambda}} \frac{  \hat\sigma_{X_{j,\mathtt{RB}}}^2- \sigma_{X_{j,\mathtt{RB}}}^2}{\sigma_{Y_j}^2}\big|+2\theta_0 \big|\sum\limits_{j\in  \mathcal{V}_{\lambda}} \frac{\nu_ju_j}{\sigma_{Y_j}^2}\big|\Big)\\&=\mathbb{P}(\mathcal{C}(\mathcal{V}^*)).
\end{align*}
}
We also have 
$$
\delta(\mathcal{V}^*)=\sum\limits_{j\in \mathcal{V}^*} \frac{(\theta_0 \cdot \beta_{X_j}+r_j-\hat{\theta}(\mathcal{V}^*) \cdot \beta_{X_j})^2  }{\sigma_{Y_j}^2}-\kappa_{n}\cdot (v^*-v_\lambda)\geq \sum\limits_{j\in \mathcal{V}^*} \frac{(\theta_0 \cdot \beta_{X_j}+r_j-\hat{\theta}(\mathcal{V}^*) \cdot \beta_{X_j})^2  }{\sigma_{Y_j}^2}\geq \sum\limits_{j\in \mathcal{V}^*}\frac{ r_j^2}{\sigma_{Y_j}^2}\cdot (1-c_0).
$$
This is because $v^*<v_\lambda$ and $\kappa_{n}\cdot (v^*-v_\lambda)<0$.\\
Then the probability $\mathbb{P}(\mathcal{C}(\mathcal{V}^*))$ is bounded by
\begin{align*}
      &\ \mathbb{P}\Big(
\big|\sum\limits_{j\in \mathcal{V}^*} \frac{u_j^2- \sigma_{X_{j,\mathtt{RB}}}^2}{\sigma_{Y_j}^2}\big|\geq \frac{1-c_0}{10}\sum\limits_{j\in \mathcal{V}^*}\frac{ r_j^2}{\sigma_{Y_j}^2}\Big)+\mathbb{P}\Big(\big|\sum\limits_{j\in \mathcal{V}_{\lambda}} \frac{  u_j^2- \sigma_{X_{j,\mathtt{RB}}}^2}{\sigma_{Y_j}^2}\big|\geq \frac{1-c_0}{10}\sum\limits_{j\in \mathcal{V}^*}\frac{ r_j^2}{\sigma_{Y_j}^2}\Big)
\\& +\mathbb{P}\Big(\big|\sum\limits_{j\in \mathcal{V}^*} \frac{\hat\sigma_{X_{j,\mathtt{RB}}}^2- \sigma_{X_{j,\mathtt{RB}}}^2}{\sigma_{Y_j}^2}\big|\geq 
\frac{1-c_0}{10}\sum\limits_{j\in \mathcal{V}^*}\frac{ r_j^2}{\sigma_{Y_j}^2}\Big)+\mathbb{P}\Big(\big|\sum\limits_{j\in \mathcal{V}_{\lambda}} \frac{  \hat\sigma_{X_{j,\mathtt{RB}}}^2- \sigma_{X_{j,\mathtt{RB}}}^2}{\sigma_{Y_j}^2}\big|\geq \frac{1-c_0}{10}\sum\limits_{j\in \mathcal{V}^*}\frac{ r_j^2}{\sigma_{Y_j}^2}\Big)\\&+\mathbb{P}\Big(\big|\sum\limits_{j\in \mathcal{V}^*} \frac{\nu_j ^2-\sigma_{Y_j}^2}{\sigma_{Y_j}^2}\big|\geq \frac{1-c_0}{10}\sum\limits_{j\in \mathcal{V}^*}\frac{ r_j^2}{\sigma_{Y_j}^2}\Big)+\mathbb{P}\Big(\big|\sum\limits_{j\in \mathcal{V}_\lambda} \frac{\nu_j ^2-\sigma_{Y_j}^2}{\sigma_{Y_j}^2}\big|\geq \frac{1-c_0}{10}\sum\limits_{j\in \mathcal{V}^*}\frac{ r_j^2}{\sigma_{Y_j}^2}\Big) \\& +\mathbb{P}\Big(2\big|\sum\limits_{j\in  \mathcal{V}_{\lambda}} \frac{\nu_ju_j}{\sigma_{Y_j}^2}\big|\geq \frac{1-c_0}{10}\sum\limits_{j\in \mathcal{V}^*}\frac{ r_j^2}{\sigma_{Y_j}^2}\Big)
+\mathbb{P}\Big(2\big| \sum\limits_{j\in \mathcal{V}^*} \frac{\nu_j u_j}{\sigma_{Y_j}^2} \big|\geq \frac{1-c_0}{10}\sum\limits_{j\in \mathcal{V}^*}\frac{ r_j^2}{\sigma_{Y_j}^2}\Big)\\& +\mathbb{P}\Big(2\big|\sum\limits_{j\in  \mathcal{V}^*} \frac{(\hat{\theta}(\mathcal{V}^*)\cdot \beta_{X_j}-\theta_0\cdot \beta_{X_j}-r_j)\cdot \nu_j}{\sigma_{Y_j}^2}\big|\geq \frac{1}{10}\sum\limits_{j\in \mathcal{V}^*} \frac{(\theta_0 \cdot \beta_{X_j}+r_j-\hat{\theta}(\mathcal{V}^*) \cdot \beta_{X_j})^2  }{\sigma_{Y_j}^2}\Big)\\&+\mathbb{P}\Big(2\big|\sum\limits_{j\in  \mathcal{V}^*} \frac{(\hat{\theta}(\mathcal{V}^*)\cdot \beta_{X_j}-\theta_0\cdot \beta_{X_j}-r_j)\cdot u_j}{\sigma_{Y_j}^2}\big|\geq \frac{1}{10}\sum\limits_{j\in \mathcal{V}^*} \frac{(\theta_0 \cdot \beta_{X_j}+r_j-\hat{\theta}(\mathcal{V}^*) \cdot \beta_{X_j})^2  }{\sigma_{Y_j}^2}\Big).
\end{align*}
Knowing that  
$$
\big|\sum\limits_{j\in  \mathcal{V}^*} \frac{(\hat{\theta}(\mathcal{V}^*)\cdot \beta_{X_j}-\theta_0\cdot \beta_{X_j}-r_j)\cdot u_j}{\sigma_{Y_j}^2}\big|\leq \sqrt{\sum\limits_{j\in  \mathcal{V}^*} \frac{(\hat{\theta}(\mathcal{V}^*)\cdot \beta_{X_j}-\theta_0\cdot \beta_{X_j}-r_j)^2}{\sigma_{Y_j}^2}}\sqrt{\sum\limits_{j\in  \mathcal{V}^*} \frac{u_j^2}{\sigma_{Y_j}^2}}
$$
$$
\big|\sum\limits_{j\in  \mathcal{V}^*} \frac{(\hat{\theta}(\mathcal{V}^*)\cdot \beta_{X_j}-\theta_0\cdot \beta_{X_j}-r_j)\cdot \nu_j}{\sigma_{Y_j}^2}\big|\leq \sqrt{\sum\limits_{j\in  \mathcal{V}^*} \frac{(\hat{\theta}(\mathcal{V}^*)\cdot \beta_{X_j}-\theta_0\cdot \beta_{X_j}-r_j)^2}{\sigma_{Y_j}^2}}\sqrt{\sum\limits_{j\in  \mathcal{V}^*} \frac{\nu_j^2}{\sigma_{Y_j}^2}}
$$
We can similarly show that there exists a $c''>0$ such that the last two terms are further bounded by
\begin{align*}
\mathbb{P}\Big(\sum\limits_{j\in  \mathcal{V}^*}\frac{\nu_j^2}{\sigma_{Y_j}^2}\geq c''\sum\limits_{j\in \mathcal{V}^*} \frac{r_j^2}{\sigma_{Y_j}^2}\Big)+\mathbb{P}\Big(\sum\limits_{j\in  \mathcal{V}^*}\frac{u_j^2}{\sigma_{Y_j}^2}\geq c''\sum\limits_{j\in \mathcal{V}^*} \frac{r_j^2  }{\sigma_{Y_j}^2}\Big).
\end{align*}
Using Lemma \ref{lem1}, we know that there exists a $c'>0$ such that all these terms are bounded by
$
 2\cdot e^{-c' \cdot \min\big\{\frac{ {v^*}^2\cdot r_{\lambda}^2(\mathcal{V}^*)}{v_\lambda},\ v^*\cdot r_{\lambda}^2(\mathcal{V}^*), \ v^* \cdot r_{\lambda}(\mathcal{V}^*)\big\}}.
$

To prove $ e^{(s_{\lambda}+1)\cdot ln(s_{\lambda})}\cdot\mathbb{P}\Big(\min\limits_{\theta \in \mathbb{R}} h(\mathcal{V}^*,  \theta)-\kappa_{n}\cdot v^* \leq  h(\mathcal{V}_{\lambda}, \theta_0)-\kappa_{n}\cdot v_{\lambda}\Big) \rightarrow 0$ for any $\mathcal{V}^*$ such that $c_1\cdot v_\lambda \leq |\mathcal{V}^*| < v_{\lambda}$ but $\rho(\mathcal{V}^*)< 1$, we only need to show
\begin{align*}
    2e^{(s_{\lambda}+1)\cdot ln(s_{\lambda})}\cdot e^{-c'\cdot min\left\{ v^*\cdot r_{\lambda}^2(\mathcal{V}^*), \ v^* \cdot r_{\lambda}(\mathcal{V}^*), \  \frac{{v^*}^2\cdot r_{\lambda}^2(\mathcal{V}^*)}{v_{\lambda}}  \right\} }\rightarrow 0.
\end{align*}
Knowing that $ |\mathcal{V}^*| < v_\lambda$, it suffices to show that 
\begin{align*}
    2e^{(s_{\lambda}+1)\cdot ln(s_{\lambda})}\cdot e^{-c'\cdot min\left\{v^* \cdot r_{\lambda}(\mathcal{V}^*), \  \frac{{v^*}^2\cdot r_{\lambda}^2(\mathcal{V}^*)}{v_{\lambda}}  \right\} }\rightarrow 0.
\end{align*}
We prove this formula in Lemma \ref{lem2} and conclude that 
\begin{align*}
    e^{(s_{\lambda}+1)\cdot ln(s_{\lambda})}\cdot\mathbb{P}\Big(\min\limits_{\theta \in \mathbb{R}} h(\mathcal{V}^*,  \theta)-\kappa_{n}\cdot v^* \leq  h(\mathcal{V}_{\lambda}, \theta_0)-\kappa_{n}\cdot v_{\lambda}\Big) \rightarrow 0.
\end{align*}
uniformly for all $\mathcal{V}^* \subseteq \mathcal{S}_\lambda$  such that $c_1\cdot v_\lambda \leq |\mathcal{V}^*| < v_{\lambda}$ but $\rho(\mathcal{V}^*)< 1$.\\
\subsubsection{Case 3: When $|\mathcal{V}^*|=v^*< c_1 \cdot v_{\lambda}$}\label{sup-sec3.4.3}
When $v^*< c_1 \cdot v_{\lambda}$, we decompose $ h(\mathcal{V}^*, \hat {\theta}(\mathcal{V}^*))$ in a different way,
\begin{align*}
    h(\mathcal{V}^*, \hat {\theta}(\mathcal{V}^*))&=\nonumber\sum\limits_{j\in \mathcal{V}^*} \frac{(\hat \beta_{Y_j}-\hat{\theta}(\mathcal{V}^*) \cdot \hat \beta_{X_{j,\mathtt{RB}}})^2  }{\sigma_{Y_j}^2}-\hat\theta(\mathcal{V}^*)^2 \sum\limits_{j\in \mathcal{V}^*} \frac{ \hat\sigma_{X_{j,\mathtt{RB}}}^2}{\sigma_{Y_j}^2}\\&=\nonumber\sum\limits_{j\in \mathcal{V}^*} \frac{(\hat \beta_{Y_j}-\hat{\theta}(\mathcal{V}^*) \cdot \hat \beta_{X_{j,\mathtt{RB}}})^2  }{\sigma_{Y_j}^2}-\hat\theta(\mathcal{V}^*)^2 \sum\limits_{j\in \mathcal{V}^*} \frac{ \hat\sigma_{X_{j,\mathtt{RB}}}^2-\sigma_{X_{j,\mathtt{RB}}}^2}{\sigma_{Y_j}^2}-\hat\theta(\mathcal{V}^*)^2 \sum\limits_{j\in \mathcal{V}^*} \frac{\sigma_{X_{j,\mathtt{RB}}}^2}{\sigma_{Y_j}^2}.
\end{align*}
Under Condition \ref{Boundedness}, there exists a $C_0>0$ such that 
{\small
\begin{align*}
 &\quad\mathbb{P}\Big(\min\limits_{\theta \in \mathbb{R}} h(\mathcal{V}^*,  \theta)-\kappa_{n}\cdot v^* \leq  h(\mathcal{V}_{\lambda}, \theta_0)-\kappa_{n}\cdot v_{\lambda}\Big)
 \\&\leq \mathbb{P}\Big(        \sum\limits_{j\in \mathcal{V}^*} \frac{(\hat \beta_{Y_j}-\hat{\theta}(\mathcal{V}^*) \cdot \hat \beta_{X_{j,\mathtt{RB}}})^2  }{\sigma_{Y_j}^2}+\kappa_{n}\cdot (v_\lambda-v^*) 
 \leq \hat\theta(\mathcal{V}^*)^2  \sum\limits_{j\in \mathcal{V}^*} \frac{\hat\sigma_{X_{j,\mathtt{RB}}}^2- \sigma_{X_{j,\mathtt{RB}}}^2}{\sigma_{Y_j}^2}+\hat\theta(\mathcal{V}^*)^2 \sum\limits_{j\in \mathcal{V}^*} \frac{\sigma_{X_{j,\mathtt{RB}}}^2}{\sigma_{Y_j}^2}
 \\&\quad +\sum\limits_{j\in \mathcal{V}_{\lambda}} \frac{\nu_j^2-\sigma_{Y_j}^2  }{\sigma_{Y_j}^2}+\theta_0^2   \sum\limits_{j\in \mathcal{V}_{\lambda}} \frac{  u_j^2- \sigma_{X_{j,\mathtt{RB}}}^2}{\sigma_{Y_j}^2}-\theta_0^2   \sum\limits_{j\in \mathcal{V}_{\lambda}} \frac{  \hat\sigma_{X_{j,\mathtt{RB}}}^2- \sigma_{X_{j,\mathtt{RB}}}^2}{\sigma_{Y_j}^2} -2\theta_0  \sum\limits_{j\in  \mathcal{V}_{\lambda}} \frac{\nu_ju_j}{\sigma_{Y_j}^2}\Big) 
   \\&\leq \mathbb{P}\Big(\sum\limits_{j\in \mathcal{V}^*} \frac{(\hat \beta_{Y_j}-\hat{\theta}(\mathcal{V}^*) \cdot \hat \beta_{X_{j,\mathtt{RB}}})^2  }{\sigma_{Y_j}^2}+\kappa_{n}\cdot (v_\lambda-v^*)
 \leq C_0^2 \big|\sum\limits_{j\in \mathcal{V}^*} \frac{\hat\sigma_{X_{j,\mathtt{RB}}}^2- \sigma_{X_{j,\mathtt{RB}}}^2}{\sigma_{Y_j}^2}\big|+C_0^2 \cdot v^*
 \\&\quad+\big|\sum\limits_{j\in \mathcal{V}_{\lambda}} \frac{\nu_j^2-\sigma_{Y_j}^2  }{\sigma_{Y_j}^2}\big|+\theta_0^2  \big|\sum\limits_{j\in \mathcal{V}_{\lambda}} \frac{  u_j^2- \sigma_{X_{j,\mathtt{RB}}}^2}{\sigma_{Y_j}^2}\big|+\theta_0^2  \big|\sum\limits_{j\in \mathcal{V}_{\lambda}} \frac{  \hat\sigma_{X_{j,\mathtt{RB}}}^2- \sigma_{X_{j,\mathtt{RB}}}^2}{\sigma_{Y_j}^2}\big|+2\theta_0 \big|\sum\limits_{j\in  \mathcal{V}_{\lambda}} \frac{\nu_ju_j}{\sigma_{Y_j}^2}\big|\Big)   \\&\leq \mathbb{P}\Big(\kappa_{n}\cdot (v_\lambda-v^*)-C_0^2 \cdot v^*
 \leq C_0^2 \big|\sum\limits_{j\in \mathcal{V}^*} \frac{\hat\sigma_{X_{j,\mathtt{RB}}}^2- \sigma_{X_{j,\mathtt{RB}}}^2}{\sigma_{Y_j}^2}\big|+\big|\sum\limits_{j\in \mathcal{V}_{\lambda}} \frac{\nu_j^2-\sigma_{Y_j}^2  }{\sigma_{Y_j}^2}\big|
 \\&\quad+\theta_0^2  \big|\sum\limits_{j\in \mathcal{V}_{\lambda}} \frac{  u_j^2- \sigma_{X_{j,\mathtt{RB}}}^2}{\sigma_{Y_j}^2}\big|+\theta_0^2  \big|\sum\limits_{j\in \mathcal{V}_{\lambda}} \frac{  \hat\sigma_{X_{j,\mathtt{RB}}}^2- \sigma_{X_{j,\mathtt{RB}}}^2}{\sigma_{Y_j}^2}\big|+2\theta_0 \big|\sum\limits_{j\in  \mathcal{V}_{\lambda}} \frac{\nu_ju_j}{\sigma_{Y_j}^2}\big|\Big).
\end{align*}
}
Denote the event
\begin{align*}
& \kappa_{n}\cdot (v_\lambda-v^*)-C_0^2 \cdot v^*
 \leq C_0^2 \cdot\big|\sum\limits_{j\in \mathcal{V}^*} \frac{\hat\sigma_{X_{j,\mathtt{RB}}}^2- \sigma_{X_{j,\mathtt{RB}}}^2}{\sigma_{Y_j}^2}\big|+\big|\sum\limits_{j\in \mathcal{V}_{\lambda}} \frac{\nu_j^2-\sigma_{Y_j}^2  }{\sigma_{Y_j}^2}\big|
 \\&\quad+\theta_0^2 \cdot \big|\sum\limits_{j\in \mathcal{V}_{\lambda}} \frac{  u_j^2- \sigma_{X_{j,\mathtt{RB}}}^2}{\sigma_{Y_j}^2}\big|+\theta_0^2 \cdot \big|\sum\limits_{j\in \mathcal{V}_{\lambda}} \frac{  \hat\sigma_{X_{j,\mathtt{RB}}}^2- \sigma_{X_{j,\mathtt{RB}}}^2}{\sigma_{Y_j}^2}\big|+2\theta_0\cdot \big|\sum\limits_{j\in  \mathcal{V}_{\lambda}} \frac{\nu_ju_j}{\sigma_{Y_j}^2}\big|
\end{align*}
as $\mathcal{D}(\mathcal{V}^*)$, we have
\begin{align*}
      \mathcal{D}(\mathcal{V}^*) \subseteq & \left\{\theta_0^2\cdot \big|\sum\limits_{j\in \mathcal{V}_{\lambda}} \frac{  u_j^2- \sigma_{X_{j,\mathtt{RB}}}^2}{\sigma_{Y_j}^2}\big|\geq \frac{1}{5}\cdot (\kappa_{n}\cdot (v_\lambda-v^*)-C_0^2 \cdot v^*)\right\}\\&\bigcup 
      \left\{2\theta_0\cdot\big|\sum\limits_{j\in  \mathcal{V}_{\lambda}} \frac{\nu_ju_j}{\sigma_{Y_j}^2}\big|\geq \frac{1}{5}\cdot (\kappa_{n}\cdot (v_\lambda-v^*)-C_0^2 \cdot v^*) \right\}\\&\bigcup\left\{\theta_0^2\cdot\big|\sum\limits_{j\in \mathcal{V}_{\lambda}} \frac{  \hat\sigma_{X_{j,\mathtt{RB}}}^2- \sigma_{X_{j,\mathtt{RB}}}^2}{\sigma_{Y_j}^2}\big|\geq \frac{1}{5}\cdot (\kappa_{n}\cdot (v_\lambda-v^*)-C_0^2 \cdot v^*)\right\}
\\& \bigcup\left\{\big|\sum\limits_{j\in \mathcal{V}_\lambda} \frac{\nu_j ^2-\sigma_{Y_j}^2}{\sigma_{Y_j}^2}\big|\geq \frac{1}{5}\cdot (\kappa_{n}\cdot (v_\lambda-v^*)-C_0^2 \cdot v^*) \right\}\\&\bigcup \left\{C_0^2\cdot\big|\sum\limits_{j\in \mathcal{V}^*} \frac{\hat\sigma_{X_{j,\mathtt{RB}}}^2- \sigma_{X_{j,\mathtt{RB}}}^2}{\sigma_{Y_j}^2}\big|\geq \frac{1}{5}\cdot (\kappa_{n}\cdot (v_\lambda-v^*)-C_0^2 \cdot v^*)\right\}.
\end{align*}
Using $v^*<c_1\cdot v_\lambda$ and Condition \ref{high dimension BIC}, we know that there must be a $c'''>0$ such that 
$$
\kappa_n\cdot(v_\lambda-v^*)-C_0^2 \cdot v^*\geq \kappa_n\cdot(1-c_1)\cdot v_\lambda -C_0^2 \cdot v^*\geq c''' \cdot \kappa_n \cdot v_\lambda.
$$
then the probability $\mathbb{P}(\mathcal{D}(\mathcal{V}^*))$ is bounded by
\begin{align*}
      &\ \mathbb{P}\Big(\theta_0^2\cdot\big|\sum\limits_{j\in \mathcal{V}_{\lambda}} \frac{  u_j^2- \sigma_{X_{j,\mathtt{RB}}}^2}{\sigma_{Y_j}^2}\big|\geq \frac{c'''}{5}\cdot \kappa_n \cdot v_\lambda\Big)+\mathbb{P}\Big(C_0^2\cdot\big|\sum\limits_{j\in \mathcal{V}^*} \frac{\hat\sigma_{X_{j,\mathtt{RB}}}^2- \sigma_{X_{j,\mathtt{RB}}}^2}{\sigma_{Y_j}^2}\big|\geq \frac{c'''}{5}\cdot \kappa_n \cdot v_\lambda\Big)
\\& +\mathbb{P}\Big(\theta_0^2\cdot\big|\sum\limits_{j\in \mathcal{V}_{\lambda}} \frac{  \hat\sigma_{X_{j,\mathtt{RB}}}^2- \sigma_{X_{j,\mathtt{RB}}}^2}{\sigma_{Y_j}^2}\big|\geq \frac{c'''}{5} \cdot \kappa_n \cdot v_\lambda\Big)+\mathbb{P}\Big(2\theta_0\cdot\big|\sum\limits_{j\in  \mathcal{V}_{\lambda}} \frac{\nu_ju_j}{\sigma_{Y_j}^2}\big|\geq \frac{c'''}{5}\cdot \kappa_n \cdot v_\lambda \Big)\\&+\mathbb{P}\Big(\big|\sum\limits_{j\in \mathcal{V}_\lambda} \frac{\nu_j ^2-\sigma_{Y_j}^2}{\sigma_{Y_j}^2}\big|\geq \frac{c'''}{5}\cdot \kappa_n \cdot v_\lambda \Big).
\end{align*}
Using Lemma \ref{lem1}, we know that there exists a $c'>0$ such that the these five terms are bounded by
$
 2\cdot e^{-c' \cdot \min\big\{\kappa_{n}^2 v_\lambda,\ \frac{\kappa_{n}^2 v_\lambda^2 }{v^*},\  \kappa_{n}\cdot v_\lambda \big\}}
$. To prove $ e^{(s_{\lambda}+1)\cdot ln(s_{\lambda})}\cdot\mathbb{P}\Big(\min\limits_{\theta \in \mathbb{R}} h(\mathcal{V}^*,  \theta)-\kappa_{n}\cdot v^* \leq  h(\mathcal{V}_{\lambda}, \theta_0)-\kappa_{n}\cdot v_{\lambda}\Big) \rightarrow 0$, we only need to show
\begin{align*}
    2e^{(s_{\lambda}+1)\cdot ln(s_{\lambda})}\cdot e^{-c' \cdot  \kappa_{n}\cdot v_\lambda }\rightarrow 0.
\end{align*}
We can prove this by Lemma \ref{lem3} and conclude that $ e^{(s_{\lambda}+1)\cdot ln(s_{\lambda})}\cdot\mathbb{P}\Big(\min\limits_{\theta \in \mathbb{R}} h(\mathcal{V}^*,  \theta)-\kappa_{n}\cdot v^* \leq  h(\mathcal{V}_{\lambda}, \theta_0)-\kappa_{n}\cdot v_{\lambda}\Big) \rightarrow 0$ uniformly for all  $\mathcal{V}^* \subseteq \mathcal{S}_\lambda$ such that $v^*<c_1\cdot v_{\lambda}$.

Therefore, we  conclude that $ e^{(s_{\lambda}+1)\cdot ln(s_{\lambda})}\cdot\mathbb{P}\Big(\min\limits_{\theta \in \mathbb{R}} h(\mathcal{V}^*,  \theta)-\kappa_{n}\cdot v^* \leq  h(\mathcal{V}_{\lambda}, \theta_0)-\kappa_{n}\cdot v_{\lambda}\Big) \rightarrow 0$ uniformly for all  $\mathcal{V}^* \subseteq \mathcal{S}_\lambda$ such that $\mathcal{V}^* \neq  \mathcal{V}_\lambda$.




\subsection{Proof of Perfect Screening Property}
Following the similar procedure in Section 3.4, for any $\mathcal{V} \subset \mathcal{{S}}_{\lambda}$, we denote a function 
\begin{align*}
h(\mathcal{V}, \theta)=\sum\limits_{j\in \mathcal{V}} \frac{(\hat \beta_{Y_j}-\theta \cdot \hat \beta_{X_{j,\mathtt{RB}}})^2-\theta^2 \cdot \hat \sigma_{X_{j,\mathtt{RB}}}^2}{\sigma_{Y_j}^2}.
\end{align*}
and show 
\begin{align*}
  \quad \mathbb{P}(\hat{\mathcal{V}}_{\lambda}\neq \mathcal{V}_{\lambda})
  \leq e^{(s_{\lambda}+1)\cdot ln(s_{\lambda})}\cdot\mathbb{P}\Big(\min\limits_{\theta \in \mathbb{R}} h(\mathcal{V}^*,  \theta)-\kappa_{n}\cdot v^* \leq  h(\mathcal{V}_{\lambda}, \theta_0)-\kappa_{n}\cdot v_{\lambda}\Big).
\end{align*}
where $\mathcal{V}^*=\argmax\limits_{\mathcal{V}\neq \mathcal{V}_{\lambda}}e^{(s_{\lambda}+1)\cdot ln(s_{\lambda})}\cdot\mathbb{P}\Big(\min\limits_{\theta \in \mathbb{R}} h(\mathcal{V},  \theta)-\kappa_{n}\cdot |\mathcal{V}| \leq  h(\mathcal{V}_{\lambda}, \theta_0)-\kappa_{n}\cdot v_{\lambda}\Big)$ and $v^*=|\mathcal{V}^*|$.

As long as we show that 
\begin{align*}
e^{(s_{\lambda}+1)\cdot ln(s_{\lambda})}\cdot\mathbb{P}\Big(\min\limits_{\theta \in \mathbb{R}} h(\mathcal{V}^*,  \theta)-\kappa_{n}\cdot v^* \leq  h(\mathcal{V}_{\lambda}, \theta_0)-\kappa_{n}\cdot v_{\lambda}\Big)\rightarrow 0.  
\end{align*} 
then $\mathbb{P}(\hat{\mathcal{V}}_{\lambda}\neq \mathcal{V}_{\lambda})\rightarrow 0$ holds. 

To prove it, we discuss $\mathcal{V}^*$ in two different cases.
\begin{itemize}
    \item $|\mathcal{V}^*|=v^* \geq v_{\lambda}$ and $\mathcal{V}^* \neq \mathcal{V}_{\lambda}$,
    \item $|\mathcal{V}^*|=v^*< v_{\lambda}$.
\end{itemize}
In each case, we show $e^{(s_{\lambda}+1)\cdot ln(s_{\lambda})}\cdot\mathbb{P}\Big(\min\limits_{\theta \in \mathbb{R}} h(\mathcal{V}^*,  \theta)-\kappa_{n}\cdot v^* \leq  h(\mathcal{V}_{\lambda}, \theta_0)-\kappa_{n}\cdot v_{\lambda}\Big)\rightarrow 0 $ 
and therefore $\mathbb{P}(\hat{\mathcal{V}}_{\lambda}\neq \mathcal{V}_{\lambda})\rightarrow 0$ holds.

\subsubsection{Case 1: $|\mathcal{V}^*|=v^* \geq v_{\lambda}$ and $\mathcal{V}^* \neq \mathcal{V}_{\lambda}$}

The proof of this section is the same as the one in Section 3.4.1.


\subsubsection{Case 2: When $|\mathcal{V}^*|=v^*< v_{\lambda}$}
When $v^*<  v_{\lambda}$, following the proof in Section 3.4.3, we show 
{\small
\begin{align*}
 &\quad\mathbb{P}\Big(\min\limits_{\theta \in \mathbb{R}} h(\mathcal{V}^*,  \theta)-\kappa_{n}\cdot v^* \leq  h(\mathcal{V}_{\lambda}, \theta_0)-\kappa_{n}\cdot v_{\lambda}\Big)
 \\&\leq \mathbb{P}\Big(\kappa_{n}\cdot (v_\lambda-v^*)-C_0^2 \cdot v^*
 \leq C_0^2 \big|\sum\limits_{j\in \mathcal{V}^*} \frac{\hat\sigma_{X_{j,\mathtt{RB}}}^2- \sigma_{X_{j,\mathtt{RB}}}^2}{\sigma_{Y_j}^2}\big|+\big|\sum\limits_{j\in \mathcal{V}_{\lambda}} \frac{\nu_j^2-\sigma_{Y_j}^2  }{\sigma_{Y_j}^2}\big|
 \\&\quad+\theta_0^2  \big|\sum\limits_{j\in \mathcal{V}_{\lambda}} \frac{  u_j^2- \sigma_{X_{j,\mathtt{RB}}}^2}{\sigma_{Y_j}^2}\big|+\theta_0^2  \big|\sum\limits_{j\in \mathcal{V}_{\lambda}} \frac{  \hat\sigma_{X_{j,\mathtt{RB}}}^2- \sigma_{X_{j,\mathtt{RB}}}^2}{\sigma_{Y_j}^2}\big|+2\theta_0 \big|\sum\limits_{j\in  \mathcal{V}_{\lambda}} \frac{\nu_ju_j}{\sigma_{Y_j}^2}\big|\Big)\\&=\mathbb{P}(\mathcal{D}(\mathcal{V}^*)).
\end{align*}
}
Where $\mathcal{D}(\mathcal{V}^*)$ is the event
\begin{align*}
&\kappa_{n}\cdot (v_\lambda-v^*)-C_0^2 \cdot v^*
 \leq C_0^2 \cdot\big|\sum\limits_{j\in \mathcal{V}^*} \frac{\hat\sigma_{X_{j,\mathtt{RB}}}^2- \sigma_{X_{j,\mathtt{RB}}}^2}{\sigma_{Y_j}^2}\big|+\big|\sum\limits_{j\in \mathcal{V}_{\lambda}} \frac{\nu_j^2-\sigma_{Y_j}^2  }{\sigma_{Y_j}^2}\big|
 \\&\quad+\theta_0^2 \cdot \big|\sum\limits_{j\in \mathcal{V}_{\lambda}} \frac{  u_j^2- \sigma_{X_{j,\mathtt{RB}}}^2}{\sigma_{Y_j}^2}\big|+\theta_0^2 \cdot \big|\sum\limits_{j\in \mathcal{V}_{\lambda}} \frac{  \hat\sigma_{X_{j,\mathtt{RB}}}^2- \sigma_{X_{j,\mathtt{RB}}}^2}{\sigma_{Y_j}^2}\big|+2\theta_0\cdot \big|\sum\limits_{j\in  \mathcal{V}_{\lambda}} \frac{\nu_ju_j}{\sigma_{Y_j}^2}\big|.
\end{align*}
We have
\begin{align*}
      \mathcal{D}(\mathcal{V}^*) \subseteq & \left\{\theta_0^2\cdot \big|\sum\limits_{j\in \mathcal{V}_{\lambda}} \frac{  u_j^2- \sigma_{X_{j,\mathtt{RB}}}^2}{\sigma_{Y_j}^2}\big|\geq \frac{1}{5}\cdot (\kappa_{n}\cdot (v_\lambda-v^*)-C_0^2 \cdot v^*)\right\}\\&\bigcup 
      \left\{2\theta_0\cdot\big|\sum\limits_{j\in  \mathcal{V}_{\lambda}} \frac{\nu_ju_j}{\sigma_{Y_j}^2}\big|\geq \frac{1}{5}\cdot (\kappa_{n}\cdot (v_\lambda-v^*)-C_0^2 \cdot v^*) \right\}\\&\bigcup\left\{\theta_0^2\cdot\big|\sum\limits_{j\in \mathcal{V}_{\lambda}} \frac{  \hat\sigma_{X_{j,\mathtt{RB}}}^2- \sigma_{X_{j,\mathtt{RB}}}^2}{\sigma_{Y_j}^2}\big|\geq \frac{1}{5}\cdot (\kappa_{n}\cdot (v_\lambda-v^*)-C_0^2 \cdot v^*)\right\}
\\& \bigcup\left\{\big|\sum\limits_{j\in \mathcal{V}_\lambda} \frac{\nu_j ^2-\sigma_{Y_j}^2}{\sigma_{Y_j}^2}\big|\geq \frac{1}{5}\cdot (\kappa_{n}\cdot (v_\lambda-v^*)-C_0^2 \cdot v^*) \right\}\\&\bigcup \left\{C_0^2\cdot\big|\sum\limits_{j\in \mathcal{V}^*} \frac{\hat\sigma_{X_{j,\mathtt{RB}}}^2- \sigma_{X_{j,\mathtt{RB}}}^2}{\sigma_{Y_j}^2}\big|\geq \frac{1}{5}\cdot (\kappa_{n}\cdot (v_\lambda-v^*)-C_0^2 \cdot v^*)\right\}.
\end{align*}
Using $v^*< v_\lambda$ and Condition \ref{cond7*}, we know that there must be a $c'''>0$ such that 
$$
\kappa_n\cdot(v_\lambda-v^*)-C_0^2 \cdot v^*\geq \kappa_n -C_0^2 \cdot v^*\geq c''' \cdot \kappa_n.
$$
then the probability $\mathbb{P}(\mathcal{D}(\mathcal{V}^*))$ is bounded by
\begin{align*}
      &\ \mathbb{P}\Big(\theta_0^2\cdot\big|\sum\limits_{j\in \mathcal{V}_{\lambda}} \frac{  u_j^2- \sigma_{X_{j,\mathtt{RB}}}^2}{\sigma_{Y_j}^2}\big|\geq \frac{c'''}{5}\cdot \kappa_n \Big)+\mathbb{P}\Big(C_0^2\cdot\big|\sum\limits_{j\in \mathcal{V}^*} \frac{\hat\sigma_{X_{j,\mathtt{RB}}}^2- \sigma_{X_{j,\mathtt{RB}}}^2}{\sigma_{Y_j}^2}\big|\geq \frac{c'''}{5}\cdot \kappa_n \Big)
\\& +\mathbb{P}\Big(\theta_0^2\cdot\big|\sum\limits_{j\in \mathcal{V}_{\lambda}} \frac{  \hat\sigma_{X_{j,\mathtt{RB}}}^2- \sigma_{X_{j,\mathtt{RB}}}^2}{\sigma_{Y_j}^2}\big|\geq \frac{c'''}{5} \cdot \kappa_n \Big)+\mathbb{P}\Big(2\theta_0\cdot\big|\sum\limits_{j\in  \mathcal{V}_{\lambda}} \frac{\nu_ju_j}{\sigma_{Y_j}^2}\big|\geq \frac{c'''}{5}\cdot \kappa_n \Big)\\&+\mathbb{P}\Big(\big|\sum\limits_{j\in \mathcal{V}_\lambda} \frac{\nu_j ^2-\sigma_{Y_j}^2}{\sigma_{Y_j}^2}\big|\geq \frac{c'''}{5}\cdot \kappa_n  \Big).
\end{align*}
Using Lemma \ref{lem1}, we know that there exists a $c'>0$ such that the these five terms are bounded by
$
 2\cdot e^{-c' \cdot \min\big\{\frac{\kappa_{n}^2}{v_\lambda} ,\ \frac{\kappa_{n}^2 }{v^*},\  \kappa_{n} \big\}}
$. To prove $ e^{(s_{\lambda}+1)\cdot ln(s_{\lambda})}\cdot\mathbb{P}\Big(\min\limits_{\theta \in \mathbb{R}} h(\mathcal{V}^*,  \theta)-\kappa_{n}\cdot v^* \leq  h(\mathcal{V}_{\lambda}, \theta_0)-\kappa_{n}\cdot v_{\lambda}\Big) \rightarrow 0$, we only need to show
\begin{align*}
    2e^{(s_{\lambda}+1)\cdot ln(s_{\lambda})}\cdot e^{-c' \cdot \min\big\{\frac{\kappa_{n}^2}{v_\lambda} ,\ \frac{\kappa_{n}^2 }{v^*},\  \kappa_{n} \big\}}=    2e^{(s_{\lambda}+1)\cdot ln(s_{\lambda})}\cdot e^{-c' \cdot \min\big\{\frac{\kappa_{n}^2}{v_\lambda} ,\   \kappa_{n} \big\}}\rightarrow 0.
\end{align*}
We can prove this by Lemma \ref{lem3} and conclude that $ e^{(s_{\lambda}+1)\cdot ln(s_{\lambda})}\cdot\mathbb{P}\Big(\min\limits_{\theta \in \mathbb{R}} h(\mathcal{V}^*,  \theta)-\kappa_{n}\cdot v^* \leq  h(\mathcal{V}_{\lambda}, \theta_0)-\kappa_{n}\cdot v_{\lambda}\Big) \rightarrow 0$ uniformly for all  $\mathcal{V}^* \subseteq \mathcal{S}_\lambda$ such that $v^*<v_{\lambda}$.

Therefore, we  conclude that $ e^{(s_{\lambda}+1)\cdot ln(s_{\lambda})}\cdot\mathbb{P}\Big(\min\limits_{\theta \in \mathbb{R}} h(\mathcal{V}^*,  \theta)-\kappa_{n}\cdot v^* \leq  h(\mathcal{V}_{\lambda}, \theta_0)-\kappa_{n}\cdot v_{\lambda}\Big) \rightarrow 0$ uniformly for all  $\mathcal{V}^* \subseteq \mathcal{S}_\lambda$ such that $\mathcal{V}^* \neq  \mathcal{V}_\lambda$.




\subsection{Lemmas}

\begin{lem}\label{lem1}
Under Condition \ref{Bound of Orlicz norm},
\begin{align*}
\mathbb{P}\Big(\big|\sum\limits_{j\in  \mathcal{V}_{\lambda}} \frac{\nu_ju_j}{\sigma_{Y_j}^2}\big|\geq t \Big)
\leq 2\cdot e^{-c \cdot \min\big\{\frac{t^2}{v_\lambda},\  t\big\}},\quad\quad \mathbb{P}\Big(\big|\sum\limits_{j\in  \mathcal{V}} \frac{\nu_ju_j}{\sigma_{Y_j}^2}\big|\geq t \Big)
\leq 2\cdot e^{-c \cdot \min\big\{\frac{t^2}{v^*},\   t\big\}}
\end{align*}
\begin{align*}
\mathbb{P}\Big(\big|\sum\limits_{j\in  \mathcal{V}_{\lambda}} \frac{\nu_j^2-\sigma_{Y_j}^2}{\sigma_{Y_j}^2}\big|\geq t \Big)\leq 2\cdot e^{-c \cdot \min\big\{\frac{t^2}{v_\lambda},\   t\big\}},\quad\quad \mathbb{P}\Big(\big|\sum\limits_{j\in  \mathcal{V}} \frac{\nu_j^2-\sigma_{Y_j}^2}{\sigma_{Y_j}^2}\big|\geq t \Big)
\leq 2\cdot e^{-c \cdot \min\big\{\frac{t^2}{v^*},\   t\big\}}
\end{align*}
\begin{align*}
\mathbb{P}\Big(    \big|\sum\limits_{j\in \mathcal{V}_{\lambda}} \frac{  u_j^2- \sigma_{X_{j,\mathtt{RB}}}^2}{\sigma_{Y_j}^2}\big|\geq t \Big)
\leq 2\cdot e^{-c \cdot \min\big\{\frac{t^2}{v_\lambda},\   t\big\}},\quad\quad \mathbb{P}\Big(    \big|\sum\limits_{j\in \mathcal{V}} \frac{  u_j^2- \sigma_{X_{j,\mathtt{RB}}}^2}{\sigma_{Y_j}^2}\big|\geq t \Big)
\leq 2\cdot e^{-c \cdot \min\big\{\frac{t^2}{v^*},\   t\big\}}
\end{align*} 
\begin{align*}
\mathbb{P}\Big(    \big|\sum\limits_{j\in \mathcal{V}_{\lambda}} \frac{ \hat\sigma_{X_{j,\mathtt{RB}}}^2- \sigma_{X_{j,\mathtt{RB}}}^2}{\sigma_{Y_j}^2}\big|\geq t \Big)
\leq 2\cdot e^{-c \cdot \min\big\{\frac{t^2}{v_\lambda},\   t\big\}},\quad\quad \mathbb{P}\Big(    \big|\sum\limits_{j\in \mathcal{V}} \frac{ \hat\sigma_{X_{j,\mathtt{RB}}}^2- \sigma_{X_{j,\mathtt{RB}}}^2}{\sigma_{Y_j}^2}\big|\geq t \Big)
\leq 2\cdot e^{-c \cdot \min\big\{\frac{t^2}{v^*},\   t\big\}}
\end{align*}  
\end{lem}

\begin{lem}\label{lem2}
Under Condition \ref{Orders of the variances and sample sizes},  \ref{Separation} and \ref{Order of the number of valid IVs}, we have 
\begin{align*}
 \frac{ ln(s_{\lambda})}{ r_{\lambda}(\mathcal{V})}\rightarrow 0, \text{ and    } \frac{ ln(s_{\lambda})}{ r_{\lambda}^2(\mathcal{V})}\rightarrow 0.
\end{align*}
uniformly holds for $|\mathcal{V}| \geq c_1\cdot v_\lambda$ and $\mathcal{V}\neq \mathcal{V}_\lambda$.\\
Therefore, when $|\mathcal{V}| \geq v_\lambda$ and $\mathcal{V}\neq \mathcal{V}_\lambda$, we have
\begin{align*}
    2e^{(s_{\lambda}+1)\cdot ln(s_{\lambda})}\cdot e^{-c'\cdot min\left\{ v\cdot r_{\lambda}^2(\mathcal{V}), \ v \cdot r_{\lambda}(\mathcal{V}) \right\} }\rightarrow 0.
\end{align*}
When $c_1\cdot v_\lambda \leq |\mathcal{V}| < v_{\lambda}$ and $\mathcal{V}\neq \mathcal{V}_\lambda$, we have
\begin{align*}
    2e^{(s_{\lambda}+1)\cdot ln(s_{\lambda})}\cdot e^{-c\cdot min\left\{ v \cdot r_{\lambda}(\mathcal{V}),\ \frac{v^2 \cdot r_\lambda^2(\mathcal{V})}{v_\lambda} \right\} }\rightarrow 0.
\end{align*}
\end{lem}

\begin{lem}\label{lem3}
    Under Condition \ref{Order of the number of valid IVs} and \ref{high dimension BIC}, 
    \begin{align*}
     \frac{ \ln(s_{\lambda})}{\kappa_{n}}\rightarrow 0,
    \end{align*}
Therefore we have
\begin{align*}
    2e^{(s_{\lambda}+1)\cdot ln(s_{\lambda})}\cdot e^{-c' \cdot  \kappa_{n}\cdot v_\lambda }\rightarrow 0.
\end{align*}
Furthermore, under Condition \ref{Order of the number of valid IVs} and Condition \ref{cond7*}, which is a stronger condition of $\kappa_n$, we have
 \begin{align*}
    2e^{(s_{\lambda}+1)\cdot ln(s_{\lambda})}\cdot e^{-c' \cdot \min\big\{\frac{\kappa_{n}^2}{v_\lambda} ,\  \kappa_{n} \big\}}\rightarrow 0.
\end{align*}
\end{lem}

\subsection{Proof of Lemmas}

\subsubsection{Proof of Lemma \ref{lem1}}

We first prove $u_j$ is a sub-Gaussian random variable. We know from the definition that the n-th moment of $u_j$ conditional on selection is 
\begin{align*}
\Expectation[u_{j}^{n}|S_j > 0] &= \frac{\sigma_{X_j}^n}{\Prob[S_j > 0]}\int_{-\infty}^\infty \left(y - \frac{1}{\eta_j}\frac{\phi\left(B_{j,+}(y)\right) - \phi\left(B_{j,-}(y)\right)}{1 - \Phi\left(B_{j,+}(y)\right) + \Phi\left(B_{j,-}(y)\right)}\right)^n\\
&\qquad\qquad\qquad \qquad\qquad  \phi\left(y \right)\left[1 - \Phi\left(B_{j,+}(y)\right)  + \Phi\left(B_{j,-}(y)\right)  \right]\diff y.
\end{align*}
where 
\begin{align*}
\Prob[S_j > 0] &= \Phi\left( \frac{-\lambda+\frac{\beta_{X_j}}{\sigma_{X_j}}}{\sqrt{1+\eta_j^2}} \right) + \Phi\left( \frac{-\lambda-\frac{\beta_{X_j}}{\sigma_{X_j}}}{\sqrt{1+\eta_j^2}} \right) .
\end{align*}
and
\begin{align*}
    B_{j,\pm}(y) &= -\left(\frac{\beta_{X_j}}{\sigma_{X_j}\eta_j}+\frac{y}{\eta_j}\right) \pm\frac{\lambda}{\eta_j}.
\end{align*}

Given that the calculations can be quite involved, we let $\eta_j = 1$ and $\gamma_j = 0$ to streamline the presentation. That is, we consider
\begin{align*}
\frac{\Expectation[u_{X_j}^{n}|S_j > 0]}{\sigma_{X_j}^n} &= \frac{1}{\Prob[S_j > 0]}\int_{-\infty}^\infty \left(y - \frac{\phi\left(-\lambda + y\right) - \phi\left(-\lambda - y\right)}{\Phi\left(-\lambda + y\right) + \Phi\left(-\lambda - y\right)}\right)^n \phi\left(y \right)\Big[\Phi\left(-\lambda +  y\right)  + \Phi\left(-\lambda -y\right)  \Big]\diff y.
\end{align*}
Here 
\begin{align*}
\Prob[S_j > 0] &= 2 \cdot \Phi\left( \frac{-\lambda}{\sqrt{2}} \right) .
\end{align*}
Then when $n$ is an odd number, $\frac{\Expectation[u_{X_j}^{n}|S_j > 0]}{\sigma_{X_j}^n}=0$.  When $n$ is an even number, we have 
\begin{align*}
\frac{\Expectation[u_{X_j}^{n}|S_j > 0]}{\sigma_{X_j}^n} &= \frac{2}{\Prob[S_j > 0]}\int_{0}^\infty \left(y - \frac{\phi\left(-\lambda + y\right) - \phi\left(-\lambda - y\right)}{\Phi\left(-\lambda + y\right) + \Phi\left(-\lambda - y\right)}\right)^n \phi\left(y \right)\Big[\Phi\left(-\lambda +  y\right)  + \Phi\left(-\lambda -y\right)  \Big]\diff y.
\end{align*}
\begin{align*}
    \frac{\phi\left(-\lambda + y\right) - \phi\left(-\lambda - y\right)}{\Phi\left(-\lambda + y\right) + \Phi\left(-\lambda - y\right)}\leq \frac{\phi\left(-\lambda + y\right) - \phi\left(-\lambda - y\right)}{\Phi\left(-\lambda\right) + \Phi\left(-\lambda\right)}\leq \frac{\phi\left(-\lambda + y\right)}{\Phi\left(-\lambda\right) + \Phi\left(-\lambda\right)}\leq\frac{1/\sqrt{2\pi}}{\Phi\left(-\lambda\right) + \Phi\left(-\lambda\right)}.
\end{align*}

\begin{align*}
\frac{\Expectation[u_{X_j}^{n}|S_j > 0]}{\sigma_{X_j}^n} &= \frac{2}{\Prob[S_j > 0]}\int_{0}^{\frac{1/\sqrt{2\pi}}{\Phi\left(-\lambda\right) + \Phi\left(-\lambda\right)}} \left(y - \frac{\phi\left(-\lambda + y\right) - \phi\left(-\lambda - y\right)}{\Phi\left(-\lambda + y\right) + \Phi\left(-\lambda - y\right)}\right)^n \phi\left(y \right)\Big[\Phi\left(-\lambda +  y\right)  + \Phi\left(-\lambda -y\right)  \Big]\diff y\\&+ \frac{2}{\Prob[S_j > 0]}\int_{\frac{1/\sqrt{2\pi}}{\Phi\left(-\lambda\right) + \Phi\left(-\lambda\right)}}^\infty \left(y - \frac{\phi\left(-\lambda + y\right) - \phi\left(-\lambda - y\right)}{\Phi\left(-\lambda + y\right) + \Phi\left(-\lambda - y\right)}\right)^n \phi\left(y \right)\Big[\Phi\left(-\lambda +  y\right)  + \Phi\left(-\lambda -y\right)  \Big]\diff y\\&\leq  \frac{2}{\Prob[S_j > 0]}\int_{0}^{\frac{1/\sqrt{2\pi}}{\Phi\left(-\lambda\right) + \Phi\left(-\lambda\right)}} \left(\frac{1/\sqrt{2\pi}}{\Phi\left(-\lambda\right) + \Phi\left(-\lambda\right)}\right)^n \phi\left(y \right)\Big[\Phi\left(-\lambda +  y\right)  + \Phi\left(-\lambda -y\right)  \Big]\diff y\\&+ \frac{2}{\Prob[S_j > 0]}\int_{\frac{1/\sqrt{2\pi}}{\Phi\left(-\lambda\right) + \Phi\left(-\lambda\right)}}^\infty y^n \phi\left(y \right)\Big[\Phi\left(-\lambda +  y\right)  + \Phi\left(-\lambda -y\right)  \Big]\diff y\\&\leq  \frac{2}{\Prob[S_j > 0]}\int_{0}^{\infty} \left(\frac{1/\sqrt{2\pi}}{\Phi\left(-\lambda\right) + \Phi\left(-\lambda\right)}\right)^n \phi\left(y \right)\Big[\Phi\left(-\lambda +  y\right)  + \Phi\left(-\lambda -y\right)  \Big]\diff y\\&+ \frac{2}{\Prob[S_j > 0]}\int_{0}^\infty y^n \phi\left(y \right)\Big[\Phi\left(-\lambda +  y\right)  + \Phi\left(-\lambda -y\right)  \Big]\diff y\\& = 2\left(\frac{1/\sqrt{2\pi}}{\Phi\left(-\lambda\right) + \Phi\left(-\lambda\right)}\right)^n + \frac{2}{\Prob[S_j > 0]}\int_{0}^\infty y^n \phi\left(y \right)\Big[\Phi\left(-\lambda +  y\right)  + \Phi\left(-\lambda -y\right)  \Big]\diff y\\&\leq \left(\frac{1/\sqrt{2\pi}}{\Phi\left(-\lambda\right) + \Phi\left(-\lambda\right)}\right)^n + \frac{1}{\Prob[S_j > 0]}\int_{-\infty}^\infty y^n \phi\left(y \right)\diff y.
\end{align*}

Then by the property of sub-Gaussian random variable, we know that there exists a $K$ such that
\begin{align*}
   \Big(\int_{-\infty}^\infty y^n \phi\left(y \right)\diff y\Big)^{\frac{1}{n}}\leq K \sqrt{n}.
\end{align*}
for any $n\geq 1$.
\begin{align*}
    \frac{\Expectation[u_{X_j}^{n}|S_j > 0]}{\sigma_{X_j}^n}\leq \left(\frac{1/\sqrt{2\pi}}{\Phi\left(-\lambda\right) + \Phi\left(-\lambda\right)}\right)^n + \frac{1}{2 \cdot \Phi\left( \frac{-\lambda}{\sqrt{2}} \right)}\cdot K^n \cdot n^\frac{n}{2}.
\end{align*}
Then we can know that there exists a $K'> 0$ such that 
\begin{align*}
    \Big(\frac{\Expectation[u_{X_j}^{n}|S_j > 0]}{\sigma_{X_j}^n}\Big)^\frac{1}{n}\leq K' \cdot \sqrt{n}.
\end{align*}
for any $n\geq 1$.

This proves that conditional on re-randomized selection, $u_j$ is a sub-Gaussian random variable. Since $\nu_j$ is a Gaussian random variable, it's also a sub-Gaussian random variable. Then we know $\nu_j u_j$, $u_j^2$ and $\nu_j^2$ are all subexponential random variables.

Also we know from Condition \ref{Bound of Orlicz norm} that $\frac{  \hat\sigma_{X_{j,\mathtt{RB}}}^2- \sigma_{X_{j,\mathtt{RB}}}^2}{\sigma_{Y_j}^2}$ is a sub-exponential random variable.

Then by Bernstein's inequality, we have
\begin{align*}
\mathbb{P}\Big(\big|\sum\limits_{j\in  \mathcal{V}_{\lambda}} \frac{\nu_ju_j}{\sigma_{Y_j}^2}\big|\geq t \Big)
&\leq 2\cdot e^{-c \cdot \min\big(\frac{t^2}{\sum\limits_{j\in  \mathcal{V}_{\lambda}}||\frac{\nu_ju_j}{\sigma_{Y_j}^2}||_{\psi_1}^2}, \frac{t}{\max_i||\frac{\nu_ju_j}{\sigma_{Y_j}^2}||_{\psi_1}}\big)}
\\&\leq 2\cdot e^{-c \cdot \min\big(\frac{t^2}{\sum\limits_{j\in  \mathcal{V}_{\lambda}}||\frac{\nu_j}{\sigma_{Y_j}}||_{\psi_2}^2\cdot||\frac{u_j}{\sigma_{Y_j}}||_{\psi_2}^2}, \frac{t}{\max_i\sqrt{||\frac{\nu_j}{\sigma_{Y_j}}||_{\psi_2}^2\cdot||\frac{u_j}{\sigma_{Y_j}}||_{\psi_2}^2}}\big)}.
\end{align*}

\begin{align*}
\mathbb{P}\Big(\big|\sum\limits_{j\in  \mathcal{V}} \frac{\nu_ju_j}{\sigma_{Y_j}^2}\big|\geq t \Big)
&\leq 2\cdot e^{-c \cdot \min\big(\frac{t^2}{\sum\limits_{j\in  \mathcal{V}}||\frac{\nu_ju_j}{\sigma_{Y_j}^2}||_{\psi_1}^2}, \frac{t}{\max_i||\frac{\nu_ju_j}{\sigma_{Y_j}^2}||_{\psi_1}}\big)}
\\&\leq 2\cdot e^{-c \cdot \min\big(\frac{t^2}{\sum\limits_{j\in  \mathcal{V}}||\frac{\nu_j}{\sigma_{Y_j}}||_{\psi_2}^2\cdot||\frac{u_j}{\sigma_{Y_j}}||_{\psi_2}^2}, \frac{t}{\max_i\sqrt{||\frac{\nu_j}{\sigma_{Y_j}}||_{\psi_2}^2\cdot||\frac{u_j}{\sigma_{Y_j}}||_{\psi_2}^2}}\big)}.
\end{align*}
\begin{align*}
\mathbb{P}\Big(\big|\sum\limits_{j\in  \mathcal{V}_{\lambda}} \frac{\nu_j^2-\sigma_{Y_j}^2}{\sigma_{Y_j}^2}\big|\geq t \Big)
&\leq 2\cdot e^{-c \cdot \min\big(\frac{t^2}{\sum\limits_{j\in  \mathcal{V}_{\lambda}}||\frac{\nu_j^2-\sigma_{Y_j}^2}{\sigma_{Y_j}^2}||_{\psi_1}^2}, \frac{t}{\max_i||\frac{\nu_j^2-\sigma_{Y_j}^2}{\sigma_{Y_j}^2}||_{\psi_1}}\big)}
\\&\leq 2\cdot e^{-c \cdot \min\big(\frac{t^2}{\sum\limits_{j\in  \mathcal{V}_{\lambda}}||\sqrt{\frac{\nu_j^2-\sigma_{Y_j}^2}{\sigma_{Y_j}^2}}||_{\psi_2}^4}, \frac{t}{\max_i||\sqrt{\frac{\nu_j^2-\sigma_{Y_j}^2}{\sigma_{Y_j}^2}}||_{\psi_2}^2}\big)}.
\end{align*}
\begin{align*}
\mathbb{P}\Big(\big|\sum\limits_{j\in  \mathcal{V}} \frac{\nu_j^2-\sigma_{Y_j}^2}{\sigma_{Y_j}^2}\big|\geq t \Big)
&\leq 2\cdot e^{-c \cdot \min\big(\frac{t^2}{\sum\limits_{j\in  \mathcal{V}}||\frac{\nu_j^2-\sigma_{Y_j}^2}{\sigma_{Y_j}^2}||_{\psi_1}^2}, \frac{t}{\max_i||\frac{\nu_j^2-\sigma_{Y_j}^2}{\sigma_{Y_j}^2}||_{\psi_1}}\big)}
\\&\leq 2\cdot e^{-c \cdot \min\big(\frac{t^2}{\sum\limits_{j\in  \mathcal{V}}||\sqrt{\frac{\nu_j^2-\sigma_{Y_j}^2}{\sigma_{Y_j}^2}}||_{\psi_2}^4}, \frac{t}{\max_i||\sqrt{\frac{\nu_j^2-\sigma_{Y_j}^2}{\sigma_{Y_j}^2}}||_{\psi_2}^2}\big)}.
\end{align*}
\begin{align*}
\mathbb{P}\Big(    \big|\sum\limits_{j\in \mathcal{V}_{\lambda}} \frac{  u_j^2- \sigma_{X_{j,\mathtt{RB}}}^2}{\sigma_{Y_j}^2}\big|\geq t \Big)
&\leq 2\cdot e^{-c \cdot \min\big(\frac{t^2}{\sum\limits_{j\in  \mathcal{V}_{\lambda}}||\frac{  u_j^2- \sigma_{X_{j,\mathtt{RB}}}^2}{\sigma_{Y_j}^2}||_{\psi_1}^2}, \frac{t}{\max_i||\frac{  u_j^2- \sigma_{X_{j,\mathtt{RB}}}^2}{\sigma_{Y_j}^2}||_{\psi_1}}\big)}
\\&\leq 2\cdot e^{-c \cdot \min\big(\frac{t^2}{\sum\limits_{j\in  \mathcal{V}_{\lambda}}||\sqrt{\frac{  u_j^2- \sigma_{X_{j,\mathtt{RB}}}^2}{\sigma_{Y_j}^2}}||_{\psi_2}^4}, \frac{t}{\max_i||\sqrt{\frac{  u_j^2- \sigma_{X_{j,\mathtt{RB}}}^2}{\sigma_{Y_j}^2}}||_{\psi_2}^2}\big)}.
\end{align*} 
\begin{align*}
\mathbb{P}\Big(    \big|\sum\limits_{j\in \mathcal{V}} \frac{  u_j^2- \sigma_{X_{j,\mathtt{RB}}}^2}{\sigma_{Y_j}^2}\big|\geq t \Big)
&\leq 2\cdot e^{-c \cdot \min\big(\frac{t^2}{\sum\limits_{j\in  \mathcal{V}}||\frac{  u_j^2- \sigma_{X_{j,\mathtt{RB}}}^2}{\sigma_{Y_j}^2}||_{\psi_1}^2}, \frac{t}{\max_i||\frac{  u_j^2- \sigma_{X_{j,\mathtt{RB}}}^2}{\sigma_{Y_j}^2}||_{\psi_1}}\big)}
\\&\leq 2\cdot e^{-c \cdot \min\big(\frac{t^2}{\sum\limits_{j\in  \mathcal{V}}||\sqrt{\frac{  u_j^2- \sigma_{X_{j,\mathtt{RB}}}^2}{\sigma_{Y_j}^2}}||_{\psi_2}^4}, \frac{t}{\max_i||\sqrt{\frac{  u_j^2- \sigma_{X_{j,\mathtt{RB}}}^2}{\sigma_{Y_j}^2}}||_{\psi_2}^2}\big)}.
\end{align*} 

\begin{align*}
\mathbb{P}\Big(    \big|\sum\limits_{j\in \mathcal{V}_{\lambda}} \frac{ \hat\sigma_{X_{j,\mathtt{RB}}}^2- \sigma_{X_{j,\mathtt{RB}}}^2}{\sigma_{Y_j}^2}\big|\geq t \Big)
&\leq 2\cdot e^{-c \cdot \min\big(\frac{t^2}{\sum\limits_{j\in  \mathcal{V}_{\lambda}}||\frac{  \hat\sigma_{X_{j,\mathtt{RB}}}^2- \sigma_{X_{j,\mathtt{RB}}}^2}{\sigma_{Y_j}^2}||_{\psi_1}^2}, \frac{t}{\max_i||\frac{  \hat\sigma_{X_{j,\mathtt{RB}}}^2- \sigma_{X_{j,\mathtt{RB}}}^2}{\sigma_{Y_j}^2}||_{\psi_1}}\big)}
\\&\leq 2\cdot e^{-c \cdot \min\big(\frac{t^2}{\sum\limits_{j\in  \mathcal{V}_{\lambda}}||\sqrt{\frac{  \hat\sigma_{X_{j,\mathtt{RB}}}^2- \sigma_{X_{j,\mathtt{RB}}}^2}{\sigma_{Y_j}^2}}||_{\psi_2}^4}, \frac{t}{\max_i||\sqrt{\frac{  \hat\sigma_{X_{j,\mathtt{RB}}}^2- \sigma_{X_{j,\mathtt{RB}}}^2}{\sigma_{Y_j}^2}}||_{\psi_2}^2}\big)}.
\end{align*} 

\begin{align*}
\mathbb{P}\Big(    \big|\sum\limits_{j\in \mathcal{V}} \frac{ \hat\sigma_{X_{j,\mathtt{RB}}}^2- \sigma_{X_{j,\mathtt{RB}}}^2}{\sigma_{Y_j}^2}\big|\geq t \Big)
&\leq 2\cdot e^{-c \cdot \min\big(\frac{t^2}{\sum\limits_{j\in  \mathcal{V}}||\frac{  \hat\sigma_{X_{j,\mathtt{RB}}}^2- \sigma_{X_{j,\mathtt{RB}}}^2}{\sigma_{Y_j}^2}||_{\psi_1}^2}, \frac{t}{\max_i||\frac{  \hat\sigma_{X_{j,\mathtt{RB}}}^2- \sigma_{X_{j,\mathtt{RB}}}^2}{\sigma_{Y_j}^2}||_{\psi_1}}\big)}
\\&\leq 2\cdot e^{-c \cdot \min\big(\frac{t^2}{\sum\limits_{j\in  \mathcal{V}}||\sqrt{\frac{  \hat\sigma_{X_{j,\mathtt{RB}}}^2- \sigma_{X_{j,\mathtt{RB}}}^2}{\sigma_{Y_j}^2}}||_{\psi_2}^4}, \frac{t}{\max_i||\sqrt{\frac{  \hat\sigma_{X_{j,\mathtt{RB}}}^2- \sigma_{X_{j,\mathtt{RB}}}^2}{\sigma_{Y_j}^2}}||_{\psi_2}^2}\big)}.
\end{align*} 
Under Condition \ref{Bound of Orlicz norm}, we can have the conclusion in Lemma 1.
\begin{align*}
\mathbb{P}\Big(\big|\sum\limits_{j\in  \mathcal{V}_{\lambda}} \frac{\nu_ju_j}{\sigma_{Y_j}^2}\big|\geq t \Big)
\leq 2\cdot e^{-c \cdot \min\big\{\frac{t^2}{v_\lambda},\  t\big\}},\quad\quad \mathbb{P}\Big(\big|\sum\limits_{j\in  \mathcal{V}} \frac{\nu_ju_j}{\sigma_{Y_j}^2}\big|\geq t \Big)
\leq 2\cdot e^{-c \cdot \min\big\{\frac{t^2}{v^*},\   t\big\}}.
\end{align*}
\begin{align*}
\mathbb{P}\Big(\big|\sum\limits_{j\in  \mathcal{V}_{\lambda}} \frac{\nu_j^2-\sigma_{Y_j}^2}{\sigma_{Y_j}^2}\big|\geq t \Big)\leq 2\cdot e^{-c \cdot \min\big\{\frac{t^2}{v_\lambda},\   t\big\}},\quad\quad \mathbb{P}\Big(\big|\sum\limits_{j\in  \mathcal{V}} \frac{\nu_j^2-\sigma_{Y_j}^2}{\sigma_{Y_j}^2}\big|\geq t \Big)
\leq 2\cdot e^{-c \cdot \min\big\{\frac{t^2}{v^*},\   t\big\}}.
\end{align*}
\begin{align*}
\mathbb{P}\Big(    \big|\sum\limits_{j\in \mathcal{V}_{\lambda}} \frac{  u_j^2- \sigma_{X_{j,\mathtt{RB}}}^2}{\sigma_{Y_j}^2}\big|\geq t \Big)
\leq 2\cdot e^{-c \cdot \min\big\{\frac{t^2}{v_\lambda},\   t\big\}},\quad\quad \mathbb{P}\Big(    \big|\sum\limits_{j\in \mathcal{V}} \frac{  u_j^2- \sigma_{X_{j,\mathtt{RB}}}^2}{\sigma_{Y_j}^2}\big|\geq t \Big)
\leq 2\cdot e^{-c \cdot \min\big\{\frac{t^2}{v^*},\   t\big\}}.
\end{align*} 
\begin{align*}
\mathbb{P}\Big(    \big|\sum\limits_{j\in \mathcal{V}_{\lambda}} \frac{ \hat\sigma_{X_{j,\mathtt{RB}}}^2- \sigma_{X_{j,\mathtt{RB}}}^2}{\sigma_{Y_j}^2}\big|\geq t \Big)
\leq 2\cdot e^{-c \cdot \min\big\{\frac{t^2}{v_\lambda},\   t\big\}},\quad\quad \mathbb{P}\Big(    \big|\sum\limits_{j\in \mathcal{V}} \frac{ \hat\sigma_{X_{j,\mathtt{RB}}}^2- \sigma_{X_{j,\mathtt{RB}}}^2}{\sigma_{Y_j}^2}\big|\geq t \Big)
\leq 2\cdot e^{-c \cdot \min\big\{\frac{t^2}{v^*},\   t\big\}}.
\end{align*}  

\subsubsection{Proof of Lemma \ref{lem2}}
\begin{proof}
 Without loss of generality, we can assume that $\sigma_{Y_j}^2=\frac{1}{n}$, then we have
\begin{align*}
    r_{\lambda}(\mathcal{V})=\frac{1}{v}\sum\limits_{j\in \mathcal{V}}\frac{r_j^2}{\sigma_{Y_j}^2}\geq \frac{1}{v}\cdot n \min\limits_{j\in\mathcal{S}_\lambda,\ r_j\neq 0} r_j^2.
\end{align*}  
Under Condition \ref{Separation}, we have
    \begin{align*}
        & \frac{ ln(s_{\lambda})}{  r_{\lambda}(\mathcal{V})}=\frac{v \cdot ln(s_{\lambda})}{n \min\limits_{j\in S_\lambda,\ r_j\neq 0} r_j^2}\leq \frac{s_\lambda \cdot ln(s_{\lambda})}{n \min\limits_{j\in S_\lambda,\ r_j\neq 0} r_j^2}\rightarrow 0.
   \\& \frac{ln(s_{\lambda})}{r_{\lambda}^2(\mathcal{V})}\leq \frac{ ln(s_{\lambda})\cdot v^2}{n^2 \min\limits_{j\in S_\lambda,\ r_j\neq0}r_j^4} \leq \frac{s_{\lambda}^2\cdot ln(s_{\lambda})}{n^2 \min\limits_{j\in S_\lambda,\ r_j\neq0}r_j^4}\rightarrow 0.
\end{align*}
Notice that when $|\mathcal{V}| \geq v_\lambda$ and $\mathcal{V}\neq \mathcal{V}_\lambda$, under Condition \ref{Order of the number of valid IVs}, we have
\begin{align*}
  \frac{(s_{\lambda}+1)\cdot ln(s_{\lambda})}{v\cdot  r_{\lambda}(\mathcal{V})}\leq   \frac{(s_{\lambda}+1)\cdot ln(s_{\lambda})}{v_\lambda \cdot  r_{\lambda}(\mathcal{V})}\rightarrow 0 ,\quad   \frac{(s_{\lambda}+1)\cdot ln(s_{\lambda})}{v\cdot r_{\lambda}^2(\mathcal{V})}\leq   \frac{(s_{\lambda}+1)\cdot ln(s_{\lambda})}{v_\lambda\cdot r_{\lambda}^2(\mathcal{V})}\rightarrow 0. 
\end{align*}
Then we can show 
\begin{align*}
    2e^{(s_{\lambda}+1)\cdot ln(s_{\lambda})}\cdot e^{-c'\cdot min\left\{ v\cdot r_{\lambda}^2(\mathcal{V}), \ v \cdot r_{\lambda}(\mathcal{V}) \right\} }\rightarrow 0.
\end{align*}
Notice that when $c_1\cdot v_\lambda \leq |\mathcal{V}| < v_{\lambda}$ and $\mathcal{V}\neq \mathcal{V}_\lambda$, under Condition \ref{Order of the number of valid IVs}, we have
\begin{align*}
  \frac{(s_{\lambda}+1)\cdot ln(s_{\lambda})}{v\cdot  r_{\lambda}(\mathcal{V})}\leq   \frac{(s_{\lambda}+1)\cdot ln(s_{\lambda})}{c_1\cdot v_\lambda \cdot  r_{\lambda}(\mathcal{V})}\rightarrow 0 ,\quad   \frac{(s_{\lambda}+1)\cdot ln(s_{\lambda})\cdot v_\lambda}{v^2\cdot r_{\lambda}^2(\mathcal{V})}\leq   \frac{(s_{\lambda}+1)\cdot ln(s_{\lambda})}{c_1^2\cdot v_\lambda\cdot r_{\lambda}^2(\mathcal{V})}\rightarrow 0. 
\end{align*}
Then we can show 
\begin{align*}
    2e^{(s_{\lambda}+1)\cdot ln(s_{\lambda})}\cdot e^{-c\cdot min\left\{ v \cdot r_{\lambda}(\mathcal{V}),  \ \frac{v^2 \cdot r_\lambda^2(\mathcal{V})}{v_\lambda} \right\} }\rightarrow 0.
\end{align*}
\end{proof}
\subsubsection{Proof of Lemma \ref{lem3}}
\begin{proof}

Under Condition \ref{Order of the number of valid IVs} and \ref{high dimension BIC} , we know that 

\begin{align*}
    \kappa_{n}\gg  
    ln(s_{\lambda}),\quad \text{ and }\quad \frac{v_\lambda}{s_\lambda} \text{ is bounded away from 0}.
\end{align*} 
Then
    \begin{align*}
 \frac{(s_{\lambda}+1)\cdot ln(s_{\lambda})}{\kappa_{n} \cdot v_{\lambda}}\rightarrow 0. 
    \end{align*}
Then we know
\begin{align*}
    2e^{(s_{\lambda}+1)\cdot ln(s_{\lambda})}\cdot e^{-c'\kappa_{n}\cdot v_\lambda }\rightarrow 0.
\end{align*}
Under Condition \ref{Order of the number of valid IVs} and Condition \ref{cond7*}, we have
\begin{align*}
 \frac{(s_{\lambda}+1)\cdot ln(s_{\lambda})}{\kappa_{n} }\rightarrow 0. 
\end{align*}
Then we know
 \begin{align*}
    2e^{(s_{\lambda}+1)\cdot ln(s_{\lambda})}\cdot e^{-c' \cdot \min\big\{\frac{\kappa_{n}^2}{v_\lambda},\  \kappa_{n} \big\}}\rightarrow 0.
\end{align*}
\end{proof}

\section{ An example that Assumption \ref{sup-assumption: Negligible invalid IV induced bias} is satisfied without perfect screening}\label{sup:sec4}

\subsection{Main results}

With a slight abuse of notation, we consider a special case where the instruments can be divided into three clusters: 
\begin{align*}
    &  \mathcal{V}_1 = \left\{j: \ r_j=0,\ \beta_{X_j}=\beta_0 \right\} \text{ with } |\mathcal{V}_1|=v_1, \\
    &  \mathcal{V}_2 = \left\{j: \ r_j=r_2,\ \beta_{X_j}=\beta_0 \right\} \text{ with } |\mathcal{V}_2|=v_2, \\
    &  \mathcal{V}_3 = \left\{j: \ r_j= r_3,\ \beta_{X_j}=\beta_0 \right\} \text{ with } |\mathcal{V}_3|=v_3. 
\end{align*}
Here, $\mathcal{V}_1$ represents the set of valid IVs with $r_j = 0$, and $\mathcal{V}_2$ represents the set of invalid IVs with vanishing pleiotropic effects with $r_2$ tending to zero at an appropriate rate (see Lemma \ref{lem: negiliable bias} and Theorem \ref{thm:check assumption 4} for its precise characterization), and $\mathcal{V}_3$ represents the set of invalid IVs with non-vanishing pleiotropic effects.  We note that it is not necessary to restrict all $\beta_{X_j}$'s to have the same magnitude, and our results presented in this section can be extended to cases where the standardized IV strength lies in a neighborhood of $\beta_0/\sigma_{X_j}$, in the sense that $\frac{\beta_{X_j}}{\sigma_{X_j}} \in \left[\frac{\beta_0}{\sigma_{X_j}} \pm \delta \times \frac{\beta_0}{\sigma_{X_j}}\right]$ with $\delta$ tending to zero.

To further simplify our theoretical derivation, we consider $\sigma_{Y_j}^2=\frac{1}{n}$ for all $j \in \mathcal{S}_\lambda$.  Next, for any subset $\mathcal{V} \subseteq \mathcal{S}_\lambda$, we further define the following quantities:
\begin{align*}
 p_1(\mathcal{V})=\frac{|\mathcal{V}_1 \cap \mathcal{V}|}{|\mathcal{V}|}, \quad    p_2(\mathcal{V})=\frac{|\mathcal{V}_2 \cap \mathcal{V}|}{|\mathcal{V}|}, \quad p_3(\mathcal{V})=\frac{|\mathcal{V}_3 \cap \mathcal{V}|}{|\mathcal{V}|}.
\end{align*}
Lastly, we denote $A_n= \big(ln(s_\lambda)\vee\kappa_n\big) \cdot \sqrt{s_\lambda/n }$. 

   
    


In what follows, we will argue that to satisfy Assumption \ref{assumption: Negligible invalid IV induced bias}, our invalid IV screening procedure does not need to have a perfect screening property. In other words, our estimator remains asymptotically unbiased even if our IV screening procedure does not select $\mathcal{V}_1$ with probability approaching one. As shall be made clear in Lemma \ref{lem: bias set screening} and Lemma \ref{lem: negiliable bias}, our method avoids the need for perfect IV screening by showing that the selected IV set $\hat{\mathcal{V}}$ can include both invalid IVs from $\mathcal{V}_2$ and a vanishing portion of invalid IVs from $\mathcal{V}_3$ in the selected set $\hat{\mathcal{V}}$. 

We impose the following conditions: 

\begin{condition}[The order of the number of valid IVs]\label{example:Order of the number of valid IVs}
The number of valid IVs $v_1$ is of the same order as $s_\lambda$. For the number of invalid IVs, there exists a positive constant $c_1 \in (0,1)$ such that $(v_2/v_1 \vee  v_3/v_1) \leq c_1$. 
\end{condition}
\noindent The above condition requires that the majority of the IVs included in MR are valid IVs. The next condition is needed so that our optimization problem does not suffer from potential over-fitting issues in high-dimensional settings: 


\begin{condition}[High-dimensional BIC]\label{example:high dimension BIC} $\kappa_{n}\gg  ln(s_{\lambda}).$
\end{condition}


Next, for any given $\epsilon>0$,  define a collection of sets 
\begin{align*}
    \bm{\mathcal{V}}_{\texttt{bias}}(\epsilon) = & \bm{\mathcal{V}}(\epsilon) \cup \bm{\mathcal{V}}_{\texttt{BIC}} \\
    = & \left\{\mathcal{V} \Big| \mathcal{V} \subseteq \mathcal{S}_\lambda, p_3(\mathcal{V})\geq \frac{\epsilon}{\sqrt{n s_\lambda }\cdot r_3} \right\} \bigcup \left\{\mathcal{V} \Big| \mathcal{V} \subseteq \mathcal{S}_\lambda, |\mathcal{V}|=v < \frac{1+c_1}{2}\cdot v_1 \right\}.
\end{align*}
$\bm{\mathcal{V}}_{\texttt{bias}}(\epsilon)$ is a union of two types of sets that will be screened out by our invalid IV screening procedure. The first set $\bm{\mathcal{V}}(\epsilon)$ consists of all possible sets with a non-vanishing proportion of IVs in $\mathcal{V}_3$, defined by the condition $p_3(\mathcal{V})\geq \frac{\epsilon}{\sqrt{n s_\lambda }\cdot r_3}$. Consequently, if our selected set $\hat{\mathcal{V}}$ belongs to $\bm{\mathcal{V}}(\epsilon)$, the resulting causal effect estimator is biased. The second set $ \bm{\mathcal{V}}_{\texttt{BIC}}$ comprises all sets containing a total number of IVs smaller than $v_1$.
As our IV screening procedure adopts $l_0$ penalty with BIC to screen out invalid IVs, our selected set $\hat{\mathcal{V}}$ tends to select an IV set with cardinality larger than $v_1$. Therefore, $\hat{\mathcal{V}}$ does not belong to  $ \bm{\mathcal{V}}_{\texttt{BIC}}$ as well. The following lemma provides rigorous statement about our selected IV set $\hat{\mathcal{V}}$:


\begin{lem}\label{lem: bias set screening}
 For any given $\epsilon>0$, if $r_3$ is sufficiently large in the sense that $A_n/(r_3\varepsilon) = o(1)$ and $|\hat{\theta}(\mathcal{V})|$ is bounded by a constant for all $\mathcal{V}\in \mathcal{S}_{\lambda}$,  then under Condition \ref{Bound of Orlicz norm}, \ref{Orders of the variances and sample sizes}, \ref{example:Order of the number of valid IVs} and \ref{example:high dimension BIC}, the selected IV set  $\hat{\mathcal{V}}$ using our procedure satisfies $ \mathbb{P}(\hat{\mathcal{V}} \in \bm{\mathcal{V}}_{\texttt{bias}}(\epsilon))\rightarrow 0.$
\end{lem}

Next, we demonstrate that when $r_2$ tends to zero at an appropriate rate and $\beta_0$ are sufficiently large, for the set $\mathcal{V}$ that does not fall into $ \bm{\mathcal{V}}_{\texttt{bias}}(\epsilon)$, the bias term described in Assumption 3 is asymptotically negligible: 

\begin{lem}\label{lem: negiliable bias} We choose \( a_\lambda \asymp s_\lambda \cdot \kappa_\lambda = n s_\lambda \cdot \beta_0^2 \) to stabilize the variance (other choices for \( a_\lambda \) can also be adopted). For any given $\epsilon>0$, whenever $r_2<\frac{\epsilon}{\sqrt{n s_\lambda }}$ and $\frac{r_3}{\beta_0^2\varepsilon\cdot \sqrt{n s_{\lambda}} } =o(1)$, under Condition \ref{Bound of Orlicz norm} and \ref{Orders of the variances and sample sizes}, we have 
    $$ |\frac{a_\lambda  }{\sqrt{s_\lambda \cdot \kappa_\lambda}}\cdot\frac{\sum_{j \in \mathcal{V}} r_j\cdot\hat\beta_{X_{j,\mathtt{RB}}}}{\sum_{j \in \mathcal{V}} \hat\beta_{X_{j,\mathtt{RB}}}^2-\hat\sigma_{X_{j,\mathtt{RB}}}^2}|= O_p(\epsilon).
    $$
for any $\mathcal{V} \notin \bm{\mathcal{V}}_{\texttt{bias}}(\epsilon)$.
\end{lem}

In Lemma \ref{lem: negiliable bias}, $\epsilon$ can tend to zero at different rates, each affecting the conditions on $r_2$, $r_3$, and $\beta_0$ differently. To cast some insights into this result, we consider a simple example. For a positive constant $\delta$, we let 
\begin{align*}
    \epsilon = \frac{1}{\ln^\delta(s_\lambda)} = o(1), \quad \kappa_n = \ln^{1+\delta}(s_\lambda) \gg \ln(s_\lambda), \quad r_3 = \ln^{1+3\delta}(s_\lambda)\cdot \sqrt{\frac{s_\lambda}{n}}. 
\end{align*}
If $r_2$ and $\beta_0$ satisfy
\begin{align*}
    r_2 < \frac{1}{\ln^\delta(s_\lambda) \cdot \sqrt{n s_\lambda}}, \quad \beta_0 \gg \sqrt{\frac{\ln^{1+4\delta}(s_\lambda)}{n}}, 
\end{align*}
the conditions of Lemma \ref{lem: negiliable bias} are met. Here, the above requirement on the magnitude of $\beta_0$ is rather mild, as the selected IV strength in $\mathcal{S}_{\lambda}$ typically has an order greater than $\sqrt{\log p / n}$, since the cut-off value $\lambda$ is often of the order $\sqrt{\log p}$. In practice, since relevant IVs often constitute only a small fraction of all candidate IVs, $s_{\lambda}$ should be a term of smaller order compared to $p$. Therefore, the condition $\beta_0 \gg \sqrt{\ln^{1+4\delta}(s_\lambda)/n}$ that we impose here is rather mild.

With these two lemmas, we are ready to show that the set $\hat{\mathcal{V}}$ selected by our proposed invalid screening procedure induces negligible bias: 

\begin{thm}\label{thm:check assumption 4}
For a vanishing number $\varepsilon>0$, we assume that 
\begin{enumerate}
    \item[(i)]  $r_3$ is sufficiently large in the sense that $A_n/(r_3\varepsilon) = o(1)$, which ensures our invalid screening procedure to effectively screen out IVs from $\mathcal{V}_3$. 
    \item[(ii)] $r_2$ is a vanishing number in the sense that $r_2<\frac{\epsilon}{\sqrt{n s_\lambda }}$, which ensures IVs from $\mathcal{V}_2$ to have vanishing pleiotrophic effects.
    \item[(iii)] $\beta_0$ is sufficiently large in the sense that  $\frac{\sqrt{s_\lambda} ln(s_\lambda)}{\sqrt{n}}\frac{r_3}{\epsilon\beta_0^2}\rightarrow 0$. 
\end{enumerate}
If $|\hat{\theta}(\mathcal{V})|$ is bounded by a constant for all $\mathcal{V}\in \mathcal{S}_{\lambda}$, under Condition \ref{Bound of Orlicz norm}, \ref{Orders of the variances and sample sizes},  \ref{example:Order of the number of valid IVs} and \ref{example:high dimension BIC}, choosing $a_\lambda \asymp s_\lambda \cdot \kappa_\lambda= n s_\lambda \cdot\beta_0^2$ to stabilize the variance, we can prove that 
\begin{align*}
        \frac{a_\lambda  }{\sqrt{s_\lambda \cdot \kappa_\lambda}}\cdot\frac{\sum_{j \in \hat{\mathcal{V}}} r_j\cdot\hat\beta_{X_{j,\mathtt{RB}}}}{\sum_{j \in \hat{\mathcal{V}}} \hat\beta_{X_{j,\mathtt{RB}}}^2-\hat\sigma_{X_{j,\mathtt{RB}}}^2}= o_p(1).
\end{align*}
\end{thm}
We note that the third condition in the above theorem is slightly stronger than what was assumed in Lemma 6, as we applied a union bound, needed to account for uniformity across all possible subsets of $\mathcal{S}_{\lambda}$. The conditions we impose here are sufficient but by no means necessary.


%

\subsection{Proof of Theorem \ref{thm:check assumption 4}}

For any given $\epsilon>0$, we have
\begin{align*} 
    \frac{a_\lambda  }{\sqrt{s_\lambda \cdot \kappa_\lambda}}\cdot\frac{\sum_{j \in \hat{\mathcal{V}}} r_j\cdot\hat\beta_{X_{j,\mathtt{RB}}}}{\sum_{j \in \hat{\mathcal{V}}} \hat\beta_{X_{j,\mathtt{RB}}}^2-\hat\sigma_{X_{j,\mathtt{RB}}}^2}& =    \sum_{\mathcal{V} \in  \bm{\mathcal{V}}_{\texttt{bias}}(\epsilon)} \frac{a_\lambda  }{\sqrt{s_\lambda \cdot \kappa_\lambda}}\cdot\frac{\sum_{j \in \mathcal{V}} r_j\cdot\hat\beta_{X_{j,\mathtt{RB}}}}{\sum_{j \in \mathcal{V}} \hat\beta_{X_{j,\mathtt{RB}}}^2-\hat\sigma_{X_{j,\mathtt{RB}}}^2} \cdot \mathrm{1}(\hat{\mathcal{V}}=\mathcal{V})\\& + \sum_{\mathcal{V} \notin  \bm{\mathcal{V}}_{\texttt{bias}}(\epsilon)} \frac{a_\lambda  }{\sqrt{s_\lambda \cdot \kappa_\lambda}}\cdot\frac{\sum_{j \in \mathcal{V}} r_j\cdot\hat\beta_{X_{j,\mathtt{RB}}}}{\sum_{j \in \mathcal{V}} \hat\beta_{X_{j,\mathtt{RB}}}^2-\hat\sigma_{X_{j,\mathtt{RB}}}^2}\cdot \mathrm{1}(\hat{\mathcal{V}}=\mathcal{V})
\end{align*}

For any $\epsilon_0>0$, the first term in the right hand side can be bounded using the following inequality:
\begin{align*}
   \mathbb{P}(|  \sum_{\mathcal{V} \in \bm{\mathcal{V}}_{\texttt{bias}}(\epsilon)} \frac{a_\lambda  }{\sqrt{s_\lambda \cdot \kappa_\lambda}}\cdot\frac{\sum_{j \in \mathcal{V}} r_j\cdot\hat\beta_{X_{j,\mathtt{RB}}}}{\sum_{j \in \mathcal{V}} \hat\beta_{X_{j,\mathtt{RB}}}^2-\hat\sigma_{X_{j,\mathtt{RB}}}^2} \cdot \mathrm{1}(\hat{\mathcal{V}}=\mathcal{V})|>\epsilon_0)\leq \mathbb{P}(\cup_{\mathcal{V} \in \bm{\mathcal{V}}_{\texttt{bias}}(\epsilon)} \{\hat{\mathcal{V}}=\mathcal{V}\})=\mathbb{P}(\hat{\mathcal{V}}\in \bm{\mathcal{V}}_{\texttt{bias}}(\epsilon))
\end{align*}

By using the result in Lemma \ref{lem: bias set screening} and letting $\epsilon_0\rightarrow 0$, we are able to show 
$$
\sum_{\mathcal{V} \in \bm{\mathcal{V}}_{\texttt{bias}}(\epsilon)} \frac{a_\lambda  }{\sqrt{s_\lambda \cdot \kappa_\lambda}}\cdot\frac{\sum_{j \in \mathcal{V}} r_j\cdot\hat\beta_{X_{j,\mathtt{RB}}}}{\sum_{j \in \mathcal{V}} \hat\beta_{X_{j,\mathtt{RB}}}^2-\hat\sigma_{X_{j,\mathtt{RB}}}^2} \cdot \mathrm{1}(\hat{\mathcal{V}}=\mathcal{V})=o_p(1).
$$

Thus it suffices to show that the second term on the right-hand side satisfies
\begin{align*}
    \sum_{\mathcal{V} \notin \bm{\mathcal{V}}_{\texttt{bias}}(\epsilon)} \frac{a_\lambda  }{\sqrt{s_\lambda \cdot \kappa_\lambda}}\cdot\frac{\sum_{j \in \mathcal{V}} r_j\cdot\hat\beta_{X_{j,\mathtt{RB}}}}{\sum_{j \in \mathcal{V}} \hat\beta_{X_{j,\mathtt{RB}}}^2-\hat\sigma_{X_{j,\mathtt{RB}}}^2}\cdot \mathrm{1}(\hat{\mathcal{V}}=\mathcal{V})=o_p(1).
\end{align*}

In Lemma \ref{lem: negiliable bias}, we show that under the event $\mathcal{A}(\mathcal{V},\epsilon)$, we have 
\begin{align*}
          |\frac{a_\lambda  }{\sqrt{s_\lambda \cdot \kappa_\lambda}}\cdot\frac{\sum_{j \in \mathcal{V}} r_j\cdot\hat\beta_{X_{j,\mathtt{RB}}}}{\sum_{j \in \mathcal{V}} \hat\beta_{X_{j,\mathtt{RB}}}^2-\hat\sigma_{X_{j,\mathtt{RB}}}^2}|< 9\epsilon
\end{align*}
for any given $\mathcal{V} \notin \bm{\mathcal{V}}_{\texttt{bias}}(\epsilon)$. 

Under the event $\bigcap_{\mathcal{V} \notin \bm{\mathcal{V}}_{\texttt{bias}}(\epsilon)} \mathcal{A}(\mathcal{V},\epsilon)$, we can show that this holds uniformly for all $\mathcal{V} \notin \bm{\mathcal{V}}_{\texttt{bias}}(\epsilon)$. Thus we have
\begin{align*}
    \sum_{\mathcal{V} \notin \bm{\mathcal{V}}_{\texttt{bias}}(\epsilon)} \frac{a_\lambda  }{\sqrt{s_\lambda \cdot \kappa_\lambda}}\cdot\frac{\sum_{j \in \mathcal{V}} r_j\cdot\hat\beta_{X_{j,\mathtt{RB}}}}{\sum_{j \in \mathcal{V}} \hat\beta_{X_{j,\mathtt{RB}}}^2-\hat\sigma_{X_{j,\mathtt{RB}}}^2}\cdot \mathrm{1}(\hat{\mathcal{V}}=\mathcal{V})<9\epsilon.
\end{align*}
We also notice that 
\begin{align*}
   \mathbb{P}\Big(\bigcap_{\mathcal{V} \notin \bm{\mathcal{V}}_{\texttt{bias}}(\epsilon)} \mathcal{A}(\mathcal{V},\epsilon)\Big)= 1-  \mathbb{P}\Big(\bigcup_{\mathcal{V} \notin \bm{\mathcal{V}}_{\texttt{bias}}(\epsilon)} \mathcal{A}^\mathcal{C}(\mathcal{V},\epsilon)\Big)\geq  1-  \sum_{\mathcal{V} \notin \bm{\mathcal{V}}_{\texttt{bias}}(\epsilon)}\mathbb{P}\Big( \mathcal{A}^\mathcal{C}(\mathcal{V},\epsilon)\Big)
\end{align*}
Here $\mathcal{A}^\mathcal{C}(\mathcal{V},\epsilon)$ is the complement of the event $\mathcal{A}(\mathcal{V},\epsilon)$.

To prove $\mathbb{P}\Big(\bigcap_{\mathcal{V} \notin \bm{\mathcal{V}}_{\texttt{bias}}(\epsilon)} \mathcal{A}(\mathcal{V},\epsilon)\Big) \rightarrow 1$, we only need to show 

$$\sum_{\mathcal{V} \notin \bm{\mathcal{V}}_{\texttt{bias}}(\epsilon)}\mathbb{P}\Big( \mathcal{A}^\mathcal{C}(\mathcal{V},\epsilon)\Big)\leq e^{(s_\lambda+1)\cdot ln(s_\lambda)} \max_{\mathcal{V} \notin \bm{\mathcal{V}}_{\texttt{bias}}(\epsilon)}\mathbb{P}\Big( \mathcal{A}^\mathcal{C}(\mathcal{V},\epsilon)\Big)\rightarrow 0.$$

In Lemma \ref{lem: negiliable bias}, we have $ \mathbb{P}(\mathcal{A}(\mathcal{V},\epsilon))\geq 1-2\cdot e^{ -\frac{c}{16} n \cdot v\cdot \beta_0^2}-4\cdot e^{ -c \cdot \min\{\frac{1}{16}\cdot n^2\cdot v\beta_0^4, \frac{1}{4}\cdot n \cdot v \beta_0^2\}}-2\cdot e^{ \frac{-c \cdot\epsilon^2\cdot \beta_0^2}{p_2(\mathcal{V})r_2^2+p_3(\mathcal{V})r_3^2} \frac{v}{ 16 s_\lambda }}$, thus $ \mathbb{P}(\mathcal{A}^\mathcal{C}(\mathcal{V},\epsilon))< 2\cdot e^{ -\frac{c}{16} n \cdot v\cdot \beta_0^2}+4\cdot e^{ -c \cdot \min\{\frac{1}{16}\cdot n^2\cdot v\beta_0^4, \frac{1}{4}\cdot n \cdot v \beta_0^2\}}+2\cdot e^{ \frac{-c \cdot\epsilon^2\cdot \beta_0^2}{p_2(\mathcal{V})r_2^2+p_3(\mathcal{V})r_3^2} \frac{v}{ 16 s_\lambda }}$. In addition, we have $v>\frac{1+c_1}{2}\cdot v_1$ for any $\mathcal{V} \notin \bm{\mathcal{V}}_{\texttt{bias}}(\epsilon)$. Under Condition \ref{example:Order of the number of valid IVs}, we have $v \asymp s_\lambda $ uniformly hold for any $\mathcal{V} \notin \bm{\mathcal{V}}_{\texttt{bias}}(\epsilon)$.

 With these results, we can prove  $e^{(s_\lambda+1)\cdot ln(s_\lambda)} \max_{\mathcal{V} \notin \bm{\mathcal{V}}_{\texttt{bias}}(\epsilon)}\mathbb{P}\Big( \mathcal{A}^\mathcal{C}(\mathcal{V},\epsilon)\Big)\rightarrow 0$ if we have  $\frac{\sqrt{s_\lambda} ln(s_\lambda)}{\sqrt{n}}\frac{r_3}{\epsilon\beta_0^2}\rightarrow 0$ and $r_2<\frac{\epsilon}{\sqrt{n s_\lambda }}$.
    
When $\epsilon$ is a vanishing number, we can show 
\begin{align*}
    \sum_{\mathcal{V} \notin \bm{\mathcal{V}}_{\texttt{bias}}(\epsilon)} \frac{a_\lambda  }{\sqrt{s_\lambda \cdot \kappa_\lambda}}\cdot\frac{\sum_{j \in \mathcal{V}} r_j\cdot\hat\beta_{X_{j,\mathtt{RB}}}}{\sum_{j \in \mathcal{V}} \hat\beta_{X_{j,\mathtt{RB}}}^2-\hat\sigma_{X_{j,\mathtt{RB}}}^2}\cdot \mathrm{1}(\hat{\mathcal{V}}=\mathcal{V})=o_p(1).
\end{align*}
Thus,
\begin{align*}
     \frac{a_\lambda  }{\sqrt{s_\lambda \cdot \kappa_\lambda}}\cdot\frac{\sum_{j \in \hat{\mathcal{V}}} r_j\cdot\hat\beta_{X_{j,\mathtt{RB}}}}{\sum_{j \in \hat{\mathcal{V}}} \hat\beta_{X_{j,\mathtt{RB}}}^2-\hat\sigma_{X_{j,\mathtt{RB}}}^2}=o_p(1).
\end{align*}

Therefore, although our selection procedure does not have perfect screening properties, the set that we select still has negligible bias.

\subsection{Proof of Lemma \ref{lem: bias set screening}} 
For any $\mathcal{V} \subseteq \mathcal{{S}}_{\lambda}$, we denote a collection of sparse vectors 
\begin{align*}
    \bm{\mathcal{R}}_{\mbox{v}} =  \left\{ \bm{a} \in \mathbb{R}^{|\mathcal{S}_{\lambda}|\times 1} :  a_j = 0,\ \text{for }j\in \mathcal{V}, \ a_k \neq 0,\ \text{for } k \in \mathcal{V}^c  \right\}
\end{align*}
and a function 
\begin{align*}
h(\mathcal{V}, \theta)&=\min\limits_{ \bm{r}\in   \bm{\mathcal{R}}_{\mbox{v}}} \sum\limits_{j\in{\mathcal{S}}_{\lambda}}  \hat l\left( \theta , \bm{r};   \hat \beta_{Y_j},\sigma_{Y_j},\hat \beta_{X_{j,\mathtt{RB}}},\hat{\sigma}_{X_{j,\mathtt{RB}}}\right)\\
&=\sum\limits_{j\in \mathcal{V}} \frac{(\hat \beta_{Y_j}-\theta \cdot \hat \beta_{X_{j,\mathtt{RB}}})^2-\theta^2 \cdot \hat \sigma_{X_{j,\mathtt{RB}}}^2}{\sigma_{Y_j}^2}.
\end{align*}

For any given $\epsilon>0$, we define the set $\bm{\mathcal{V}}_{\texttt{bias}}(\epsilon)=\{\mathcal{V} | \mathcal{V} \subseteq \mathcal{S}_\lambda, p_3(\mathcal{V})\geq \frac{\epsilon}{\sqrt{n s_\lambda }\cdot r_3} \} \cup \{\mathcal{V} | \mathcal{V} \subseteq \mathcal{S}_\lambda, |\mathcal{V}|=v < \frac{1+c_1}{2}\cdot v_1 \}.$
Now we want to analyze $\mathbb{P}(    \hat{\mathcal{V}}\in \bm{\mathcal{V}}_{\texttt{bias}}(\epsilon)) $ by utilizing the following inequality: 
\begin{equation}\label{eq:selection-inconsistency-inequality}
\begin{aligned}
  \quad \mathbb{P}(\hat{\mathcal{V}}\in  \bm{\mathcal{V}}_{\texttt{bias}}(\epsilon))
  &\leq \mathbb{P}\Big(\min\limits_{v\in\mathbb{N}_{+}, v \leq s_\lambda}\big[\min\limits_{|\mathcal{V}|=v, \mathcal{V}\in \bm{\mathcal{V}}_{\texttt{bias}}(\epsilon)} \min\limits_{\theta \in \mathbb{R}} h(\mathcal{V},  \theta)-\kappa_{n}\cdot v\big] \leq\min\limits_{\theta \in \mathbb{R}} h(\mathcal{V}_1, \theta)-\kappa_{n}\cdot v_1\Big) \\&\leq \bigcup\limits_{ \mathcal{V}\in \bm{\mathcal{V}}_{\texttt{bias}}(\epsilon) }\mathbb{P}\Big( \min\limits_{\theta \in \mathbb{R}} h(\mathcal{V},  \theta)-\kappa_{n}\cdot |\mathcal{V}|\leq\min\limits_{\theta \in \mathbb{R}} h(\mathcal{V}_1, \theta)-\kappa_{n}\cdot v_1\Big) 
    \\&\leq \sum\limits_{v=1}^{s_\lambda} \tbinom{s_\lambda}{v}\max\limits_{|\mathcal{V}|=v, \mathcal{V}\in \bm{\mathcal{V}}_{\texttt{bias}}(\epsilon)}\mathbb{P}\Big(\min\limits_{\theta \in \mathbb{R}} h(\mathcal{V},  \theta)-\kappa_{n}\cdot v \leq  h(\mathcal{V}_1, \theta_0)-\kappa_{n}\cdot v_1\Big) \\&\leq \sum\limits_{v=1}^{s_\lambda} {s_\lambda}^{v}\max\limits_{|\mathcal{V}|=v, \mathcal{V}\in\bm{\mathcal{V}}_{\texttt{bias}}(\epsilon)}\mathbb{P}\Big(\min\limits_{\theta \in \mathbb{R}} h(\mathcal{V},  \theta)-\kappa_{n}\cdot v \leq  h(\mathcal{V}_1, \theta_0)-\kappa_{n}\cdot v_1\Big)  \\&\leq  \max\limits_{\mathcal{V}\in\bm{\mathcal{V}}_{\texttt{bias}}(\epsilon)}e^{(s_{\lambda}+1)\cdot ln(s_{\lambda})}\cdot\mathbb{P}\Big(\min\limits_{\theta \in \mathbb{R}} h(\mathcal{V},  \theta)-\kappa_{n}\cdot |\mathcal{V}| \leq  h(\mathcal{V}_1, \theta_0)-\kappa_{n}\cdot v_1\Big)\\&=e^{(s_{\lambda}+1)\cdot ln(s_{\lambda})}\cdot\mathbb{P}\Big(\min\limits_{\theta \in \mathbb{R}} h(\mathcal{V}^*,  \theta)-\kappa_{n}\cdot v^* \leq  h(\mathcal{V}_1, \theta_0)-\kappa_{n}\cdot v_1\Big).
\end{aligned}
\end{equation}
where $\mathcal{V}^*=\argmax\limits_{\mathcal{V}\in\bm{\mathcal{V}}_{\texttt{bias}}(\epsilon)}e^{(s_{\lambda}+1)\cdot ln(s_{\lambda})}\cdot\mathbb{P}\Big(\min\limits_{\theta \in \mathbb{R}} h(\mathcal{V},  \theta)-\kappa_{n}\cdot |\mathcal{V}| \leq  h(\mathcal{V}_1, \theta_0)-\kappa_{n}\cdot v_1\Big)$ and $v^*=|\mathcal{V}^*|$.

As we show that 
\begin{align}\label{eq:target}
e^{(s_{\lambda}+1)\cdot ln(s_{\lambda})}\cdot\mathbb{P}\Big(\min\limits_{\theta \in \mathbb{R}} h(\mathcal{V}^*,  \theta)-\kappa_{n}\cdot v^* \leq  h(\mathcal{V}_1, \theta_0)-\kappa_{n}\cdot v_1\Big)\rightarrow 0,
\end{align} 
then $\mathbb{P}(\hat{\mathcal{V}}\in  \bm{\mathcal{V}}_{\texttt{bias}}(\epsilon))\rightarrow 0$ holds. Here, we also note that the first equation in Eq \eqref{eq:selection-inconsistency-inequality} follows from the definition of the optimization problem defined in Equation 2 in the manuscript, the second to the fifth inequalities in Eq \eqref{eq:selection-inconsistency-inequality} hold following $\min\limits_{ \theta \in \mathbb{R}} h(\mathcal{V}_1,\theta)\leq h(\mathcal{V}_1, \theta_0)$, $\tbinom{s_\lambda}{v}\leq s_\lambda^v $ and some basic calculations. 

To prove formula (\ref{eq:target}), we need to analyze the asymptotic properties of $ h(\mathcal{V}_1, \theta_0)$, $ \min\limits_{\theta \in \mathbb{R}} h(\mathcal{V}^*,  \theta)$ and $\kappa_{n}$.

We start with $h(\mathcal{V}_1, \theta_0)$ and decompose it below following our notation defined in Section \ref{notation} 
\begin{align}
   h(\mathcal{V}_1, \theta_0)&\nonumber=\sum\limits_{j\in \mathcal{V}_1} \frac{(\hat \beta_{Y_j}-\theta_0 \cdot \beta_{X_j})^2  }{\sigma_{Y_j}^2}+\theta_0^2 \cdot \sum\limits_{j\in \mathcal{V}_1} \frac{ ( \hat \beta_{X_{j,\mathtt{RB}}}-\beta_{X_j})^2- \hat\sigma_{X_{j,\mathtt{RB}}}^2}{\sigma_{Y_j}^2}\\
   &\nonumber-2\theta_0\cdot \sum\limits_{j\in  \mathcal{V}_1} \frac{(\hat \beta_{Y_j}-\theta_0 \cdot \beta_{X_j})( \hat \beta_{X_{j,\mathtt{RB}}}-\beta_{X_j})}{\sigma_{Y_j}^2} \\
   & \label{eq:expansion-of-h}= \sum\limits_{j\in \mathcal{V}_1} \frac{\nu_j^2  }{\sigma_{Y_j}^2}+\theta_0^2 \cdot \sum\limits_{j\in \mathcal{V}_1} \frac{  u_j^2- \hat\sigma_{X_{j,\mathtt{RB}}}^2}{\sigma_{Y_j}^2}-2\theta_0\cdot \sum\limits_{j\in  \mathcal{V}_1} \frac{\nu_ju_j}{\sigma_{Y_j}^2}.
\end{align}

Next, we study the asymptotic property of $ \min\limits_{\theta \in \mathbb{R}} h(\mathcal{V}^*,  \theta)$.
We denote $\hat \theta(\mathcal{V}^*)=\arg\min\limits_{\theta\in \mathbb{R}}h(\mathcal{V}^*,\theta)$ and decompose $h(\mathcal{V}^*,\hat{\theta}(\mathcal{V}^*))$ in a similar way as $h(\mathcal{V}_1,\theta_0)$, following our notation in Section \ref{notation}.

\begin{align*}
    h(\mathcal{V}^*, \hat {\theta}(\mathcal{V}^*))&=\nonumber\sum\limits_{j\in \mathcal{V}^*} \frac{(\hat \beta_{Y_j}-\hat{\theta}(\mathcal{V}^*) \cdot \beta_{X_j})^2  }{\sigma_{Y_j}^2}+\hat\theta(\mathcal{V}^*)^2 \sum\limits_{j\in \mathcal{V}^*} \frac{( \hat \beta_{X_{j,\mathtt{RB}}}-\beta_{X_j})^2- \hat\sigma_{X_{j,\mathtt{RB}}}^2}{\sigma_{Y_j}^2}\\
    &+2\hat{\theta}(\mathcal{V}^*) \sum\limits_{j\in  \mathcal{V}^*} \frac{(\hat{\theta}(\mathcal{V}^*)\cdot \beta_{X_j}-\theta_0\cdot \beta_{X_j}-r_j)\cdot( \hat \beta_{X_{j,\mathtt{RB}}}-\beta_{X_j})}{\sigma_{Y_j}^2}\\
    &-2\hat{\theta}(\mathcal{V}^*)  \sum\limits_{j\in \mathcal{V}^*} \frac{(\hat \beta_{Y_j}-\theta_0 \cdot \beta_{X_j}-r_j)( \hat \beta_{X_{j,\mathtt{RB}}}-\beta_{X_j})}{\sigma_{Y_j}^2}\\
    &=\sum\limits_{j\in \mathcal{V}^*} \frac{(\theta_0 \cdot \beta_{X_j}+r_j+\nu_j -\hat{\theta}(\mathcal{V}^*) \cdot \beta_{X_j})^2  }{\sigma_{Y_j}^2}+\hat\theta(\mathcal{V}^*)^2  \sum\limits_{j\in \mathcal{V}^*} \frac{u_j^2- \hat\sigma_{X_{j,\mathtt{RB}}}^2}{\sigma_{Y_j}^2}\\&+2\hat{\theta}(\mathcal{V}^*) \sum\limits_{j\in  \mathcal{V}^*} \frac{(\hat{\theta}(\mathcal{V}^*)\cdot \beta_{X_j}-\theta_0\cdot \beta_{X_j}-r_j)\cdot u_j}{\sigma_{Y_j}^2}-2\hat{\theta}(\mathcal{V}^*)  \sum\limits_{j\in \mathcal{V}^*} \frac{\nu_j u_j}{\sigma_{Y_j}^2}\\
    &=\sum\limits_{j\in \mathcal{V}^*} \frac{(\theta_0 \cdot \beta_{X_j}+r_j-\hat{\theta}(\mathcal{V}^*) \cdot \beta_{X_j})^2  }{\sigma_{Y_j}^2}+\sum\limits_{j\in \mathcal{V}^*} \frac{\nu_j ^2}{\sigma_{Y_j}^2}+2\sum\limits_{j\in \mathcal{V}^*} \frac{(\theta_0 \cdot \beta_{X_j}+r_j -\hat{\theta}(\mathcal{V}^*)\cdot \beta_{X_j}) \cdot \nu_j }{\sigma_{Y_j}^2}\\&+\hat\theta(\mathcal{V}^*)^2  \sum\limits_{j\in \mathcal{V}^*} \frac{u_j^2- \hat\sigma_{X_{j,\mathtt{RB}}}^2}{\sigma_{Y_j}^2}+2\hat{\theta}(\mathcal{V}^*) \sum\limits_{j\in  \mathcal{V}^*} \frac{(\hat{\theta}(\mathcal{V}^*)\cdot \beta_{X_j}-\theta_0\cdot \beta_{X_j}-r_j)\cdot u_j}{\sigma_{Y_j}^2}-2\hat{\theta}(\mathcal{V}^*)  \sum\limits_{j\in \mathcal{V}^*} \frac{\nu_j u_j}{\sigma_{Y_j}^2}.
\end{align*}

With these decompositions, the Equation (2) can be rewritten as 
\begin{align*}
 &\quad\mathbb{P}\Big(\min\limits_{\theta \in \mathbb{R}} h(\mathcal{V}^*,  \theta)-\kappa_{n}\cdot v^* \leq  h(\mathcal{V}_1, \theta_0)-\kappa_{n}\cdot v_1\Big)
 \\&=\mathbb{P}\Big(\sum\limits_{j\in \mathcal{V}^*} \frac{(\theta_0 \cdot \beta_{X_j}+r_j-\hat{\theta}(\mathcal{V}^*) \cdot \beta_{X_j})^2  }{\sigma_{Y_j}^2}+\kappa_{n}\cdot 
(v_1- v^*)
 \leq -\sum\limits_{j\in \mathcal{V}^*} \frac{\nu_j ^2}{\sigma_{Y_j}^2}-2\sum\limits_{j\in \mathcal{V}^*} \frac{(\theta_0 \cdot \beta_{X_j}+r_j -\hat{\theta}(\mathcal{V}^*)\cdot \beta_{X_j}) \cdot \nu_j }{\sigma_{Y_j}^2}
 \\&\quad-\hat\theta(\mathcal{V}^*)^2  \sum\limits_{j\in \mathcal{V}^*} \frac{u_j^2- \sigma_{X_{j,\mathtt{RB}}}^2}{\sigma_{Y_j}^2}+\hat\theta(\mathcal{V}^*)^2  \sum\limits_{j\in \mathcal{V}^*} \frac{\hat\sigma_{X_{j,\mathtt{RB}}}^2- \sigma_{X_{j,\mathtt{RB}}}^2}{\sigma_{Y_j}^2}-2\hat{\theta}(\mathcal{V}^*) \sum\limits_{j\in  \mathcal{V}^*} \frac{(\hat{\theta}(\mathcal{V}^*)\cdot \beta_{X_j}-\theta_0\cdot \beta_{X_j}-r_j)\cdot u_j}{\sigma_{Y_j}^2}
 \\&\quad+2\hat{\theta}(\mathcal{V}^*)  \sum\limits_{j\in \mathcal{V}^*} \frac{\nu_j u_j}{\sigma_{Y_j}^2}+\sum\limits_{j\in \mathcal{V}_1} \frac{\nu_j^2  }{\sigma_{Y_j}^2}+\theta_0^2   \sum\limits_{j\in \mathcal{V}_1} \frac{  u_j^2- \sigma_{X_{j,\mathtt{RB}}}^2}{\sigma_{Y_j}^2}-\theta_0^2   \sum\limits_{j\in \mathcal{V}_1} \frac{  \hat\sigma_{X_{j,\mathtt{RB}}}^2- \sigma_{X_{j,\mathtt{RB}}}^2}{\sigma_{Y_j}^2} -2\theta_0  \sum\limits_{j\in  \mathcal{V}_1} \frac{\nu_ju_j}{\sigma_{Y_j}^2}\Big)  .
 \end{align*}

By some calculations,  we can see that 
\begin{align*}
    \sum\limits_{j\in \mathcal{V}^*} \frac{(\theta_0 \cdot \beta_{X_j}+r_j-\hat{\theta}(\mathcal{V}^*) \cdot \beta_{X_j})^2  }{\sigma_{Y_j}^2}\geq \min\limits_{\theta \in \mathbb{R}}\sum\limits_{j\in \mathcal{V}^*} \frac{(\theta_0 \cdot \beta_{X_j}+r_j-\theta \cdot \beta_{X_j})^2 }{\sigma_{Y_j}^2} =\sum\limits_{j\in \mathcal{V}^*}\frac{ r_j^2}{\sigma_{Y_j}^2}-\frac{(\sum\limits_{j\in \mathcal{V}^*}\frac{
r_j\cdot\beta_{X_j}}{\sigma_{Y_j}^2})^2}{\sum\limits_{j\in \mathcal{V}^*}\frac{ \beta_{X_j}^2}{\sigma_{Y_j}^2}}.
\end{align*}
with probability 1.

Let 
$$
\Delta(\mathcal{V}^*)=\frac{1}{v^*}(\sum\limits_{j\in \mathcal{V}^*}\frac{ r_j^2}{\sigma_{Y_j}^2}-\frac{(\sum\limits_{j\in \mathcal{V}^*}\frac{
r_j\cdot\beta_{X_j}}{\sigma_{Y_j}^2})^2}{\sum\limits_{j\in \mathcal{V}^*}\frac{ \beta_{X_j}^2}{\sigma_{Y_j}^2}})  \text{ , } \quad \sum\limits_{j\in \mathcal{V}^*}\frac{ r_j^2}{\sigma_{Y_j}^2}-\frac{(\sum\limits_{j\in \mathcal{V}^*}\frac{
r_j\cdot\beta_{X_j}}{\sigma_{Y_j}^2})^2}{\sum\limits_{j\in \mathcal{V}^*}\frac{ \beta_{X_j}^2}{\sigma_{Y_j}^2}}= v^*\cdot \Delta(\mathcal{V}^*).
$$

 When $\mathcal{V}^* \in \bm{\mathcal{V}}_{\texttt{bias}}(\epsilon)$ such that $|\mathcal{V}^*|=v^*\geq \frac{1+c_1}{2} \cdot v_1$ and $\mathcal{V}^* \neq \mathcal{V}_1$, under Condition \ref{Boundedness}, there exists a $C_0>0$ such that 
{\small
\begin{align*}
 &\quad\mathbb{P}\Big(\min\limits_{\theta \in \mathbb{R}} h(\mathcal{V}^*,  \theta)-\kappa_{n}\cdot v^* \leq  h(\mathcal{V}_1, \theta_0)-\kappa_{n}\cdot v_1\Big)
 \\&\leq \mathbb{P}\Big(        \sum\limits_{j\in \mathcal{V}^*} \frac{(\theta_0 \cdot \beta_{X_j}+r_j-\hat{\theta}(\mathcal{V}^*) \cdot \beta_{X_j})^2  }{\sigma_{Y_j}^2}-\kappa_{n}\cdot (v^*-v_\lambda)
 \leq -\sum\limits_{j\in \mathcal{V}^*} \frac{\nu_j ^2-\sigma_{Y_j}^2}{\sigma_{Y_j}^2}-\hat\theta(\mathcal{V}^*)^2  \sum\limits_{j\in \mathcal{V}^*} \frac{u_j^2- \sigma_{X_{j,\mathtt{RB}}}^2}{\sigma_{Y_j}^2}
 \\&\quad +\hat\theta(\mathcal{V}^*)^2  \sum\limits_{j\in \mathcal{V}^*} \frac{\hat\sigma_{X_{j,\mathtt{RB}}}^2- \sigma_{X_{j,\mathtt{RB}}}^2}{\sigma_{Y_j}^2}-2 \sum\limits_{j\in  \mathcal{V}^*} \frac{(\hat{\theta}(\mathcal{V}^*)\cdot \beta_{X_j}-\theta_0\cdot \beta_{X_j}-r_j)\cdot \nu_j}{\sigma_{Y_j}^2}-2\hat{\theta}(\mathcal{V}^*) \sum\limits_{j\in  \mathcal{V}^*} \frac{(\hat{\theta}(\mathcal{V}^*)\cdot \beta_{X_j}-\theta_0\cdot \beta_{X_j}-r_j)\cdot u_j}{\sigma_{Y_j}^2}
 \\&\quad+2\hat{\theta}(\mathcal{V}^*)  \sum\limits_{j\in \mathcal{V}^*} \frac{\nu_j u_j}{\sigma_{Y_j}^2}+\sum\limits_{j\in \mathcal{V}_1} \frac{\nu_j^2-\sigma_{Y_j}^2  }{\sigma_{Y_j}^2}+\theta_0^2   \sum\limits_{j\in \mathcal{V}_1} \frac{  u_j^2- \sigma_{X_{j,\mathtt{RB}}}^2}{\sigma_{Y_j}^2}-\theta_0^2   \sum\limits_{j\in \mathcal{V}_1} \frac{  \hat\sigma_{X_{j,\mathtt{RB}}}^2- \sigma_{X_{j,\mathtt{RB}}}^2}{\sigma_{Y_j}^2} -2\theta_0  \sum\limits_{j\in  \mathcal{V}_1} \frac{\nu_ju_j}{\sigma_{Y_j}^2}\Big) 
  \\&\leq \mathbb{P}\Big(\sum\limits_{j\in \mathcal{V}^*} \frac{(\theta_0 \cdot \beta_{X_j}+r_j-\hat{\theta}(\mathcal{V}^*) \cdot \beta_{X_j})^2  }{\sigma_{Y_j}^2}-\kappa_{n}\cdot (v^*-v_\lambda)
 \leq \big|\sum\limits_{j\in \mathcal{V}^*} \frac{\nu_j ^2-\sigma_{Y_j}^2}{\sigma_{Y_j}^2}\big|+ \hat\theta(\mathcal{V}^*)^2 \big|\sum\limits_{j\in \mathcal{V}^*} \frac{u_j^2- \sigma_{X_{j,\mathtt{RB}}}^2}{\sigma_{Y_j}^2}\big|
 \\&\quad +2\big|\hat{\theta}(\mathcal{V}^*) \sum\limits_{j\in  \mathcal{V}^*} \frac{(\hat{\theta}(\mathcal{V}^*)\cdot \beta_{X_j}-\theta_0\cdot \beta_{X_j}-r_j)\cdot u_j}{\sigma_{Y_j}^2}\big|+2\big| \sum\limits_{j\in  \mathcal{V}^*} \frac{(\hat{\theta}(\mathcal{V}^*)\cdot \beta_{X_j}-\theta_0\cdot \beta_{X_j}-r_j)\cdot \nu_j}{\sigma_{Y_j}^2}\big|+2\big|\hat{\theta}(\mathcal{V}^*) \sum\limits_{j\in \mathcal{V}^*} \frac{\nu_j u_j}{\sigma_{Y_j}^2} \big|
 \\&\quad+\hat\theta(\mathcal{V}^*)^2 \big|\sum\limits_{j\in \mathcal{V}^*} \frac{\hat\sigma_{X_{j,\mathtt{RB}}}^2- \sigma_{X_{j,\mathtt{RB}}}^2}{\sigma_{Y_j}^2}\big|+\big|\sum\limits_{j\in \mathcal{V}_1} \frac{\nu_j^2-\sigma_{Y_j}^2  }{\sigma_{Y_j}^2}\big|+\theta_0^2  \big|\sum\limits_{j\in \mathcal{V}_1} \frac{  u_j^2- \sigma_{X_{j,\mathtt{RB}}}^2}{\sigma_{Y_j}^2}\big|+\theta_0^2  \big|\sum\limits_{j\in \mathcal{V}_1} \frac{  \hat\sigma_{X_{j,\mathtt{RB}}}^2- \sigma_{X_{j,\mathtt{RB}}}^2}{\sigma_{Y_j}^2}\big|+2\theta_0 \big|\sum\limits_{j\in  \mathcal{V}_1} \frac{\nu_ju_j}{\sigma_{Y_j}^2}\big|\Big) 
   \\&\leq \mathbb{P}\Big(\sum\limits_{j\in \mathcal{V}^*} \frac{(\theta_0 \cdot \beta_{X_j}+r_j-\hat{\theta}(\mathcal{V}^*) \cdot \beta_{X_j})^2  }{\sigma_{Y_j}^2}-\kappa_{n}\cdot (v^*-v_\lambda)
 \leq \big|\sum\limits_{j\in \mathcal{V}^*} \frac{\nu_j ^2-\sigma_{Y_j}^2}{\sigma_{Y_j}^2}\big|+C_0^2 \big|\sum\limits_{j\in \mathcal{V}^*} \frac{u_j^2- \sigma_{X_{j,\mathtt{RB}}}^2}{\sigma_{Y_j}^2}\big|
 \\&\quad+C_0^2 \big|\sum\limits_{j\in \mathcal{V}^*} \frac{\hat\sigma_{X_{j,\mathtt{RB}}}^2- \sigma_{X_{j,\mathtt{RB}}}^2}{\sigma_{Y_j}^2}\big|+2C_0\big| \sum\limits_{j\in  \mathcal{V}^*} \frac{(\hat{\theta}(\mathcal{V}^*)\cdot \beta_{X_j}-\theta_0\cdot \beta_{X_j}-r_j)\cdot u_j}{\sigma_{Y_j}^2}\big|+2\big| \sum\limits_{j\in  \mathcal{V}^*} \frac{(\hat{\theta}(\mathcal{V}^*)\cdot \beta_{X_j}-\theta_0\cdot \beta_{X_j}-r_j)\cdot \nu_j}{\sigma_{Y_j}^2}\big|\\&\quad+2C_0\big| \sum\limits_{j\in \mathcal{V}^*} \frac{\nu_j u_j}{\sigma_{Y_j}^2} \big|+\big|\sum\limits_{j\in \mathcal{V}_1} \frac{\nu_j^2-\sigma_{Y_j}^2  }{\sigma_{Y_j}^2}\big|+\theta_0^2  \big|\sum\limits_{j\in \mathcal{V}_1} \frac{  u_j^2- \sigma_{X_{j,\mathtt{RB}}}^2}{\sigma_{Y_j}^2}\big|+\theta_0^2  \big|\sum\limits_{j\in \mathcal{V}_1} \frac{  \hat\sigma_{X_{j,\mathtt{RB}}}^2- \sigma_{X_{j,\mathtt{RB}}}^2}{\sigma_{Y_j}^2}\big|+2\theta_0 \big|\sum\limits_{j\in  \mathcal{V}_1} \frac{\nu_ju_j}{\sigma_{Y_j}^2}\big|\Big).
\end{align*}

For simplicity, we can assume $C_0=1$ and $\theta_0=1$.  We also know that 

$$
\sum\limits_{j\in \mathcal{V}^*} \frac{(\theta_0 \cdot \beta_{X_j}+r_j-\hat{\theta}(\mathcal{V}^*) \cdot \beta_{X_j})^2  }{\sigma_{Y_j}^2}-\kappa_{n}\cdot (v^*-v_1)\geq \sum\limits_{j\in \mathcal{V}^*}\frac{ r_j^2}{\sigma_{Y_j}^2}-\frac{(\sum\limits_{j\in \mathcal{V}^*}\frac{
r_j\cdot\beta_{X_j}}{\sigma_{Y_j}^2})^2}{\sum\limits_{j\in \mathcal{V}^*}\frac{ \beta_{X_j}^2}{\sigma_{Y_j}^2}}-\kappa_{n}\cdot (v^*-v_1) .
$$

If we have 

\begin{align}\label{eq:underfitting}
    \frac{\kappa_{n}\cdot (v^*-v_1)}{\sum\limits_{j\in \mathcal{V}^*}\frac{ r_j^2}{\sigma_{Y_j}^2}-\frac{(\sum\limits_{j\in \mathcal{V}^*}\frac{
r_j\cdot\beta_{X_j}}{\sigma_{Y_j}^2})^2}{\sum\limits_{j\in \mathcal{V}^*}\frac{ \beta_{X_j}^2}{\sigma_{Y_j}^2}}}=\frac{\kappa_{n}\cdot (v^*-v_1)}{v^*\Delta(\mathcal{V^*})} \rightarrow 0.
\end{align}
uniformly holds, there should be a $c$ such that 

$$
\sum\limits_{j\in \mathcal{V}}\frac{ r_j^2}{\sigma_{Y_j}^2}-\frac{(\sum\limits_{j\in \mathcal{V}}\frac{
r_j\cdot\beta_{X_j}}{\sigma_{Y_j}^2})^2}{\sum\limits_{j\in \mathcal{V}}\frac{ \beta_{X_j}^2}{\sigma_{Y_j}^2}}-\kappa_{n}\cdot (v^*-v_1) \geq c\cdot (\sum\limits_{j\in \mathcal{V}}\frac{ r_j^2}{\sigma_{Y_j}^2}-\frac{(\sum\limits_{j\in \mathcal{V}}\frac{
r_j\cdot\beta_{X_j}}{\sigma_{Y_j}^2})^2}{\sum\limits_{j\in \mathcal{V}}\frac{ \beta_{X_j}^2}{\sigma_{Y_j}^2}}).
$$

We prove the Equation (\ref{eq:underfitting}) in Lemma \ref{lem7}.

With this result, the above probability is bounded by

\begin{align*}
      &\ \mathbb{P}\Big(
\big|\sum\limits_{j\in \mathcal{V}^*} \frac{u_j^2- \sigma_{X_{j,\mathtt{RB}}}^2}{\sigma_{Y_j}^2}\big|\geq \frac{c}{10}(\sum\limits_{j\in \mathcal{V}}\frac{ r_j^2}{\sigma_{Y_j}^2}-\frac{(\sum\limits_{j\in \mathcal{V}}\frac{ r_j\cdot\beta_{X_j}}{\sigma_{Y_j}^2})^2}{\sum\limits_{j\in \mathcal{V}}\frac{ \beta_{X_j}^2}{\sigma_{Y_j}^2}})\Big)+\mathbb{P}\Big(\big|\sum\limits_{j\in \mathcal{V}_1} \frac{  u_j^2- \sigma_{X_{j,\mathtt{RB}}}^2}{\sigma_{Y_j}^2}\big|\geq \frac{c}{10}(\sum\limits_{j\in \mathcal{V}}\frac{ r_j^2}{\sigma_{Y_j}^2}-\frac{(\sum\limits_{j\in \mathcal{V}}\frac{ r_j\cdot\beta_{X_j}}{\sigma_{Y_j}^2})^2}{\sum\limits_{j\in \mathcal{V}}\frac{ \beta_{X_j}^2}{\sigma_{Y_j}^2}})\Big)
\\& +\mathbb{P}\Big(\big|\sum\limits_{j\in \mathcal{V}^*} \frac{\hat\sigma_{X_{j,\mathtt{RB}}}^2- \sigma_{X_{j,\mathtt{RB}}}^2}{\sigma_{Y_j}^2}\big|\geq 
\frac{c}{10}(\sum\limits_{j\in \mathcal{V}}\frac{ r_j^2}{\sigma_{Y_j}^2}-\frac{(\sum\limits_{j\in \mathcal{V}}\frac{ r_j\cdot\beta_{X_j}}{\sigma_{Y_j}^2})^2}{\sum\limits_{j\in \mathcal{V}}\frac{ \beta_{X_j}^2}{\sigma_{Y_j}^2}})\Big)+\mathbb{P}\Big(\big|\sum\limits_{j\in \mathcal{V}_1} \frac{  \hat\sigma_{X_{j,\mathtt{RB}}}^2- \sigma_{X_{j,\mathtt{RB}}}^2}{\sigma_{Y_j}^2}\big|\geq \frac{c}{10}(\sum\limits_{j\in \mathcal{V}}\frac{ r_j^2}{\sigma_{Y_j}^2}-\frac{(\sum\limits_{j\in \mathcal{V}}\frac{ r_j\cdot\beta_{X_j}}{\sigma_{Y_j}^2})^2}{\sum\limits_{j\in \mathcal{V}}\frac{ \beta_{X_j}^2}{\sigma_{Y_j}^2}})\Big)\\&+\mathbb{P}\Big(\big|\sum\limits_{j\in \mathcal{V}^*} \frac{\nu_j ^2-\sigma_{Y_j}^2}{\sigma_{Y_j}^2}\big|\geq \frac{c}{10}(\sum\limits_{j\in \mathcal{V}}\frac{ r_j^2}{\sigma_{Y_j}^2}-\frac{(\sum\limits_{j\in \mathcal{V}}\frac{ r_j\cdot\beta_{X_j}}{\sigma_{Y_j}^2})^2}{\sum\limits_{j\in \mathcal{V}}\frac{ \beta_{X_j}^2}{\sigma_{Y_j}^2}})\Big)+\mathbb{P}\Big(\big|\sum\limits_{j\in \mathcal{V}_\lambda} \frac{\nu_j ^2-\sigma_{Y_j}^2}{\sigma_{Y_j}^2}\big|\geq \frac{c}{10}(\sum\limits_{j\in \mathcal{V}}\frac{ r_j^2}{\sigma_{Y_j}^2}-\frac{(\sum\limits_{j\in \mathcal{V}}\frac{ r_j\cdot\beta_{X_j}}{\sigma_{Y_j}^2})^2}{\sum\limits_{j\in \mathcal{V}}\frac{ \beta_{X_j}^2}{\sigma_{Y_j}^2}})\Big) \\& +\mathbb{P}\Big(2\big|\sum\limits_{j\in  \mathcal{V}_1} \frac{\nu_ju_j}{\sigma_{Y_j}^2}\big|\geq \frac{c}{10}(\sum\limits_{j\in \mathcal{V}}\frac{ r_j^2}{\sigma_{Y_j}^2}-\frac{(\sum\limits_{j\in \mathcal{V}}\frac{ r_j\cdot\beta_{X_j}}{\sigma_{Y_j}^2})^2}{\sum\limits_{j\in \mathcal{V}}\frac{ \beta_{X_j}^2}{\sigma_{Y_j}^2}})\Big)
+\mathbb{P}\Big(2\big| \sum\limits_{j\in \mathcal{V}^*} \frac{\nu_j u_j}{\sigma_{Y_j}^2} \big|\geq \frac{c}{10}(\sum\limits_{j\in \mathcal{V}}\frac{ r_j^2}{\sigma_{Y_j}^2}-\frac{(\sum\limits_{j\in \mathcal{V}}\frac{ r_j\cdot\beta_{X_j}}{\sigma_{Y_j}^2})^2}{\sum\limits_{j\in \mathcal{V}}\frac{ \beta_{X_j}^2}{\sigma_{Y_j}^2}})\Big)\\& +\mathbb{P}\Big(2\big|\sum\limits_{j\in  \mathcal{V}^*} \frac{(\hat{\theta}(\mathcal{V}^*)\cdot \beta_{X_j}-\theta_0\cdot \beta_{X_j}-r_j)\cdot \nu_j}{\sigma_{Y_j}^2}\big|\geq \frac{1}{10}\big(\sum\limits_{j\in \mathcal{V}^*} \frac{(\theta_0 \cdot \beta_{X_j}+r_j-\hat{\theta}(\mathcal{V}^*) \cdot \beta_{X_j})^2  }{\sigma_{Y_j}^2}-\kappa_{n}\cdot (v^*-v_1)\big)\Big)\\&+\mathbb{P}\Big(2\big|\sum\limits_{j\in  \mathcal{V}^*} \frac{(\hat{\theta}(\mathcal{V}^*)\cdot \beta_{X_j}-\theta_0\cdot \beta_{X_j}-r_j)\cdot u_j}{\sigma_{Y_j}^2}\big|\geq \frac{1}{10}\big(\sum\limits_{j\in \mathcal{V}^*} \frac{(\theta_0 \cdot \beta_{X_j}+r_j-\hat{\theta}(\mathcal{V}^*) \cdot \beta_{X_j})^2  }{\sigma_{Y_j}^2}-\kappa_{n}\cdot (v^*-v_1)\big)\Big).
\end{align*}
}

Also we have 
$$
\big|\sum\limits_{j\in  \mathcal{V}^*} \frac{(\hat{\theta}(\mathcal{V}^*)\cdot \beta_{X_j}-\theta_0\cdot \beta_{X_j}-r_j)\cdot u_j}{\sigma_{Y_j}^2}\big|\leq \sqrt{\sum\limits_{j\in  \mathcal{V}^*} \frac{(\hat{\theta}(\mathcal{V}^*)\cdot \beta_{X_j}-\theta_0\cdot \beta_{X_j}-r_j)^2}{\sigma_{Y_j}^2}}\sqrt{\sum\limits_{j\in  \mathcal{V}^*} \frac{u_j^2}{\sigma_{Y_j}^2}}
$$
$$
\big|\sum\limits_{j\in  \mathcal{V}^*} \frac{(\hat{\theta}(\mathcal{V}^*)\cdot \beta_{X_j}-\theta_0\cdot \beta_{X_j}-r_j)\cdot \nu_j}{\sigma_{Y_j}^2}\big|\leq \sqrt{\sum\limits_{j\in  \mathcal{V}^*} \frac{(\hat{\theta}(\mathcal{V}^*)\cdot \beta_{X_j}-\theta_0\cdot \beta_{X_j}-r_j)^2}{\sigma_{Y_j}^2}}\sqrt{\sum\limits_{j\in  \mathcal{V}^*} \frac{\nu_j^2}{\sigma_{Y_j}^2}}
$$
That means there exists a $c''>0$ such that the last two terms are further bounded by
\begin{align*}
    &\quad \mathbb{P}\Big(\big|\sum\limits_{j\in  \mathcal{V}^*} \frac{(\hat{\theta}(\mathcal{V}^*)\cdot \beta_{X_j}-\theta_0\cdot \beta_{X_j}-r_j)\cdot \nu_j}{\sigma_{Y_j}^2}\big|\geq \frac{1}{10}\big(\sum\limits_{j\in \mathcal{V}^*} \frac{(\theta_0 \cdot \beta_{X_j}+r_j-\hat{\theta}(\mathcal{V}^*) \cdot \beta_{X_j})^2  }{\sigma_{Y_j}^2}-\kappa_{n}\cdot (v^*-v_1)\big)\Big)\\&\leq    \mathbb{P}\Big(\sqrt{\sum\limits_{j\in  \mathcal{V}^*}\frac{\nu_j^2}{\sigma_{Y_j}^2}}\geq \frac{\frac{1}{10}\big(\sum\limits_{j\in \mathcal{V}^*} \frac{(\theta_0 \cdot \beta_{X_j}+r_j-\hat{\theta}(\mathcal{V}^*) \cdot \beta_{X_j})^2  }{\sigma_{Y_j}^2}-\kappa_{n}\cdot (v^*-v_1)\big)}{\sqrt{\sum\limits_{j\in \mathcal{V}^*} \frac{(\theta_0 \cdot \beta_{X_j}+r_j-\hat{\theta}(\mathcal{V}^*) \cdot \beta_{X_j})^2  }{\sigma_{Y_j}^2}}}\Big)\\& \leq \mathbb{P}\Big(\sum\limits_{j\in  \mathcal{V}^*}\frac{\nu_j^2}{\sigma_{Y_j}^2}\geq c''(\sum\limits_{j\in \mathcal{V}}\frac{ r_j^2}{\sigma_{Y_j}^2}-\frac{(\sum\limits_{j\in \mathcal{V}}\frac{
r_j\cdot\beta_{X_j}}{\sigma_{Y_j}^2})^2}{\sum\limits_{j\in \mathcal{V}}\frac{ \beta_{X_j}^2}{\sigma_{Y_j}^2}})\Big).
\end{align*}
Similarly, 
\begin{align*}
    &\quad \mathbb{P}\Big(\big|\sum\limits_{j\in  \mathcal{V}^*} \frac{(\hat{\theta}(\mathcal{V}^*)\cdot \beta_{X_j}-\theta_0\cdot \beta_{X_j}-r_j)\cdot u_j}{\sigma_{Y_j}^2}\big|\geq \frac{1}{10}\big(\sum\limits_{j\in \mathcal{V}^*} \frac{(\theta_0 \cdot \beta_{X_j}+r_j-\hat{\theta}(\mathcal{V}^*) \cdot \beta_{X_j})^2  }{\sigma_{Y_j}^2}-\kappa_{n}\cdot (v^*-v_1)\big)\Big)\\& \leq \mathbb{P}\Big(\sum\limits_{j\in  \mathcal{V}^*}\frac{u_j^2}{\sigma_{Y_j}^2}\geq c''(\sum\limits_{j\in \mathcal{V}}\frac{ r_j^2}{\sigma_{Y_j}^2}-\frac{(\sum\limits_{j\in \mathcal{V}}\frac{
r_j\cdot\beta_{X_j}}{\sigma_{Y_j}^2})^2}{\sum\limits_{j\in \mathcal{V}}\frac{ \beta_{X_j}^2}{\sigma_{Y_j}^2}})\Big).
\end{align*}
Using Lemma \ref{lem1}, we can show that these ten probabilities are bounded by
$
 2\cdot e^{-c' \cdot \min\big\{\frac{ {v^*}^2\cdot \Delta^2(\mathcal{V}^*)}{v_1},\ v^*\cdot \Delta^2(\mathcal{V}^*), \ v^* \cdot \Delta(\mathcal{V}^*)\big\}}.
$

To prove $ e^{(s_{\lambda}+1)\cdot ln(s_{\lambda})}\cdot\mathbb{P}\Big(\min\limits_{\theta \in \mathbb{R}} h(\mathcal{V}^*,  \theta)-\kappa_{n}\cdot v^* \leq  h(\mathcal{V}_1, \theta_0)-\kappa_{n}\cdot v_1\Big) \rightarrow 0$ uniformly for any $\mathcal{V}^* \in \bm{\mathcal{V}}_{\texttt{bias}}(\epsilon)$ such that $|\mathcal{V}^*|=v^*\geq \frac{1+c_1}{2} \cdot v_1$ and $\mathcal{V}^* \neq \mathcal{V}_1$, we only need to show
\begin{align*}
    2e^{(s_{\lambda}+1)\cdot ln(s_{\lambda})}\cdot e^{-c'\cdot min\left\{ v^*\cdot \Delta(\mathcal{V}^*), \ v^* \cdot \Delta^2(\mathcal{V}^*), \  \frac{{v^*}^2\cdot \Delta      (\mathcal{V}^*)}{v_1}  \right\} }\rightarrow 0.
\end{align*}

The above formula can be converted into
\begin{align}\label{eq:screening}
    2e^{(s_{\lambda}+1)\cdot ln(s_{\lambda})}\cdot e^{-c'\cdot min\left\{ v^* \cdot \Delta (\mathcal{V}^*)\right\} }\rightarrow 0.
\end{align}

We prove the Equation (\ref{eq:screening}) in Lemma \ref{lem7}.





When $v^*< \frac{1+c_1}{2} \cdot v_1$, we decompose $ h(\mathcal{V}^*, \hat {\theta}(\mathcal{V}^*))$ in a different way,
\begin{align*}
    h(\mathcal{V}^*, \hat {\theta}(\mathcal{V}^*))&=\nonumber\sum\limits_{j\in \mathcal{V}^*} \frac{(\hat \beta_{Y_j}-\hat{\theta}(\mathcal{V}^*) \cdot \hat \beta_{X_{j,\mathtt{RB}}})^2  }{\sigma_{Y_j}^2}-\hat\theta(\mathcal{V}^*)^2 \sum\limits_{j\in \mathcal{V}^*} \frac{ \hat\sigma_{X_{j,\mathtt{RB}}}^2}{\sigma_{Y_j}^2}\\&=\nonumber\sum\limits_{j\in \mathcal{V}^*} \frac{(\hat \beta_{Y_j}-\hat{\theta}(\mathcal{V}^*) \cdot \hat \beta_{X_{j,\mathtt{RB}}})^2  }{\sigma_{Y_j}^2}-\hat\theta(\mathcal{V}^*)^2 \sum\limits_{j\in \mathcal{V}^*} \frac{ \hat\sigma_{X_{j,\mathtt{RB}}}^2-\sigma_{X_{j,\mathtt{RB}}}^2}{\sigma_{Y_j}^2}-\hat\theta(\mathcal{V}^*)^2 \sum\limits_{j\in \mathcal{V}^*} \frac{\sigma_{X_{j,\mathtt{RB}}}^2}{\sigma_{Y_j}^2}
\end{align*}

Under Condition \ref{Boundedness}, there exists a $C_0>0$ such that 

{\small
\begin{align*}
 &\quad\mathbb{P}\Big(\min\limits_{\theta \in \mathbb{R}} h(\mathcal{V}^*,  \theta)-\kappa_{n}\cdot v^* \leq  h(\mathcal{V}_1, \theta_0)-\kappa_{n}\cdot v_1\Big)
 \\&\leq \mathbb{P}\Big(        \sum\limits_{j\in \mathcal{V}^*} \frac{(\hat \beta_{Y_j}-\hat{\theta}(\mathcal{V}^*) \cdot \hat \beta_{X_{j,\mathtt{RB}}})^2  }{\sigma_{Y_j}^2}+\kappa_{n}\cdot (v_1-v^*) 
 \leq \hat\theta(\mathcal{V}^*)^2  \sum\limits_{j\in \mathcal{V}^*} \frac{\hat\sigma_{X_{j,\mathtt{RB}}}^2- \sigma_{X_{j,\mathtt{RB}}}^2}{\sigma_{Y_j}^2}+\hat\theta(\mathcal{V}^*)^2 \sum\limits_{j\in \mathcal{V}^*} \frac{\sigma_{X_{j,\mathtt{RB}}}^2}{\sigma_{Y_j}^2}
 \\&\quad +\sum\limits_{j\in \mathcal{V}_1} \frac{\nu_j^2-\sigma_{Y_j}^2  }{\sigma_{Y_j}^2}+\theta_0^2   \sum\limits_{j\in \mathcal{V}_1} \frac{  u_j^2- \sigma_{X_{j,\mathtt{RB}}}^2}{\sigma_{Y_j}^2}-\theta_0^2   \sum\limits_{j\in \mathcal{V}_1} \frac{  \hat\sigma_{X_{j,\mathtt{RB}}}^2- \sigma_{X_{j,\mathtt{RB}}}^2}{\sigma_{Y_j}^2} -2\theta_0  \sum\limits_{j\in  \mathcal{V}_1} \frac{\nu_ju_j}{\sigma_{Y_j}^2}\Big) 
   \\&\leq \mathbb{P}\Big(\sum\limits_{j\in \mathcal{V}^*} \frac{(\hat \beta_{Y_j}-\hat{\theta}(\mathcal{V}^*) \cdot \hat \beta_{X_{j,\mathtt{RB}}})^2  }{\sigma_{Y_j}^2}+\kappa_{n}\cdot (v_1-v^*)
 \leq C_0^2 \big|\sum\limits_{j\in \mathcal{V}^*} \frac{\hat\sigma_{X_{j,\mathtt{RB}}}^2- \sigma_{X_{j,\mathtt{RB}}}^2}{\sigma_{Y_j}^2}\big|+C_0^2 \cdot v^*
 \\&\quad+\big|\sum\limits_{j\in \mathcal{V}_1} \frac{\nu_j^2-\sigma_{Y_j}^2  }{\sigma_{Y_j}^2}\big|+\theta_0^2  \big|\sum\limits_{j\in \mathcal{V}_1} \frac{  u_j^2- \sigma_{X_{j,\mathtt{RB}}}^2}{\sigma_{Y_j}^2}\big|+\theta_0^2  \big|\sum\limits_{j\in \mathcal{V}_1} \frac{  \hat\sigma_{X_{j,\mathtt{RB}}}^2- \sigma_{X_{j,\mathtt{RB}}}^2}{\sigma_{Y_j}^2}\big|+2\theta_0 \big|\sum\limits_{j\in  \mathcal{V}_1} \frac{\nu_ju_j}{\sigma_{Y_j}^2}\big|\Big)   \\&\leq \mathbb{P}\Big(\kappa_{n}\cdot (v_1-v^*)-C_0^2 \cdot v^*
 \leq C_0^2 \big|\sum\limits_{j\in \mathcal{V}^*} \frac{\hat\sigma_{X_{j,\mathtt{RB}}}^2- \sigma_{X_{j,\mathtt{RB}}}^2}{\sigma_{Y_j}^2}\big|+\big|\sum\limits_{j\in \mathcal{V}_1} \frac{\nu_j^2-\sigma_{Y_j}^2  }{\sigma_{Y_j}^2}\big|
 \\&\quad+\theta_0^2  \big|\sum\limits_{j\in \mathcal{V}_1} \frac{  u_j^2- \sigma_{X_{j,\mathtt{RB}}}^2}{\sigma_{Y_j}^2}\big|+\theta_0^2  \big|\sum\limits_{j\in \mathcal{V}_1} \frac{  \hat\sigma_{X_{j,\mathtt{RB}}}^2- \sigma_{X_{j,\mathtt{RB}}}^2}{\sigma_{Y_j}^2}\big|+2\theta_0 \big|\sum\limits_{j\in  \mathcal{V}_1} \frac{\nu_ju_j}{\sigma_{Y_j}^2}\big|\Big).
\end{align*}
}

For simplicity, we can assume $C_0=1$ and $\theta_0=1$.

Using $v^*< \frac{1+c_1}{2} \cdot v_1$, we know that there must be a $c>0$ such that 
$$
\kappa_n\cdot(v_1-v^*)-C_0^2 \cdot v^*\geq c \cdot \kappa_n \cdot v_1.
$$
then the above probability is bounded by
\begin{align*}
      &\ \mathbb{P}\Big(\big|\sum\limits_{j\in \mathcal{V}_1} \frac{  u_j^2- \sigma_{X_{j,\mathtt{RB}}}^2}{\sigma_{Y_j}^2}\big|\geq \frac{c}{5}\cdot \kappa_n \cdot v_1\Big)+\mathbb{P}\Big(\big|\sum\limits_{j\in \mathcal{V}^*} \frac{\hat\sigma_{X_{j,\mathtt{RB}}}^2- \sigma_{X_{j,\mathtt{RB}}}^2}{\sigma_{Y_j}^2}\big|\geq \frac{c}{5}\cdot \kappa_n \cdot v_1\Big)
\\& +\mathbb{P}\Big(\big|\sum\limits_{j\in \mathcal{V}_1} \frac{  \hat\sigma_{X_{j,\mathtt{RB}}}^2- \sigma_{X_{j,\mathtt{RB}}}^2}{\sigma_{Y_j}^2}\big|\geq \frac{c}{5} \cdot \kappa_n \cdot v_1\Big)+\mathbb{P}\Big(2\big|\sum\limits_{j\in  \mathcal{V}_1} \frac{\nu_ju_j}{\sigma_{Y_j}^2}\big|\geq \frac{c}{5}\cdot \kappa_n \cdot v_1 \Big)+\mathbb{P}\Big(\big|\sum\limits_{j\in \mathcal{V}_\lambda} \frac{\nu_j ^2-\sigma_{Y_j}^2}{\sigma_{Y_j}^2}\big|\geq \frac{c}{5}\cdot \kappa_n \cdot v_1 \Big).
\end{align*}

Using Lemma \ref{lem1}, we know that there exists a $c'>0$ such that the these five terms are bounded by
$
 2\cdot e^{-c' \cdot \min\big\{\kappa_{n}^2 v_1,\ \frac{\kappa_{n}^2 v_1^2 }{v^*},\  \kappa_{n}\cdot v_1 \big\}}
$.

To prove $ e^{(s_{\lambda}+1)\cdot ln(s_{\lambda})}\cdot\mathbb{P}\Big(\min\limits_{\theta \in \mathbb{R}} h(\mathcal{V}^*,  \theta)-\kappa_{n}\cdot v^* \leq  h(\mathcal{V}_1, \theta_0)-\kappa_{n}\cdot v_1\Big) \rightarrow 0$, we only need to show
\begin{align*}
    2e^{(s_{\lambda}+1)\cdot ln(s_{\lambda})}\cdot e^{-c' \cdot  \kappa_{n}\cdot v_1 }\rightarrow 0.
\end{align*}

This can be easily verified by Condition  \ref{example:Order of the number of valid IVs} and \ref{example:high dimension BIC}. So we can conclude that $ e^{(s_{\lambda}+1)\cdot ln(s_{\lambda})}\cdot\mathbb{P}\Big(\min\limits_{\theta \in \mathbb{R}} h(\mathcal{V}^*,  \theta)-\kappa_{n}\cdot v^* \leq  h(\mathcal{V}_1, \theta_0)-\kappa_{n}\cdot v_1\Big) \rightarrow 0$ uniformly for $\mathcal{V}^* \in \bm{\mathcal{V}}_{\texttt{bias}}(\epsilon)$ such that $v^*< \frac{1+c_1}{2} \cdot v_1$.

Therefore, we  conclude that $ e^{(s_{\lambda}+1)\cdot ln(s_{\lambda})}\cdot\mathbb{P}\Big(\min\limits_{\theta \in \mathbb{R}} h(\mathcal{V}^*,  \theta)-\kappa_{n}\cdot v^* \leq  h(\mathcal{V}_1, \theta_0)-\kappa_{n}\cdot v_1\Big) \rightarrow 0$ uniformly for all  $\mathcal{V}^* \in \bm{\mathcal{V}}_{\texttt{bias}}(\epsilon)$.

\subsection{Proof of Lemma \ref{lem: negiliable bias}}

To prove this lemma, we first define the event:
\begin{align*}
    \mathcal{A}(\mathcal{V},\epsilon)&=\left\{\big|\sum\limits_{j\in \mathcal{V}} \frac{ \hat\sigma_{X_{j,\mathtt{RB}}}^2-\sigma_{X_{j,\mathtt{RB}}}^2}{\sigma_{Y_j}^2}\big|< \frac{1}{4}\cdot\sum_{j \in \mathcal{V}}\frac{\beta_{X_j}^2}{\sigma_{Y_j}^2}\right\} \bigcap \left\{\big|\sum\limits_{j\in \mathcal{V}} \frac{ u_{j}^2-\sigma_{X_{j,\mathtt{RB}}}^2}{\sigma_{Y_j}^2}\big|\geq \frac{1}{4}\cdot\sum_{j \in \mathcal{V}}\frac{\beta_{X_j}^2}{\sigma_{Y_j}^2}\right\} \\&\quad \bigcap \left\{     \big|\sum\limits_{j\in \mathcal{V}} \frac{\beta_{X_j} u_j }{\sigma_{Y_j}^2}\big|< \frac{1}{4}\cdot\sum_{j \in \mathcal{V}}\frac{\beta_{X_j}^2}{\sigma_{Y_j}^2}\right\} \bigcap \left\{\frac{4 a_\lambda  }{\sqrt{s_\lambda \cdot \kappa_\lambda}} \big|\sum\limits_{j\in \mathcal{V}} \frac{r_j u_j}{\sigma_{Y_j}^2} \big|< \epsilon \cdot \sum_{j \in \mathcal{V}}\frac{\beta_{X_j}^2}{\sigma_{Y_j}^2}\right\}.
\end{align*}
We want to show for any $\epsilon>0$ and any given $\mathcal{V} \notin \bm{\mathcal{V}}_{\texttt{bias}}(\epsilon)$, under event $ \mathcal{A}(\mathcal{V},\epsilon)$ 
\begin{align*}
    \frac{a_\lambda  }{\sqrt{s_\lambda \cdot \kappa_\lambda}}\cdot\frac{\sum_{j \in \mathcal{V}} r_j\cdot\hat\beta_{X_{j,\mathtt{RB}}}}{\sum_{j \in \mathcal{V}} \hat\beta_{X_{j,\mathtt{RB}}}^2-\hat\sigma_{X_{j,\mathtt{RB}}}^2}< 9 \cdot \epsilon.
\end{align*}
and $\mathbb{P}(\mathcal{A}(\mathcal{V},\epsilon))\rightarrow 1$.

To do this, we make the following decomposition,
\begin{align*}
    \frac{a_\lambda  }{\sqrt{s_\lambda \cdot \kappa_\lambda}}\frac{\sum_{j \in \mathcal{V}} r_j\cdot\hat\beta_{X_{j,\mathtt{RB}}}}{\sum_{j \in \mathcal{V}} \hat\beta_{X_{j,\mathtt{RB}}}^2-\hat\sigma_{X_{j,\mathtt{RB}}}^2}=     \frac{a_\lambda  }{\sqrt{s_\lambda \cdot \kappa_\lambda}}\frac{\sum_{j \in \mathcal{V}} r_j\cdot\beta_{X_{j}}+\sum_{j \in \mathcal{V}} r_j\cdot u_{j}}{\sum_{j \in \mathcal{V}} \beta_{X_{j}}^2+\sum_{j \in \mathcal{V}} 2 \beta_{X_{j}}u_{j}+\sum_{j \in \mathcal{V}} (\hat\sigma_{X_{j,\mathtt{RB}}}^2-\sigma_{X_{j,\mathtt{RB}}}^2)+\sum_{j \in \mathcal{V}} (u_{j}^2-\sigma_{X_{j,\mathtt{RB}}}^2)}
\end{align*}
and notice that under $\mathcal{A}(\mathcal{V},\epsilon)$ we have 

\begin{align*}
        \frac{a_\lambda  }{\sqrt{s_\lambda \cdot \kappa_\lambda}}\frac{\sum_{j \in \mathcal{V}} r_j\cdot\hat\beta_{X_{j,\mathtt{RB}}}}{\sum_{j \in \mathcal{V}} \hat\beta_{X_{j,\mathtt{RB}}}^2-\hat\sigma_{X_{j,\mathtt{RB}}}^2} &\leq \frac{a_\lambda  }{\sqrt{s_\lambda \cdot \kappa_\lambda}}\frac{\sum_{j \in \mathcal{V}} r_j\cdot\beta_{X_{j}}+\sum_{j \in \mathcal{V}} r_j\cdot u_{j}}{\frac{1}{4}\cdot\sum_{j \in \mathcal{V}} \beta_{X_{j}}^2}
        \\&=     \frac{4 \cdot a_\lambda  }{\sqrt{s_\lambda \cdot \kappa_\lambda}}\frac{\sum_{j \in \mathcal{V}} r_j\cdot\beta_{X_{j}}}{\sum_{j \in \mathcal{V}} \beta_{X_{j}}^2}+\frac{4 \cdot a_\lambda  }{\sqrt{s_\lambda \cdot \kappa_\lambda}}\frac{\sum_{j \in \mathcal{V}} r_j\cdot u_{j}}{\sum_{j \in \mathcal{V}} \beta_{X_{j}}^2}
        \\&<     \frac{4 \cdot a_\lambda  }{\sqrt{s_\lambda \cdot \kappa_\lambda}}\frac{\sum_{j \in \mathcal{V}} r_j\cdot\beta_{X_{j}}}{\sum_{j \in \mathcal{V}} \beta_{X_{j}}^2}+\epsilon.
\end{align*}

For the first term on the right-hand side
$$
 \frac{4\cdot a_\lambda}{\sqrt{s_\lambda\cdot\kappa_\lambda}} \cdot \frac{\sum_{j \in \mathcal{V}} r_j\cdot\beta_{X_{j}}}{\sum_{j \in \mathcal{V}} \beta_{X_{j}}^2}
$$, we can rewrite it as
\begin{align*}
    4\cdot \sqrt{n s_\lambda \cdot\beta_0^2} \cdot \frac{\sum_{j \in \mathcal{V}} r_j\cdot\beta_{X_{j}}}{\sum_{j \in \mathcal{V}} \beta_{X_{j}}^2}&=4\cdot \sqrt{n s_\lambda \cdot\beta_0^2}\cdot\frac{(v_2'k+v_3')\cdot r_3\beta_0}{v_1' \beta_0^2+v_2'\beta_0^2+v_3'\beta_0^2}\\&=4\cdot \sqrt{n s_\lambda \cdot\beta_0^2}\cdot\frac{v_2'k+v_3'}{(v_1' +v_2'+v_3')}\cdot\frac{r_3}{\beta_0}\\&=4\cdot \sqrt{n s_\lambda \cdot\beta_0^2}\cdot(p_2(\mathcal{V})\cdot k+p_3(\mathcal{V}))\frac{r_3}{\beta_0}\\&=4\cdot \sqrt{n s_\lambda }\cdot(p_2(\mathcal{V})\cdot k+p_3(\mathcal{V}))\cdot r_3.
\end{align*}

where $k=\frac{r_2}{r_3}$. For any given $\epsilon>0$, we have $p_3(\mathcal{V})<\frac{\epsilon}{\sqrt{n s_\lambda }\cdot r_3}$ for all $\mathcal{V} \notin \bm{\mathcal{V}}_{\texttt{bias}}(\epsilon)$.
If we have $
r_2<\frac{\epsilon}{\sqrt{n s_\lambda }},$
then we know 
$$
    4\cdot \sqrt{n s_\lambda \cdot\beta_0^2} \cdot \frac{\sum_{j \in \mathcal{V}} r_j\cdot\beta_{X_{j}}}{\sum_{j \in \mathcal{V}} \beta_{X_{j}}^2}=4\cdot \sqrt{n s_\lambda }\cdot(p_2(\mathcal{V})\cdot k+p_3(\mathcal{V}))\cdot r_3<8\epsilon.
$$
holds for any given $\mathcal{V} \notin \bm{\mathcal{V}}_{\texttt{bias}}(\epsilon)$.

Now we show that under $\mathcal{A}(\mathcal{V},\epsilon)$,
\begin{align*}
    \frac{a_\lambda  }{\sqrt{s_\lambda \cdot \kappa_\lambda}}\cdot\frac{\sum_{j \in \mathcal{V}} r_j\cdot\hat\beta_{X_{j,\mathtt{RB}}}}{\sum_{j \in \mathcal{V}} \hat\beta_{X_{j,\mathtt{RB}}}^2-\hat\sigma_{X_{j,\mathtt{RB}}}^2}< 9 \cdot \epsilon.
\end{align*}

It suffices to show that $  \mathbb{P}(\mathcal{A}(\mathcal{V},\epsilon))\rightarrow 1$. We have 
\begin{align*}
    \mathbb{P}(\mathcal{A}(\mathcal{V},\epsilon))&\geq 1-\mathbb{P}\Big(      \big|\sum\limits_{j\in \mathcal{V}} \frac{ \hat\sigma_{X_{j,\mathtt{RB}}}^2-\sigma_{X_{j,\mathtt{RB}}}^2}{\sigma_{Y_j}^2}\big|\geq \frac{1}{4}\cdot\sum_{j \in \mathcal{V}}\frac{\beta_{X_j}^2}{\sigma_{Y_j}^2} \Big)-\mathbb{P}\Big(      \big|\sum\limits_{j\in \mathcal{V}} \frac{ u_{j}^2-\sigma_{X_{j,\mathtt{RB}}}^2}{\sigma_{Y_j}^2}\big|\geq \frac{1}{4}\cdot\sum_{j \in \mathcal{V}}\frac{\beta_{X_j}^2}{\sigma_{Y_j}^2} \Big)\\&\ \ \ -\mathbb{P}\Big(      \big|\sum\limits_{j\in \mathcal{V}} \frac{\beta_{X_j} u_j }{\sigma_{Y_j}^2}\big|\geq \frac{1}{4}\cdot\sum_{j \in \mathcal{V}}\frac{\beta_{X_j}^2}{\sigma_{Y_j}^2} \Big)-\mathbb{P}\Big(         \frac{4  a_\lambda  }{\sqrt{s_\lambda \cdot \kappa_\lambda}} \big|\sum\limits_{j\in \mathcal{V}} \frac{r_j u_j}{\sigma_{Y_j}^2} \big|\geq \epsilon \cdot \sum_{j \in \mathcal{V}}\frac{\beta_{X_j}^2}{\sigma_{Y_j}^2} \Big)
\end{align*}
Under Condition \ref{Bound of Orlicz norm}, we have 
\begin{align*}
     &\quad \mathbb{P}\Big(      \big|\sum\limits_{j\in \mathcal{V}} \frac{ \hat\sigma_{X_{j,\mathtt{RB}}}^2-\sigma_{X_{j,\mathtt{RB}}}^2}{\sigma_{Y_j}^2}\big|\geq \frac{1}{4}\cdot\sum_{j \in \mathcal{V}}\frac{\beta_{X_j}^2}{\sigma_{Y_j}^2} \Big) \leq  2\cdot e^{ -c \cdot \min\{\frac{(\sum_{j \in \mathcal{V}}\frac{\beta_{X_j}^2}{\sigma_{Y_j}^2})^2}{16\cdot v}, \frac{1}{4}\cdot\sum_{j \in \mathcal{V}}\frac{\beta_{X_j}^2}{\sigma_{Y_j}^2}\}} =2\cdot e^{ -c \cdot \min\{\frac{1}{16}\cdot n^2\cdot v\beta_0^4, \frac{1}{4}\cdot n \cdot v \beta_0^2\}}   
     \\&\quad \mathbb{P}\Big(      \big|\sum\limits_{j\in \mathcal{V}} \frac{ u_{j}^2-\sigma_{X_{j,\mathtt{RB}}}^2}{\sigma_{Y_j}^2}\big|\geq \frac{1}{4}\cdot\sum_{j \in \mathcal{V}}\frac{\beta_{X_j}^2}{\sigma_{Y_j}^2} \Big) \leq  2\cdot e^{ -c \cdot \min\{\frac{(\sum_{j \in \mathcal{V}}\frac{\beta_{X_j}^2}{\sigma_{Y_j}^2})^2}{16\cdot v}, \frac{1}{4}\cdot\sum_{j \in \mathcal{V}}\frac{\beta_{X_j}^2}{\sigma_{Y_j}^2}\}} =2\cdot e^{ -c \cdot \min\{\frac{1}{16}\cdot n^2\cdot v\beta_0^4, \frac{1}{4}\cdot n \cdot v \beta_0^2\}}  
     \\& \quad \mathbb{P}\Big(      \big|\sum\limits_{j\in \mathcal{V}} \frac{\beta_{X_j} u_j }{\sigma_{Y_j}^2}\big|\geq \frac{1}{4}\cdot\sum_{j \in \mathcal{V}}\frac{\beta_{X_j}^2}{\sigma_{Y_j}^2} \Big)
\leq 2\cdot e^{ \frac{-c \cdot (\sum_{j \in \mathcal{V}}\frac{\beta_{X_j}^2}{\sigma_{Y_j}^2})^2}{ 16\cdot \sum_{j \in \mathcal{V}}\frac{\beta_{X_j}^2}{\sigma_{Y_j}^2}}}=2\cdot e^{ -\frac{c}{16}\cdot \sum_{j \in \mathcal{V}}\frac{\beta_{X_j}^2}{\sigma_{Y_j}^2}}= 2\cdot e^{ -\frac{c}{16} n \cdot \sum_{j \in \mathcal{V}}\beta_{X_j}^2}= 2\cdot e^{ -\frac{c}{16} n \cdot v\cdot \beta_0^2}.
    \end{align*}  
\begin{align*}   \mathbb{P}\Big(         \frac{4a_\lambda  }{\sqrt{s_\lambda \cdot \kappa_\lambda}} \big|\sum\limits_{j\in \mathcal{V}} \frac{r_j u_j}{\sigma_{Y_j}^2} \big|\geq \epsilon \cdot \sum_{j \in \mathcal{V}}\frac{\beta_{X_j}^2}{\sigma_{Y_j}^2} \Big)\leq 2\cdot e^{ \frac{-c \cdot \epsilon^2 \cdot (\sum_{j \in \mathcal{V}}\frac{\beta_{X_j}^2}{\sigma_{Y_j}^2} )^2}{16 \cdot \sum_{j \in \mathcal{V}} \frac{r_j^2}{\sigma_{Y_j}^2}} \cdot \frac{s_\lambda\cdot \kappa_\lambda}{a_\lambda^2}}=2\cdot e^{ \frac{-c \cdot\epsilon^2\cdot n^2\cdot v^2 \beta_0^4}{16 n \sum_{j \in \mathcal{V}} r_j^2} \frac{1}{n s_\lambda \beta_0^2}}=2\cdot e^{ \frac{-c \cdot\epsilon^2\cdot \beta_0^2}{p_2(\mathcal{V})r_2^2+p_3(\mathcal{V})r_3^2} \frac{v}{ 16 s_\lambda }}.
   \end{align*}
 Thus we have  $    \mathbb{P}(\mathcal{A}(\mathcal{V},\epsilon))\geq 1-2\cdot e^{ -\frac{c}{16} n \cdot v\cdot \beta_0^2}-4\cdot e^{ -c \cdot \min\{\frac{1}{16}\cdot n^2\cdot v\beta_0^4, \frac{1}{4}\cdot n \cdot v \beta_0^2\}}-2\cdot e^{ \frac{-c \cdot\epsilon^2\cdot \beta_0^2}{p_2(\mathcal{V})r_2^2+p_3(\mathcal{V})r_3^2} \frac{v}{ 16 s_\lambda }}$.

   Since $|\mathcal{V}|=v\geq \frac{1+c_1}{2}s_\lambda$ and $p_3(\mathcal{V})<\frac{\epsilon}{\sqrt{n s_\lambda }\cdot r_3}$ for $\mathcal{V} \notin \bm{\mathcal{V}}_{\texttt{bias}}(\epsilon)$, if we have $k<\frac{\epsilon}{\sqrt{n s_\lambda }\cdot r_3}$ then 
\begin{align*} \frac{p_2(\mathcal{V})r_2^2+p_3(\mathcal{V})r_3^2}{\epsilon^2\cdot \beta_0^2} &\leq \frac{2\epsilon}{\sqrt{n s_\lambda}r_3}\frac{r_3^2}{\epsilon^2 \beta_0^2}\\&=\frac{2\epsilon}{\sqrt{n s_\lambda}}\frac{r_3}{\epsilon^2\beta_0^2}\\&= \frac{2}{\sqrt{n s_\lambda}}\frac{r_3}{\epsilon\beta_0^2}.
\end{align*}
If $\frac{1}{\sqrt{n s_\lambda}}\frac{r_3}{\epsilon\beta_0^2}\rightarrow 0$ and $ \frac{1}{n^2\cdot s_\lambda \beta_0^4}\rightarrow 0$, we then can show $\mathbb{P}(\mathcal{A}(\mathcal{V},\epsilon))\rightarrow 1$.

Thus we have 
    $$
      \frac{a_\lambda  }{\sqrt{s_\lambda \cdot \kappa_\lambda}}\cdot\frac{\sum_{j \in \mathcal{V}} r_j\cdot\hat\beta_{X_{j,\mathtt{RB}}}}{\sum_{j \in \mathcal{V}} \hat\beta_{X_{j,\mathtt{RB}}}^2-\hat\sigma_{X_{j,\mathtt{RB}}}^2}= O_p(\epsilon),
    $$
 for any given $\mathcal{V} \notin \bm{\mathcal{V}}_{\texttt{bias}}(\epsilon)$.

\subsection{Additional Lemmas}

\begin{lem}\label{lem7}
    Under Condition \ref{Orders of the variances and sample sizes} and \ref{example:Order of the number of valid IVs}, if
    $A_n/(r_3\varepsilon) = o(1)$, we have 
\begin{align*}
    2e^{(s_{\lambda}+1)\cdot ln(s_{\lambda})}\cdot e^{-c'\cdot min\left\{ v \cdot \Delta (\mathcal{V})\right\} }\rightarrow 0 \text{ and }  \frac{\kappa_{n}\cdot (v-v_1)}{v\cdot\Delta(\mathcal{V})} \rightarrow 0.    
\end{align*}
uniformly hold for all $\mathcal{V} \in \bm{\mathcal{V}}_{\texttt{bias}}(\epsilon)$ and $v=|\mathcal{V}|\geq \frac{1+c_1}{2} v_1$.
\end{lem}

\subsubsection{Proof of Lemma \ref{lem7}}

For a given set $\mathcal{V} \in \bm{\mathcal{V}}_{\texttt{bias}}(\epsilon)$, we let  $v_1'=|\mathcal{V}_1 \cap \mathcal{V}|$, $v_2'=|\mathcal{V}_2 \cap \mathcal{V}|$ and $v_3'=|\mathcal{V}_3 \cap \mathcal{V}|$. Then 
\begin{align*}
    \Delta(\mathcal{V})=\frac{1}{v_1'+v_2'+v_3'}(\sum\limits_{j\in \mathcal{V}}\frac{ r_j^2}{\sigma_{Y_j}^2}-\frac{(\sum\limits_{j\in \mathcal{V}}\frac{
r_j\cdot\beta_{X_j}}{\sigma_{Y_j}^2})^2}{\sum\limits_{j\in \mathcal{V}}\frac{ \beta_{X_j}^2}{\sigma_{Y_j}^2}})&=\frac{v_2'k^2+v_3'}{v_1'+v_2'+v_3'}\cdot n r_3^2-n\cdot \frac{(v_2'k\cdot r_3\beta_0+v_3'\cdot r_3\beta_0)^2}{v_1'\beta_0^2+v_2'\beta_0^2+v_3'\beta_0^2}\cdot\frac{1}{v_1'+v_2'+v_3'}\\&=\frac{v_2'k^2+v_3'}{v_1'+v_2'+v_3'}\cdot n r_3^2-\frac{(v_2'k+v_3')^2\cdot nr_3^2}{(v_1'+v_2'+v_3')^2}.
\end{align*}
Using the definition of  $p_1(\mathcal{V})$, $p_2(\mathcal{V})$ and $p_3(\mathcal{V})$, we have 
\begin{align*}
    \Delta(\mathcal{V})=\frac{1}{v_1'+v_2'+v_3'}(\sum\limits_{j\in \mathcal{V}}\frac{ r_j^2}{\sigma_{Y_j}^2}-\frac{(\sum\limits_{j\in \mathcal{V}}\frac{
r_j\cdot\beta_{X_j}}{\sigma_{Y_j}^2})^2}{\sum\limits_{j\in \mathcal{V}}\frac{ \beta_{X_j}^2}{\sigma_{Y_j}^2}})=(p_2(\mathcal{V})k^2+p_3(\mathcal{V}))\cdot n r_3^2-(p_2(\mathcal{V})k+p_3(\mathcal{V}))^2\cdot n r_3^2.
\end{align*}
where $k=\frac{r_2}{r_3}$.

We have
\begin{align*}
    &\quad p_2(\mathcal{V})k^2+p_3(\mathcal{V})-(p_2(\mathcal{V})k+p_3(\mathcal{V}))^2\\&=(p_2(\mathcal{V})-p_2(\mathcal{V})^2)k^2-2 p_2(\mathcal{V})p_3(\mathcal{V})\cdot k+p_3(\mathcal{V})-p_3(\mathcal{V})^2\\&=p_2(\mathcal{V})(p_1(\mathcal{V})+p_3(\mathcal{V}))k^2-2 p_2(\mathcal{V})p_3(\mathcal{V})\cdot k+p_3(\mathcal{V})(p_1(\mathcal{V})+p_2(\mathcal{V}))\\&=p_2(\mathcal{V})p_3(\mathcal{V})\cdot k^2-2 p_2(\mathcal{V})p_3(\mathcal{V})\cdot k+p_2(\mathcal{V})p_3(\mathcal{V})+p_1(\mathcal{V})p_3(\mathcal{V})+p_2(\mathcal{V})p_1(\mathcal{V})\cdot k^2\\&=p_2(\mathcal{V})p_3(\mathcal{V})\cdot (k-1)^2+p_1(\mathcal{V})p_3(\mathcal{V})+p_2(\mathcal{V})p_1(\mathcal{V})\cdot k^2
\end{align*}

Note that we have 
$$
p_3(\mathcal{V})< \frac{v_3'}{v_1'+v_2'+v_3'} < \frac{c_1\cdot v_1}{\frac{1+c_1}{2}\cdot v_1}=\frac{2c_1}{1+c_1}.
$$

Let $c_1'=\frac{2c_1}{1+c_1}$.
If $p_3(\mathcal{V})\geq \frac{\epsilon}{\sqrt{n s_\lambda }\cdot r_3}$, we have $1-c_1'<p_1(\mathcal{V})+p_2(\mathcal{V})\leq 1-\frac{\epsilon}{\sqrt{n s_\lambda }\cdot r_3}$. Then we consider two different situations:
\begin{itemize}
    \item $p_2(\mathcal{V}) \geq \frac{1-c_1'}{2}$: 
    by choosing $k<\frac{1}{2}$, we have
    $$
     (p_2(\mathcal{V})k^2+p_3(\mathcal{V}))-(p_2(\mathcal{V})k+p_3(\mathcal{V}))^2\geq p_2(\mathcal{V})p_3(\mathcal{V})\cdot (k-1)^2> \frac{(1-c_1')}{8}\frac{\epsilon}{\sqrt{n s_\lambda }\cdot r_3}.
    $$
    \item $p_2(\mathcal{V}) < \frac{1-c_1'}{2}$: 
    we have $p_1(\mathcal{V}) > \frac{1-c_1'}{2}$,
    $$
     (p_2(\mathcal{V})k^2+p_3(\mathcal{V}))-(p_2(\mathcal{V})k+p_3(\mathcal{V}))^2\geq p_1(\mathcal{V})p_3(\mathcal{V})> \frac{1-c_1'}{2} \cdot \frac{\epsilon}{\sqrt{n s_\lambda }\cdot r_3} .
    $$
\end{itemize}

So we have when $p_3(\mathcal{V})\geq \frac{\epsilon}{\sqrt{n s_\lambda }\cdot r_3}$, 
$$
(p_2(\mathcal{V})k^2+p_3(\mathcal{V}))-(p_2(\mathcal{V})k+p_3(\mathcal{V}))^2> \frac{(1-c_1')}{8}\cdot \frac{\epsilon}{\sqrt{n s_\lambda }\cdot r_3}.
$$

Now we consider

$$\frac{(s_\lambda+1)\cdot ln(s_\lambda)}{v \cdot
\Delta(\mathcal{V})}=\frac{s_\lambda+1}{v}\frac{ln(s_\lambda)}{((p_2(\mathcal{V})k^2+p_3(\mathcal{V}))-(p_2(\mathcal{V})k+p_3(\mathcal{V}))^2)\cdot nr_3^2}\leq \frac{s_\lambda+1}{\frac{c_1+1}{2}\cdot v_1} \frac{8\sqrt{s_\lambda } \max\{ln(s_\lambda),\kappa_n\}}{(1-c_1')\cdot \epsilon\cdot \sqrt{n} r_3}.$$

$$\frac{\kappa_{n}\cdot (v-v_1)}{v\cdot\Delta(\mathcal{V})} \leq \frac{\kappa_n}{((p_2(\mathcal{V})k^2+p_3(\mathcal{V}))-(p_2(\mathcal{V})k+p_3(\mathcal{V}))^2)\cdot nr_3^2}\leq \frac{8\sqrt{s_\lambda } \max\{ln(s_\lambda),\kappa_n\}}{(1-c_1')\cdot \epsilon\cdot \sqrt{n} r_3}.$$

Using Condition \ref{example:Order of the number of valid IVs} we have $\frac{s_\lambda+1}{\frac{c_1+1}{2}\cdot v_1}= O(1)$. Then if we further have 
$$
\frac{A_n}{\epsilon r_3}= o(1), 
$$
we then can show 
\begin{align*}
    2e^{(s_{\lambda}+1)\cdot ln(s_{\lambda})}\cdot e^{-c'\cdot min\left\{ v \cdot \Delta (\mathcal{V})\right\} }\rightarrow 0 \text{ and }  \frac{\kappa_{n}\cdot (v-v_1)}{v\cdot\Delta(\mathcal{V})} \rightarrow 0.    
\end{align*}

\subsection{Sufficient conditions for the Boundness condition}

\begin{condition}[Boundedness]
For any $\mathcal{V} \in S_{\lambda}$, $|\hat\theta(\mathcal{V})|$ is uniformly bounded away from $\infty$ with probability goes to 1.
\end{condition}

To see this, we can decompose $\hat{\theta}(\mathcal{V})$ as follows:
\begin{align*}
\hat\theta(\mathcal{V})&=\frac{\sum_{j \in \mathcal{V}}\frac{(\theta_0\beta_{X_j}+r_j)\beta_{X_j}}{\sigma^2_{Y_j}}+\sum_{j \in \mathcal{V}}\frac{\beta_{X_j}\nu_j}{\sigma^2_{Y_j}}+\sum_{j \in \mathcal{V}}\frac{(\theta_0\beta_{X_j}+r_j)u_j}{\sigma^2_{Y_j}}+\sum_{j \in \mathcal{V}}\frac{u_j\nu_j}{\sigma^2_{Y_j}}}{\sum_{j \in \mathcal{V}}\frac{\beta^2_{X_{j}}}{\sigma^2_{Y_j}}+2\sum_{j \in \mathcal{V}}\frac{\beta_{X_{j}}u_j}{\sigma^2_{Y_j}}+\sum_{j \in \mathcal{V}}\frac{u^2_{j}-\sigma^2_{X_{j,\mathtt{RB}}}}{\sigma^2_{Y_j}}-\sum_{j \in \mathcal{V}}\frac{\hat\sigma^2_{X_{j,\mathtt{RB}}}-\sigma^2_{X_{j,\mathtt{RB}}}}{\sigma^2_{Y_j}}}\\&=\frac{(\theta_0+\max\limits_{j\in \mathcal{S}_\lambda}|\frac{r_j}{\beta_{X_j}}|)\sum_{j \in \mathcal{V}}\frac{\beta_{X_j}^2}{\sigma^2_{Y_j}}+\sum_{j \in \mathcal{V}}\frac{\beta_{X_j}\nu_j}{\sigma^2_{Y_j}}+(\theta_0+\max\limits_{j\in \mathcal{S}_\lambda}|\frac{r_j}{\beta_{X_j}}|)\sum_{j \in \mathcal{V}}\frac{\beta_{X_j}u_j}{\sigma^2_{Y_j}}+\sum_{j \in \mathcal{V}}\frac{u_j\nu_j}{\sigma^2_{Y_j}}}{\sum_{j \in \mathcal{V}}\frac{\beta^2_{X_{j}}}{\sigma^2_{Y_j}}+2\sum_{j \in \mathcal{V}}\frac{\beta_{X_{j}}u_j}{\sigma^2_{Y_j}}+\sum_{j \in \mathcal{V}}\frac{u^2_{j}-\sigma^2_{X_{j,\mathtt{RB}}}}{\sigma^2_{Y_j}}-\sum_{j \in \mathcal{V}}\frac{\hat\sigma^2_{X_{j,\mathtt{RB}}}-\sigma^2_{X_{j,\mathtt{RB}}}}{\sigma^2_{Y_j}}}    
\end{align*}

Define the event $\mathcal{B}(\mathcal{V})$
\begin{align*}
    \mathcal{B}(\mathcal{V})&=\left\{\big|\sum\limits_{j\in \mathcal{V}} \frac{ \hat\sigma_{X_{j,\mathtt{RB}}}^2-\sigma_{X_{j,\mathtt{RB}}}^2}{\sigma_{Y_j}^2}\big|< \frac{1}{4}\cdot\sum_{j \in \mathcal{V}}\frac{\beta_{X_j}^2}{\sigma_{Y_j}^2}\right\} \bigcap \left\{\big|\sum\limits_{j\in \mathcal{V}} \frac{ u_{j}^2-\sigma_{X_{j,\mathtt{RB}}}^2}{\sigma_{Y_j}^2}\big|<\frac{1}{4}\cdot\sum_{j \in \mathcal{V}}\frac{\beta_{X_j}^2}{\sigma_{Y_j}^2}\right\} \\&\quad \bigcap \left\{ 2   \big|\sum\limits_{j\in \mathcal{V}} \frac{\beta_{X_j} u_j }{\sigma_{Y_j}^2}\big|< \frac{1}{4}\cdot\sum_{j \in \mathcal{V}}\frac{\beta_{X_j}^2}{\sigma_{Y_j}^2}\right\} \bigcap \left\{\big|\sum\limits_{j\in \mathcal{V}} \frac{ u_{j} \nu_j}{\sigma_{Y_j}^2}\big|< \frac{1}{4}\cdot\sum_{j \in \mathcal{V}}\frac{\beta_{X_j}^2}{\sigma_{Y_j}^2}\right\}\bigcap \left\{\big|\sum\limits_{j\in \mathcal{V}} \frac{\beta_{X_j} \nu_j }{\sigma_{Y_j}^2}\big|< \frac{1}{4}\cdot\sum_{j \in \mathcal{V}}\frac{\beta_{X_j}^2}{\sigma_{Y_j}^2}\right\}.
\end{align*}

Under $\mathcal{B}(\mathcal{V})$, we have 
\begin{align*}
\hat\theta(\mathcal{V})&=\frac{(\theta_0+\max\limits_{j\in \mathcal{S}_\lambda}|\frac{r_j}{\beta_{X_j}}|)\sum_{j \in \mathcal{V}}\frac{\beta_{X_j}^2}{\sigma^2_{Y_j}}+\sum_{j \in \mathcal{V}}\frac{\beta_{X_j}\nu_j}{\sigma^2_{Y_j}}+(\theta_0+\max\limits_{j\in \mathcal{S}_\lambda}|\frac{r_j}{\beta_{X_j}}|)\sum_{j \in \mathcal{V}}\frac{\beta_{X_j}u_j}{\sigma^2_{Y_j}}+\sum_{j \in \mathcal{V}}\frac{u_j\nu_j}{\sigma^2_{Y_j}}}{\sum_{j \in \mathcal{V}}\frac{\beta^2_{X_{j}}}{\sigma^2_{Y_j}}+2\sum_{j \in \mathcal{V}}\frac{\beta_{X_{j}}u_j}{\sigma^2_{Y_j}}+\sum_{j \in \mathcal{V}}\frac{u^2_{j}-\sigma^2_{X_{j,\mathtt{RB}}}}{\sigma^2_{Y_j}}-\sum_{j \in \mathcal{V}}\frac{\hat\sigma^2_{X_{j,\mathtt{RB}}}-\sigma^2_{X_{j,\mathtt{RB}}}}{\sigma^2_{Y_j}}} \\& \leq    \frac{(\theta_0+\max\limits_{j\in \mathcal{S}_\lambda}|\frac{r_j}{\beta_{X_j}}|)\sum_{j \in \mathcal{V}}\frac{\beta_{X_j}^2}{\sigma^2_{Y_j}}+\frac{1}{4}\cdot\sum_{j \in \mathcal{V}}\frac{\beta^2_{X_{j}}}{\sigma^2_{Y_j}}+(\theta_0+\max\limits_{j\in \mathcal{S}_\lambda}|\frac{r_j}{\beta_{X_j}}|)\cdot \frac{1}{8}\cdot\sum_{j \in \mathcal{V}}\frac{\beta^2_{X_{j}}}{\sigma^2_{Y_j}}+\frac{1}{4}\cdot\sum_{j \in \mathcal{V}}\frac{\beta^2_{X_{j}}}{\sigma^2_{Y_j}}}{\frac{1}{4}\cdot\sum_{j \in \mathcal{V}}\frac{\beta^2_{X_{j}}}{\sigma^2_{Y_j}}}\\& =    \frac{9}{2}\cdot (\theta_0+|\frac{r_3}{\beta_0}|)+2.
\end{align*}

Notice that
\begin{align*}
    \mathbb{P}(\mathcal{B}(\mathcal{V}))&\geq 1-\mathbb{P}\Big(      \big|\sum\limits_{j\in \mathcal{V}} \frac{ \hat\sigma_{X_{j,\mathtt{RB}}}^2-\sigma_{X_{j,\mathtt{RB}}}^2}{\sigma_{Y_j}^2}\big|\geq \frac{1}{4}\cdot\sum_{j \in \mathcal{V}}\frac{\beta_{X_j}^2}{\sigma_{Y_j}^2} \Big)-\mathbb{P}\Big(      \big|\sum\limits_{j\in \mathcal{V}} \frac{ u_{j}^2-\sigma_{X_{j,\mathtt{RB}}}^2}{\sigma_{Y_j}^2}\big|\geq \frac{1}{4}\cdot\sum_{j \in \mathcal{V}}\frac{\beta_{X_j}^2}{\sigma_{Y_j}^2} \Big)\\&\ \ \ -\mathbb{P}\Big(      2\big|\sum\limits_{j\in \mathcal{V}} \frac{\beta_{X_j} u_j }{\sigma_{Y_j}^2}\big|\geq \frac{1}{4}\cdot\sum_{j \in \mathcal{V}}\frac{\beta_{X_j}^2}{\sigma_{Y_j}^2} \Big)-\mathbb{P}\Big(\big|\sum\limits_{j\in \mathcal{V}} \frac{ u_{j} \nu_j}{\sigma_{Y_j}^2}\big|\geq \frac{1}{4}\cdot\sum_{j \in \mathcal{V}}\frac{\beta_{X_j}^2}{\sigma_{Y_j}^2}\Big)- \mathbb{P}\Big(\big|\sum\limits_{j\in \mathcal{V}} \frac{\beta_{X_j} \nu_j }{\sigma_{Y_j}^2}\big|\geq \frac{1}{4}\cdot\sum_{j \in \mathcal{V}}\frac{\beta_{X_j}^2}{\sigma_{Y_j}^2}\Big).
\end{align*}

Under Condition \ref{Bound of Orlicz norm}, we have 
\begin{align*}
     &\quad \mathbb{P}\Big(      \big|\sum\limits_{j\in \mathcal{V}} \frac{ \hat\sigma_{X_{j,\mathtt{RB}}}^2-\sigma_{X_{j,\mathtt{RB}}}^2}{\sigma_{Y_j}^2}\big|\geq \frac{1}{4}\cdot\sum_{j \in \mathcal{V}}\frac{\beta_{X_j}^2}{\sigma_{Y_j}^2} \Big) \leq  2\cdot e^{ -c \cdot \min\{\frac{(\sum_{j \in \mathcal{V}}\frac{\beta_{X_j}^2}{\sigma_{Y_j}^2})^2}{16\cdot v}, \frac{1}{4}\cdot\sum_{j \in \mathcal{V}}\frac{\beta_{X_j}^2}{\sigma_{Y_j}^2}\}} =2\cdot e^{ -c \cdot \min\{\frac{1}{16}\cdot n^2\cdot v\beta_0^4, \frac{1}{4}\cdot n \cdot v \beta_0^2\}}   
     \\&\quad \mathbb{P}\Big(      \big|\sum\limits_{j\in \mathcal{V}} \frac{ u_{j}^2-\sigma_{X_{j,\mathtt{RB}}}^2}{\sigma_{Y_j}^2}\big|\geq \frac{1}{4}\cdot\sum_{j \in \mathcal{V}}\frac{\beta_{X_j}^2}{\sigma_{Y_j}^2} \Big) \leq  2\cdot e^{ -c \cdot \min\{\frac{(\sum_{j \in \mathcal{V}}\frac{\beta_{X_j}^2}{\sigma_{Y_j}^2})^2}{16\cdot v}, \frac{1}{4}\cdot\sum_{j \in \mathcal{V}}\frac{\beta_{X_j}^2}{\sigma_{Y_j}^2}\}} =2\cdot e^{ -c \cdot \min\{\frac{1}{16}\cdot n^2\cdot v\beta_0^4, \frac{1}{4}\cdot n \cdot v \beta_0^2\}}  
     \\& \quad \mathbb{P}\Big(      2 \big|\sum\limits_{j\in \mathcal{V}} \frac{\beta_{X_j} u_j }{\sigma_{Y_j}^2}\big|\geq \frac{1}{4}\cdot\sum_{j \in \mathcal{V}}\frac{\beta_{X_j}^2}{\sigma_{Y_j}^2} \Big)
\leq 2\cdot e^{ \frac{-c \cdot (\sum_{j \in \mathcal{V}}\frac{\beta_{X_j}^2}{\sigma_{Y_j}^2})^2}{ 64\cdot \sum_{j \in \mathcal{V}}\frac{\beta_{X_j}^2}{\sigma_{Y_j}^2}}}=2\cdot e^{ -\frac{c}{64}\cdot \sum_{j \in \mathcal{V}}\frac{\beta_{X_j}^2}{\sigma_{Y_j}^2}}= 2\cdot e^{ -\frac{c}{64} n \cdot \sum_{j \in \mathcal{V}}\beta_{X_j}^2}= 2\cdot e^{ -\frac{c}{64} n \cdot v\cdot \beta_0^2} \\& \quad \mathbb{P}\Big(\big|\sum\limits_{j\in \mathcal{V}} \frac{\beta_{X_j} \nu_j }{\sigma_{Y_j}^2}\big|\geq \frac{1}{4}\cdot\sum_{j \in \mathcal{V}}\frac{\beta_{X_j}^2}{\sigma_{Y_j}^2} \Big)
\leq 2\cdot e^{ \frac{-c \cdot (\sum_{j \in \mathcal{V}}\frac{\beta_{X_j}^2}{\sigma_{Y_j}^2})^2}{ 16\cdot \sum_{j \in \mathcal{V}}\frac{\beta_{X_j}^2}{\sigma_{Y_j}^2}}}=2\cdot e^{ -\frac{c}{16}\cdot \sum_{j \in \mathcal{V}}\frac{\beta_{X_j}^2}{\sigma_{Y_j}^2}}= 2\cdot e^{ -\frac{c}{16} n \cdot \sum_{j \in \mathcal{V}}\beta_{X_j}^2}= 2\cdot e^{ -\frac{c}{16} n \cdot v\cdot \beta_0^2}\\&\quad \mathbb{P}\Big(      \big|\sum\limits_{j\in \mathcal{V}} \frac{ u_{j} \nu_j}{\sigma_{Y_j}^2}\big|\geq \frac{1}{4}\cdot\sum_{j \in \mathcal{V}}\frac{\beta_{X_j}^2}{\sigma_{Y_j}^2} \Big) \leq  2\cdot e^{ -c \cdot \min\{\frac{(\sum_{j \in \mathcal{V}}\frac{\beta_{X_j}^2}{\sigma_{Y_j}^2})^2}{16\cdot v}, \frac{1}{4}\cdot\sum_{j \in \mathcal{V}}\frac{\beta_{X_j}^2}{\sigma_{Y_j}^2}\}} =2\cdot e^{ -c \cdot \min\{\frac{1}{16}\cdot n^2\cdot v\beta_0^4, \frac{1}{4}\cdot n \cdot v \beta_0^2\}}  .
    \end{align*}
    
 Thus we have  $    \mathbb{P}(\mathcal{B}(\mathcal{V}))\geq 1-2\cdot e^{ -\frac{c}{16} n \cdot v\cdot \beta_0^2}-2\cdot e^{ -\frac{c}{64} n \cdot v\cdot \beta_0^2}-6\cdot e^{ -c \cdot \min\{\frac{1}{16}\cdot n^2\cdot v\beta_0^4, \frac{1}{4}\cdot n \cdot v \beta_0^2\}}$.

 Under the event $\bigcap_{\mathcal{V} \subseteq \mathcal{S}_\lambda} \mathcal{B}(\mathcal{V})$, we can show that
 \begin{align*}
\hat\theta(\mathcal{V}) \leq   \frac{9}{2}\cdot (\theta_0+|\frac{r_3}{\beta_{0}}|)+2.
\end{align*}
 holds uniformly for all subset $\mathcal{V} \subseteq \mathcal{S}_\lambda$. We also notice that 
\begin{align*}
   \mathbb{P}\Big(\bigcap_{\mathcal{V} \subseteq \mathcal{S}_\lambda} \mathcal{B}(\mathcal{V})\Big)= 1-  \mathbb{P}\Big(\bigcup_{\mathcal{V} \subseteq \mathcal{S}_\lambda} \mathcal{B}^\mathcal{C}(\mathcal{V})\Big)\geq  1-  \sum_{\mathcal{V} \subseteq \mathcal{S}_\lambda}\mathbb{P}\Big( \mathcal{B}^\mathcal{C}(\mathcal{V})\Big).
\end{align*}
Here $\mathcal{B}^\mathcal{C}(\mathcal{V})$ is the complement of the event $\mathcal{B}(\mathcal{V})$.

To prove $\mathbb{P}\Big(\bigcap_{\mathcal{V} \subseteq \mathcal{S}_\lambda} \mathcal{B}(\mathcal{V})\Big) \rightarrow 1$, we only need to show 

$$\sum_{\mathcal{V} \subseteq \mathcal{S}_\lambda}\mathbb{P}\Big( \mathcal{B}^\mathcal{C}(\mathcal{V})\Big)\leq e^{(s_\lambda+1)\cdot ln(s_\lambda)} \max_{\mathcal{V} \subseteq \mathcal{S}_\lambda}\mathbb{P}\Big( \mathcal{B}^\mathcal{C}(\mathcal{V})\Big)\rightarrow 0.$$

We have shown  $    \mathbb{P}(\mathcal{B}(\mathcal{V}))\geq 1-2\cdot e^{ -\frac{c}{16} n \cdot v\cdot \beta_0^2}-2\cdot e^{ -\frac{c}{64} n \cdot v\cdot \beta_0^2}-6\cdot e^{ -c \cdot \min\{\frac{1}{16}\cdot n^2\cdot v\beta_0^4, \frac{1}{4}\cdot n \cdot v \beta_0^2\}}$, thus $ \mathbb{P}(\mathcal{B}^\mathcal{C}(\mathcal{V}))< 2\cdot e^{ -\frac{c}{16} n \cdot v\cdot \beta_0^2}+2\cdot e^{ -\frac{c}{64} n \cdot v\cdot \beta_0^2}+6\cdot e^{ -c \cdot \min\{\frac{1}{16}\cdot n^2\cdot v\beta_0^4, \frac{1}{4}\cdot n \cdot v \beta_0^2\}}$.

With these results, we can prove  $e^{(s_\lambda+1)\cdot ln(s_\lambda)} \max_{\mathcal{V} \subseteq \mathcal{S}_\lambda}\mathbb{P}\Big( \mathcal{B}^\mathcal{C}(\mathcal{V})\Big)\rightarrow 0$ if we have  $\frac{s_\lambda ln(s_\lambda)}{n\beta_0^2}\rightarrow 0$.

Thus we have 
\begin{align*}
\hat\theta(\mathcal{V}) \leq   \frac{9}{2}\cdot (\theta_0+|\frac{r_3}{\beta_0}|)+2.
\end{align*}
 holds uniformly for all subset $\mathcal{V} \subseteq \mathcal{S}_\lambda$ with probability approaching one.

If there exists a constant $C>0$ such that $|\frac{r_3}{\beta_0}|<C$, we then can verify that 
\begin{align*}
\hat\theta(\mathcal{V}) \leq   \frac{9}{2}\cdot (\theta_0+C)+2.
\end{align*}
for all subset $\mathcal{V} \subseteq \mathcal{S}_\lambda$ with probability going to one.



\section{ Connections and differences with \cite{bowden2009unbiased}} \label{sup:connection}

    \noindent\textbf{A summary of the proposed method in \cite{bowden2009unbiased}.} The authors consider a setup with an initial GWAS scan and a replication study. In the initial scan, they let the $\{X_1, \ldots, X_K\}$ be the estimated effect size for $K$ SNPs. They also assume these effect sizes follow normal distributions:
    $$
    X_i \sim \mathcal{N}(\mu_i, \sigma_{1,i}^2).
    $$
    This initial scan is used for selecting the strong SNPs. They ordered these effect size as $X_{(1)}, \ldots, X_{(K)}$ and denote the corresponding means and variances as $\mu_{(1)}, \ldots, \mu_{(K)}$ and $\sigma^2_{1,(1)}, \ldots, \sigma^2_{1,(K)}$, and then perform the following selection:
    $$
    \frac{|X_{(1)}|}{\sigma_{1,(1)}}\geq \frac{|X_{(2)}|}{\sigma_{1,(2)}} \geq \ldots  \frac{|X_{(1,k)}|}{\sigma_{(k)}} \geq \Phi^{-1}(1-c_{crit})=\lambda.
    $$
    Due to the selection step, the distribution of $X_{(i)}$ becomes truncated normal, and therefore, $X_{(i)}$ is a biased estimator of $\mu_{(i)}$. To get an unbiased estimation, the authors leverage the replication study, where $Y_i$ is the effect size of the $i$-th ranked SNP such that 
    $$
    Y_i \sim \mathcal{N}(\mu_{(i)}, \sigma^2_{2,i})
    $$

    For simplicity, we let $\sigma^2_{1,(i)}=\sigma^2_{2,i}=\sigma_i^2$.  Obviously, $Y_i$ is an unbiased estimator of $\mu_{(i)}$.  However, it often has a large variance. For this reason, the authors proposed a weighed version of estimator, 
    $$
    \hat{\mu}_{(i)}=\frac{\sigma_{1,(i)}^2 Y_i+\sigma_{2,i}^2 X_{(i)}}{\sigma_{1,(i)}^2+\sigma_{2,i}^2}=\frac{ Y_i+X_{(i)}}{2},
    $$
    which can effectively combine the data from both the initial scan and the replication study.

    Based on this estimator $\hat{\mu}_{(i)}$, they then construct an unbiased estimator of $\mu_{(i)}$ and further use the Rao-Blackwellization to obtain an unbiased estimator with the minimum variance.

    This estimator takes the form:
    $$
    \tilde \mu_{(i)}=\hat{\mu}_{(i)}-\frac{\sigma_{i}}{\sqrt{2}}\cdot\frac{\phi(W^{(0)}_{i,i+1})-\phi(W^{(0)}_{i,i-1})-\phi(W^{(1)}_{i,i+1})+\phi(W^{(1)}_{i,i-1})}{\Phi(W^{(0}_{i,i+1})-\Phi(W^{(0}_{i,i-1})-\Phi(W^{(0}_{i,i+1})+\Phi(W^{(0}_{i,i-1})},
    $$
    where
    $
    W^{(p)}_{s,t}=\frac{\sqrt{2}}{\sigma_{s}} \cdot (\hat \mu_{(s)}-(-1)^{p} \frac{\sigma_{s}|X_{(t)}|}{\sigma_{t}})$,    
    $\frac{|X_{(0)}|}{\sigma_{(0)}}=\infty$, and $\frac{|X_{(k+1)}|}{\sigma_{(k+1)}} = \Phi^{-1}(1-c_{crit})=\lambda$.

    \medskip
      \textbf{Connections and differences with our approach.} When customizing the proposed approach in \cite{bowden2009unbiased} to our problem for SNP selection, we may perform the selection following: 
    $$
    \frac{|X_{(1)}|}{\sigma_{1}}\geq \Phi^{-1}(1-c_{crit})=\lambda.
    $$
    Then, the corresponding unbiased estimator for $\mu_{(1)}$ can then be given by 
\begin{align*}
        \tilde \mu_{(1)}&=\hat{\mu}_{(1)}-\frac{\sigma_{1}}{\sqrt{2}}\cdot\frac{\phi(W^{(0)}_{1,2})-\phi(W^{(0)}_{1,0})-\phi(W^{(1)}_{1,2})+\phi(W^{(1)}_{1,0})}{\Phi(W^{(0)}_{1,2})-\Phi(W^{(0)}_{1,0})-\Phi(W^{(1)}_{1,2})+\Phi(W^{(1)}_{1,0})}\\&=\hat{\mu}_{(1)}-\frac{\sigma_{1}}{\sqrt{2}}\cdot\frac{\phi(W^{(0)}_{1,2})-\phi(W^{(1)}_{1,2})}{\Phi(W^{(0)}_{1,2})+1-\Phi(W^{(1)}_{1,2})}\\&=\hat{\mu}_{(1)}-\frac{\sigma_{1}}{\sqrt{2}}\cdot\frac{\phi(\frac{\sqrt{2}}{\sigma_{1}} \cdot ( \hat \mu_{(1)}-\lambda))-\phi(\frac{\sqrt{2}}{\sigma_{1}} \cdot ( \hat \mu_{(1)}+\lambda))}{\Phi(\frac{\sqrt{2}}{\sigma_{1}} \cdot ( \hat \mu_{(1)}-\lambda))+1-\Phi(\frac{\sqrt{2}}{\sigma_{1}} \cdot ( \hat \mu_{(1)}+\lambda))}.
\end{align*}
Although this estimator appears very similar to the one proposed in our manuscript, it requires summary statistics from a replication study (as \( W_{s,t}^{(p)} \) depends on \( Y_i \)). In other words, to provide an unbiased estimator of \( \mu_{(1)} \), \cite{bowden2009unbiased} requires two GWAS: one initial GWAS for selection and another replication study for unbiased estimator construction. Therefore, the key difference between \cite{bowden2009unbiased} and our approach is that our method can construct an unbiased estimator of \( \mu_{(1)} \) without a replication study. In other words, the setting considered in \cite{bowden2009unbiased} is aligned with the three-sample MR setting, where SNP selection is performed on a third independent exposure GWAS sample. In constrast, we focus on the two-sample MR, where SNP selection and parameter estimation are carried out in the same exposure GWAS sample. From a different perspective, our approach is indeed connected to \cite{bowden2009unbiased}
as a Rao-Blackwellization step is applied to improve the estimation efficiency.

\section{ Simulation settings and additional simulation results}\label{sup:sec6}

\subsection{Simulation settings}\label{sup:setup}
Note that the total effects of SNP $j$ on exposure $X$ and outcome $Y$ can be written as:
	\begin{align*}
	    \beta_{X_j} = \gamma_j + \beta_{XU}\phi_j;\quad \beta_{Y_j} = \theta \beta_{X_j} +\alpha_j + \beta_{YU} \phi_j,
	\end{align*}
	where $\gamma_j$, $\phi_j$, and $\alpha_j$ is the true direct effect of SNP $j$ on $X$, confounding factor $U$, and $Y$, respectively (see Figure~\ref{fig:causal-diagram}). Following \cite{qi2021comprehensive}, we simulate summary-level association statistics $\hat{\beta}_{X_j}$ and $\hat{\beta}_{Y_j}$ directly. Specifically, we generate 
		\begin{align*}
		\begin{bmatrix}
		\hat{\beta}_{X_j}\\
		\hat{\beta}_{Y_j}
		\end{bmatrix} \sim \mathcal{N}\left(\begin{bmatrix}
		\beta_{X_j} \\
		\beta_{Y_j}
		\end{bmatrix}\ ,\ \begin{bmatrix}
		{\sigma}_{X_j} & 0 \\
		0 & {\sigma}_{Y_j}
		\end{bmatrix} \right), 
		\end{align*}
where $\sigma_{X_j} = \sqrt{1/n_X}$ and $\sigma_{X_j} = \sqrt{1/n_Y}$.  

To save space and make the simulations representative of real GWAS data, we focus on general simulation settings where both directional correlated pleiotropy and balanced uncorrelated pleiotropy are considered simultaneously. Other specific simulation settings have also been briefly considered. Specifically, we generate the underlying parameters from a mixture of distributions, a setup that has been widely used for modeling the effect sizes of complex traits in GWAS \citep{bulik2015ld,stephens2017false,zeng2018signatures}:
\begin{align}\label{eq:main simulation setup normal}
    \begin{pmatrix}
		\gamma_j \\
		\alpha_j \\
		\phi_j
		\end{pmatrix} \sim \underbrace{\pi_1 \begin{pmatrix}
		\mathsf{N}(0,\sigma_x^2)\\
		\delta_0 \\
		\delta_0
		\end{pmatrix}}_{\text{Valid IVs}} & + \underbrace{\pi_2 \begin{pmatrix}
		\mathsf{N}(0,{\sigma}_{x}^2)\\
		\mathsf{N}(0.015,{\sigma}_{u}^2)\\
		\mathsf{N}(0,\sigma_{u}^2)
		\end{pmatrix}}_{\text{correlated pleiotropy}} + \underbrace{ \pi_3 \begin{pmatrix}
		\mathsf{N}(0,{\sigma}_x^2)\\
		\mathsf{N}(0,{\sigma}_{y}^2)\\
		\delta_0
		\end{pmatrix} }_{\text{uncorrelated pleiotropy}} 
		 + \underbrace{\pi_4 \begin{pmatrix}
		\delta_0\\
		\mathsf{N}(0,\sigma_{y}^2)\\
		\delta_0
		\end{pmatrix} + \pi_5 \begin{pmatrix}
		\delta_0\\
		\delta_0\\
		\delta_0
		\end{pmatrix}}_{\text{IVs fail the relevance assumption}},
\end{align}
where $\delta_0$ is a Dirac measure centered on zero (i.e., the point mass at $0$), $\pi_1$ controls the proportion of valid IVs (where $\gamma_j\neq 0$ and both $\alpha_j$ and $\phi_j$ are equal to zero), $\pi_2$ controls the proportion of invalid IVs due to correlated pleiotropy, $\pi_3$ controls the proportion of invalid IVs due to uncorrelated pleiotropy, $\pi_4$ controls the proportion of SNPs that are only associated with $Y$, and $\pi_5 = 1 - \sum_{j=1}^{3}\pi_j$ controls the proportion of SNPs that have no association with both $X$ and $Y$. Note that when $\phi_j\neq 0$, the Instrument Strength Independent on Direct Effect (InSIDE) assumption is violated for SNP $j$ because the exposure effect is correlated with their pleiotropic effects on the come due to mediation by common confounding factor $U$.  InSIDE assumption is popular in MR literature and requires that the exposure effects of individual SNPs are independent of their pleiotropic effects on the outcome \citep{burgess2017interpreting}.


Following \cite{qi2021comprehensive}, we generate 200,000 independent SNPs to represent all underlying common variants and set $\sigma_x^2 = \sigma_y^2 = \sigma_u^2 = 1\times 10^{-5}$, $\beta_{XU} = \beta_{YU}= 0.3$. We set $n_{X} = n_{Y} = 500,000$ to reflect the sample size of a typical GWAS in our real data analyses. We further set $\pi_1 + \pi_2 + \pi_3 = 0.02$,  $\pi_2 = \pi_3$, $\pi_4  = 0.01$, and $\pi_5 = 0.97$. We vary the proportion of invalid IVs, which is defined as $(\pi_2 + \pi_3)/ (\pi_1 + \pi_2 + \pi_3)$, to simulate different situations. 

Our proposed CARE estimator is compared with widely used  IVW method \citep{burgess2013mendelian} and seven other popular, recently proposed robust MR methods, including cML and cML-DP \citep{xue2021constrained}, MR-Egger \citep{bowden2015mendelian}, Weighted-Median \citep{bowden2016consistent}, MR-mix \citep{qi2019mendelian}, Weighted-Mode \citep{hartwig2017robust}, MR-APSS \citep{hu2022mendelian}, RAPS \citep{zhao2020statistical},  contamination mixture  \citep[ContMix;][]{burgess2020robust}, and MR-Lasso \citep{rees2019robust}. For IVW, we use the random effects version, which accounts for invalid IVs by allowing over-dispersion in the regression model. For CARE and MR-APSS, we set the significance threshold at $5\times 10^{-5}$. Following common practice, for other benchmark methods, we set the cut-off value $\lambda$ at 5.45 (corresponding to the significance threshold $5\times 10^{-8}$). In our numerical studies, we used $\eta = 0.5$ as the default value in the winner's curse removal step. We simulate 500 Monte  Carlo repetitions to evaluate empirical statistical power ($\theta \neq 0$) and 1,000 Monte  Carlo repetitions to evaluate Type 1 error rates ($\theta =0$).

We report our simulation results with five measures: Type 1 error rates (proportions of mistaken rejection under $\theta= 0$), power (proportions of p-values less than the significance threshold 0.05 under $\theta\neq 0$), absolute bias (the absolute difference between the estimated $\hat{\theta}$ and the true $\theta$), mean squared error (the average squared difference between the estimated $\hat{\theta}$ and the true $\theta$), and coverage probability (average coverage probability of the 95\% confidence interval). 

\subsection{Different proportions of invalid IVs, CARE without winner's curse, and running time}\label{sup:sec_proportion}
\begin{figure}[!htbp]
	\centering	\includegraphics[width=\linewidth]{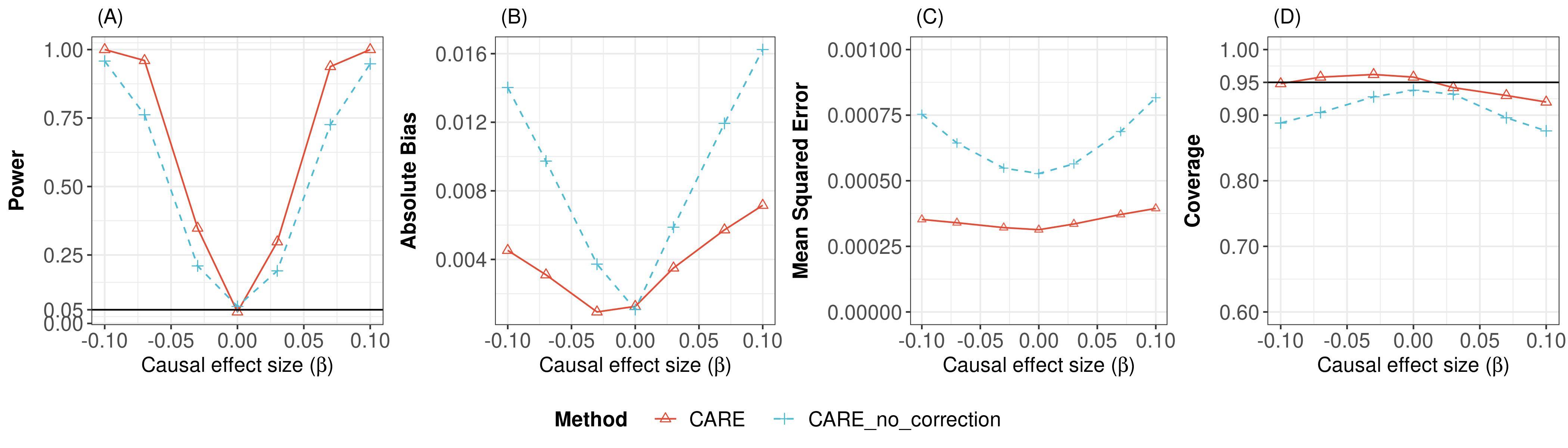}
	\caption{	\label{supfig:simulation_no_correction} Type 1 error rates, power, absolute bias, mean squared error, and coverage of the CARE estimator and CARE estimator without winner's curse bias correction (CARE\_no\_correction) under the setting with 50\% invalid IVs. Power is the empirical power estimated by the proportion of p-values less than the significance threshold 0.05.  Coverage is the empirical coverage probability of the 95\% confidence interval.} 
\end{figure}

\begin{figure}[!htbp]
	\centering	\includegraphics[width=\linewidth]{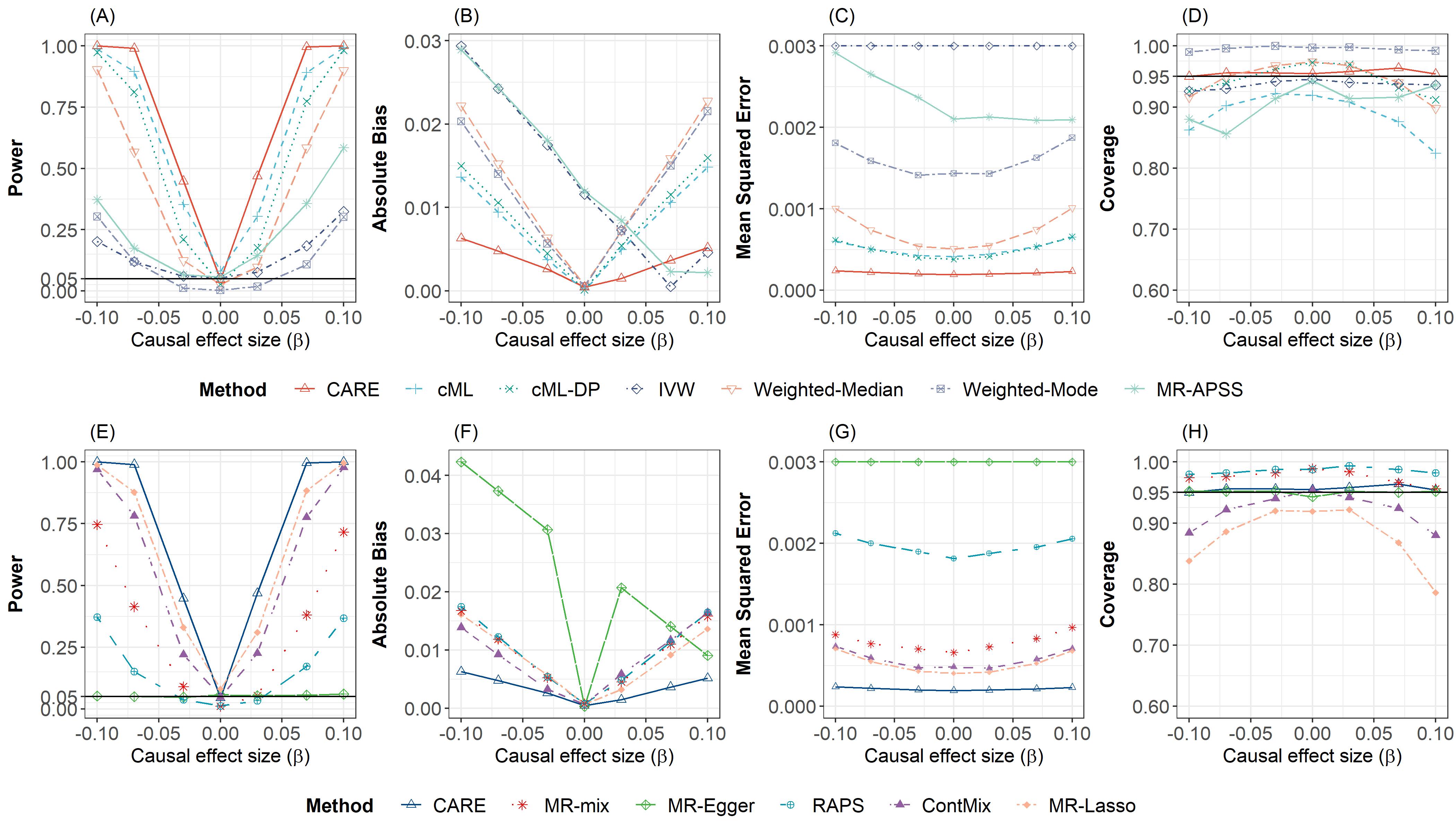}
	\caption{	\label{supfig:simulation_main03} Power, absolute bias, mean squared error, and coverage of the CARE estimator and several robust MR methods under the main setting with 30\% invalid IVs. Power is the empirical power estimated by the proportion of p-values less than the significance threshold 0.05.  Coverage is the empirical coverage probability of the 95\% confidence interval.} 
\end{figure}

\begin{figure}[!htbp]
	\centering	\includegraphics[width=\linewidth]{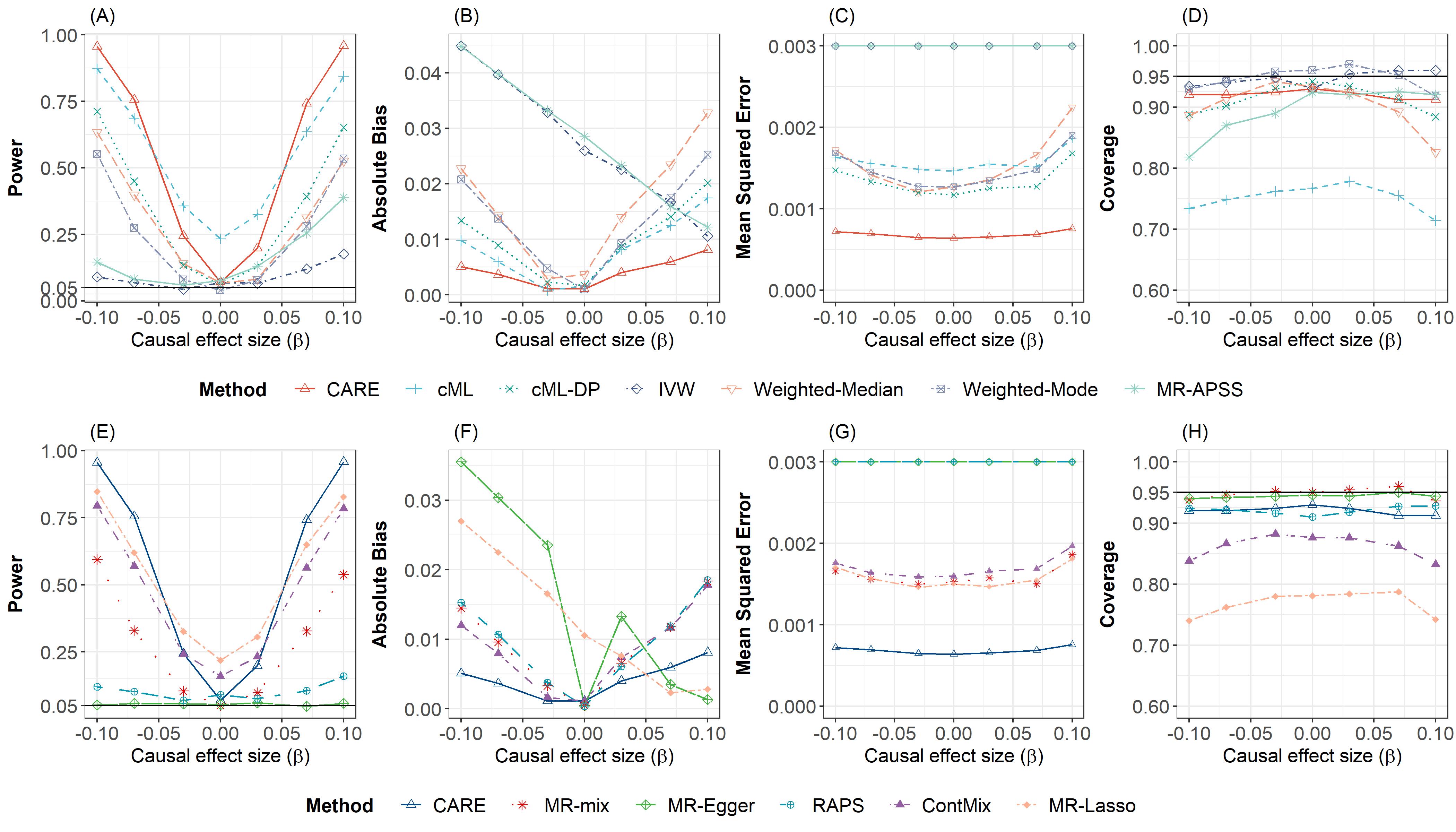}
	\caption{	\label{supfig:simulation_main07} Power, absolute bias, mean squared error, and coverage of the CARE estimator and several robust MR methods under the main setting with 70\% invalid IVs. Power is the empirical power estimated by the proportion of p-values less than the significance threshold 0.05.  Coverage is the empirical coverage probability of the 95\% confidence interval.} 
\end{figure}

\begin{figure}[!htbp]
	\centering	\includegraphics[width=\linewidth]{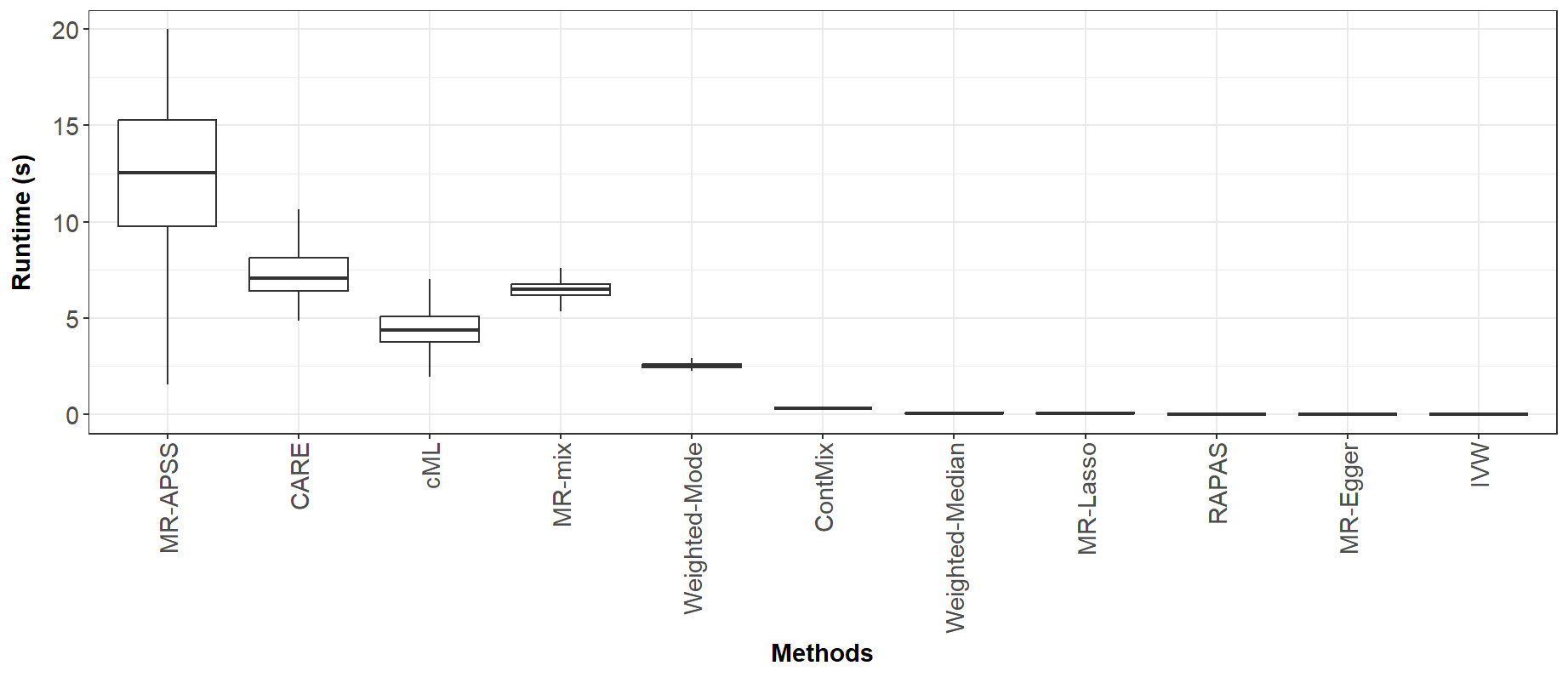}
	\caption{	\label{supfig:simulation_runtime} Runtime (in seconds) of the CARE estimator and several robust MR methods under the main setting (12,000 simulations in total). The box limits represent the lower and upper quartiles, the central line represents the median, and the whiskers represent all samples lying within 1.5 times the interquartile range (IQR).} 
\end{figure}

We conduct several additional simulations. We generate the parameters using the following distribution:
\begin{align}\label{eq:simulation setup normal}
    \begin{pmatrix}
		\gamma_j \\
		\alpha_j \\
		\phi_j
		\end{pmatrix} \sim \underbrace{\pi_1 \begin{pmatrix}
		\mathsf{N}(0,\sigma_x^2)\\
		\delta_0 \\
		\delta_0
		\end{pmatrix}}_{\text{Valid IVs}} & + \underbrace{\pi_2 \begin{pmatrix}
		\mathsf{N}(0,{\sigma}_{x}^2)\\
		\mathsf{N}(0.015,{\sigma}_{u}^2)\\
		\mathsf{N}(0,\sigma_{u}^2)
		\end{pmatrix}}_{\text{correlated pleiotropy}} + \underbrace{ \pi_3 \begin{pmatrix}
		\mathsf{N}(0,{\sigma}_x^2)\\
		\mathsf{N}(0,{\sigma}_{y}^2)\\
		\delta_0
		\end{pmatrix} }_{\text{uncorrelated pleiotropy}} 
		 + \underbrace{\pi_4 \begin{pmatrix}
		\delta_0\\
		\mathsf{N}(0,\sigma_{y}^2)\\
		\delta_0
		\end{pmatrix} + \pi_5 \begin{pmatrix}
		\delta_0\\
		\delta_0\\
		\delta_0
		\end{pmatrix}}_{\text{IVs fail the relevance assumption}},
\end{align}
We follow the main simulation setting and set $\pi_1 + \pi_2 + \pi_3 = 0.02$,  $\pi_4  = 0.01$, and $\pi_5 = 0.97$. We vary the proportion of invalid IVs, which is defined as $(\pi_2 + \pi_3)/ (\pi_1 + \pi_2 + \pi_3)$, to simulate different situations. Figure~\ref{supfig:simulation_no_correction} summarizes the result to compare the performance of the CARE estimator and CARE estimator without winner's curse bias correction under the setting with 50\% invalid IVs. Figures~\ref{supfig:simulation_main03} and~\ref{supfig:simulation_main07}  summarize the results for the settings with 30\% and 70\% invalid IVs. Figure~\ref{supfig:simulation_runtime} summarizes the runtime of the CARE estimator and several robust MR methods for the setting with 50\% invalid IVs.

\newpage
\subsection{Uniform distributed effects in correlated pleiotropy} \label{sup:sec_uniform}
Under the setting using uniform distributed effects in correlated pleiotropy, $\alpha_{j}$ follows the uniform distribution. We generate $\gamma_j, \alpha_j$ using the following distribution:

\begin{align}\label{eq:simulation setup uniform}
    \begin{pmatrix}
		\gamma_j \\
		\alpha_j \\
		\phi_j
		\end{pmatrix} \sim \underbrace{\pi_1 \begin{pmatrix}
		\mathsf{N}(0,\sigma_x^2)\\
		\delta_0 \\
		\delta_0
		\end{pmatrix}}_{\text{Valid IVs}} & + \underbrace{\pi_2 \begin{pmatrix}
		\mathsf{N}(0,{\sigma}_{x}^2)\\
		U(0.01,0.03) \\
		\mathsf{N}(0,\sigma_{u}^2)
		\end{pmatrix}}_{\text{correlated pleiotropy}} + \underbrace{ \pi_3 \begin{pmatrix}
		\mathsf{N}(0,{\sigma}_x^2)\\
		\mathsf{N}(0,{\sigma}_{y}^2)\\
		\delta_0
		\end{pmatrix} }_{\text{uncorrelated pleiotropy}} 
		 + \underbrace{\pi_4 \begin{pmatrix}
		\delta_0\\
		\mathsf{N}(0,\sigma_{y}^2)\\
		\delta_0
		\end{pmatrix} + \pi_5 \begin{pmatrix}
		\delta_0\\
		\delta_0\\
		\delta_0
		\end{pmatrix}}_{\text{IVs fail the relevance assumption}},
\end{align}

\begin{figure}[!htbp]
	\centering	\includegraphics[width=\linewidth]{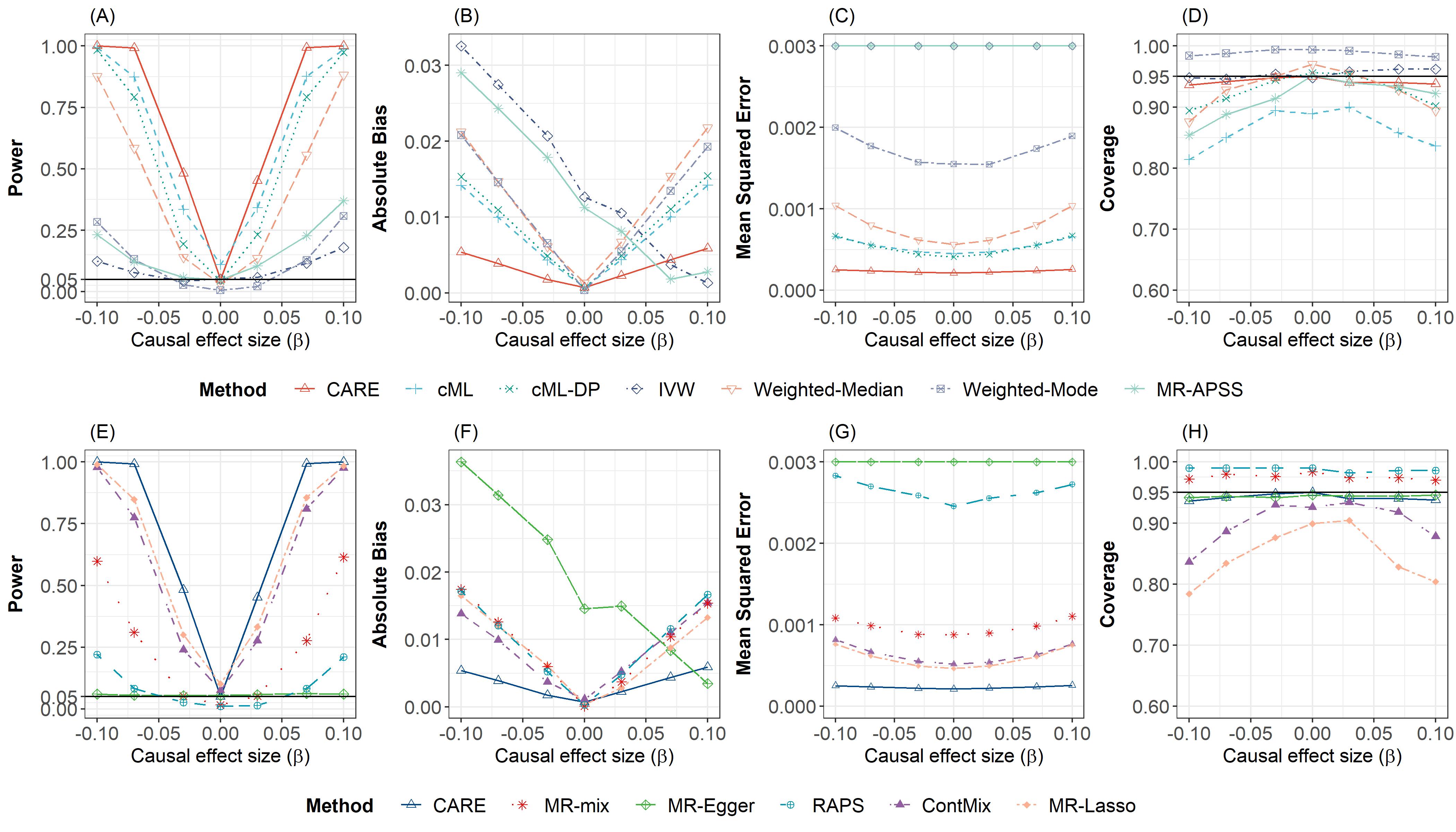}
	\caption{	\label{supfig:simulation_uniform03} Power, absolute bias, mean squared error, and coverage of the CARE estimator and several robust MR methods under the setting of uniformly distributed effects in correlated pleiotropy with 30\% invalid IVs. Power is the empirical power estimated by the proportion of p-values less than the significance threshold 0.05.  Coverage is the empirical coverage probability of the 95\% confidence interval.} 
\end{figure}

\begin{figure}[!htbp]
	\centering	\includegraphics[width=\linewidth]{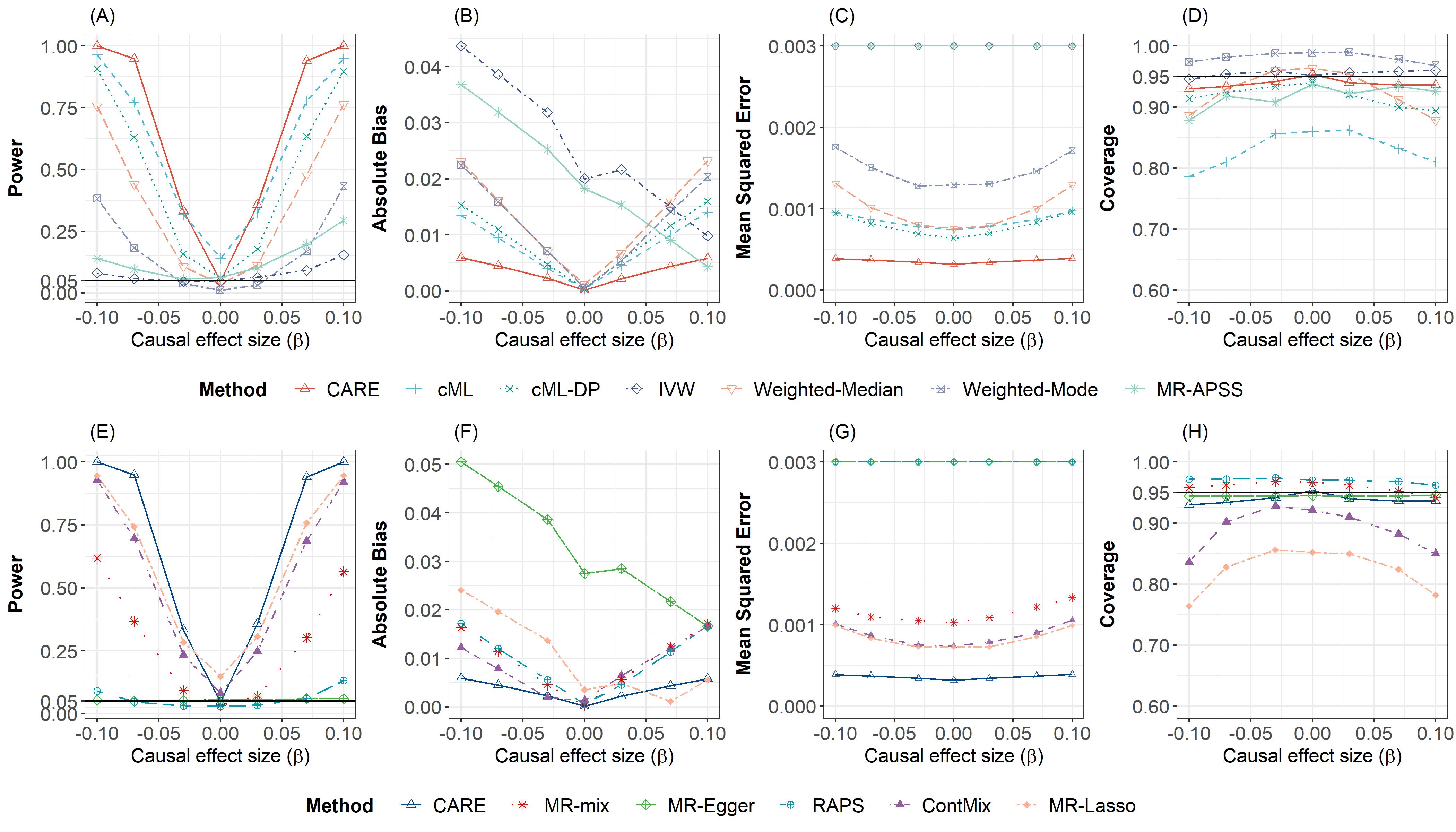}
	\caption{	\label{fig:simulation_uniform05} Power, absolute bias, mean squared error, and coverage of the CARE estimator and several robust MR methods under the setting of uniform distributed effects in correlated pleiotropy with 50\% invalid IVs. Power is the empirical power estimated by the proportion of p-values less than the significance threshold 0.05.  Coverage is the empirical coverage probability of the 95\% confidence interval.} 
\end{figure}

\begin{figure}[!htbp]
	\centering	\includegraphics[width=\linewidth]{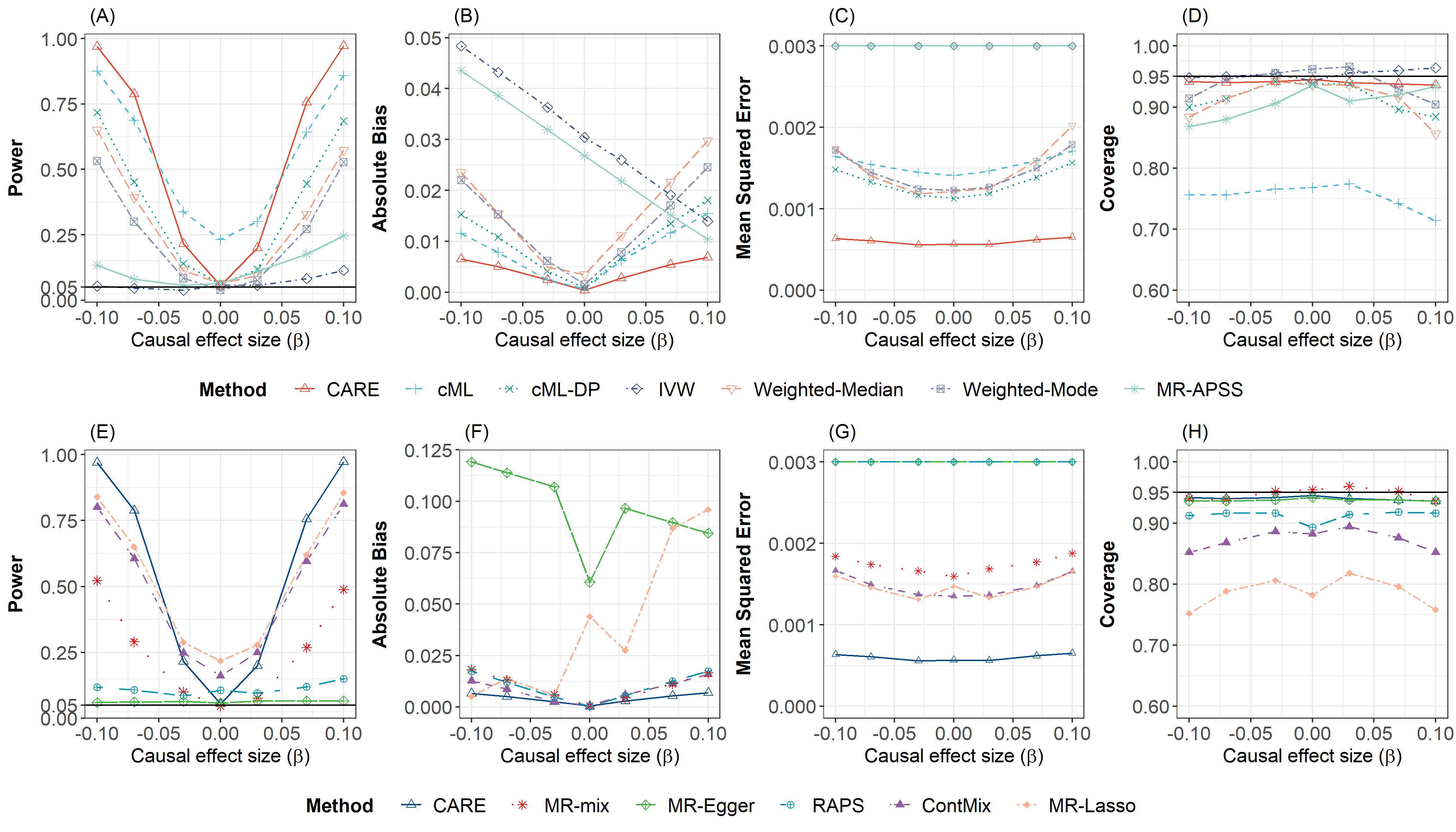}
	\caption{	\label{supfig:simulation_uniform07} Power, absolute bias, mean squared error, and coverage of the CARE estimator and several robust MR methods under the setting of uniformly distributed effects in correlated pleiotropy with 70\% invalid IVs. Power is the empirical power estimated by the proportion of p-values less than the significance threshold 0.05.  Coverage is the empirical coverage probability of the 95\% confidence interval.} 
\end{figure}

We follow the main simulation setting and set $\pi_1 + \pi_2 + \pi_3 = 0.02$,  $\pi_4  = 0.01$, and $\pi_5 = 0.97$. We vary the proportion of invalid IVs, which is defined as $(\pi_2 + \pi_3)/ (\pi_1 + \pi_2 + \pi_3)$, to simulate different situations. Figures~\ref{supfig:simulation_uniform03} to \ref{supfig:simulation_uniform07} summarize the results for the settings with 30\%, 50\%, and 70\% invalid IVs.

\subsection{Balanced horizontal pleiotropy with InSIDE assumption satisfied}\label{sup:sec_bal_pleio}
\begin{figure}[!htbp]
	\centering	\includegraphics[width=\linewidth]{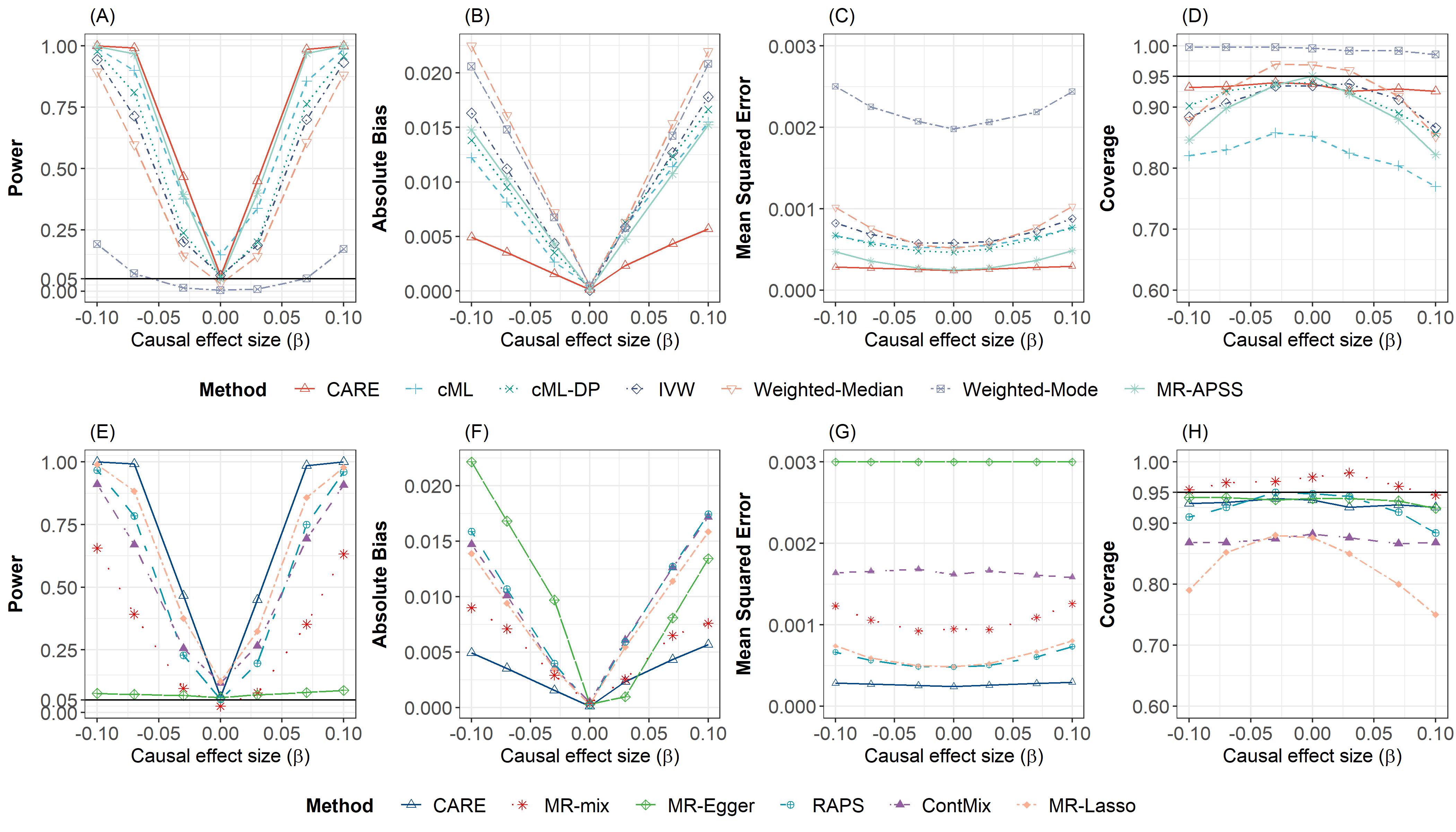}
	\caption{	\label{supfig:simulation_bal_plei_03} Power, absolute bias, mean squared error, and coverage of the CARE estimator and several robust MR methods under the setting of balanced horizontal pleiotropy with InSIDE assumption satisfied with 30\% invalid IVs. Power is the empirical power estimated by the proportion of p-values less than the significance threshold 0.05.  Coverage is the empirical coverage probability of the 95\% confidence interval.} 
\end{figure}

\begin{figure}[!htbp]
	\centering	\includegraphics[width=\linewidth]{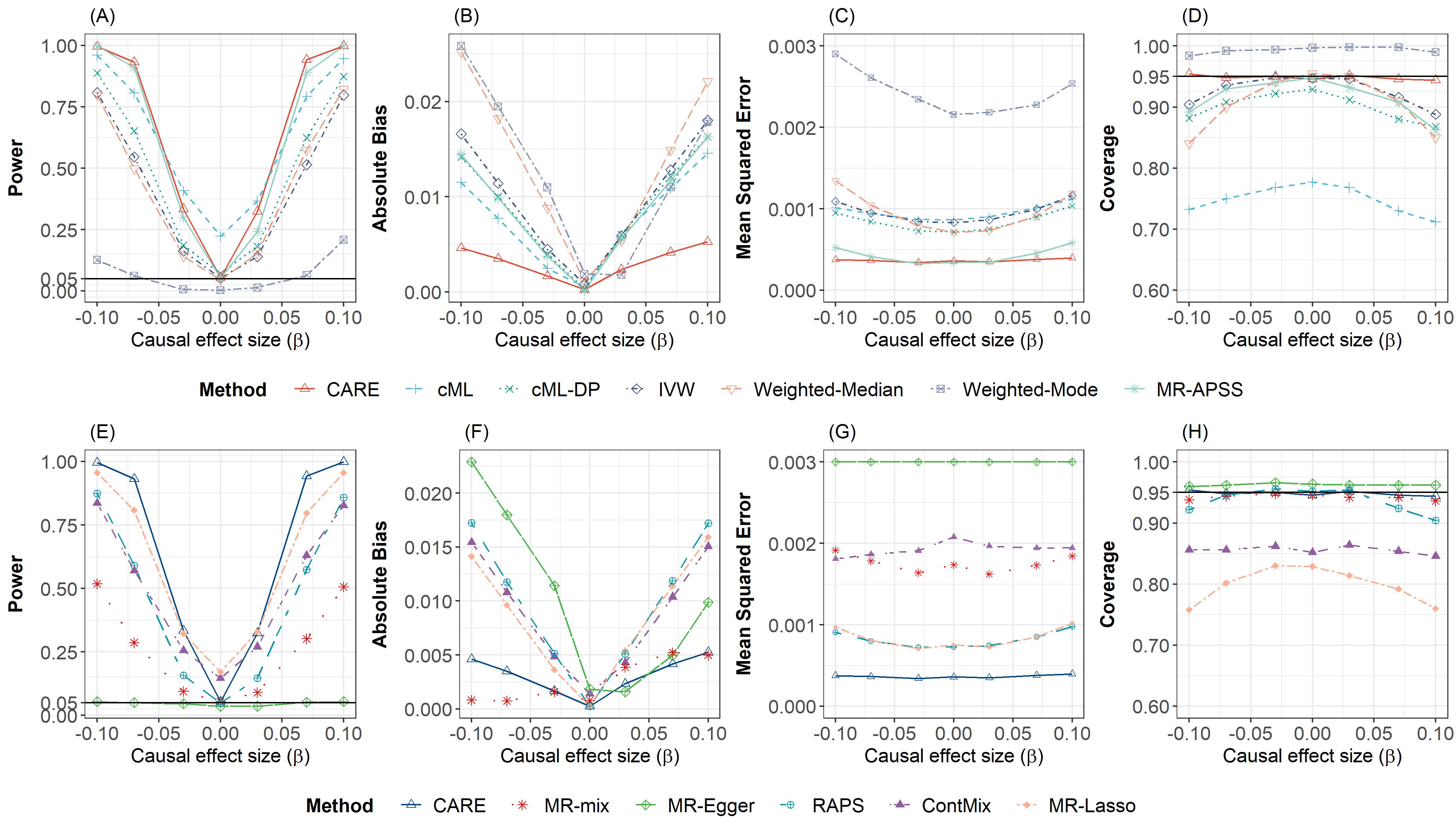}
	\caption{	\label{supfig:simulation_bal_plei_05} Power, absolute bias, mean squared error, and coverage of the CARE estimator and several robust MR methods under the setting of balanced horizontal pleiotropy with InSIDE assumption satisfied with 50\% invalid IVs. Power is the empirical power estimated by the proportion of p-values less than the significance threshold 0.05.  Coverage is the empirical coverage probability of the 95\% confidence interval.} 
\end{figure}

\begin{figure}[!htbp]
	\centering	\includegraphics[width=\linewidth]{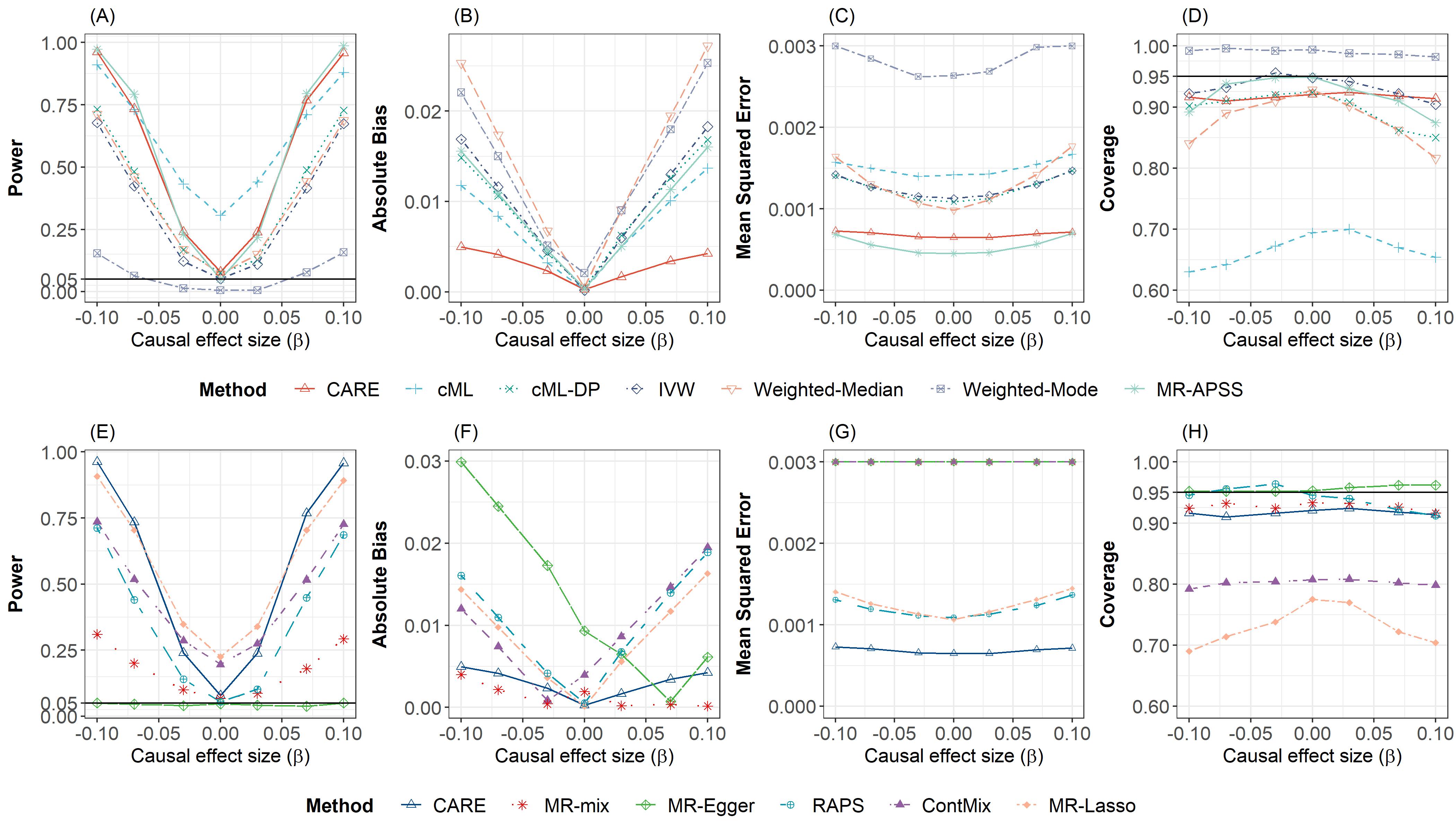}
	\caption{	\label{supfig:simulation_bal_plei_07} Power, absolute bias, mean squared error, and coverage of the CARE estimator and several robust MR methods under the setting of balanced horizontal pleiotropy with InSIDE assumption satisfied with 70\% invalid IVs. Power is the empirical power estimated by the proportion of p-values less than the significance threshold 0.05.  Coverage is the empirical coverage probability of the 95\% confidence interval.} 
\end{figure}

Under the setting of balanced horizontal pleiotropy with the InSIDE assumption satisfied, we allow the InSIDE assumption to be satisfied by setting $\phi_j = 0$. We generate $\gamma_j, \alpha_j$ using the following distribution:
\begin{align*}
    \begin{pmatrix}
		\gamma_j \\
		\alpha_j \\
		\end{pmatrix} \sim \underbrace{\pi_1 \begin{pmatrix}
		\mathsf{N}(0,\sigma_x^2)\\
		\delta_0 
		\end{pmatrix}}_{\text{Valid IVs}}  + \underbrace{ \pi_3 \begin{pmatrix}
		\mathsf{N}(0,{\sigma}_x^2)\\
		\mathsf{N}(0,{\sigma}_{y}^2)
		\end{pmatrix} }_{\text{uncorrelated pleiotropy}} 
		 + \underbrace{\pi_4 \begin{pmatrix}
		\delta_0\\
		\mathsf{N}(0,\sigma_{y}^2)
		\end{pmatrix} + \pi_5 \begin{pmatrix}
		\delta_0\\
		\delta_0
		\end{pmatrix}}_{\text{IVs fail the relevance assumption}}.
\end{align*}
We follow the main simulation setting and set $\pi_1 + \pi_3 = 0.02$,  $\pi_4  = 0.01$, and $\pi_5 = 0.97$. We vary the proportion of invalid IVs, which is defined as $(\pi_3)/ (\pi_1 + \pi_3)$, to simulate different situations. Figures~\ref{supfig:simulation_bal_plei_03} to \ref{supfig:simulation_bal_plei_07} summarize the results for the settings with 30\%, 50\%, and 70\% invalid IVs.

\subsection{Directional horizontal pleiotropy with InSIDE assumption violated}\label{sup:sec_dir_pleio}
\begin{figure}[!htbp]
	\centering	\includegraphics[width=\linewidth]{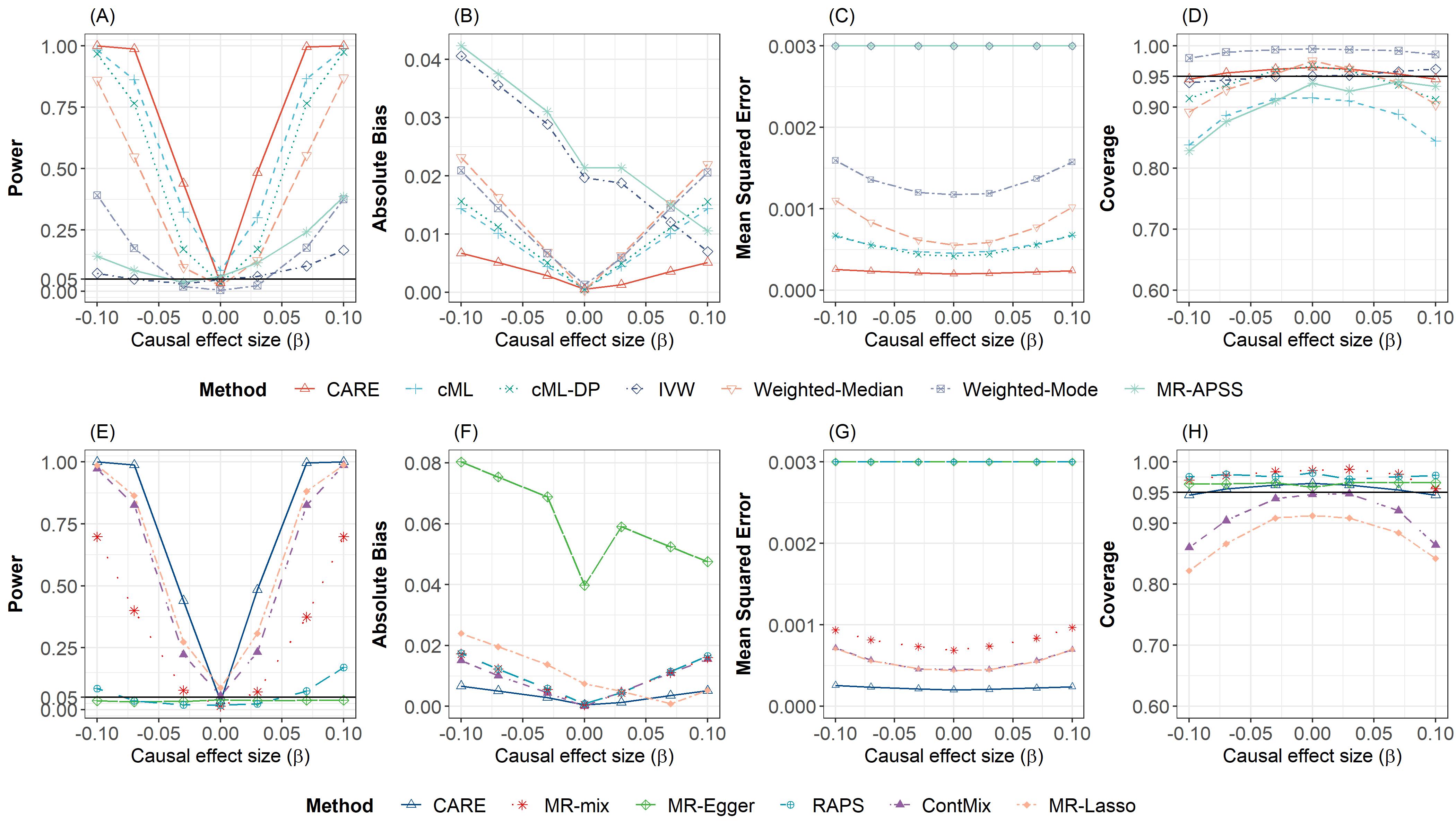}
	\caption{	\label{supfig:simulation_dir_plei_03} Power, absolute bias, mean squared error, and coverage of the CARE estimator and several robust MR methods under the setting of directional horizontal pleiotropy with InSIDE assumption violated with 30\% invalid IVs. Power is the empirical power estimated by the proportion of p-values less than the significance threshold 0.05.  Coverage is the empirical coverage probability of the 95\% confidence interval.} 
\end{figure}

\begin{figure}[!htbp]
	\centering	\includegraphics[width=\linewidth]{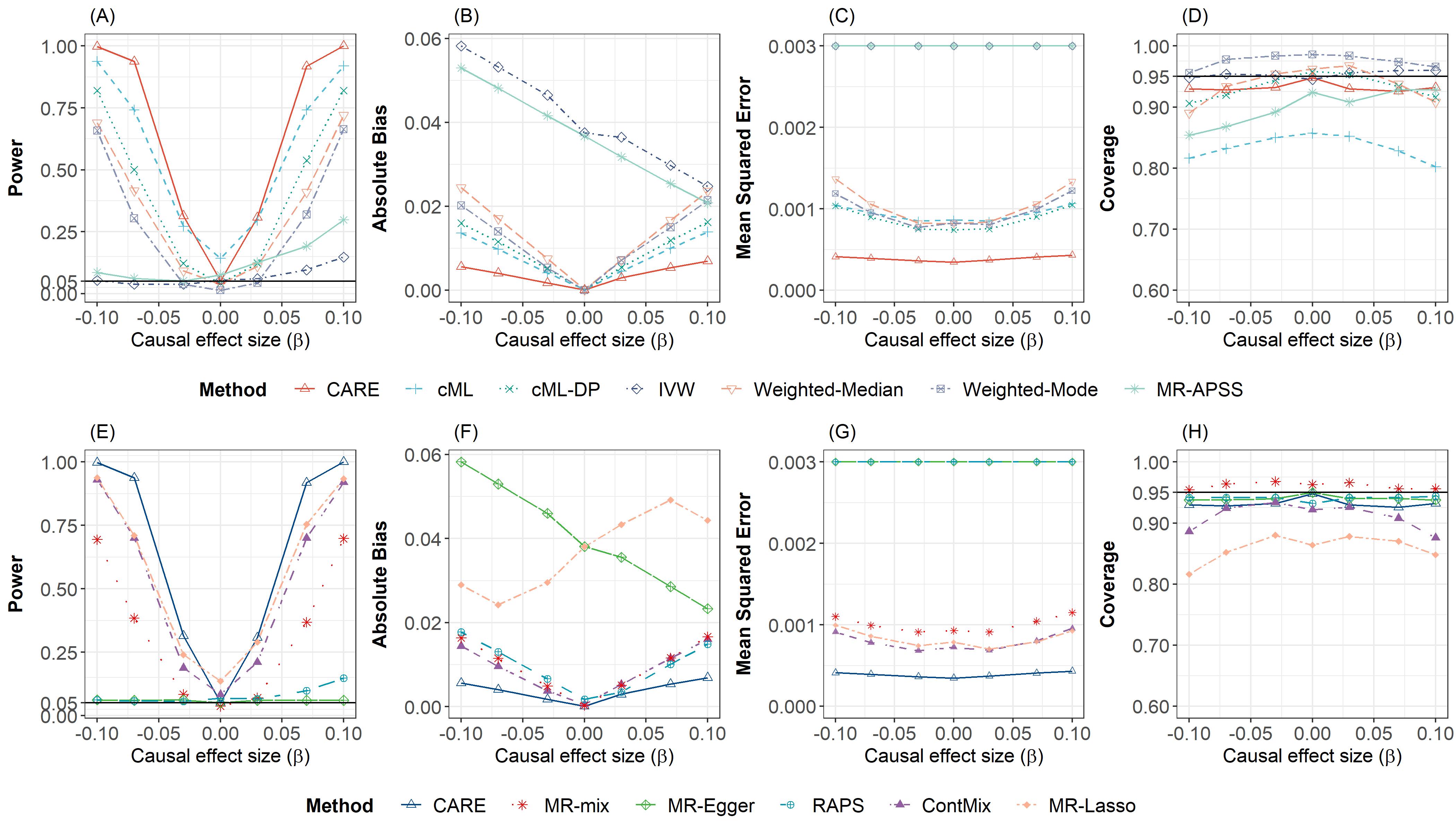}
	\caption{	\label{supfig:simulation_dir_plei_05} Power, absolute bias, mean squared error, and coverage of the CARE estimator and several robust MR methods under the setting of directional horizontal pleiotropy with InSIDE assumption violated with 50\% invalid IVs. Power is the empirical power estimated by the proportion of p-values less than the significance threshold 0.05.  Coverage is the empirical coverage probability of the 95\% confidence interval.} 
\end{figure}

\begin{figure}[!htbp]
	\centering	\includegraphics[width=\linewidth]{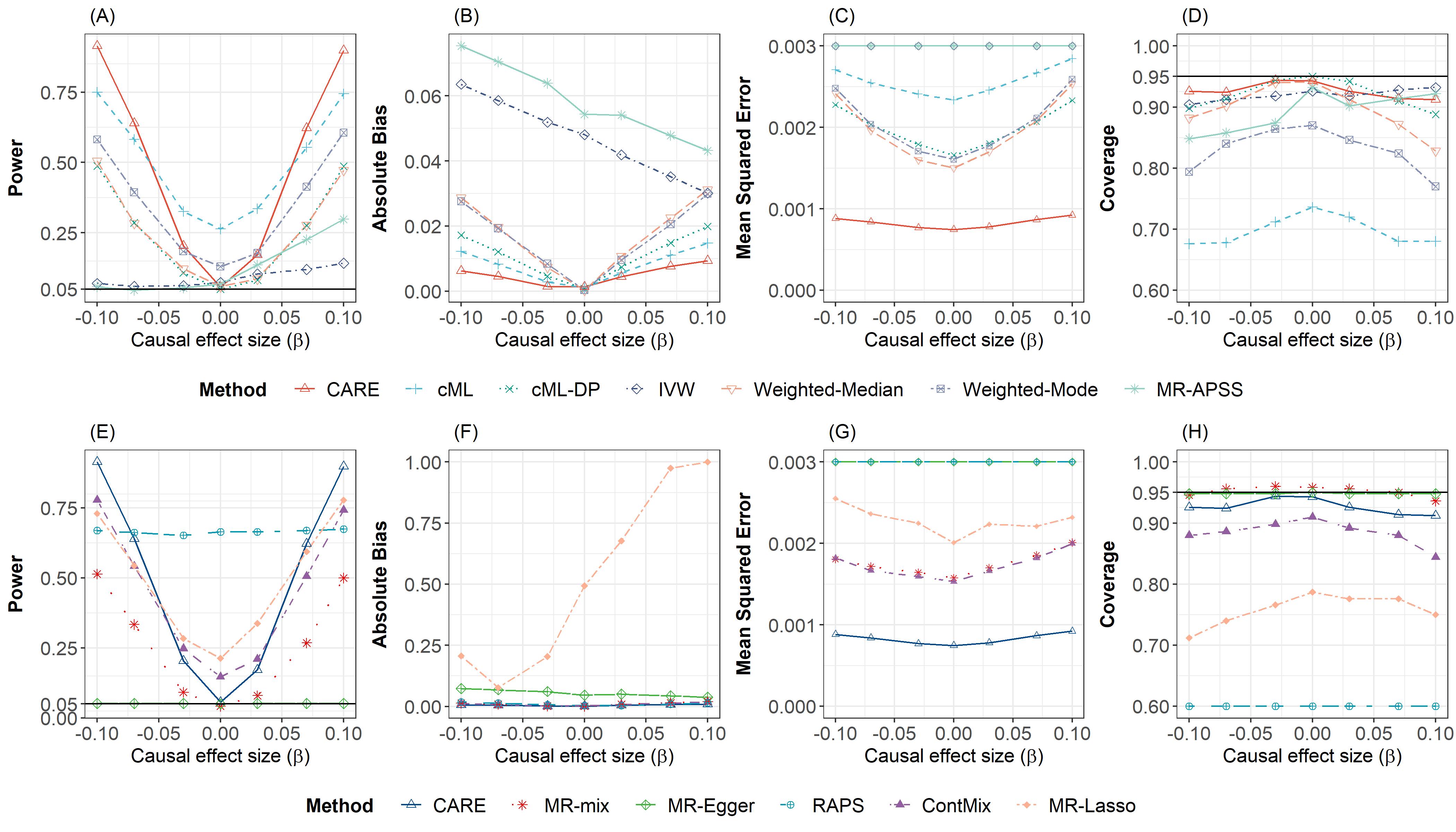}
	\caption{	\label{supfig:simulation_dir_plei_07} Power, absolute bias, mean squared error, and coverage of the CARE estimator and several robust MR methods under the setting of directional horizontal pleiotropy with InSIDE assumption violated with 70\% invalid IVs. Power is the empirical power estimated by the proportion of p-values less than the significance threshold 0.05.  Coverage is the empirical coverage probability of the 95\% confidence interval.} 
\end{figure}

Under the setting of directional horizontal pleiotropy with InSIDE assumption violated, we generate the underlying parameters using the following distribution:
\begin{align*}
    \begin{pmatrix}
		\gamma_j \\
		\alpha_j \\
		\phi_j
		\end{pmatrix} \sim \underbrace{\pi_1 \begin{pmatrix}
		\mathsf{N}(0,\sigma_x^2)\\
		\delta_0 \\
		\delta_0
		\end{pmatrix}}_{\text{Valid IVs}} & + \underbrace{\pi_2 \begin{pmatrix}
		\mathsf{N}(0,{\sigma}_{x}^2)\\
		U(0.01,0.03) \\
		\mathsf{N}(0,\sigma_{u}^2)
		\end{pmatrix}}_{\text{correlated pleiotropy}}
		 + \underbrace{\pi_4 \begin{pmatrix}
		\delta_0\\
		\mathsf{N}(0,\sigma_{y}^2)\\
		\delta_0
		\end{pmatrix} + \pi_5 \begin{pmatrix}
		\delta_0\\
		\delta_0\\
		\delta_0
		\end{pmatrix}}_{\text{IVs fail the relevance assumption}},
\end{align*}
We follow the main simulation setting and set $\pi_1 + \pi_2 = 0.02$,  $\pi_4  = 0.01$, and $\pi_5 = 0.97$. We vary the proportion of invalid IVs, which is defined as $(\pi_2)/ (\pi_1 + \pi_2)$, to simulate different situations. Figures~\ref{supfig:simulation_dir_plei_03} to \ref{supfig:simulation_dir_plei_07} summarize the results for the settings with 30\%, 50\%, and 70\% invalid IVs.

\subsection{Sensitivity analysis using different values of $\eta$}\label{sup:sec_eta}
We conducted sensitivity analyses using different values of $\eta$ (0.1, 0.3, 0.5, 0.7, 0.9) in our main setting. We generate the underlying parameters using the following distribution:
\begin{align*}
    \begin{pmatrix}
		\gamma_j \\
		\alpha_j \\
		\phi_j
		\end{pmatrix} \sim \underbrace{\pi_1 \begin{pmatrix}
		\mathsf{N}(0,\sigma_x^2)\\
		\delta_0 \\
		\delta_0
		\end{pmatrix}}_{\text{Valid IVs}} & + \underbrace{\pi_2 \begin{pmatrix}
		\mathsf{N}(0,{\sigma}_{x}^2)\\
		\mathsf{N}(0.015,{\sigma}_{u}^2)\\
		\mathsf{N}(0,\sigma_{u}^2)
		\end{pmatrix}}_{\text{correlated pleiotropy}} + \underbrace{ \pi_3 \begin{pmatrix}
		\mathsf{N}(0,{\sigma}_x^2)\\
		\mathsf{N}(0,{\sigma}_{y}^2)\\
		\delta_0
		\end{pmatrix} }_{\text{uncorrelated pleiotropy}} 
		 + \underbrace{\pi_4 \begin{pmatrix}
		\delta_0\\
		\mathsf{N}(0,\sigma_{y}^2)\\
		\delta_0
		\end{pmatrix} + \pi_5 \begin{pmatrix}
		\delta_0\\
		\delta_0\\
		\delta_0
		\end{pmatrix}}_{\text{IVs fail the relevance assumption}},
\end{align*}

\begin{figure}[!htbp]
	\centering	\includegraphics[width=\linewidth]{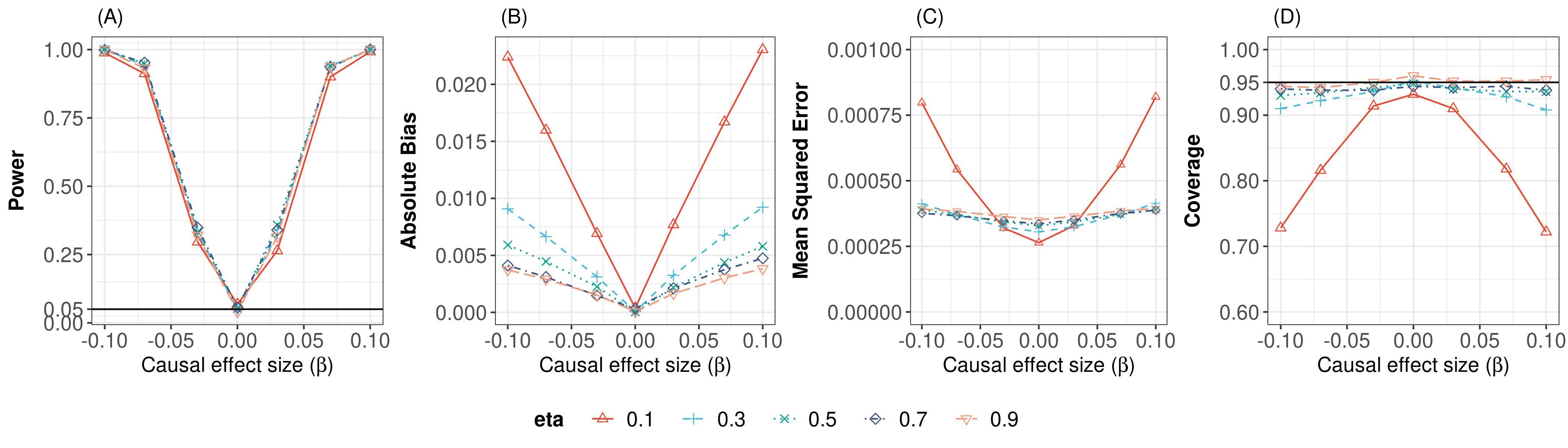}
	\caption{	\label{supfig:eta} 
Power, absolute bias, mean squared error, and coverage of the CARE estimator with different $\eta$ under the main setting. Power is the empirical power estimated by the proportion of p-values less than the significance threshold 0.05. Coverage is the empirical coverage probability of the 95\% confidence interval.} 
\end{figure}

We follow the main simulation setting and set $\pi_1 + \pi_2 + \pi_3 = 0.02$,  $\pi_4  = 0.01$, and $\pi_5 = 0.97$. Figures~\ref{supfig:eta} summarize the results for the settings with 50\% invalid IVs.

\subsection{Consistency of using GBIC with different choices of $\kappa_n$ as model selection methods}\label{sup:gic}
We discuss the adjustment of BIC when $s_{\lambda}$ tends to infinity with generalized BIC (GBIC) of the following form: 
\[
    \text{GBIC}(v) = - 2 \hat{l}\big(\hat\theta(v), \{ \hat{r}_{j}(v)\}_{j\in \hat{\mathcal{V}}} \big)+\kappa_n \cdot(s_{\lambda}-v), \quad s_\lambda=|\mathcal{S}_\lambda|. 
\]
We tested two choices of \(\kappa_n\): (i) $\kappa_n = \log n$ and (ii)  $\kappa_n = \log(s_\lambda) \cdot \log(\log(n))$, both satisfying  \(\kappa_{n} \gg \log(s_{\lambda})\). 
We generate the underlying parameters using the following distribution:
\begin{align*}
    \begin{pmatrix}
		\gamma_j \\
		\alpha_j \\
		\phi_j
		\end{pmatrix} \sim \underbrace{\pi_1 \begin{pmatrix}
		\mathsf{N}(0,\sigma_x^2)\\
		\delta_0 \\
		\delta_0
		\end{pmatrix}}_{\text{Valid IVs}} & + \underbrace{\pi_2 \begin{pmatrix}
		\mathsf{N}(0,{\sigma}_{x}^2)\\
		\mathsf{N}(0.015,{\sigma}_{u}^2)\\
		\mathsf{N}(0,\sigma_{u}^2)
		\end{pmatrix}}_{\text{correlated pleiotropy}} + \underbrace{ \pi_3 \begin{pmatrix}
		\mathsf{N}(0,{\sigma}_x^2)\\
		\mathsf{N}(0,{\sigma}_{y}^2)\\
		\delta_0
		\end{pmatrix} }_{\text{uncorrelated pleiotropy}} 
		 + \underbrace{\pi_4 \begin{pmatrix}
		\delta_0\\
		\mathsf{N}(0,\sigma_{y}^2)\\
		\delta_0
		\end{pmatrix} + \pi_5 \begin{pmatrix}
		\delta_0\\
		\delta_0\\
		\delta_0
		\end{pmatrix}}_{\text{IVs fail the relevance assumption}},
\end{align*}

\begin{figure}[!htbp]
	\centering	\includegraphics[width=\linewidth]{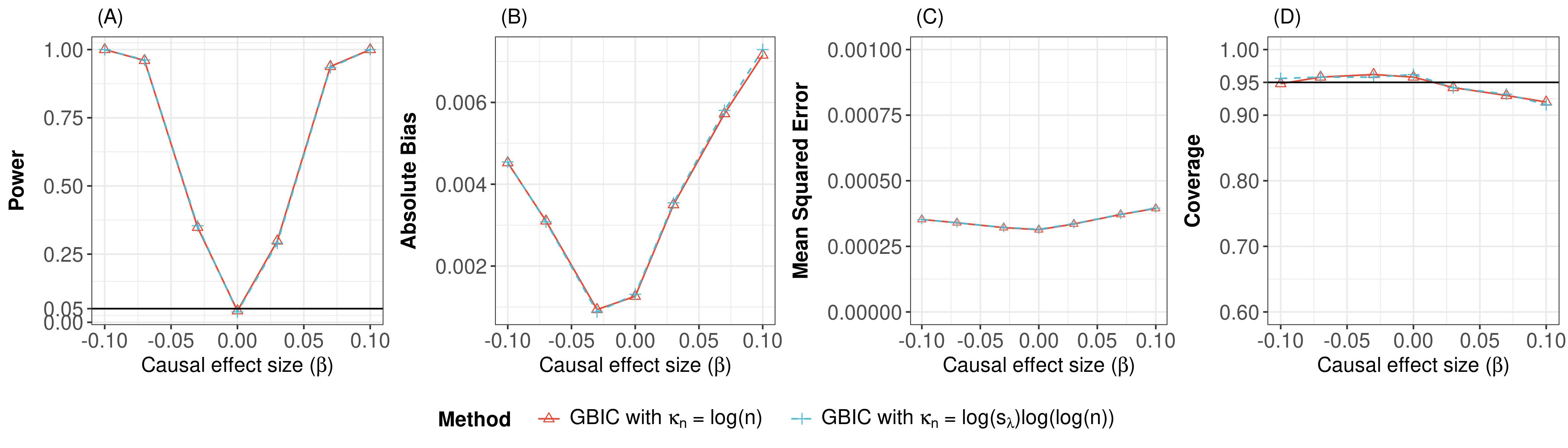}
	\caption{	\label{supfig:gic} 
Power, absolute bias, mean squared error, and coverage of the CARE estimator using GBIC with $\kappa_n = \log n$ and $\kappa_n = \log(s_\lambda) \cdot \log(\log(n))$ under the main setting. Power is the empirical power estimated by the proportion of p-values less than the significance threshold 0.05. Coverage is the empirical coverage probability of the 95\% confidence interval.
    } 
\end{figure}

\begin{figure}[H]
    \centering
    \includegraphics[width=1\linewidth]{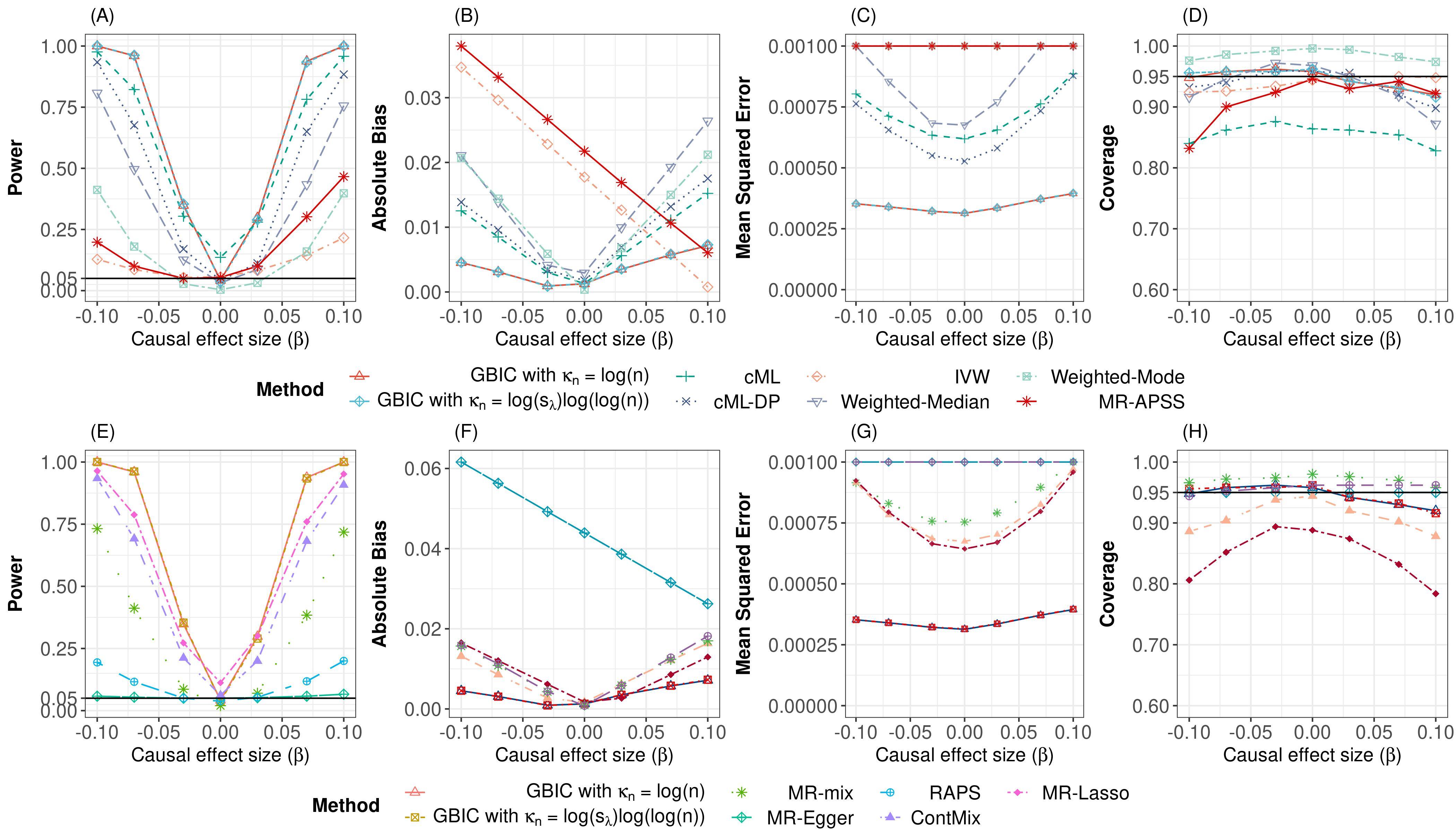}
    \caption{Comparison of Power, absolute bias, mean squared error, and coverage of the CARE estimator using GBIC with $\kappa_n = \log n$ and $\kappa_n = \log(s_\lambda) \cdot \log(\log(n))$  and other benchmark methods under the main setting with 50\% invalid IVs. Power is the empirical power estimated by the proportion of p-values less than the significance threshold 0.05. Coverage is the empirical coverage probability of the 95\% confidence interval. \label{subfig:GBIC}} 
\end{figure}

We follow the main simulation setting and set $\pi_1 + \pi_2 + \pi_3 = 0.02$,  $\pi_4  = 0.01$, and $\pi_5 = 0.97$. Figures~\ref{supfig:gic} and \ref{subfig:GBIC} summarize the results for the settings with 50\% invalid IVs.

\subsection{Nonlinear settings}\label{sup:nonlinear}
In our non-linear simulation settings, we implement a four-step process to model complex genetic relationships. First, we simulate $p$ mutually independent single nucleotide polymorphisms (SNPs), denoted as $\mathbf{G} = (G_1, ..., G_p)^T$. Each SNP $G_j$ follows a $\text{Binomial}(2, \text{MAF}_j)$ distribution, where $\text{MAF}_j$ represents the minor allele frequency drawn from a $\text{Uniform}(0.01, 0.5)$ distribution. Next, we simulate an unmeasured confounder $U$ as $U = \sum_{j=1}^{p}\phi_j G_j + E_U$. The risk factor $X$ is then simulated as $X = \sum_{j=1}^{p} f(G_j) + \beta_{XU} U + E_X$, and finally, the outcome $Y$ is modeled as $Y = \theta X + \beta_{YU} U + \sum_{j=1}^{p} \alpha_j G_j + E_Y$. In these equations, $E_U$, $E_X$, and $E_Y$ represent mutually independent random noise terms, distributed as $E_U \sim \mathcal{N}(0, \sigma_U^2)$, $E_X \sim \mathcal{N}(0, \sigma_X^2)$, and $E_Y \sim \mathcal{N}(0, \sigma_Y^2)$, respectively. These distributions are consistent with the main setting. Similarly, the coefficients $\gamma_j$, $\alpha_j$, and $\phi_j$ are generated from the same mixture of distributions as described in the main setting. To explore different non-linear relationships, we consider three scenarios. In the first, we focus on non-linearity in $X$ with a linear $Y$, where $f(G_j) = \gamma_{1j} G_j^2 + \gamma_{2j} G_j$, with $\gamma_{1j} = \gamma_{2j} = \gamma_j$. The second scenario introduces additional complexity by incorporating interaction terms between SNPs in the model for $X$, such that $X = \sum_{j=1}^{p} f(G_j) + \sum_{i,j \in S} \gamma_{ij} G_i G_j + \beta_{XU} U + E_X$, where $f(G_j)$ remains as in the first scenario, and $S$ represents a randomly selected set of 20 SNP pairs for which interaction effects are modeled. The third scenario introduces non-linearity in $Y$ with $Y = \theta^2 X + \beta_{YU} U + \sum_{j=1}^{p} \alpha_j G_j + E_Y$.

\begin{figure}[!htbp]
	\centering	\includegraphics[width=\linewidth]{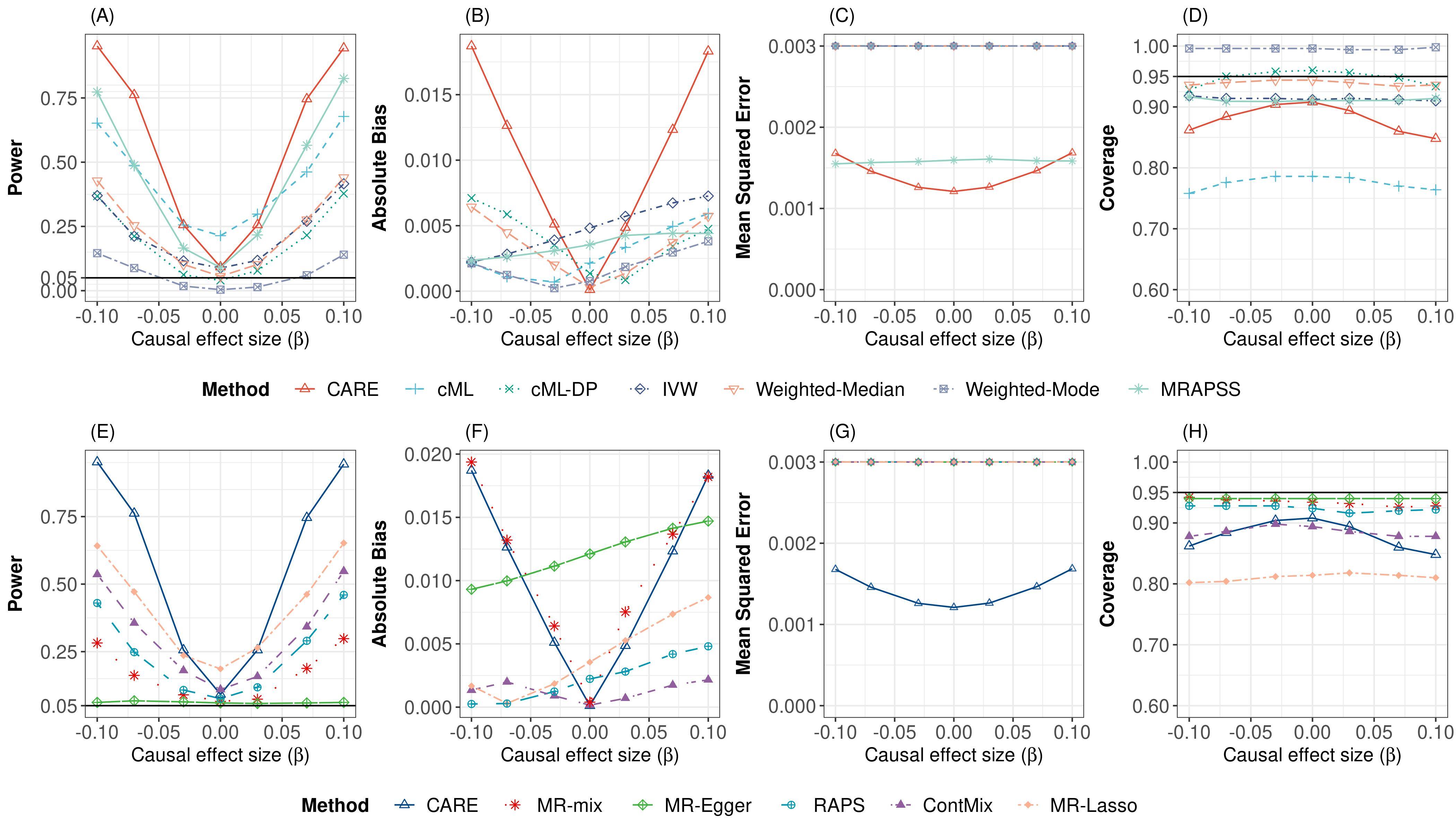}
	\caption{	\label{supfig:nonlinear} Power, absolute bias, mean squared error, and coverage of the CARE estimator and several robust MR methods under the setting of non-linearity in exposure without interaction terms with 50\% invalid IVs. Power is the empirical power estimated by the proportion of p-values less than the significance threshold 0.05.  Coverage is the empirical coverage probability of the 95\% confidence interval.} 
\end{figure}

\begin{figure}[!htbp]
	\centering	\includegraphics[width=\linewidth]{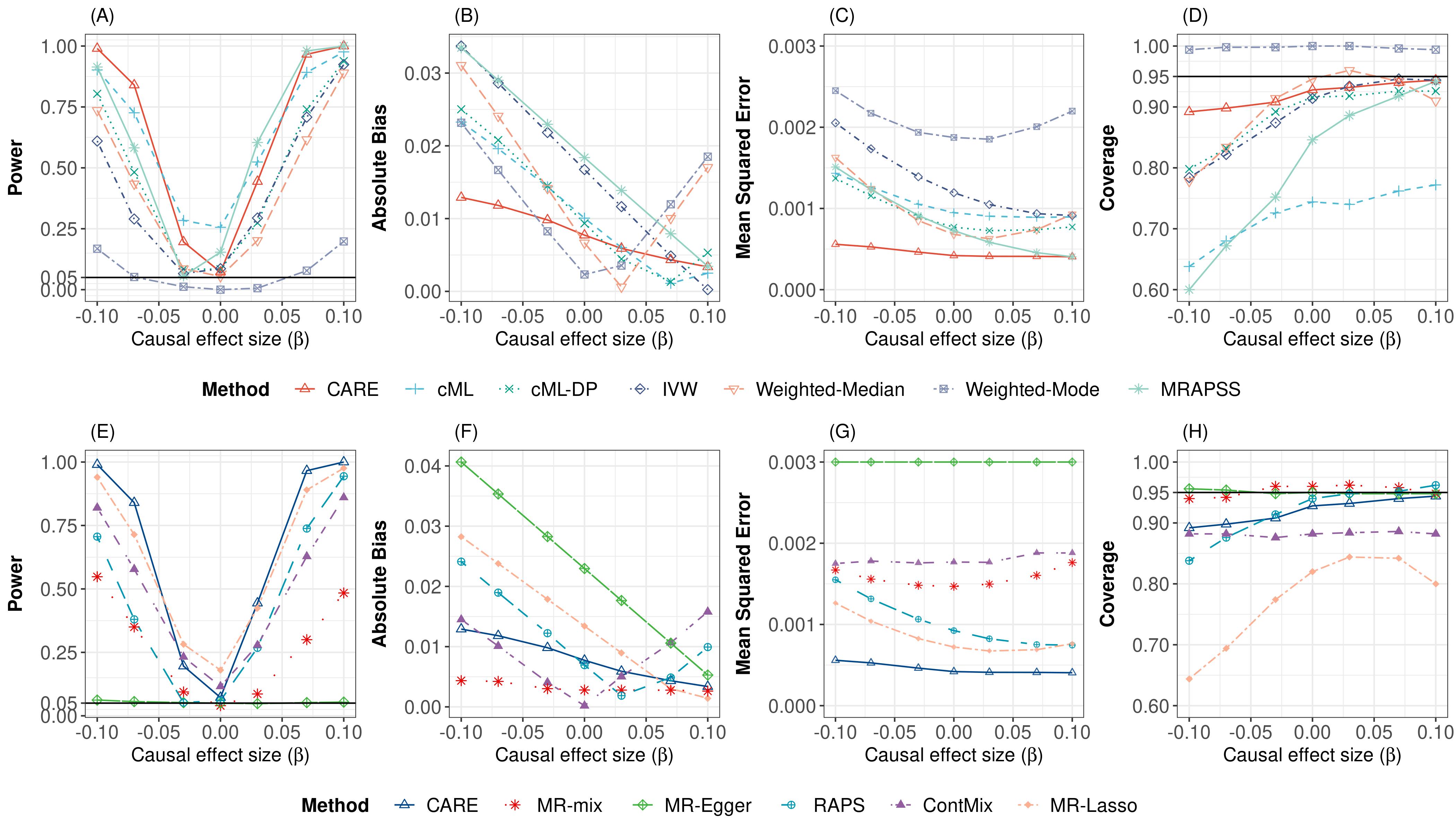}
	\caption{	\label{supfig:nonlinear_interaction} Power, absolute bias, mean squared error, and coverage of the CARE estimator and several robust MR methods under the setting of non-linearity in exposure with interaction terms with 50\% invalid IVs. Power is the empirical power estimated by the proportion of p-values less than the significance threshold 0.05.  Coverage is the empirical coverage probability of the 95\% confidence interval.} 
\end{figure}

\begin{figure}[!htbp]
	\centering	\includegraphics[width=\linewidth]{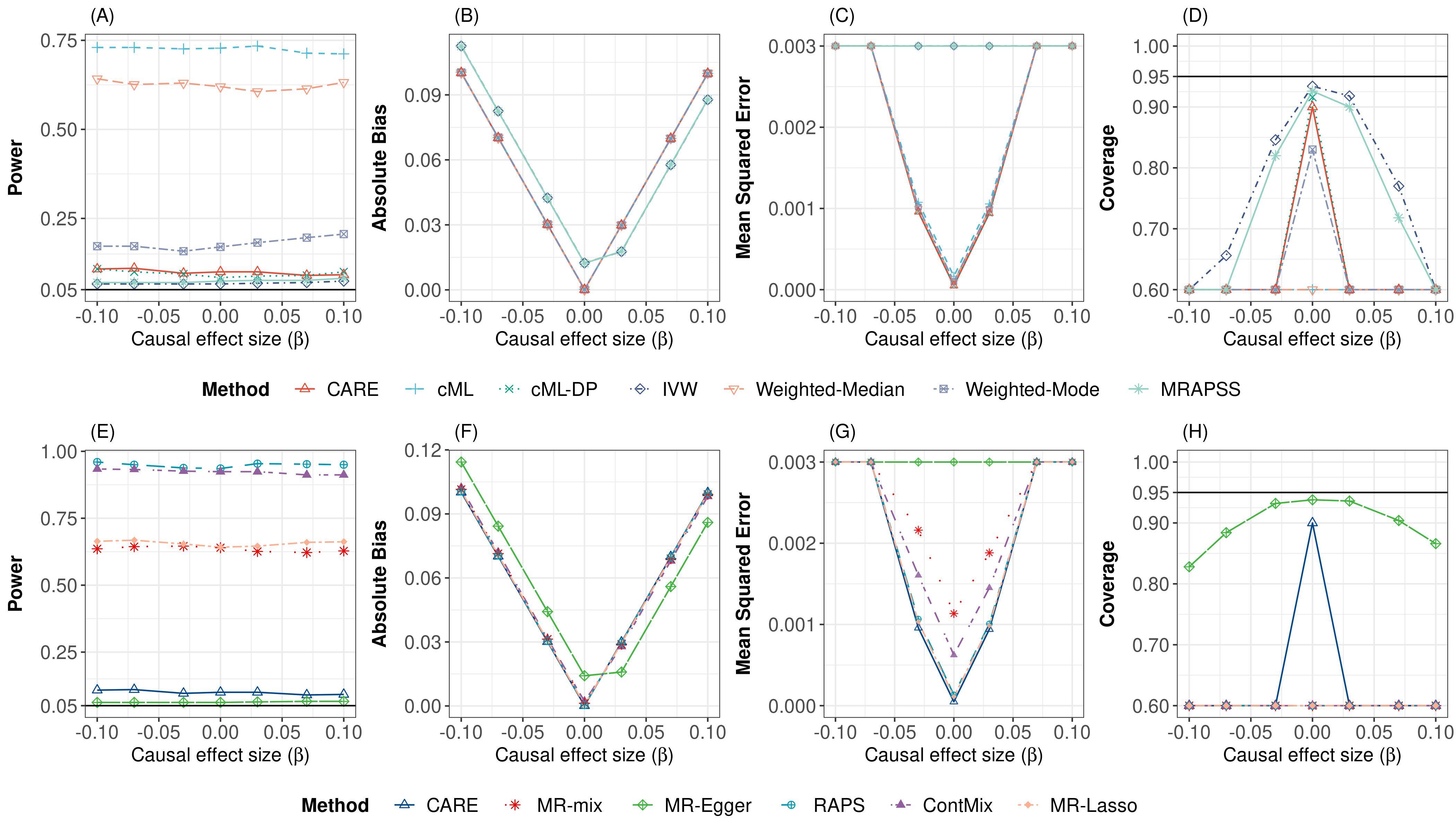}
	\caption{	\label{supfig:nonlinear_Y} Power, absolute bias, mean squared error, and coverage of the CARE estimator and several robust MR methods under the  setting of non-linearity in both exposure and outcome without interaction terms with 50\% invalid IVs. Power is the empirical power estimated by the proportion of p-values less than the significance threshold 0.05.  Coverage is the empirical coverage probability of the 95\% confidence interval.} 
\end{figure}

Supplementary Figures~\ref{supfig:nonlinear} to~\ref{supfig:nonlinear_Y} summarize the results for the three scenarios with 50\% invalid IVs. In the first two scenarios with non-linear $X$ on $G$, CARE showed slightly inflated Type 1 error rates, larger bias, and worse coverage (Supplementary Figures~\ref{supfig:nonlinear} and \ref{supfig:nonlinear_interaction}). The third scenario revealed that CARE demonstrated diminished power, larger bias, and poor coverage (Supplementary Figure~\ref{supfig:nonlinear_Y}).

\subsection{Sample size variation of GWAS}\label{sup:sample_size}
We evaluate the performance of CARE estimator and benchmark MR methods with different sample sizes of both exposure and outcome GWAS (100000, 50000, 10000, 5000). We generate the underlying parameters using the same distribution as the main setting. 

\begin{figure}[!htbp]
	\centering	\includegraphics[width=\linewidth]{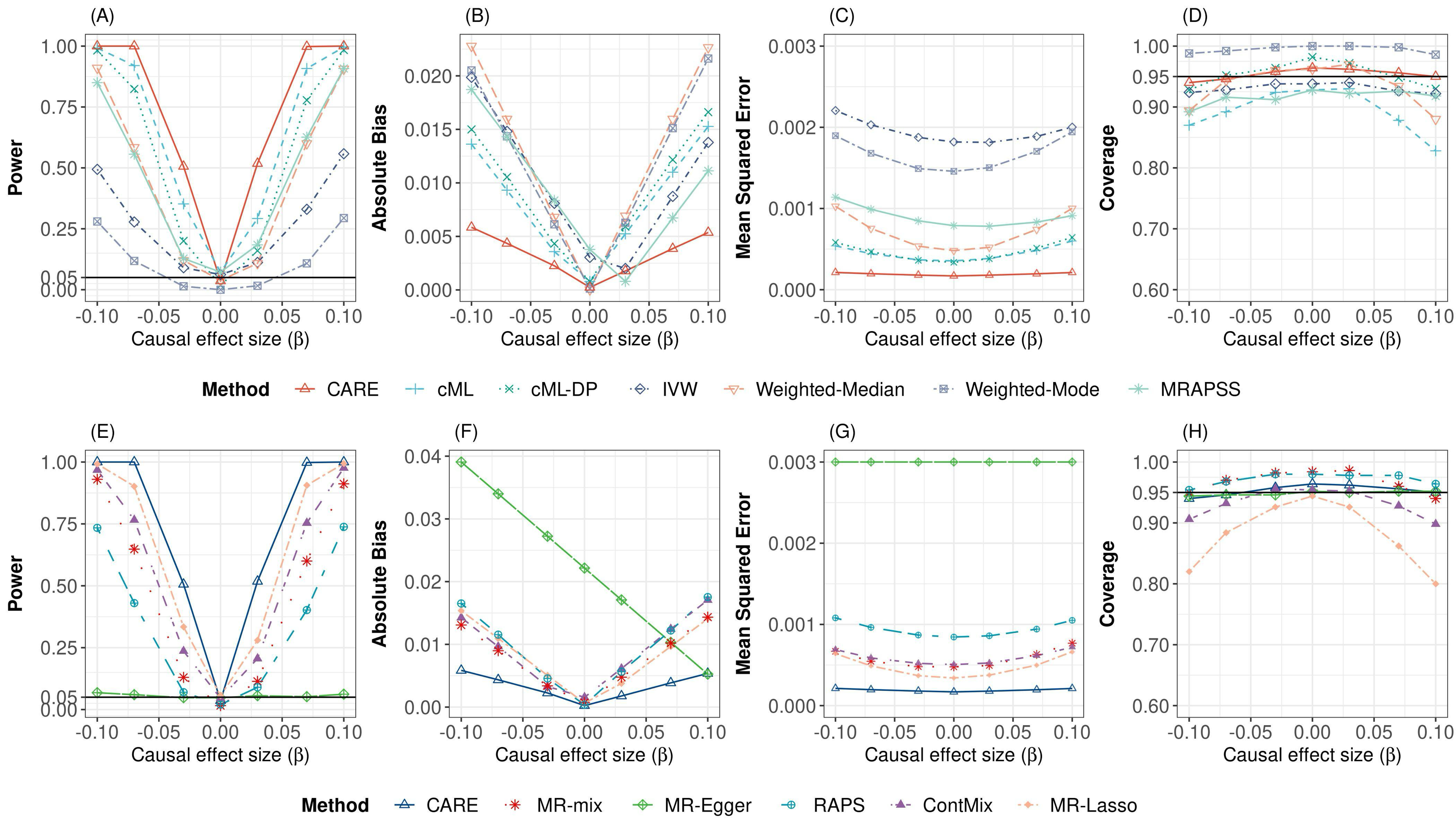}
	\caption{	\label{supfig:samplesize_100000} Power, absolute bias, mean squared error, and coverage of the CARE estimator and several robust MR methods with sample size = 100000, $\sigma^2_x = 1\times10^{-5}$ and  50\% invalid IVs. Power is the empirical power estimated by the proportion of p-values less than the significance threshold 0.05.  Coverage is the empirical coverage probability of the 95\% confidence interval.} 
\end{figure}

\begin{figure}[!htbp]
	\centering	\includegraphics[width=\linewidth]{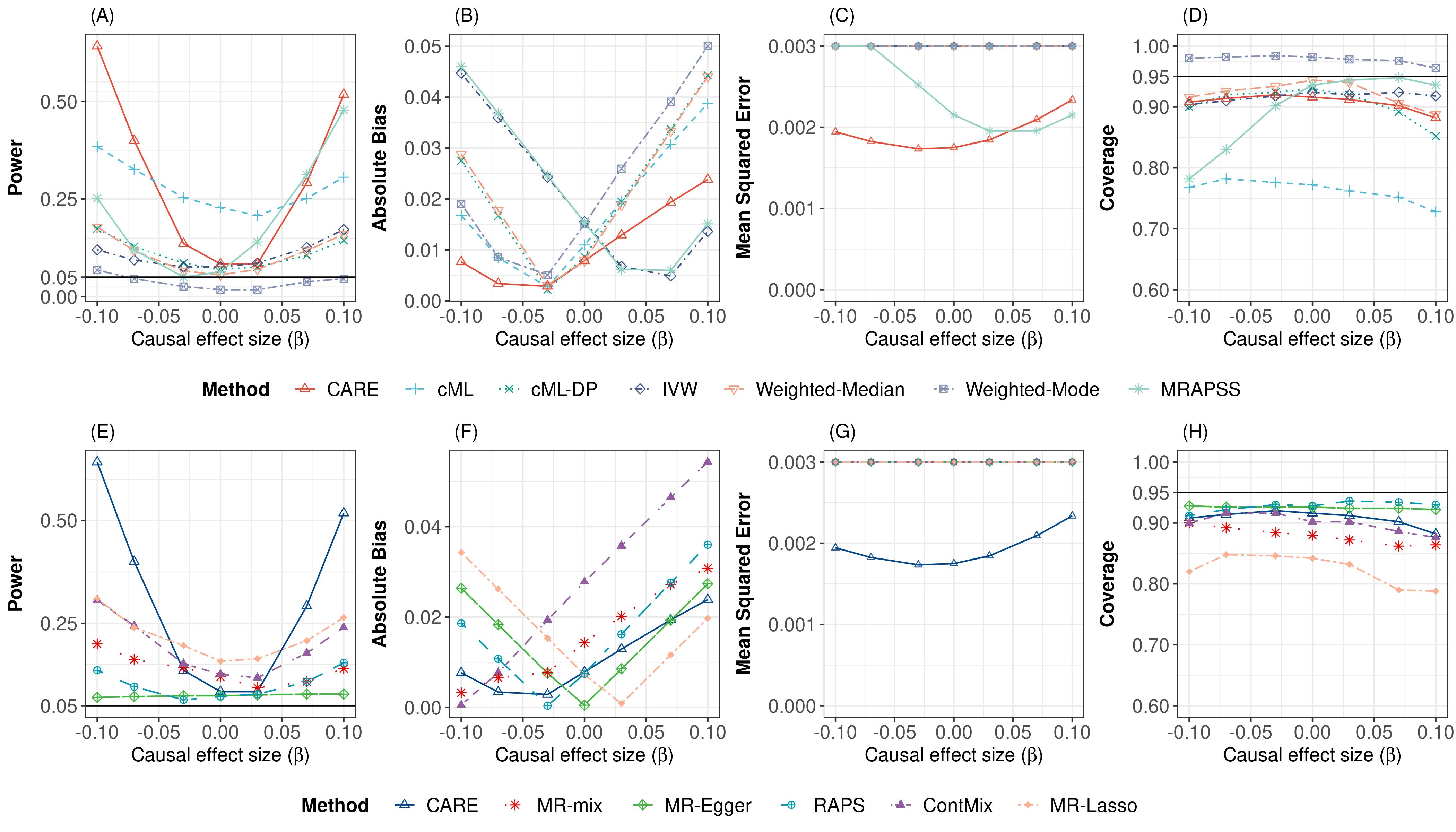}
	\caption{	\label{supfig:samplesize_50000} Power, absolute bias, mean squared error, and coverage of the CARE estimator and several robust MR methods with sample size = 50000, $\sigma^2_x = 5\times10^{-5}$ and  50\% invalid IVs. Power is the empirical power estimated by the proportion of p-values less than the significance threshold 0.05.  Coverage is the empirical coverage probability of the 95\% confidence interval.} 
\end{figure}

\begin{figure}[!htbp]
	\centering	\includegraphics[width=\linewidth]{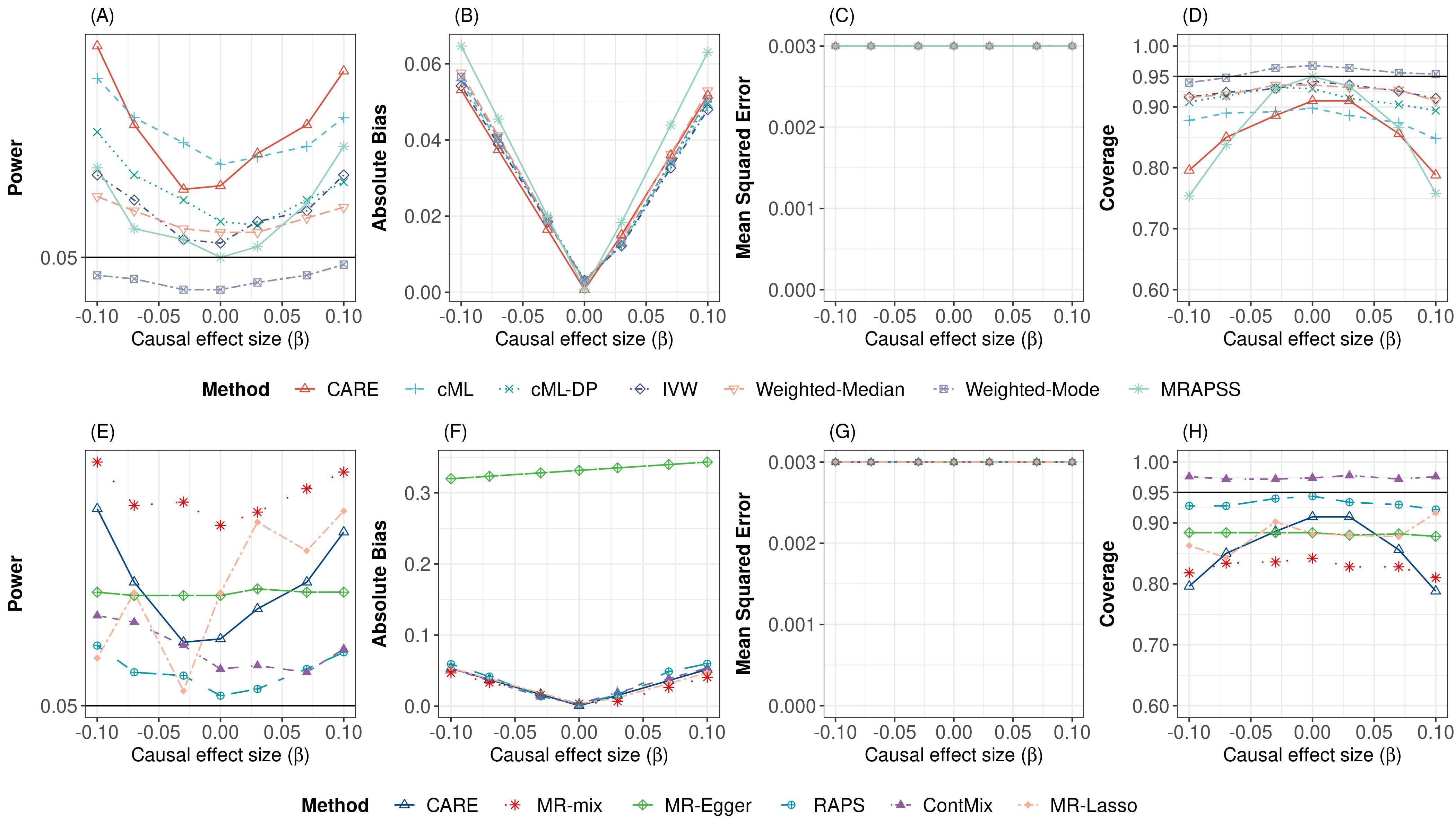}
	\caption{	\label{supfig:samplesize_10000} Power, absolute bias, mean squared error, and coverage of the CARE estimator and several robust MR methods with sample size = 10000, $\sigma^2_x = 8\times10^{-5}$ and  50\% invalid IVs. Power is the empirical power estimated by the proportion of p-values less than the significance threshold 0.05.  Coverage is the empirical coverage probability of the 95\% confidence interval.} 
\end{figure}

\begin{figure}[!htbp]
	\centering	\includegraphics[width=\linewidth]{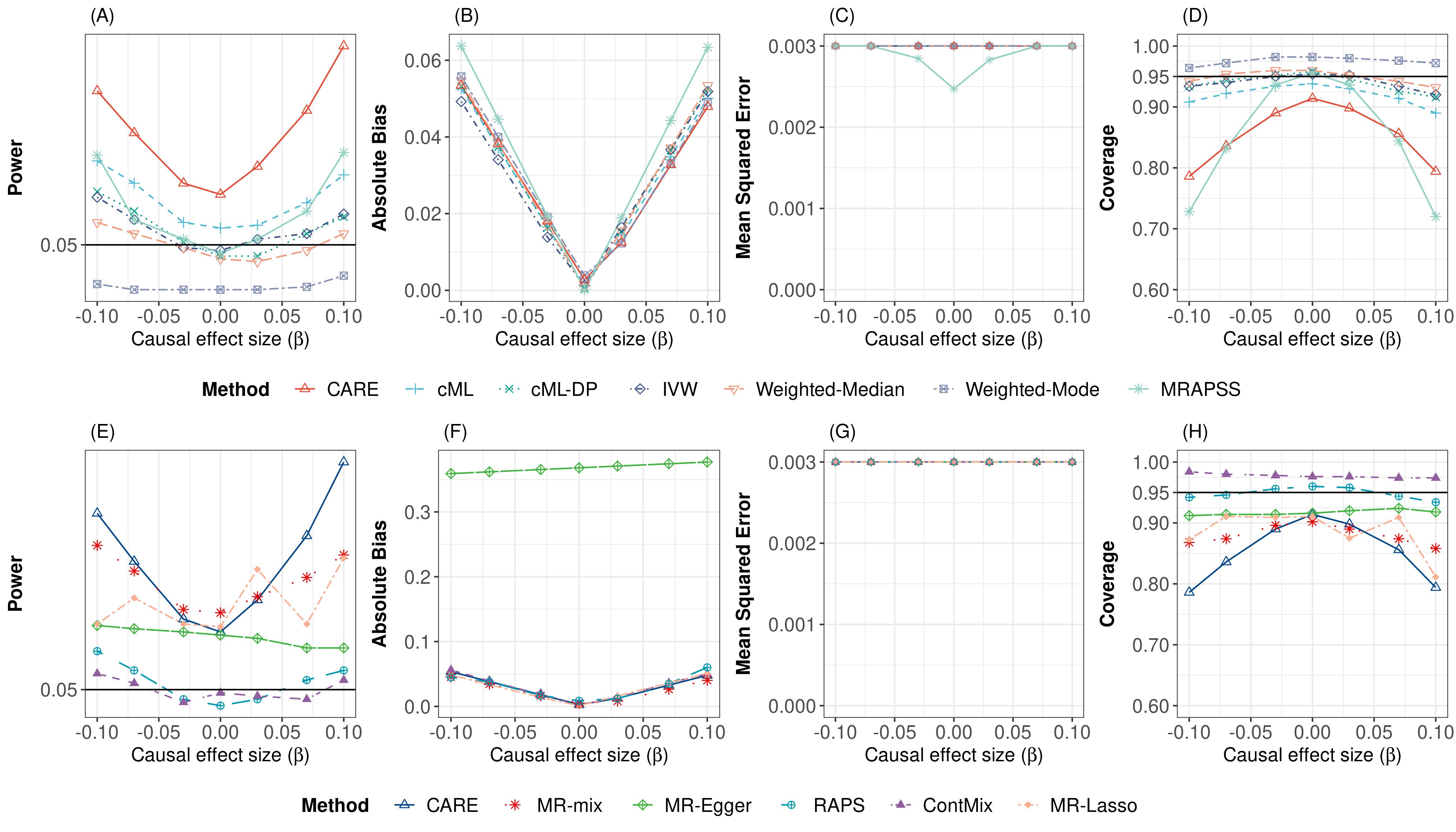}
	\caption{	\label{supfig:samplesize_5000} Power, absolute bias, mean squared error, and coverage of the CARE estimator and several robust MR methods with sample size = 5000, $\sigma^2_x = 1\times10^{-4}$ and  50\% invalid IVs. Power is the empirical power estimated by the proportion of p-values less than the significance threshold 0.05.  Coverage is the empirical coverage probability of the 95\% confidence interval.} 
\end{figure}

We follow the main simulation setting and set $\pi_1 + \pi_2 + \pi_3 = 0.02$,  $\pi_4  = 0.01$, and $\pi_5 = 0.97$. We vary the proportion of invalid IVs, which is defined as $(\pi_2 + \pi_3)/ (\pi_1 + \pi_2 + \pi_3)$, to simulate different situations. To maintain heritability within a biologically plausible range, we adjust the variance of the risk factor, denoted as $\sigma_{X}^2$, across different simulation scenarios to maintain reasonable heritability. Figure~\ref{supfig:samplesize_100000} to~\ref{supfig:samplesize_5000} summarize the results for different sample sizes. The findings indicate that CARE's performance deteriorates as the sample size of GWAS decreases.

\newpage
\subsection{   Variations in number of SNPs}\label{sup:sample_size_SNP}
We evaluate the performance of CARE estimator and benchmark MR methods with different sample sizes of SNPs (100000, 50000, 10000, 5000, 1000). We generate the underlying parameters using the following distribution:
\begin{align*}
    \begin{pmatrix}
		\gamma_j \\
		\alpha_j \\
		\phi_j
		\end{pmatrix} \sim \underbrace{\pi_1 \begin{pmatrix}
		\mathsf{N}(0,\sigma_x^2)\\
		\delta_0 \\
		\delta_0
		\end{pmatrix}}_{\text{Valid IVs}} & + \underbrace{\pi_2 \begin{pmatrix}
		\mathsf{N}(0,{\sigma}_{x}^2)\\
		\mathsf{N}(0.015,{\sigma}_{u}^2)\\
		\mathsf{N}(0,\sigma_{u}^2)
		\end{pmatrix}}_{\text{correlated pleiotropy}} + \underbrace{ \pi_3 \begin{pmatrix}
		\mathsf{N}(0,{\sigma}_x^2)\\
		\mathsf{N}(0,{\sigma}_{y}^2)\\
		\delta_0
		\end{pmatrix} }_{\text{uncorrelated pleiotropy}} 
		 + \underbrace{\pi_4 \begin{pmatrix}
		\delta_0\\
		\mathsf{N}(0,\sigma_{y}^2)\\
		\delta_0
		\end{pmatrix} + \pi_5 \begin{pmatrix}
		\delta_0\\
		\delta_0\\
		\delta_0
		\end{pmatrix}}_{\text{IVs fail the relevance assumption}},
\end{align*}

\begin{figure}[!htbp]
	\centering	\includegraphics[width=\linewidth]{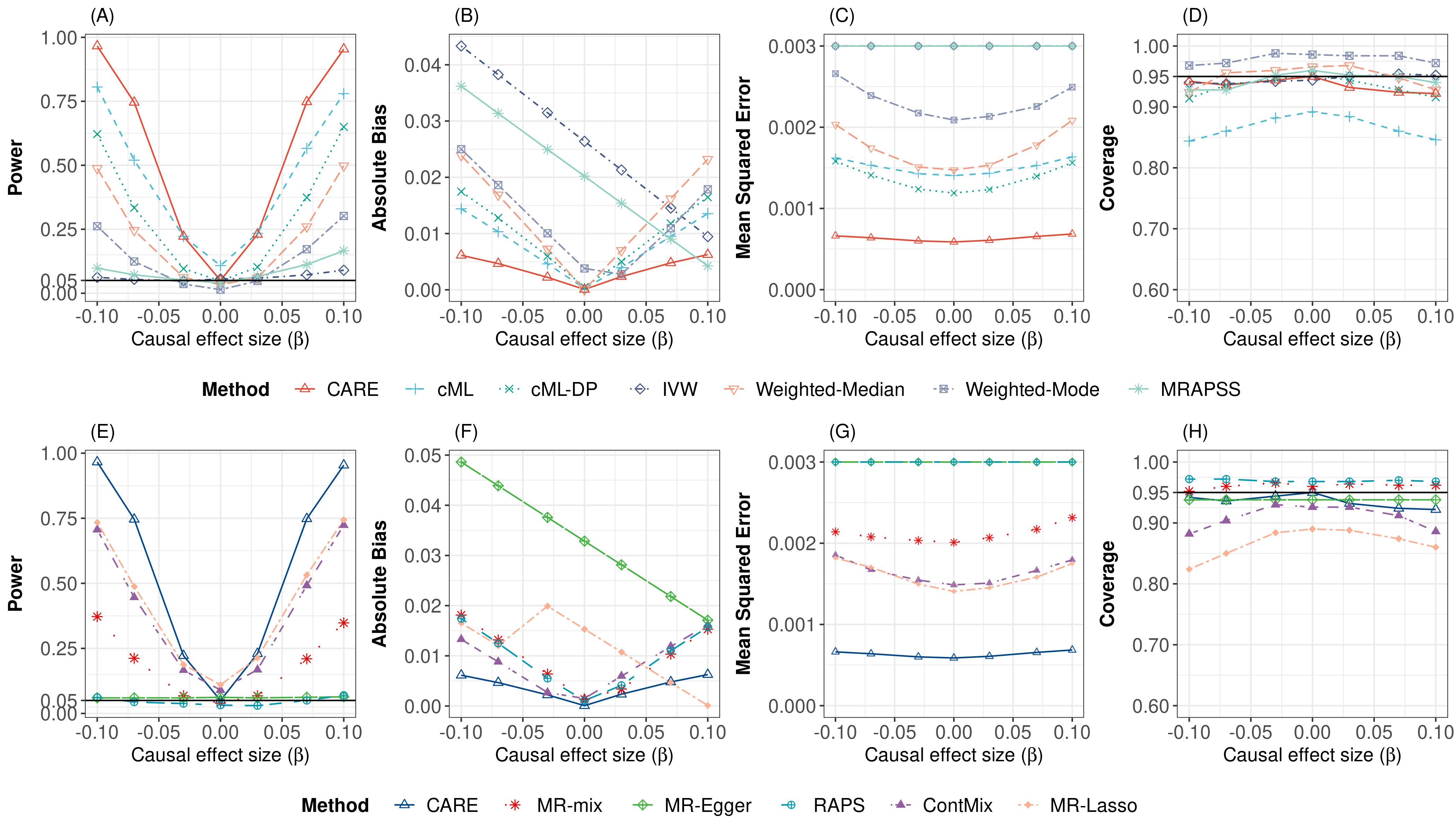}
	\caption{	\label{supfig:SNP_number_100000} Power, absolute bias, mean squared error, and coverage of the CARE estimator and several robust MR methods with the number of SNPs equal to 100,000, and  50\% invalid IVs. Power is the empirical power estimated by the proportion of p-values less than the significance threshold 0.05.  Coverage is the empirical coverage probability of the 95\% confidence interval.} 
\end{figure}

\begin{figure}[!htbp]
	\centering	\includegraphics[width=\linewidth]{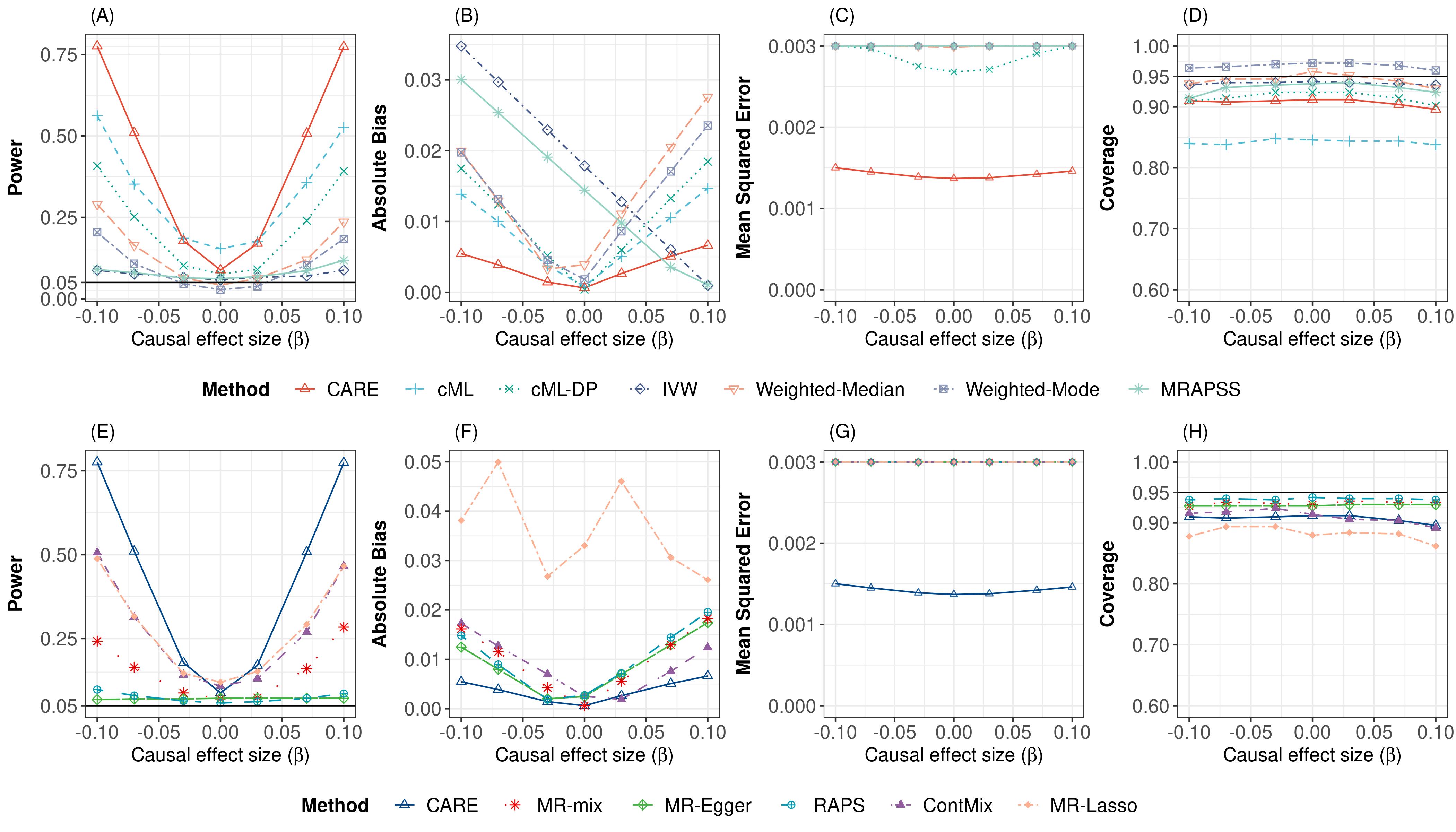}
	\caption{	\label{supfig:SNP_number_50000} Power, absolute bias, mean squared error, and coverage of the CARE estimator and several robust MR methods with the number of SNPs equal to 50,000, and  50\% invalid IVs. Power is the empirical power estimated by the proportion of p-values less than the significance threshold 0.05.  Coverage is the empirical coverage probability of the 95\% confidence interval.} 
\end{figure}

\begin{figure}[!htbp]
	\centering	\includegraphics[width=\linewidth]{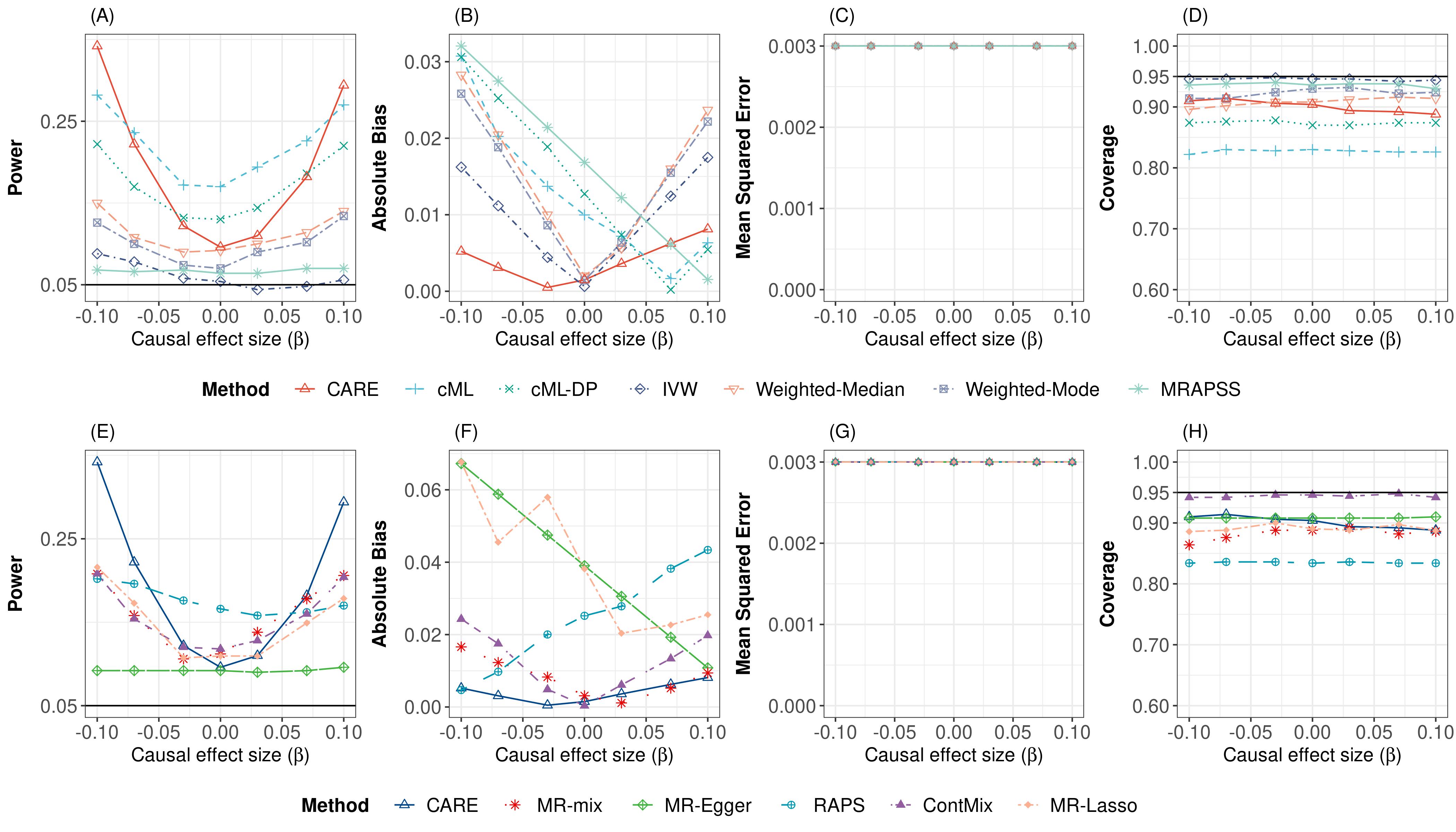}
	\caption{	\label{supfig:SNP_number_10000} Power, absolute bias, mean squared error, and coverage of the CARE estimator and several robust MR methods with the number of SNPs equal to 10,000, and  50\% invalid IVs. Power is the empirical power estimated by the proportion of p-values less than the significance threshold 0.05.  Coverage is the empirical coverage probability of the 95\% confidence interval.} 
\end{figure}

\begin{figure}[!htbp]
	\centering	\includegraphics[width=\linewidth]{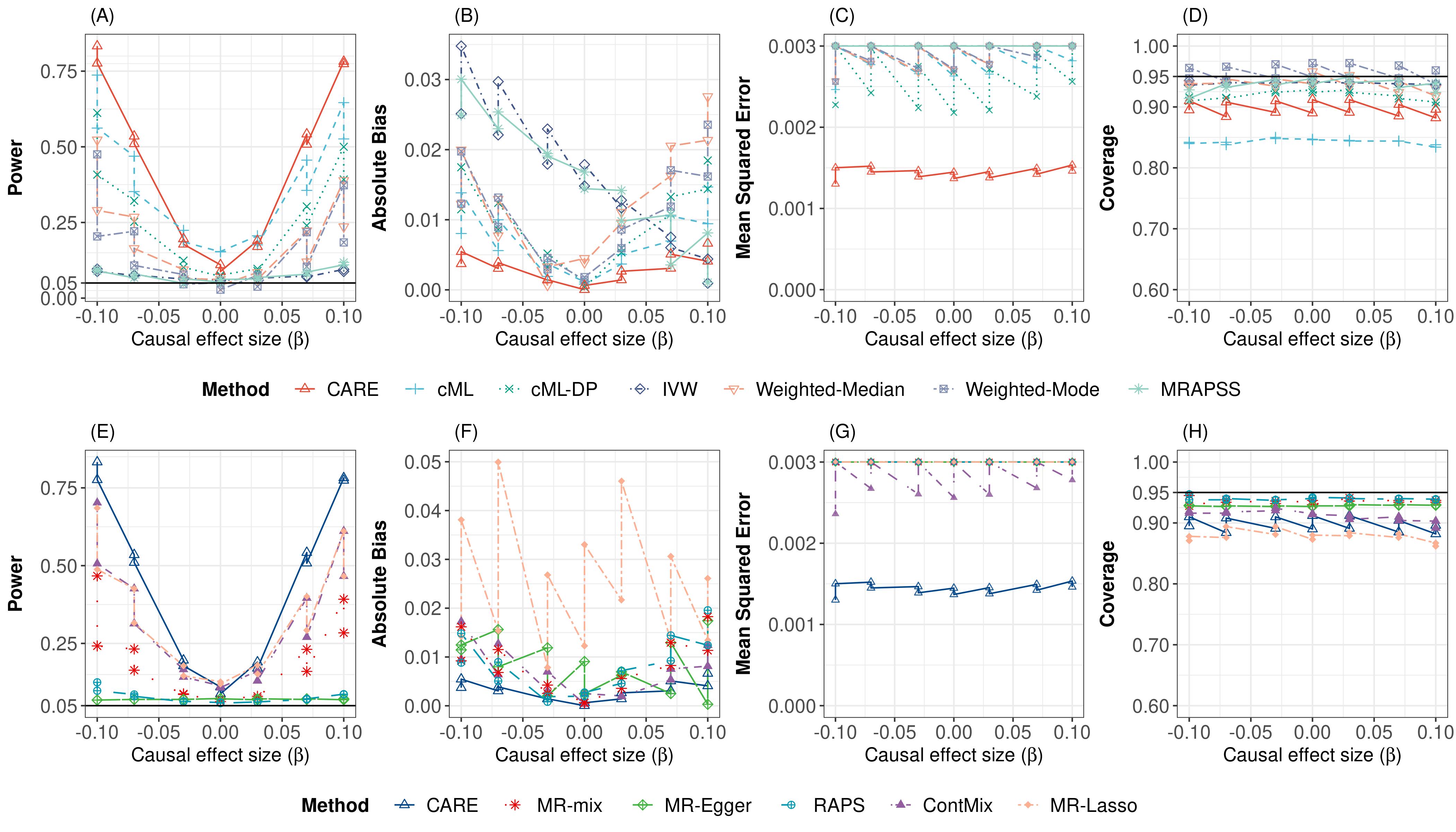}
	\caption{	\label{supfig:SNP_number_5000} Power, absolute bias, mean squared error, and coverage of the CARE estimator and several robust MR methods with the number of SNPs equal to 5,000, and  50\% invalid IVs. Power is the empirical power estimated by the proportion of p-values less than the significance threshold 0.05.  Coverage is the empirical coverage probability of the 95\% confidence interval.} 
\end{figure}

We follow the main simulation setting and set $\pi_1 + \pi_2 + \pi_3 = 0.02$,  $\pi_4  = 0.01$, and $\pi_5 = 0.97$. Supplementary Figure~\ref{supfig:SNP_number_100000} to \ref{supfig:SNP_number_5000} summarize the results for different sample sizes of SNPs. The findings indicate that CARE's performance deteriorates as the sample size of SNPs decreases.

\subsection{  Using the same liberal threshold}\label{sup:same threshold}
\begin{figure}[!http]
    \centering \includegraphics[width=\linewidth]{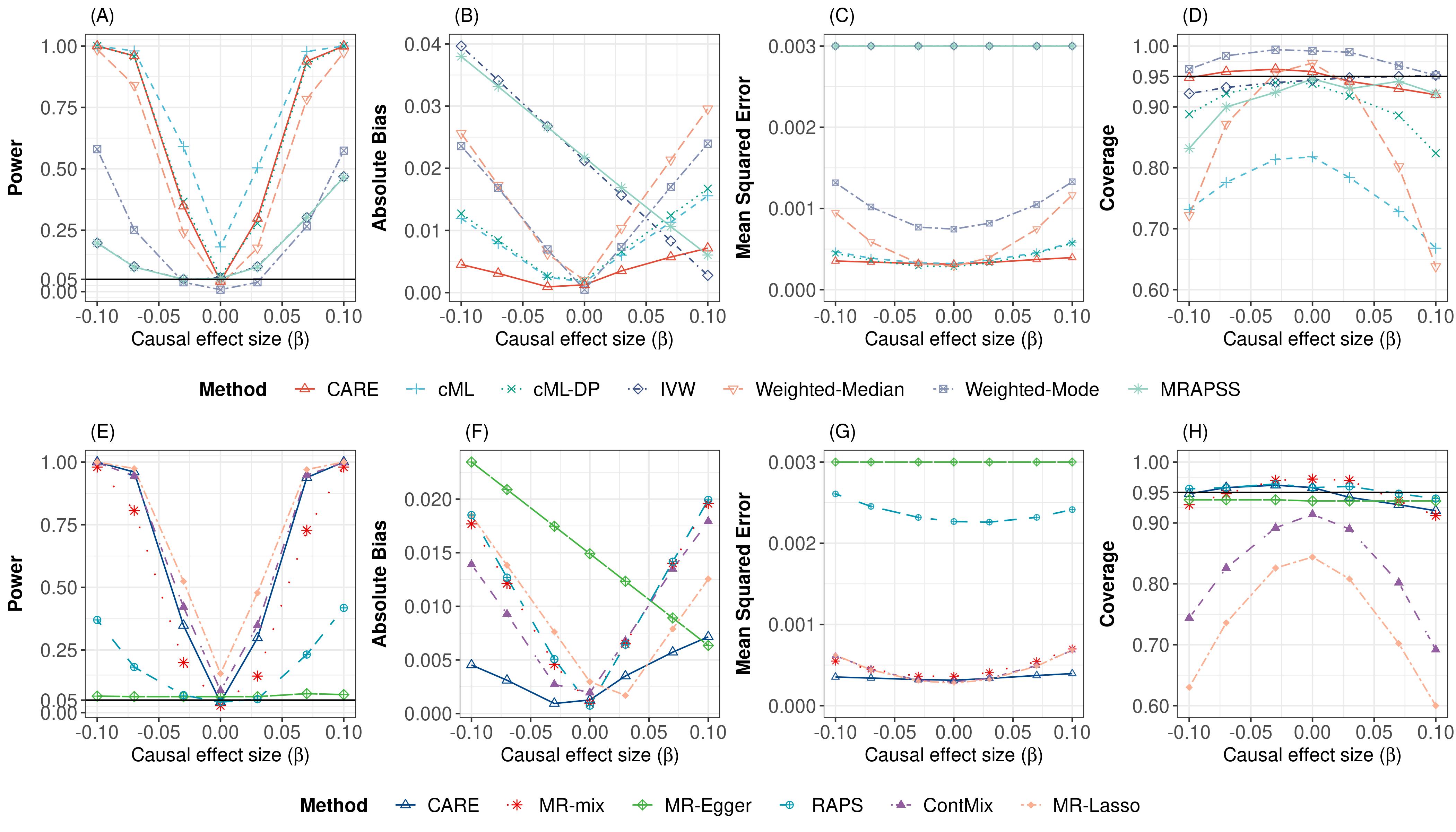}
    \caption{   \label{fig:simulation_threshold_5e-5} Power, absolute bias, mean squared error, and coverage of the CARE estimator and several robust MR methods under the main setting with 50\% invalid IVs. The significant threshold is $5\times 10^{-5}$ for all methods. Power is the empirical power estimated by the proportion of p-values less than the significance threshold of 0.05. Coverage is the empirical coverage probability of the 95\% confidence interval.}
\end{figure}

To assess the influence of the IV selection threshold, we compared all methods using the same liberal threshold of $p < 5 \times 10^{-5}$ under the main setting. We generate 200,000 independent SNPs to represent all underlying common variants and set $\sigma_x^2 = \sigma_y^2 = \sigma_u^2 = 1\times 10^{-5}$, $\beta_{XU} = \beta_{YU}= 1$. We set $n_{X} = n_{Y} = 500,000$ to reflect the sample size of a typical GWAS in our real data analyses. We further set $\pi_1 + \pi_2  = 0.02$,   $\pi_4  = 0.01$, and $\pi_5 = 0.97$. We let the proportion of invalid IVs, which is defined as $\pi_2/ (\pi_1 + \pi_2)$ be equal to 50\%. While some competing methods showed increased power, this often came at the cost of inflated Type I error rates and poor confidence interval coverage. CARE maintained its advantages in terms of bias, mean squared error, and valid inference. Figure~\ref{fig:simulation_threshold_5e-5} summarize the results for this setting.

\subsection{  Comparison of $l_0$ and $l_1$ algorithms}\label{sup:l0 and l1}
We conduct a series of simulations to compare the performances of these two methods with the \( l_0 \) constraint approach adopted in this manuscript. Firstly, we varied the proportion of invalid IVs (30\%, 50\%). We also tested the performance under the setting of uniform distributed effects in correlated pleiotropy with 50\% invalid IVs (See \ref{sup:sec_uniform} for the details of the setting). 

\begin{figure}[!http]
    \centering \includegraphics[width=\linewidth]{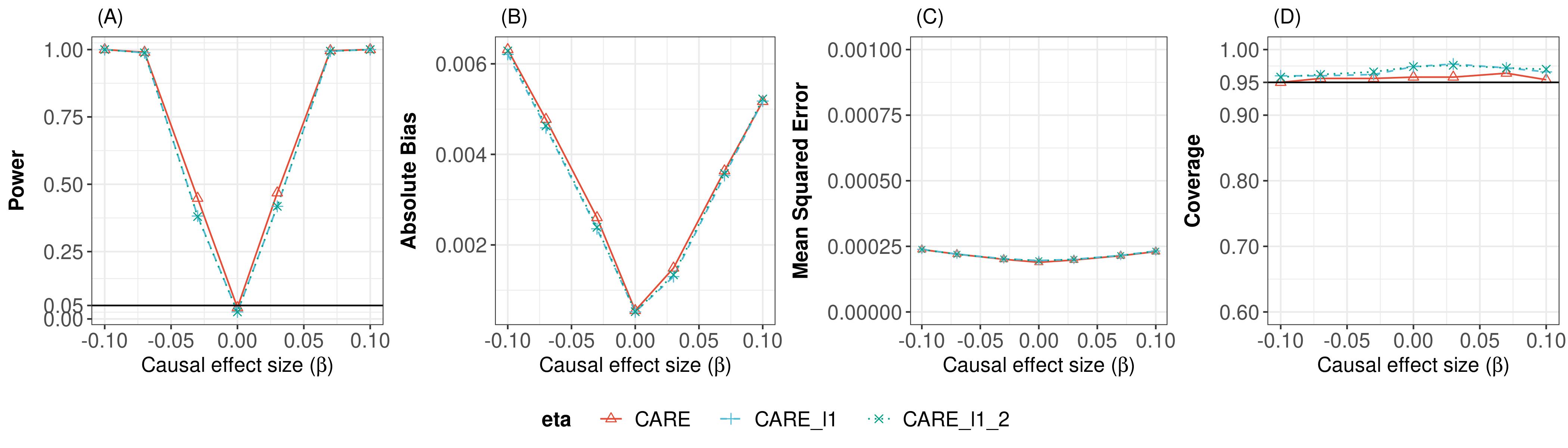}
    \caption{   \label{fig:simulation_l1_0.3} Power, absolute bias, mean squared error, and coverage of the CARE estimator with $l_0$ and two $l_1$ algorithms  under the main setting with 30\% invalid IVs. Power is the empirical power estimated by the proportion of p-values less than the significance threshold of 0.05. Coverage is the empirical coverage probability of the 95\% confidence interval.}
\end{figure}

\begin{figure}[!http]
    \centering \includegraphics[width=\linewidth]{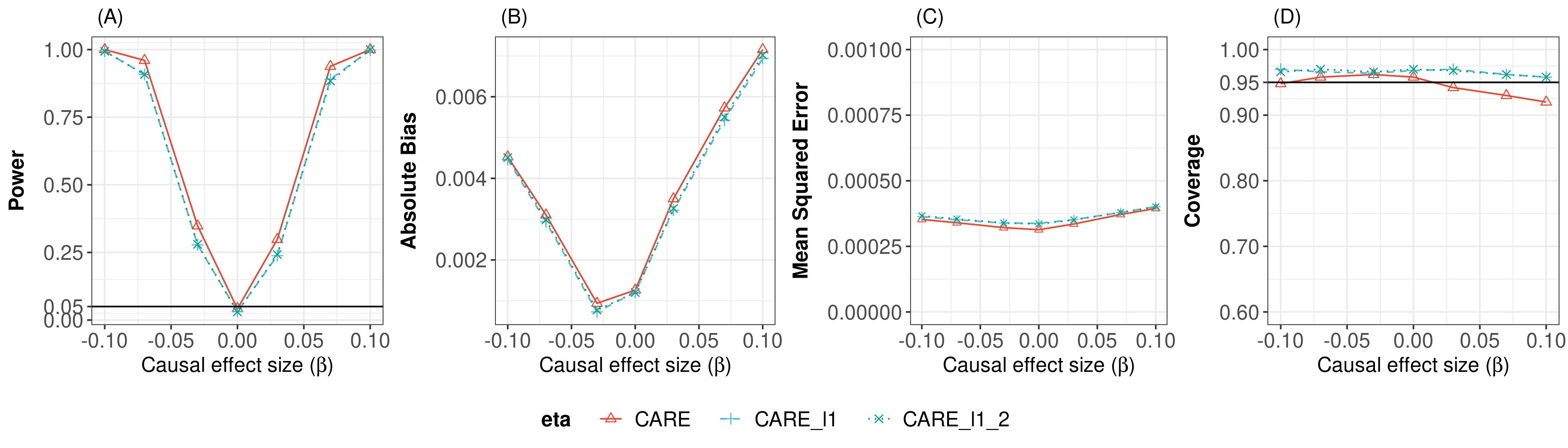}
    \caption{   \label{fig:simulation_l1_0.5} Power, absolute bias, mean squared error, and coverage of the CARE estimator with $l_0$ and two $l_1$ algorithms  under the main setting with 50\% invalid IVs. Power is the empirical power estimated by the proportion of p-values less than the significance threshold of 0.05. Coverage is the empirical coverage probability of the 95\% confidence interval.}
\end{figure}

\begin{figure}[!http]
    \centering \includegraphics[width=\linewidth]{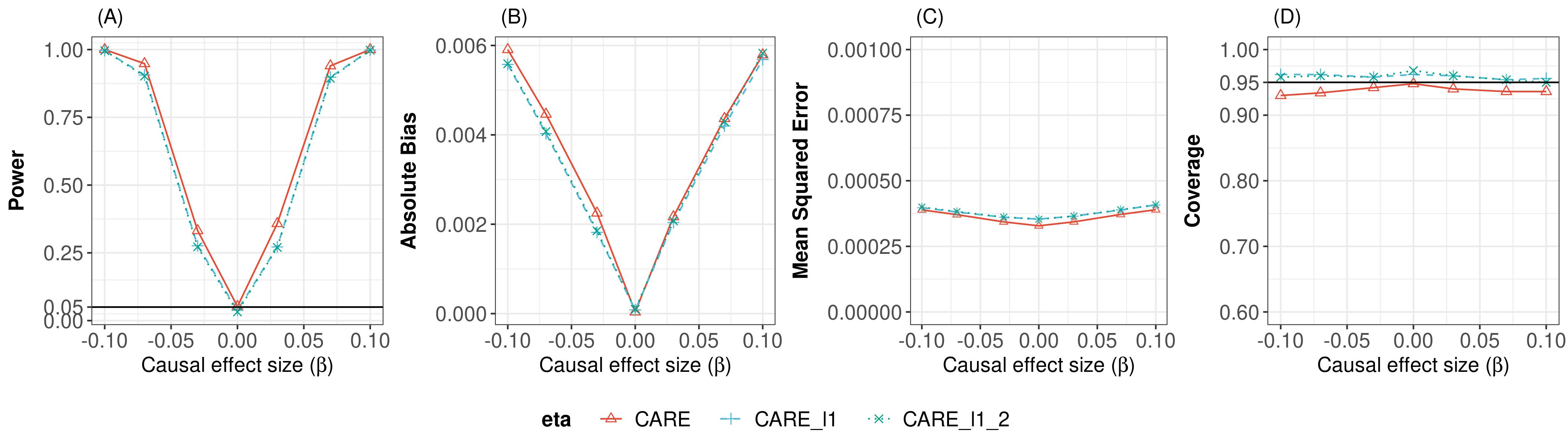}
    \caption{   \label{fig:simulation_l1_uniform} Power, absolute bias, mean squared error, and coverage of the CARE estimator with $l_0$ and two $l_1$ algorithms under the setting of uniform distributed effects in correlated pleiotropy with 50\% invalid IVs. Power is the empirical power estimated by the proportion of p-values less than the significance threshold 0.05.  Coverage is the empirical coverage probability of the 95\% confidence interval.}
\end{figure}

We generate 200,000 independent SNPs to represent all underlying common variants and set $\sigma_x^2 = \sigma_y^2 = \sigma_u^2 = 1\times 10^{-5}$, $\beta_{XU} = \beta_{YU}= 1$. We set $n_{X} = n_{Y} = 500,000$ to reflect the sample size of a typical GWAS in our real data analyses. We further set $\pi_1 + \pi_2  = 0.02$,   $\pi_4  = 0.01$, and $\pi_5 = 0.97$. Figure~\ref{fig:simulation_l1_0.3} to \ref{fig:simulation_l1_uniform} summarize the results for these settings. Our findings consistently demonstrate that while both approaches maintain comparable Type I error control, absolute bias, mean squared error (MSE), and coverage probability across various scenarios, the $l_0$-based CARE method achieves noticeably higher statistical power.

\subsection{ Third sample for selecting IVs}\label{sup:three samples}
Our method—CARE—effectively integrates winner’s curse correction via Rao-Blackwellization with robust handling of both measurement error and pleiotropy. However, in scenarios where the winner’s curse is no longer a concern—for example, when a third independent sample is available for IV selection based on association strength—some alternative methods may outperform CARE.

To investigate this, we conducted an additional simulation study using a three-sample MR design. The data generating process is the same as the main setting in our manuscript, which favors other methods with parametric assumptions (See details in Section \ref{sup:setup}) In this design, a third independent sample is used exclusively for IV selection based on association strength, thereby eliminating the need for winner’s curse correction in all methods. We uniformly apply a liberal IV selection threshold of $p < 5 \times 10^{-5}$ to this third sample across all methods for fair comparison.

\begin{figure}[!http]
    \centering \includegraphics[width=\linewidth]{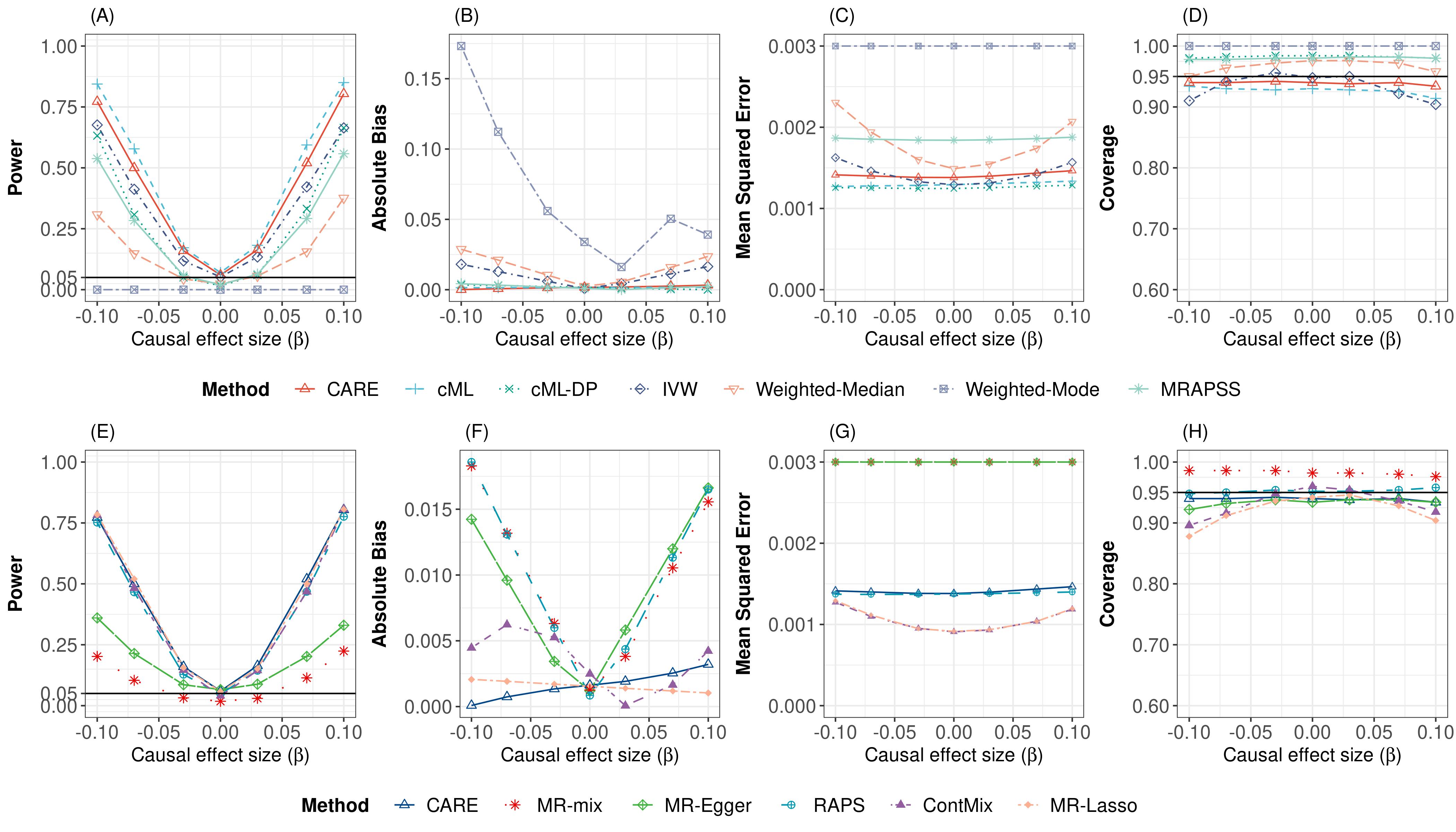}
    \caption{   \label{fig:three samples 2} Power, absolute bias, mean squared error, and coverage of the CARE estimator and several robust MR methods under the main setting with 50\% invalid IVs, all using a third sample for IV selection based on association strength. Power is the empirical power estimated by the proportion of p-values less than the significance threshold 0.05.  Coverage is the empirical coverage probability of the 95\% confidence interval.}
\end{figure}

As shown in Figure~\ref{fig:three samples 2}, cML outperforms CARE in terms of both power and mean squared error (MSE), while maintaining comparable empirical coverage. Other methods, such as cML-DP and IVW, also exhibit competitive performances. These results highlight that when a third sample is available and winner’s curse correction is unnecessary, CARE may not be the optimal choice.

\clearpage
\section{Additional Real Data Results}\label{sup:sec5}

\subsection{Data harmonization}\label{Sec:extension-pruning}

We harmonize GWAS summary data through the following steps. First, we exclude genetic variants that are not available in the outcome GWAS dataset. Second, we select independent genetic variants that have no linkage disequilibrium with other selected genetic variants. No linkage disequilibrium is defined as R-squared $ < 0.001$ with an extension of 10,000 Kb in the genome, which has been widely adopted in applied MR studies \cite{hemani2018mr}. For the benchmark methods,  in line with the current practice  \cite{hemani2018mr}, we employ standard clumping, selecting the variant with the smallest $p$-value of the SNP-exposure association when genetic variants are in linkage disequilibrium. For the proposed method CARE, we employ a revised sigma-based pruning procedure and select the variant with the smallest standard deviation of the SNP-exposure association when genetic variants are in linkage disequilibrium \citep{ma2023breaking}. We employ this revised sigma-based pruning procedure because standard clumping introduces a different type of selection bias; see \cite{robertson2016accounting} for related discussion. Third, by leveraging allele frequency information, we infer the strand direction of ambiguous SNPs and harmonize exposure-outcome datasets using the \texttt{twosampleMR} package. We use the default setting with $\lambda=4.06$ and $\eta = 0.5$ for our proposed CARE estimator, and set $\lambda=4.06$ and $\lambda = 5.45$ for MR-APSS and other benchmark methods, respectively.

\subsection{Comparative analysis of four MR methods for assessing COVID-19 severity}
\label{sup-sec7.2}
\begin{figure}[!htbp]
 \centering
       \includegraphics[width=0.7\linewidth]{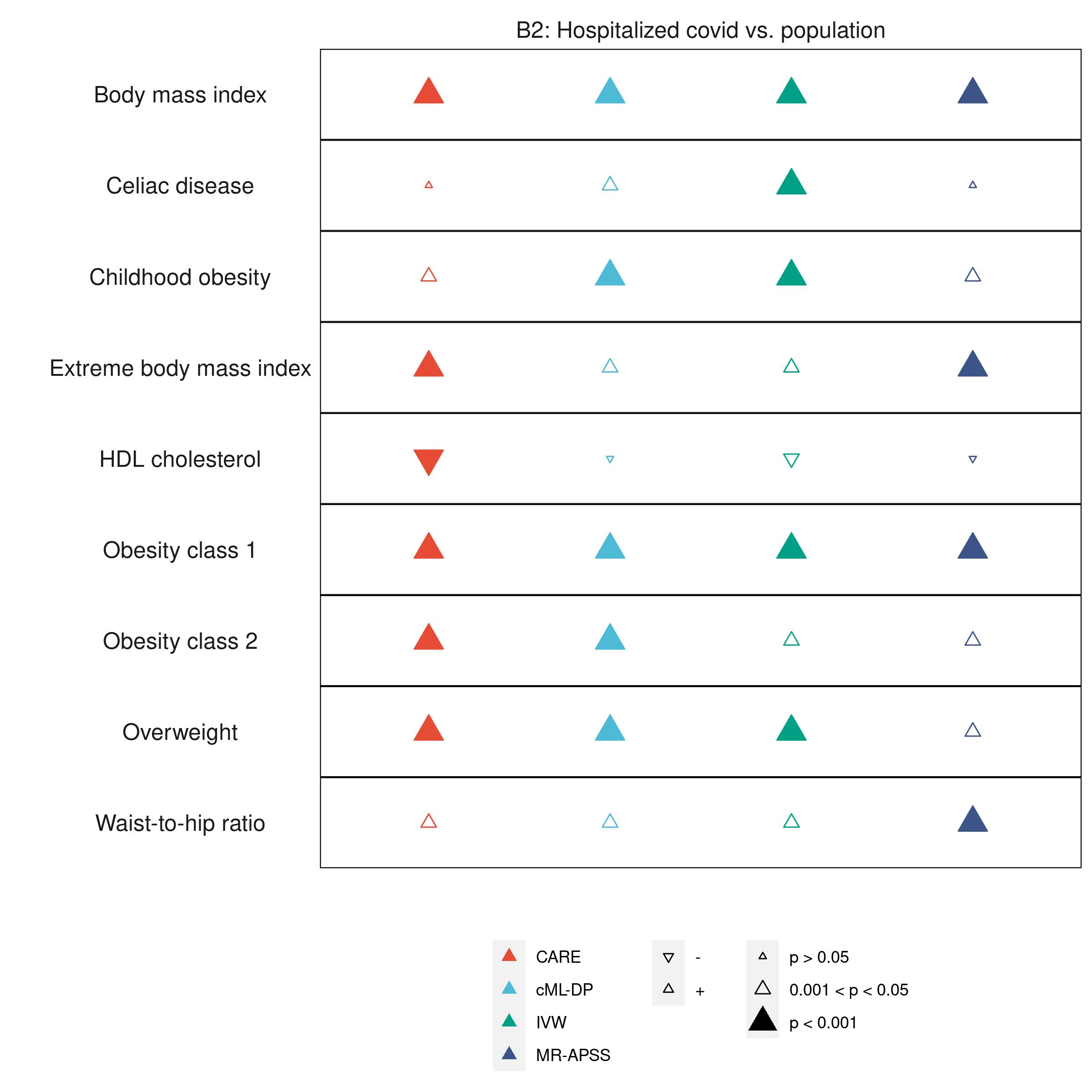}
        \caption{Significant causal exposure COVID-19 severity pairs identified by CARE, cML-DP, IVW, and MR-APSS. We summarize the significant causal exposure identified by at least one method under Bonferroni correction.}	\label{fig:COVIDsig}
\end{figure}

Second, we focus on four methods with relatively good performance under our negative control outcome analysis to alleviate the concerns of false positives. Figure~\ref{fig:COVIDsig} summarizes the results. First,  CARE identifies body mass index (BMI), obesity class 1, obesity class 2, overweight, and extreme BMI are causally associated with COVID-19 severity. According to the Centers for Disease Control and Prevention (CDC), the risk of severe illness (i.e., hospitalization) from COVID-19 increases sharply with higher BMI, indicating that extreme BMI may be a likely causal risk factor for COVID-19 severity.  Second, CARE identifies that HDL cholesterol (present in the blood, associated with a lower risk of coronary heart disease) is causally associated with COVID-19 severity. Low HDL level in the blood is reported to be associated with COVID-19 severity and most COVID-19 patients (65\%) exhibit severely low HDL levels \citep{nain2022high}. In comparison, the competing methods fail to identify HDL. Third, competing methods such as IVW and cML-DP identify childhood obesity, and celiac disease as causally associated with COVID-19 severity, while CARE does not. However, limited evidence supports their roles in COVID-19 severity as these risk factors are not listed on the CDC website, and hard to find support from the literature, suggesting that these two risk factors identified by competing methods may be false positives. Fourth, MR-APSS identifies the waist-to-hip ratio, which has been missed by the other three methods; however, the waist-to-hip ratio has been reported to have no association with COVID-19 severity \citep{gao2022associations}.

\subsection{Supplementary tables and figures for real data results}
\begin{longtable}{lc|ll|c} 
   \hline\hline
GWAS ID & Trait & \# SNP & N & PMID \\
  \hline
  \endhead 

  \href{https://gwas.mrcieu.ac.uk/datasets/ieu-a-298/}{ieu-a-298} & Alzheimer's disease* & 11,633 & 74,046 & 24162737 \\
  \href{https://gwas.mrcieu.ac.uk/datasets/ieu-a-45/}{ieu-a-45} & Anorexia nervosa & 1,149,254 & 17,767 & 24514567 \\
  \href{https://gwas.mrcieu.ac.uk/datasets/ieu-a-44/}{ieu-a-44} & Asthma* & 546,183 & 26,475 & 20860503 \\
  \href{https://gwas.mrcieu.ac.uk/datasets/ieu-a-806/}{ieu-a-806} & Autism* & 9,499,590 & 10,263 & 23453885 \\
  \href{https://gwas.mrcieu.ac.uk/datasets/ieu-a-801/}{ieu-a-801} & Bipolar disorder* & 2,427,221 & 16,731 & 21926972 \\
  \href{https://gwas.mrcieu.ac.uk/datasets/ieu-a-29/}{ieu-a-29} & Birth length & 2,201,972 & 28,459 & 25281659 \\
  \href{https://gwas.mrcieu.ac.uk/datasets/ieu-a-1083/}{ieu-a-1083} & Birth weight & 16,245,524 & 143,677 & 27680694 \\
  \href{https://gwas.mrcieu.ac.uk/datasets/ieu-a-2/}{ieu-a-2} & Body Mass Index* & 2,555,511 & 339,224 & 25673413 \\
  \href{https://gwas.mrcieu.ac.uk/datasets/ieu-a-1109/}{ieu-a-1109} & Cardioembolic stroke& 2,421,920 & 21,185 & 26935894 \\
  \href{https://gwas.mrcieu.ac.uk/datasets/ieu-a-1058/}{ieu-a-1058} & Celiac disease & 38,037 & 24,269 & 22057235 \\
  \href{https://gwas.mrcieu.ac.uk/datasets/ieu-a-1096/}{ieu-a-1096} & Childhood obesity & 2,442,739 & 13,848 & 22494627 \\
  \href{https://gwas.mrcieu.ac.uk/datasets/ieu-a-1102/}{ieu-a-1102} & Chronic kidney disease* & 2,191,877 & 117,165 & 26831199 \\
  \href{https://gwas.mrcieu.ac.uk/datasets/ieu-a-7/}{ieu-a-7} & Coronary heart disease* & 9,455,779 & 123,504 & 26343387 \\
  \href{https://gwas.mrcieu.ac.uk/datasets/ieu-a-12/}{ieu-a-12} & Crohn's disease* & 124,888 & 51,874 & 26192919 \\
  \href{https://gwas.mrcieu.ac.uk/datasets/ieu-a-1000/}{ieu-a-1000} & Depressive symptoms* & 6,524,475 & 161,460 & 27089181 \\
  \href{https://gwas.mrcieu.ac.uk/datasets/ieu-a-1040/}{ieu-a-1040} & Difference in height between adolescence and adulthood & 2,401,290 & 9,228 & 23449627 \\
  \href{https://gwas.mrcieu.ac.uk/datasets/ieu-a-1037/}{ieu-a-1037} & Difference in height between childhood and adulthood & 2,384,832 & 10,799 & 23449627 \\
  \href{https://gwas.mrcieu.ac.uk/datasets/ieu-a-85/}{ieu-a-85} & Extreme body mass index* & 1,984,814 & 16,068 & 23563607 \\
  \href{https://gwas.mrcieu.ac.uk/datasets/ieu-a-86/}{ieu-a-86} & Extreme height & 1,966,557 & 16,196 & 23563607 \\
  \href{https://gwas.mrcieu.ac.uk/datasets/ieu-a-87/}{ieu-a-87} & Extreme waist-to-hip ratio & 1,939,901 & 10,255 & 23563607 \\
  \href{https://gwas.mrcieu.ac.uk/datasets/ieu-a-1054/}{ieu-a-1054} & Gout & 2,450,548 & 69,374 & 23263486 \\
  \href{https://gwas.mrcieu.ac.uk/datasets/ieu-a-299/}{ieu-a-299} & HDL cholesterol* & 2,447,442 & 187,167 & 24097068 \\
  \href{https://gwas.mrcieu.ac.uk/datasets/ieu-a-89/}{ieu-a-89} & Height & 2,550,859 & 253,288 & 25282103 \\
  \href{https://gwas.mrcieu.ac.uk/datasets/ieu-a-31/}{ieu-a-31} & inflammatory bowel disease* & 12,716,084 & 34,652 & 26192919 \\
  \href{https://gwas.mrcieu.ac.uk/datasets/ieu-a-814/}{ieu-a-814} & Ischaemic stroke & 393,465 & 517 & 17434096 \\
  \href{https://gwas.mrcieu.ac.uk/datasets/ieu-a-300/}{ieu-a-300} & LDL cholesterol & 2,437,752 & 173,082 & 24097068 \\
  \href{https://gwas.mrcieu.ac.uk/datasets/ieu-a-965/}{ieu-a-965} & Lung adenocarcinoma* & 8,881,354 & 18,336 & 24880342 \\
  \href{https://gwas.mrcieu.ac.uk/datasets/ieu-a-966/}{ieu-a-966} & Lung cancer* & 8,945,893 & 27,209 & 24880342 \\
  \href{https://gwas.mrcieu.ac.uk/datasets/ieu-a-1025/}{ieu-a-1025} & Multiple sclerosis* & 156,632 & 38,589 & 24076602 \\
  \href{https://gwas.mrcieu.ac.uk/datasets/ieu-a-798/}{ieu-a-798} & Mycocardial infarction* & 9,289,492 & 171,875 & 26343387 \\
  \href{https://gwas.mrcieu.ac.uk/datasets/ieu-a-1007/}{ieu-a-1007} & Neuroticism & 6,524,433 & 170,911 & 27089181 \\
  \href{https://gwas.mrcieu.ac.uk/datasets/ieu-a-90/}{ieu-a-90} & Obesity class 1* & 2,380,428 & 98,697 & 23563607 \\
  \href{https://gwas.mrcieu.ac.uk/datasets/ieu-a-91/}{ieu-a-91} & Obesity class 2* & 2,331,456 & 72,546 & 23563607 \\
  \href{https://gwas.mrcieu.ac.uk/datasets/ieu-a-92/}{ieu-a-92} & Obesity class 3* & 2,250,779 & 50,364 & 23563607 \\
  \href{https://gwas.mrcieu.ac.uk/datasets/ieu-a-93/}{ieu-a-93} & Overweight* & 2,435,045 & 158,855 & 23563607 \\
  \href{https://gwas.mrcieu.ac.uk/datasets/ieu-a-975/}{ieu-a-975} & Paget's disease & 2,479,235& 3,440 & 21623375 \\
  \href{https://gwas.mrcieu.ac.uk/datasets/ieu-a-812/}{ieu-a-812} & Parkinson's disease* & 453,218 & 5,691 & 19915575 \\
  \href{https://gwas.mrcieu.ac.uk/datasets/ieu-a-833/}{ieu-a-833} & Rheumatoid arthritis* & 9,739,304 & 80,799 & 24390342 \\
  \href{https://gwas.mrcieu.ac.uk/datasets/ieu-a-22/}{ieu-a-22} & Schizophrenia* & 9,444,231 & 82,315 & 25056061 \\
  \href{https://gwas.mrcieu.ac.uk/datasets/ieu-a-967/}{ieu-a-967} & Squamous cell lung cancer & 8,893,750 & 18,313 & 24880342 \\
  \href{https://gwas.mrcieu.ac.uk/datasets/ieu-a-1009/}{ieu-a-1009} & Subjective well being & 2,268,675 & 298,420 & 27089181 \\
  \href{https://gwas.mrcieu.ac.uk/datasets/ieu-a-301/}{ieu-a-301} & Total cholesterol & 2,446,982 & 187,365 & 24097068 \\ 
  \href{https://gwas.mrcieu.ac.uk/datasets/ieu-a-26/}{ieu-a-26} & Type 2 diabetes* & 2,473,442 & 69,033 & 22885922 \\
  \href{https://gwas.mrcieu.ac.uk/datasets/ieu-a-970/}{ieu-a-970} & Ulcerative colitis & 156,116 & 47,745 & 26192919 \\
  \href{https://gwas.mrcieu.ac.uk/datasets/ieu-a-72/}{ieu-a-72} & Waist-to-hip ratio & 2,562,516 & 224,459 & 25673412 \\
  \href{https://gwas.mrcieu.ac.uk/datasets/ukb-b-533/}{ukb-b-553} & Ease of skin tanning & 9,851,867 & 453,065 & \\
  \href{https://gwas.mrcieu.ac.uk/datasets/ukb-d-1747_1/}{ukb-d-1747\_1} & Hair colour (natural, before greying): Blonde & 13,586,531 & 360,270 & \\
  \href{https://gwas.mrcieu.ac.uk/datasets/ukb-d-1747_2/}{ukb-d-1747\_2} & Hair colour (natural, before greying): Red & 13,586,531 & 360,270 & \\
  \href{https://gwas.mrcieu.ac.uk/datasets/ukb-d-1747_3/}{ukb-d-1747\_3} & Hair colour (natural, before greying): Light brown & 13,586,531 & 360,270 & \\
  \href{https://gwas.mrcieu.ac.uk/datasets/ukb-d-1747_4/}{ukb-d-1747\_4} & Hair colour (natural, before greying): Dark brown & 13,586,531 & 360,270 & \\
  \href{https://gwas.mrcieu.ac.uk/datasets/ukb-d-1747_5/}{ukb-d-1747\_5} & Hair colour (natural, before greying): Black & 13,586,531 & 360,270 & \\

   \hline
   \hline
   \caption{45 exposures and six negative control outcomes included in the current study. GWAS ID, Trait, \# SNP, N, and PMID stand GWAS ID used in IEU OpenGWAS database, exposure name, number of SNPs in the corresponding full GWAS summary data, sample size of the corresponding study, and PMID used in PubMed, respectively. Traits with stars represent those have been reported by the CDC or in peer-reviewed literature as risk factors for COVID-19 severity.}
\label{sup:Table1_Exposure_detail}\\
\end{longtable}

\begin{figure}[H]
 \centering
       \includegraphics[width=\linewidth]{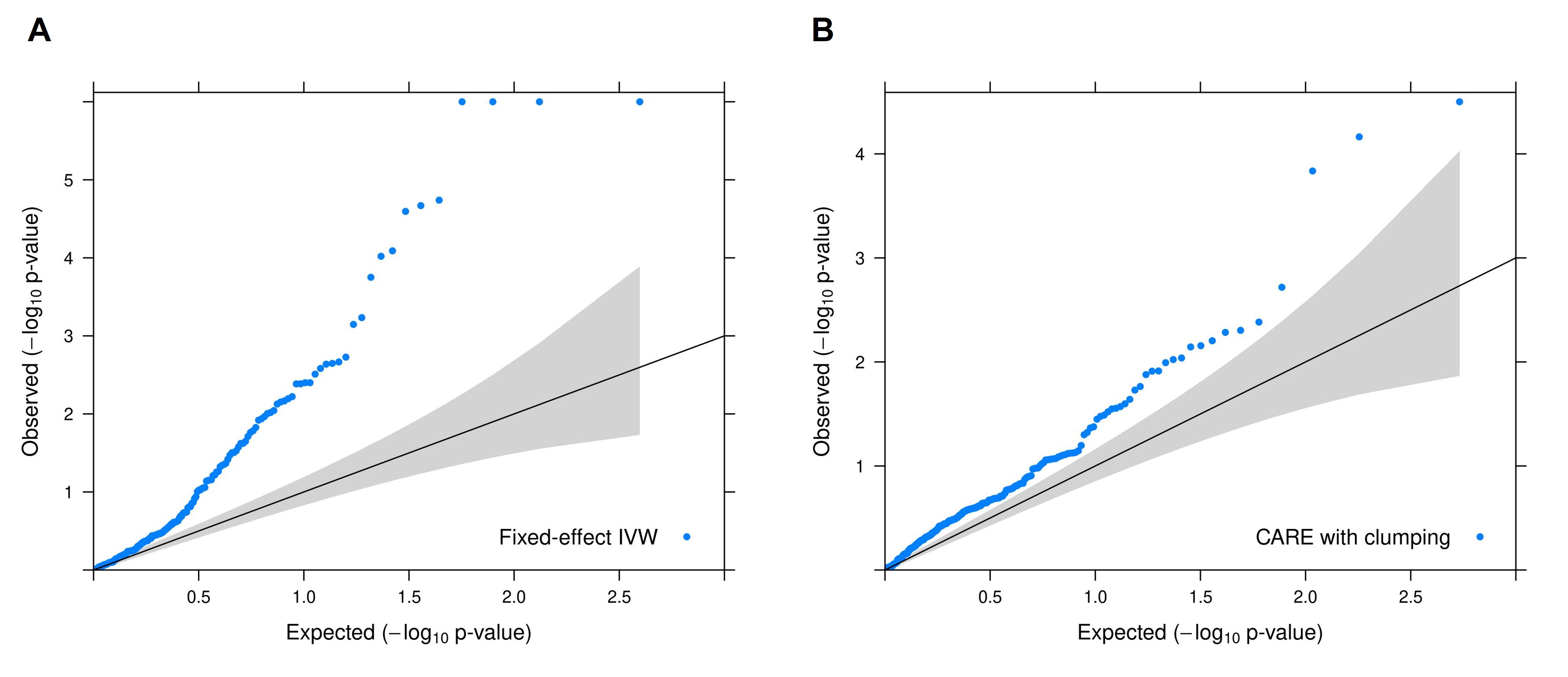}
        \caption{QQ plots of p-values in negative control outcome analysis for fixed-effect IVW (panel A) and CARE using clumping to select candidate IVs (panel B). The gray-shaded part is 95\% confidence interval.}	\label{supfig:negative_control_qqplot}
\end{figure} 

\begin{table}[htbp]
\centering
\label{tab:health_metrics}
\resizebox{\textwidth}{!}{
\begin{tabular}{l|ccc|ccc|ccc|ccc}
\hline\hline
\multirow{2}{*}{Condition} & \multicolumn{3}{c|}{CARE} & \multicolumn{3}{c|}{cML-DP} & \multicolumn{3}{c|}{IVW} & \multicolumn{3}{c}{MR-APSS} \\
\cline{2-13}
& $\beta$ & SE & p-value & $\beta$ & SE & p-value & $\beta$ & SE & p-value & $\beta$ & SE & p-value \\
\hline
Body mass index & 0.3893 & 0.0595 & 5.96E-11 & 0.3952 & 0.0533 & 1.22E-13 & 0.4024 & 0.0580 & 4.11E-12 & 0.4006 & 0.1008 & 7.00E-05 \\
Celiac disease & 0.0213 & 0.0189 & 0.2603 & 0.0293 & 0.0089 & 0.0011 & 0.0299 & 0.0086 & 0.0005 & 0.0200 & 0.0166 & 0.2291 \\
Childhood obesity & 0.0749 & 0.0280 & 0.0074 & 0.0915 & 0.0226 & 5.49E-05 & 0.0946 & 0.0230 & 3.88E-05 & 0.0540 & 0.0231 & 0.0192 \\
Extreme body mass index & 0.0746 & 0.0194 & 0.0001 & 0.0561 & 0.0212 & 0.0081 & 0.0545 & 0.0191 & 0.0042 & 0.0622 & 0.0180 & 0.0005 \\
HDL cholesterol & -0.1840 & 0.0509 & 0.0003 & -0.0598 & 0.0315 & 0.0573 & -0.0809 & 0.0359 & 0.0244 & -0.1177 & 0.0836 & 0.1591 \\
Obesity class 1 & 0.1916 & 0.0379 & 4.27E-07 & 0.1312 & 0.0254 & 2.47E-07 & 0.1288 & 0.0257 & 5.61E-07 & 0.1461 & 0.0307 & 2.01E-06 \\
Obesity class 2 & 0.0924 & 0.0266 & 0.0005 & 0.0805 & 0.0222 & 0.0003 & 0.0793 & 0.0245 & 0.0012 & 0.0549 & 0.0203 & 0.0069 \\
Overweight & 0.2184 & 0.0602 & 0.0003 & 0.1475 & 0.0407 & 0.0003 & 0.1487 & 0.0443 & 0.0008 & 0.1621 & 0.0514 & 0.0016 \\
Waist-to-hip ratio & 0.3279 & 0.1139 & 0.0040 & 0.1980 & 0.0841 & 0.0186 & 0.2134 & 0.0842 & 0.0113 & 0.4114 & 0.0990 & 3.27E-05 \\
\hline\hline
\end{tabular}
}
\caption{Association between significant exposure COVID-19 severity pairs using four methods: CARE, cML-DP, IVW, and MR-APSS. Values represent effect sizes ($\beta$), standard errors (SE), and p-values.}
\end{table}

\end{document}